\documentclass[
11pt, 
oneside, 
english, 
singlespacing, 
liststotoc, 
headsepline, 
]{MastersDoctoralThesis} 

\usepackage[utf8]{inputenc} 
\usepackage[T1]{fontenc} 

\newcommand\varmp{\mathbin{\vcenter{\hbox{%
  \oalign{\hfil$\scriptstyle-$\hfil\cr
          \noalign{\kern-.3ex}
          $\scriptscriptstyle({+})$\cr}%
}}}}

\usepackage{amsmath,amssymb}
\usepackage{graphicx}
\usepackage[colorlinks=true, allcolors=blue]{hyperref}
\usepackage{upgreek}
\newcommand{\varbeta}{\text{\fontfamily{lmss}\selectfont{\ss}}}
\usepackage{float}
\usepackage[countmax]{subfloat}
\usepackage{empheq}
\usepackage{bm}
\usepackage{color}
\usepackage{tensor}
\usepackage[scr = dutchcal]{mathalpha} 

\newcommand{\hl}{}

\DeclareMathOperator{\Mass}{M}
\DeclareMathOperator{\Length}{L}
\DeclareMathOperator{\Time}{T}
\usepackage{comment}
\usepackage{subcaption}
\usepackage{tabularx}
\usepackage{tablefootnote}
\usepackage[stable]{footmisc}

\newcounter{daggerfootnote}
\newcounter{ddaggerfootnote}

\newcommand*{\daggerfootnote}[1]{%
    \setcounter{daggerfootnote}{\value{footnote}}%
    \renewcommand*{\thefootnote}{\fnsymbol{footnote}}%
    \footnote[2]{#1}%
    \setcounter{footnote}{\value{daggerfootnote}}%
    \renewcommand*{\thefootnote}{\arabic{footnote}}%
    }

 \newcommand*{\ddaggerfootnote}[1]{%
    \setcounter{ddaggerfootnote}{\value{footnote}}%
    \renewcommand*{\thefootnote}{\fnsymbol{footnote}}%
    \footnote[3]{#1}%
    \setcounter{footnote}{\value{ddaggerfootnote}}%
    \renewcommand*{\thefootnote}{\arabic{footnote}}%
}

\renewcommand{\thefootnote}{\arabic{footnote}}

\newcommand{\gfootnote}[1]{%
\let\oldthefootnote=\thefootnote%
\setcounter{footnote}{0}%
\renewcommand{\thefootnote}{\fnsymbol{footnote}}%
\footnote{#1}%
\let\thefootnote=\oldthefootnote%
}
\usepackage[backend=biber, style=alphabetic]{biblatex}
\DeclareNolabel{
  \nolabel{\regexp{[\p{Z}\p{P}\p{S}\p{C}]+}}
}
\renewbibmacro*{doi+eprint+url}{%
    \printfield{doi}%
    \newunit\newblock%
    \iftoggle{bbx:eprint}{%
        \usebibmacro{eprint}%
    }{}%
    \newunit\newblock%
    \iffieldundef{eprint}{%
        \usebibmacro{url+urldate}}%
        {}%
    }
\usepackage{pdfpages}

\newcommand{\customtilde}{$\sim$}
\DeclareSourcemap{
  \maps[datatype=bibtex]{
    \map{
      \step[fieldsource=title,
            match=\regexp{\\textasciitilde}, 
            replace=\regexp{\\customtilde}]
    }
  }
}
\usepackage[autostyle=true]{csquotes} 

\makeatletter

\makeatletter

\thesistitle{Cosmology using Strong Gravitational Lensing} 
\supervisor{Prof. C\'{e}line \textsc{B{\oe}hm}} 
\examiner{} 
\degree{Doctor of Philosophy} 
\author{Angela \textsc{Ng}} 
\addresses{} 

\subject{} 
\university{The University of Sydney} 
\department{School of Physics} 
\faculty{Faculty of Science} 

\AtBeginDocument{
\hypersetup{pdftitle=\ttitle} 
\hypersetup{pdfauthor=\authorname} 
\hypersetup{pdfkeywords=\keywordnames} 
}

\begin{document}

\frontmatter 

\pagestyle{plain} 


\begin{titlepage}
\begin{center}

\vspace*{.06\textheight}
{\scshape\LARGE \univname\par}\vspace{1.5cm} 

\textsc{\Large Doctoral Thesis}\\[0.5cm] 

\HRule \\[0.4cm] 
{\huge \bfseries \ttitle\par}\vspace{0.4cm} 
\HRule \\[1.5cm] 
 
\begin{minipage}[t]{0.4\textwidth}
\begin{flushleft} \large
\emph{Author:}\\
\authorname 
\end{flushleft}
\end{minipage}
\begin{minipage}[t]{0.4\textwidth}
\begin{flushright} \large
\emph{Supervisor:} \\
\supname\\ 
\vspace{1cm}
\emph{Examiners:} \\
Prof. Rachel \textsc{Webster}\\
Dr Pierre \textsc{Fleury}
\end{flushright}
\end{minipage}\\[3cm]
 
\vfill

\large \textit{A thesis submitted in fulfilment of the requirements\\ for the degree of \degreename}\\[0.3cm] 
\textit{in the}\\[0.4cm]
\deptname \\[0.5\baselineskip]
Faculty of Science \\[2cm] 


{\large 2023}\\[4cm] 
\vfill
\end{center}
\end{titlepage}


\begin{originality}
\addchaptertocentry{\originalityname} 

\bigskip

\noindent I, \authorname, declare that this thesis titled, \enquote{\ttitle} and the work presented in it are my own. I confirm that:

\begin{itemize} 
\item This work was done wholly or mainly while in candidature for a research degree at this University.
\item Where any part of this thesis has previously been submitted for a degree or any other qualification at this University or any other institution, this has been clearly stated.
\item Where I have consulted the published work of others, this is always clearly attributed.
\item Where I have quoted from the work of others, the source is always given. With the exception of such quotations, this thesis is entirely my own work.
\item I have acknowledged all main sources of help.
\item Where the thesis is based on work done by myself jointly with others, I have made clear exactly what was done by others and what I have contributed myself.
\end{itemize}
 
\noindent Signed:\\
\rule[0.5em]{25em}{0.5pt} 
 
\noindent Date:\\
\rule[0.5em]{25em}{0.5pt} 
\end{originality}

\clearpage




\begin{authorship}
\addchaptertocentry{\authorshipname} 

\bigskip

\begin{itemize}

\item \autoref{chap:p1} of this thesis is published as \fullcite{Ng2020}. The approach was spearheaded by GL.

\item \autoref{chap:p2} of this thesis is published as \fullcite{Ng2023}.

\end{itemize}



\bigskip

\noindent In addition to the statements above, in cases where I am not the corresponding author of a published item, permission to include the published material has been granted by the corresponding author.

\bigskip
\medskip

\noindent Signed:

\noindent \rule[0.5em]{25em}{0.5pt}
\newline
\null \qquad \qquad (Author)

\medskip

\noindent Date:

\noindent \rule[0.5em]{25em}{0.5pt} 

\bigskip
\bigskip

\noindent As supervisor for the candidature upon which this thesis is based, I can confirm that the authorship attribution statements above are correct.

\bigskip
\medskip

\noindent Signed:

\noindent \rule[0.5em]{25em}{0.5pt}
\newline
\null \qquad \qquad (Supervisor)

\medskip

\noindent Date:

\noindent \rule[0.5em]{25em}{0.5pt} 

\end{authorship}

\begin{abstract}
\addchaptertocentry{\abstractname} 

The light we observe from distant astrophysical objects including supernovae and quasars allows us to determine large distances in terms of a cosmological model. 
Despite the success of the standard cosmological model in fitting the data, there remains no underlying explanation for the accelerated expansion and dark matter. Furthermore, there is a current tension between early- and late-universe determinations of the Hubble constant. New techniques may offer the possibility of measuring out to larger distances, provide complementary information, or be able to side-step current limitations. After reviewing the fundamentals of standard cosmology and gravitational lensing, this thesis investigates a novel method of cosmography based on combining the techniques of strong gravitational lensing time delay measurements and quasar reverberation mapping. The motivation for this method was the possibility of avoiding lens modelling challenges, such as the mass-sheet degeneracy, typically associated with time delay cosmography. It suggested that differential time delays originating from spatially separated signals in the Broad Line Region of a quasar could be distinguished and measured from the spectroscopy of the images, and utilised to provide a ratio of cosmological distances independent of the lensing potential. An analytic description of the effect of the differential lensing on the emission line spectral flux for axisymmetric Broad Line Region geometries is given, with the inclined ring or disk, spherical shell, and double cone as examples. This critical examination shows that the proposed method is unable to recover cosmological information, as the observed time delay and inferred line-of-sight velocity do not uniquely map to the three-dimensional position within the quasar.
\end{abstract}


\begin{acknowledgements}
\addchaptertocentry{\acknowledgementname} 
\bigskip
\bigskip
I am grateful to C\'{e}line B{\oe}hm for taking on the role of advisor in the past year. Thank you sincerely for supporting my academic agency, and permitting me the time and space to think.

My heartfelt and unreserved thanks go to Mark Wardle. I can only endeavour to pay forward your altruism. I also extend my appreciation to Shari Breen, Nic Scott, Caro Foster and particularly Kevin Varvell.

To my partner, Ed McDonald -- for your unwavering love, for bearing witness to it all, and all our mathematical discussions.

I thank for their friendship Joseph, Tiphaine, Mat, Jiro, Gurashish, and my many other academic brethren; outside of academia: Avi, Chere, Oz, the rest of DTR, and Alex; the Bobbi's community; and Jenny and Andrew for their love and support.

I am grateful to C\'{e}line, Joseph, Ed and Mark; as well as my thesis examiners, Pierre Fleury and Rachel Webster, for taking time to carefully read and provide feedback on this manuscript at different stages -- my examiners' comments will be cherished always. I especially thank Pierre for his meticulous examination which caught many errors, as well as many useful and interesting exchanges.

Specifically regarding Chapter 6 of this thesis, I additionally thank Mat Varidel, Josh Speagle, Oz Brent, and the anonymous reviewer. I would also like to thank the Artemis HPC support staff and Gillian Flack for technical assistance during the very early period of my PhD, although the work on cosmic voids did not make it into this manuscript. I acknowledge funding from an Australian Government Research Training Program Scholarship, the University of Sydney Hunstead Merit Award, and University of Sydney travel scholarships.

The theme of this thesis may perhaps also be interpreted as a metaphor: light in the cosmos is deflected and delayed by unseen matter only inferred by a distant observer, even as it travels on the straightest possible path. During an exchange at a conference with a fellow student regarding senseless tribulations, I attempted an encouraging ``we're still here''. Her paraphrased reply was ``but many more are not''. This thesis is dedicated to those many students.

\end{acknowledgements}


\tableofcontents 

\begin{symbols}{lll} 


\addlinespace 
$(-, +, +, +)$ &  metric signature & \\
$x^\mu = (ct, x^i)$ & choice of the time coordinate $x^0=ct$ & \\
& \\
bold & tuples, vectors and tensors &\\
Greek indices & temporal-spatial components & $\{ 0, 1, 2,3 \}$  \\
Latin indices & &\\
-- lowercase  & spatial components &$\{1, 2, 3 \}$ \\
-- uppercase  & transverse or angular spatial components &$\{1, 2 \}$\\ 
& & \\
repeated indices & Einstein summation notation & \\
&(unless fixed) &\\
$T_{[\mu_1 ... \mu_n]}$ & anti-symmetric tensor notation  &\\
$ \equiv \frac{1}{n!} \epsilon_{\mu_1 ... \mu_n}\sum\limits_{\text{permut.}} T_{\mu_1 ... \mu_n} $ & & \\
$T_{(\mu_1 ... \mu_n)} $ & symmetric tensor notation & \\
$\equiv \frac{1}{n!} \sum\limits_{\text{permut.}} T_{\mu_1 ... \mu_n} $ & & \\

&&\\
$\nabla_{\nu} f = f_{;\nu}$ &  covariant derivative of $f$ & \\
{\hl $\frac{D f}{d \sigma} $} & covariant derivative of $f$ along a curve\\
$\partial_{\nu} f = f_{,\nu}$ &  partial derivative of $f$ & \\
$\bm{\nabla}$ & gradient & \\
$\Box \equiv \nabla^{\nu} \nabla_{\nu}$ & covariant d'Alembertian & \\
$d_tf$ & total derivative of $f$ & \\
&&\\
$\delta f$ & perturbation of $f$ & \\
& OR variation  of $f$ & \\
$\delta(\,)$ or $\delta_D(\,)$ & Dirac delta & \\
$\delta\indices{^i_j}$ & Kronecker delta & \\ 
\end{symbols}
\begin{longtable}{lll}
\hline
&&\\
$c$ & \textit{in vacuo} speed of light in absence of gravity  & [$\Length \Time^{-1}$]\daggerfootnote{\label{dagfoot}Natural units are frequently and sometimes exclusively used in the literature, where c and $\hbar$ are dimensionless and set to 1, such that quantities with dimensions in terms of mass, length and time $[\Mass^\alpha \Length^\beta \Time^\gamma]$ have dimensions of energy $[\text{E}^{\alpha - \beta - \gamma}]$. The conversion factor from natural units to MLT units is multiplication by $\hbar^{\beta + \gamma} c^{\beta - 2 \alpha}$. Appendix C of \cite{Baumann2022} has details and examples. We refrain from using natural units, to attempt consistency throughout when covering observations and varied sub-fields with their own preferred systems of units: we find, following e.g. \cite{Landau1951}, setting coordinates $x^0 = ct$ which render velocities dimensionless is sufficient to remove cumbersome factors of $c$ in most calculations. (Note \cite{Landau1951} uses $u^\mu = \frac{d x^\mu}{ds}$ from $ds^2 = d(c\uptau)^2$ due to the $(+,-,-,-)$ signature convention.)}\\
$\hbar$ & reduced Planck constant & [$\Mass \Length^2 \Time^{-1}$]$^{\dagger}$\\
$G$ & gravitational constant & [$\Mass^{-1} \Length^{3} \Time^{-2}$]\\
$S$ & action & [$\Mass \Length^2 \Time^{-1}$]\\
$\uptau$ & proper time &  [$\Time$] \\
$g_{\mu\nu}$ & metric tensor & [ ]\ddaggerfootnote{\label{ddagfoot}The dimensional analysis of tensors as geometric objects is non-trivial, see for example \cite{Post1982}. Furthermore, when written in coordinate form, the dimensions of each tensor component changes and can differ from each other depending on the choice of coordinates. Consider that we make the standard choice that $ds^2 = g_{\mu \nu} dx^{\mu} dx^{\nu}$ has units of length squared, and an angular element $d\theta$ is dimensionless, whereas a radial coordinate $dr$ is usually taken to have dimensions of length -- swapping from spherical to (quasi-)Cartesian coordinates changes the dimensions of $g_{\mu \nu}$. The important thing is we are clear on the physical interpretation when tensors contract into (observable) scalar quantities. As another example, we are free to choose $x^0$ as either $ct$ or $t$, and the $T^{00}= \bm{T}(\bm{e}_0, \bm{e}_0)$ component of the stress energy tensor can be interpreted as either an energy density or mass density respectively. Here we quote dimensions, of all components, assuming all the $x^{\mu}$ coordinates have units of length, which may sometimes not be the case, but is nonetheless useful.} \\
&&\\
$h$ & Planck constant & [$\Mass \Length^2 \Time^{-1}$]$^{\dagger}$\\
 & OR reduced Hubble constant & [ ]\\
$H_0$ & Hubble constant & [$\Time^{-1}$]\\
& (present-day Hubble parameter) & \\
$H \equiv a^{-1}\frac{da}{dt}$ & Hubble parameter & [$\Time^{-1}$]\\
$\mathscr{H}\equiv a^{-1}\frac{da}{d(c \eta)}$ & conformal Hubble parameter & [$\Length^{-1}$]\\
&&\\
$z$ & redshift & [ ]\\
$D_A$ & angular size distance & [$\Length$]\\
$D_d$, $D_s$, $D_{ds}$ & angular size distance: from observer to lens,  & [$\Length$], [$\Length$], [$\Length$]\\
 & from observer to source, and from lens to source & \\
$D_\tau \equiv \frac{D_d D_s}{D_{ds}}$ & lensing time-delay distance & [$\Length$]\\
$D_\textsc{r}$ & radar distance & [$\Length$]\\
$D_\textsc{p}$ & parallax distance & [$\Length$]\\
$D_L$ & luminosity distance & [$\Length$]\\
&& \\
$ds$ & line element of spacetime & [$\Length$]\\
$t$ & cosmic (coordinate) time & [$\Time$] \\
$\eta$ & conformal time & [$\Time$] \\
$a$ & scale factor & [ ]\\
$K$ & spatial curvature & [$\Length^{-2}$]\\
$f_K \equiv \sin_K$ & comoving areal distance & [$\Length$]\\
$\Phi$ & Newtonian gravitational potential  & [$\Length^2\Time^{-2}$]\\
$\hat{\Phi}$, $\hat{\Psi}$ & Bardeen potentials &  [ ] \\
 & $\Phi = \hat{\Phi}c^2 = \hat{\Psi}c^2$ in non-relativistic limit \\
&&\\

$u^\mu \equiv \frac{d x^{\mu}}{d (c \uptau)}$ & four-velocity & [ ]$^\ddagger$\\
$p^\mu \equiv m u^\mu$ & four-momentum & [$\Mass$]$^\ddagger$\\
$v^i \equiv \frac{d x^i}{d(c \eta)}$ & peculiar velocity & [ ]$^\ddagger$\\
$\varepsilon \equiv \rho c^2$ & energy density & [$\Mass\Length^{-1}\Time^{-2}$]\\
$\rho$ & relativistic mass density & [$\Mass \Length^{-3}$] \\
$\varrho$ & rest mass density & [$\Mass \Length^{-3}$] \\
$P$ & pressure & [$\Mass \Length^{-1} \Time^{-2}$] \\
$T^{\mu \nu}$ & stress-energy tensor & [$\Mass \Length^{-1} \Time^{-2}$]$^\ddagger$ \\
&&\\

$j^{\mu}$ & four-current density & [$\Mass^{\frac{1}{2}}\Length^{- \frac{1}{2}}\Time^{-2}$]\gfootnote{All electromagnetic quantities are given with dimensions according to Gaussian units, here marked with an asterisk. Gaussian units employ one fewer number of base dimensions than SI units, as charge [Q] = [$\Mass^{\frac{1}{2}}\Length^\frac{3}{2}\Time^{-1}$]. This is perhaps not a very enlightened choice, but nonetheless it is sometimes conventional and convenient when performing dimensional analysis with other quantities mostly in [MLT]. As a note, there is a strong case \cite{Hehl2005, Post1982} for using the dimensions of the action as a fundamental dimension to replace mass $M$ which has drawbacks such as being non-additive; in addition to using charge as a dimension.}\\
$A_{\mu}$ & four-potential & [$\Mass^\frac{1}{2}\Length^\frac{1}{2}\Time^{-1}$]$^*$\\
$F^{\mu \nu}$ & electromagnetic field tensor & [$\Mass^{\frac{1}{2}}\Length^{-\frac{1}{2}}\Time^{-1}$]$^*$\\
&&\\
$\omega$ & angular frequency & [$\Time^{-1}$]\\
$\varphi$ & phase function & [ ]\\
$k^{\mu}$ & four-wavevector & [$\Length^{-1}$]$^\ddagger$\\
$\sigma$ & (possibly affine) parameter along a path & [ ]\\
$\lambda$ & wavelength & [$\Length$]\\
 & OR eigenvalues of the Jacobian of the lens map & [ ]\\
&&\\

$\mathscr{Q}$ & time-like spacetime path &\\
$\mathscr{P}$ & null spacetime path & \\

$T$ & lensing arrival time functional & [$\Time$] \\
$\tau$ & lensing time delay & [$\Time$] \\
$\phi$ & Fermat potential & [ ] \\
$\hat{\psi}$, $\psi$ &  lensing potential, scaled & [$\Length$], [ ] \\
$\hat{\bm{\alpha}}$, $\bm{\alpha}$ & deflection angle, scaled & [ ], [ ]\\
$\bm{\beta}$ & angular position of source & [ ]\\
$\bm{\theta}$ & angular position of image & [ ]\\
$\kappa$ & dimensionless surface mass density   & [ ] \\
 & equiv. convergence of the lens & \\
& OR Einstein gravitational constant & $[\Mass^{-1} \Length^{-1} \Time^2]$\\
&&\\
\end{longtable}


\begin{abbreviations}{ll} 
\textbf{AGN} & \textbf{A}ctive \textbf{G}alactic \textbf{N}ucleus\\
\textbf{BAO} & \textbf{B}aryon \textbf{A}coustic \textbf{O}scillation \\
\textbf{BBN} & \textbf{B}ig \textbf{B}ang \textbf{N}ucleosynthesis \\
\textbf{BLR} & \textbf{B}road \textbf{L}ine \textbf{R}egion\\
\textbf{CDM} & \textbf{C}old \textbf{D}ark \textbf{M}atter \\
\textbf{CMB} & \textbf{C}osmic \textbf{M}icrowave \textbf{B}ackground \\
\textbf{DM} & \textbf{D}ark \textbf{M}atter\\
\textbf{EEP} & \textbf{E}instein \textbf{E}quivalence \textbf{P}rinciple\\
\textbf{EFE} & \textbf{E}instein \textbf{F}ield \textbf{E}quations \\
\textbf{FLRW} & \textbf{F}riedmann-\textbf{L}ema\^itre-\textbf{R}obertson-\textbf{W}alker \\
\textbf{GDE} & \textbf{G}eodesic \textbf{D}eviation \textbf{E}quation\\
\textbf{GR} & \textbf{G}eneral \textbf{R}elativity\\
\textbf{ISW} & \textbf{I}ntegrated \textbf{S}achs-\textbf{W}olfe \\
\textbf{MOND} & \textbf{Mo}dified \textbf{N}ewtonian \textbf{D}ynamics\\
\textbf{ODE} & \textbf{O}rdinary \textbf{D}ifferential \textbf{E}quation\\
\textbf{PDE} & \textbf{P}artial \textbf{D}ifferential \textbf{E}quation\\
\textbf{QSO} & \textbf{Q}uasi-\textbf{S}tellar \textbf{O}bject (Quasar)\\
\textbf{RM} & \textbf{R}everberation \textbf{M}apping\\
\textbf{SARM} & \textbf{S}pectro\textbf{a}strometry-\textbf{R}everberation \textbf{M}apping\\
\textbf{SA} & \textbf{S}pectro\textbf{a}strometry\\
\textbf{SEP} & \textbf{S}trong \textbf{E}quivalence \textbf{P}rinciple\\
\textbf{SN(e) Ia} & \textbf{S}uper\textbf{n}ova(e) Type \textbf{Ia}\\
\textbf{WEP} & \textbf{W}eak \textbf{E}quivalence \textbf{P}rinciple\\

\end{abbreviations}

\mainmatter 

\pagestyle{thesis} 



\chapter{Overview} 

\label{Introduction} 


\paragraph{Theoretical Background}

This thesis contains a presentation of theoretical aspects of cosmology and gravitational lensing -- the deflection of light from a distant source by intervening matter -- with a focus on distance measures which arise from the phenomenon of strong gravitational lensing. The overall flavour of this thesis aims to be systematic and motivated from first principles, rather than heuristic.

Chapters \ref{chap:standardcosmo} and \ref{chap:lightlensaction} respectively review in detail the fundamentals of standard cosmology and gravitational lensing. Chapter \ref{chap:lightlensaction} strives to address a particular pedagogical gap in {\hl some of the literature on} strong lensing, in which the relationship between the quasi-Newtonian approximation (considering a background FLRW cosmology) and its general relativistic foundations is not always given either detailed exposition, or made explicitly clear. We therefore utilise a relativistic, perturbative approach to derive the equations of the quasi-Newtonian formalism, taking care to enumerate the precise assumptions and approximations before taking each assumption one-by-one. A connection is thereby also made with the typical weak lensing formalism.

Chapter \ref{chap:cosmoglqso} introduces the usage of gravitational lensing and quasars in cosmology, providing the context for the original publications included in Chapters \ref{chap:p1} and \ref{chap:p2}.

\paragraph{Research}

Our ability to infer and interpret cosmological distances, such as to distant astrophysical objects including supernovae and quasars, forms one of the main foundations of contemporary cosmology. The increase in our observational capabilities over the past century has resulted in our current era being labelled as that of precision cosmology. Yet despite this precision fit of our standard cosmological model to the data, there remains no fundamental explanation for two of its main components -- the accelerated expansion assumed to be driven by a dark energy, and dark matter which appears to interact only gravitationally. In addition, the measurement of cosmological distances is difficult and therefore has been controversial both in the past and currently. For example, it is a matter of current debate whether the tension of the local ``distance ladder'' determination of the Hubble constant $H_0$ with the determination from the CMB data is due to unknown or under-estimated systematics, or new physics. The development of different techniques, such as those which can either measure out to yet larger distances with increasing precision or which do not suffer from the same limitations as current ones, have the potential to offer new insights or complementary information. This thesis centres on one such novel method based on combining the techniques of strong gravitational lensing time delay measurements and quasar reverberation mapping, which was proposed to measure a particular ratio of cosmological distances.

An observer may see multiple images of a distant source, as its light is deflected en route by the gravitational field of massive objects (such as a galaxy or cluster of galaxies) near the line-of-sight. This is the regime of strong gravitational lensing. If the source is additionally time-variable -- such as a supernova or quasar -- then the observer is able to detect a delay in the arrival times of light from each image. This time delay arises from both the difference in the path length travelled and the difference in the gravitational potential experienced by photons propagating in different directions. The measurement of these strong lensing time delays as a tool to determine cosmological distances and distance ratios, and thus cosmological parameters (chiefly the Hubble constant $H_0$) is known as time-delay cosmography. Time-delay cosmography is now a well-established method capable of providing independent determinations of $H_0$, important in the context of the current tension between {\hl the early-} and late-universe measurements. However, its limitations include the uncertainties in the assumed mass distribution of the lens.  

The method presented in Chapter \ref{chap:p1} aimed to circumvent these limitations, as it is independent of the lensing potential. It suggested a means to determine a particular ratio of cosmological distances (separate from the usual dimensionful distance ratio which is strongly dependent on $H_0$) by considering differential time delays over images, originating from spatially-separated photometric signals within a strongly lensed quasar. The difference in the light travel time within the quasar also contributes to the total time delay. This corresponds to the reverberation mapping time delay, and is able to give constraints on the geometry; information on the kinematics (i.e. the line-of-sight velocity) of these regions, is furthermore provided by spectroscopic data.

Critical examination of this method however, presented in Chapter \ref{chap:p2}, shows it to be unsound. An analytic description of the effect of the differential lensing on the emission line spectral flux for axisymmetric Broad Line Region geometries is also given, with the inclined ring or disk, spherical shell, and double cone as examples. The proposed method is unable to recover cosmological information as the observed time delay and inferred line-of-sight velocity do not uniquely map to the three-dimensional position within the source.

\chapter{Standard Cosmology} 
\label{chap:standardcosmo}
\captionsetup{width=.9\linewidth}
Rapid advances in observational cosmology have led to the establishment of the standard or concordance cosmological model, $\Lambda$CDM (defined in Section \ref{lcdm}). This includes measurements of the Cosmic Microwave Background anisotropies, with the highest precision observations from the Planck Satellite \cite{Planck2020}; as well as supernovae Ia data \cite[e.g.][]{Rest2014, Campbell2013, Guy2010}. Many key cosmological parameters have now been determined to one or two significant figure accuracy \cite{PDG2022}. Despite the precision fit of the $\Lambda$CDM model to the data, however, we lack explanations for fundamental physics of the accelerated expansion and for the nature of dark matter (matter which clusters gravitationally, but otherwise does not appear to interact). In this introductory chapter, we review the fundamentals of standard cosmology and  discuss the various issues unresolved by standard cosmology. We recommend the interested reader to \cite{DInverno1992, Baumann2022, Mukhanov2005, Bertschinger1995, Peebles2020} as pedagogical texts on theoretical aspects of cosmology and \cite{PDG2022, Abdalla2022} for thorough and up-to-date reviews of cosmology. This work also assumes familiarity with General Relativity (GR); standard textbooks include \cite{DInverno1992, Hartle2003, Carroll2004, Misner1973} amongst others -- a more recent textbook is \cite{Guidry2019}.

\section{Distances in Cosmology}

Since the 1920s, cosmology has transformed from a branch of philosophy to a precision science. At the core of this huge change has been our ability to measure or determine cosmological distances, an endeavour which is called \textit{cosmography}. By cosmological distances, we refer to large distances scales over which the Universe is on average homogeneous and isotropic. This addresses two fundamental questions: what there is in the Universe; and how it is distributed and moving on large scales. These questions have led to unexpected discoveries, notably the expansion of the Universe \cite{Lemaitre1931, Hubble1929} and the acceleration of this expansion \cite{Perlmutter1997, Riess1998}; as well as corroborating the evidence on galactic scales (e.g. \cite{Borriello2001}) for the existence of dark matter (for a review see e.g. \cite{Young2017}).

\subsection{Distance Measures in Curved Spacetime} \label{sec:distmeasuresincurvedspacetime}

As we relinquish the Newtonian concept of absolute time and space in both Special and General Relativity, a spatial distance is in general an observer-dependent quantity. The relativistic generalisation of the three-dimensional spatial position vector of Newtonian physics might na\"ively be assumed to be a four-vector position, but on this point we must also be careful. In a flat (Minkowski) spacetime, a vector may be defined by several equivalent definitions, including bi-locally as an algebraic difference between two points or spacetime events. Any bi-local concept of a vector is invalid in curved spacetime, as there is no concept of a unique straight path between two events (consider attempting to define a vector along the North and South poles of a sphere in this manner). On a curved manifold, a vector is strictly defined only locally, as a tangent to a curve at a point.

Fundamentally, there is no such thing as a \textit{position}, \textit{distance} or \textit{displacement vector} in general relativity without first defining a local coordinate system. The most basic kind of vector is a velocity or tangent vector. Vectors do not exist in a curved manifold itself, but live instead in a {\hl Minkowski} tangent space attached to the manifold at each point. However, since a curved manifold is \textit{locally flat}, it is nonetheless valid to consider \textit{infinitesimal} displacement vectors, equivalent to tangent vectors, which represent the displacement between two neighbouring points or events.

The metric tensor $\bm{g}$ on a curved manifold or spacetime is an important mathematical object which allows us to compute the inner product of two input vectors. By inputting two infinitesimal displacement vectors to the metric, we are able to define a \textit{line element} $ds$ which is invariant under coordinate transformations. The fundamental notion of a (temporal-spatial) distance, or more precisely the \textit{invariant interval} between spacetime events, is therefore defined in general relativity interchangeably by the metric or the line element
\begin{equation}
    ds^2 = g_{\mu \nu} (x^\alpha) dx^\mu dx^\nu
\end{equation}
and in some sense, this is a generalised Pythagorean theorem. The metric is our clock-and-ruler. The spacetime interval is called null, time-like and space-like for $ds^2 \{=0, <0, > 0 \}$ respectively. We notice that when two events $\mathscr{A}$ and $\mathscr{B}$ are space-like separated, then there exists a reference frame in which $\mathscr{A}$ and $\mathscr{B}$ occur simultaneously at different spatial locations, in which case $ds$ coincides with a purely spatial distance. However, this does not lead to a very pragmatic notion of a spatial distance, since actually observing an event (or some physical object) corresponds to photons arriving from an event (or the worldline of the object) to our own worldline.

In fact, although there are a number of such mathematically well-motivated definitions for spatial distances \cite{Fleury2015}, what is of greater practical relevance to cosmology are notions of spatial distance which are defined by physical observables. We will explore these in Section \ref{sec:observationaldistmeasures}, but we first need to write a metric to describe the Universe.

\subsection{A First Approximation of the Universe}

\subsubsection{The Cosmological Principle} \label{sec:cosmologicalprinciple}

The \textit{cosmological principle} posits that no observer, including ourselves, holds a privileged spatial position within the Universe; nor do there exist any privileged spatial directions. Its validity allows our own observations at a single location to be taken as a representative sample of the Universe, and utilised to test cosmological models. Historically the cosmological principle was a mere assumption generalising the Copernican principle, implying that statistically\footnote{Ensemble averaging is the same as spatial averaging assuming ergodicity, see Section \ref{sec:lambdacdm}.} the Universe is everywhere homogeneous and isotropic. Note if the Universe is isotropic or independent of direction around all points, then it is also homogeneous or independent of position; but the converse need not be true (see e.g. \cite{Peacock1999}). Observations since the end of the twentieth century have generally upheld the principle, showing the Universe to be approximately homogeneous and isotropic when averaged over large scales; and inhomogeneous on small scales --  we are of course not actually in a perfectly homogeneous Universe; certainly not at all scales, otherwise our galaxy would not exist and nor would we. The length scale at the transition from clumpy to smooth is 10 to 100Mpc, where $1\text{Mpc} = 3.1 \times 10^{22} \text{m} = 3.3 \times 10^6$ light years, whereas the length scale of the observable Universe is the \textit{Hubble distance} or \textit{Hubble radius} of order 4000Mpc \cite{Amendola2010}.

\paragraph{Weyl's Postulate} 

To relate the cosmological principle to our Universe, we need to apply Weyl's Postulate \cite{DInverno1992} as a complement on the kind of matter allowed. This states that
\begin{displayquote}
Galaxies behave like fundamental particles in a cosmological fluid, and these fundamental particles follow  timelike geodesics or worldlines which only ever intersect at a point in the finite or infinite past, and possibly in the future\footnote{That is, besides during the Big Bang and possibly during a Big Crunch.}. 
\end{displayquote}
These timelike geodesics are then orthogonal to a family of spacelike hypersurfaces; and a distance on these surfaces is called a foliation distance or physical distance. The family of timelike curves corresponds to a \textit{threading} of spacetime, and the spacelike hypersurfaces form a \textit{foliation} of spacetime; this is explored for more general metrics or spacetimes in Section \ref{sec:threading}.

 Any set of curves filling spacetime (or a given open region in spacetime) and which is non-intersecting is called a \textit{congruence} of curves \cite{Poisson2004}. {\hl If the geodesics are non-intersecting, there is one and only one geodesic passing through each point of spacetime such that there exists a unique four-velocity at every point and a hydrodynamical (fluid) description of the matter of the Universe is adequate.\footnote{\hl If the four-velocity field is multiply-valued and hence ill-defined, the fluid description breaks down and a phase-space description is necessary. The statement in \cite{DInverno1992} is misleading as Weyl's postulate by itself does not imply a \textit{perfect fluid} description of matter.}} This is not quite true in reality, e.g. Andromeda is on a collision course with the Milky Way and worldlines cross as structures form; but it is a good approximation. The relative peculiar velocities of galaxies are random and much, much smaller than the speed of light, to which the general velocity of the bulk motion on cosmological scales is comparable.

Thus when we say that the cosmological principle applies statistically, it means that it applies in a given \textit{preferred frame} only -- defined by a privileged family of freely falling observers that move with the average velocity of typical galaxies in their respective neighbourhoods. These are called \textit{fundamental} or \textit{comoving} observers. For example, on Earth we are not quite comoving and therefore do not see complete isotropy; we see a \textit{dipole} in the Cosmic Microwave Background (CMB). This is assumed to be kinematic in origin, so we may perform a special relativistic boost to the so-called \textit{CMB rest frame} in which the Universe is isotropic. Ultimately, the cosmological principle or spatial isotropy, and the existence of preferred frames which thread spacetime are rooted in the Weak and Einstein Equivalence Principles respectively (see Section \ref{sec:modifiedgrav}).

\subsubsection{The Friedmann-Lema\^itre-Robertson-Walker Metric}

Applying spatial homogeneity and isotropy to the Universe means that it can be represented by a time-ordered sequence of three-dimensional spatial slices $\Sigma_t$, each of which is homogeneous and isotropic and therefore can only possess \textit{constant} 3-curvature. These are therefore global hypersurfaces of simultaneity: there is a global coordinate time (so-called cosmic time) $t$ which coincides with the proper time $\uptau$ of all privileged observers at fixed $x^i$. The idea of global simultaneity may seem strange; after all, in Special and General Relativity we grappled with the idea of simultaneity being observer-dependent -- but the cosmological principle is a very strict condition.
This singles out a unique form for the spacetime geometry, given by the Friedmann-Lema\^itre-Robertson-Walker (FLRW) metric whose line element may be expressed as
\begin{equation}
\boxed{
ds^2 = g_{\mu \nu}dx^\mu dx^\nu = -d(ct)^2 + a^2(t)\tilde{\gamma}_{ij} dx^i dx^j = a^2(\eta) (-d(c \eta)^2 + \tilde{\gamma}_{ij} dx^i dx^j). }\label{FLRW}
\end{equation}
The time dependence of the position of a fundamental particle or galaxy is factored out of the coordinate system and into the scale factor $a$ -- the physical or proper distance between points are expanding in time proportional to $a$, and the comoving coordinates $x^i$ are such that the coordinate separation remains constant in time. Factoring out the scale factor from the whole metric in the last equality in Equation \eqref{FLRW} is a particular case of a \textit{conformal transformation}, so we may use either the cosmic time $t$ or conformal time $\eta \equiv a^{-1} t$. The scale factor is the normalised three-dimensional spatial hypersurface analogue of the radius of a two-dimensional sphere.

It may be convenient in different circumstances to use different comoving coordinates for the conformal spatial metric $\tilde{\gamma}_{ij}$. For example, in spherical and quasi-Cartesian coordinates respectively $$\tilde{\gamma}_{ij} dx^i dx^j = \frac{d r^2}{1- K r^2} + r^2  (d\theta^2 + \sin^2 \theta d \phi ^2) =  \left( \delta_{i j} + K \frac{x_i x_j}{1-K x_l x^l} \right) d x^i d x^j.$$ When we consider light, as in a large portion of this work, we choose to work in \textit{modified  spherical} or hyperspherical spatial coordinates 
\begin{equation}
\boxed{
    \tilde{\gamma}_{ij} dx^i dx^j = d \chi^2 + f_K^2 (\chi) ( d \theta^2 + \sin^2 \theta d \phi^2)} \label{flrwconformalspatialmetric}
\end{equation}
since photon paths in the background can be assumed, without loss of generality given the isotropy of space, to travel on radial paths along $\chi$. This defines a notion of a line-of-sight comoving distance (set $ds^2=0$ and take the negative solution for a photon towards the observer at $\chi = 0$), which may be written as a function of time
\begin{equation}
    \chi(t) = - \int_{t_0}^{t} \frac{c dt'}{a(t')}. \label{chi_integral}
\end{equation}
Since it is a line-of-sight distance, it is sometimes called a radial coordinate, but it is independent of spatial curvature (unlike $r$) and analogous to the polar angle of a 2-sphere (but here has dimensions of length, see footnote \ref{footnote_scalefactor}).

The function $f_{K}(\chi)$ may be called a \textit{comoving areal radius} \cite{Fleury2015}, \textit{metric distance} \cite{Baumann2022} (although we find this latter terminology imprecise), or \textit{comoving transverse distance} \cite{Hogg1999}. It relates the spherical radial coordinate $r = f_K(\chi)$ to the modified radial coordinate $\chi$, and is defined\footnote{The unified definition \eqref{fK} is just as justified as the more common definition by the individual cases for the sign of K, as the strict mathematical definition of a sine is given by a Taylor series expansion \cite{Abramowitz1970}. Written out explicitly, this is $f_K(\chi) \equiv K^{-1/2}\sin(K^{1/2} \chi) = \sum_{n=0}^\infty \frac{(-1)^{n}}{(2n+1)!} K^n \chi^{2n+1}$. The $K=0$ case is defined formally by analytic continuation, using the limit $K \to 0$. The unified expression is much more useful to us for manipulating trigonometric identities.} according to the constant spatial curvature or Gaussian curvature $K$
\begin{equation}
f_K(\chi) \equiv \sin_K(\chi) \equiv K^{-1/2} \sin(K^{1/2}\chi) \label{fK}
\end{equation}
where $K$ is $\{>0, =0, < 0\}$ for closed, flat and open spatial geometries, with corresponding $f_K(\chi) = \{|K|^{-1/2} \sin(|K|^{1/2} \chi)$, $\chi$, $|K|^{-1/2} \sinh(|K|^{1/2} \chi) \}$.

The present-day scale factor $a_0 \equiv a(t_0)$ is usually normalised to $1$. We note there are different normalisation conventions for the FLRW metric: it is important to keep in mind that if we normalise $K$ as the dimensionless $k = \{0, \pm 1\}$, then the metric uses an alternative scale factor\footnote{\label{footnote_scalefactor}The length scale $R(t)$ is the (\textit{unnormalised}) analogue in a curved 3-space to the radius of a 2-sphere, see e.g. \cite{Peacock1999, Baumann2022}. To be explicit, the metric presented here is invariant under the rescaling: $R= R_0 a$, $r' = \frac{r}{R_0}$ or $\chi' = \frac{\chi}{R_0}$, $k = K R_0^2$. Then $\chi'$ is a polar angle which ranges from $[0, \infty)$ in flat and open spacetimes and $[0, \pi]$ in closed spacetimes.} $R(t)$ -- or includes its present-day value $R_0$ -- with units of length, which cannot be simultaneously normalised to $1$ except when $K = 0$.

As we will later see, the field equations of general relativity relate the energy and matter content of the Universe to its geometry. A spatially flat or Euclidean universe is one in which the energy density is equal to a critical value; if the density is higher or lower than this value, then the Universe is closed (spherical) or open (hyperbolic) respectively. The Universe is observed to be very close to spatially flat, and is often assumed to be exactly so. This observation is popularly framed as the flatness problem (found in many textbooks e.g. \cite{Baumann2022}), in some versions stated as a coincidence or fine-tuning such that the energy density of the Universe should be precisely the critical density -- therefore proposed to be a result of inflation, a period of very rapid expansion, in the early Universe. It has been argued that the flatness problem, at least in all of its versions, is not truly a problem \cite{Helbig2020b, Holman2018}.

The different possibilities for spatial curvature may be understood by considering two free test particles which are initially moving in parallel. A spatially flat universe is Euclidean, and the particles remain parallel. The particles converge in a spatially closed universe, analogous to how aeroplanes following lines of longitude (i.e. great circles or geodesics of a 2-sphere) would converge at either pole on the globe; and diverge in a spatially open universe. Although an open or negatively curved 3D space is often illustrated by a 2D saddle, such a surface does not have constant negative curvature. It is possible, however, to think of the saddle as local representation rather than a global one. 

\subsection{Observational Distance Measures} \label{sec:observationaldistmeasures}

Metric distances, in particular those defined by space-like separations, are not directly observable -- we see distant objects or events as a consequence of light\footnote{Up until recently, astronomy was dominated by electromagnetic observations, mostly optical as well as some other bands e.g. radio, microwave. Non-electromagnetic observations involve neutrinos and cosmic rays. With the advent of gravitational wave detection in 2015, there is now a non-electromagnetic field of observational cosmology revealing new information \cite{Bailes2021}. The coordinated observation and interpretation of all these types of signals or particles, predicted for the upcoming era, is called \textit{multi-messenger} astronomy.} travelling from them. As such, what is of practical relevance to cosmology are notions of spatial distance which are defined by physical observables. These observables include the measured time differences of light signals (e.g. radar distances and strong lensing time delay distances), angles on the sky (e.g. parallax distances and angular diameter distances), the intensity of light (e.g. luminosity distance) and the spectra of light (i.e. redshift).

The measurement of cosmological distances is difficult and therefore has been controversial both in the past and currently. This is exemplified respectively by the so-called \textit{Hubble wars} ($\sim$1960-80) \cite{Narlikar1996}, and \textit{Hubble tension} of the local \textit{distance ladder} determination of the present day Hubble parameter $H_0$ with the determination from the CMB data. Whether this current Hubble tension is due to unknown systematics or new physics is the subject of much discussion \cite{Alam2017, DiValentino2021, Riess2018, Riess2019}. The main line of empirical attack is to measure the history of expansion with increasing precision over an increasing range of redshifts \cite{Ellis2012}.

There are broadly two classes of objects with which we may use to measure cosmological distances: standard candles which provide a luminosity distance $D_{L}$, and standard rulers which provide an angular diameter distance $D_A$. The classic standard candles used in cosmology are Cepheid variable stars and Supernovae Ia; and examples of standard rulers are the Baryon Acoustic Oscillations (BAO) observed as a preferred distance scale in the galaxy distribution, and the acoustic peaks of the CMB power spectrum. These standard candles and rulers of cosmology need to be calibrated. For example, we may first find distances to nearby stars, e.g. Cepheids within the galaxy, using parallax measurements. These measurements in turn are used to calibrate Cepheid stars in nearby galaxies, which are then further used to calibrate Type Ia supernovae to higher and higher redshifts. This chain of measurements forms the local \textit{distance ladder} \cite{Riess2019}. Alternatively, we may use the \textit{inverse distance ladder}, by calibrating supernovae Ia from BAO data calibrated using the CMB power spectrum \cite{Aubourg2015}. However these methods involve complicated astrophysics, such as supernovae explosions and structure formation. Furthermore, systematic errors accumulate along each step of the distance ladder \cite{DeGrijs2011, Riess2018}. This section is a non-exhaustive presentation of some of the main observational distance measures from first principles.

\subsubsection{Cosmological Redshift}

A spectral shift, commonly described as a blueshift or redshift, is the change of a wavelength of a wave from a value $\lambda_s$ measured at its source to a value $\lambda_o$ measured by an observer. It is defined by a fractional wavelength (or equivalently frequency via the dispersion relation or angular frequency $\omega$) change parameter, usually simply called the \textit{redshift} $z$ which for a given source is
\begin{equation}
z_s \equiv \frac{\lambda_o - \lambda_s}{ \lambda_s} = \frac{\omega_s -\omega_o}{\omega_o}, 
\end{equation}
hence a value $z<0$ implies a blueshift and $z>0$ implies a redshift. In the context of astrophysics and cosmology, the observed emission lines from distant light sources such as galaxies may be compared to a known rest-frame pattern, allowing us to measure a spectral shift.\footnote{When considering large galaxy surveys, it is not practical to perform spectroscopy on the entire survey. Instead, the redshift of a galaxy is estimated using multi-band photometry and the difference in the intensity of the object in several colours, calibrated using subsamples which do have spectroscopic data. The redshift estimate is called a photometric redshift, or a \textit{photo-z}.}

\paragraph{Cosmological Redshift as a Measure of Time and Distance}

In relativity, time and space are intertwined as a single manifold described using the metric. In addition, galaxies may be considered to be roughly fixed at comoving coordinates (see Weyl's Postulate, discussed in Section \ref{sec:cosmologicalprinciple}). This allows redshift, which is usually interpreted only kinematically, to be used as a measure of both time and spatial distance in FLRW cosmology.

\begin{figure}
    \centering
    \includegraphics[width=.8\linewidth]{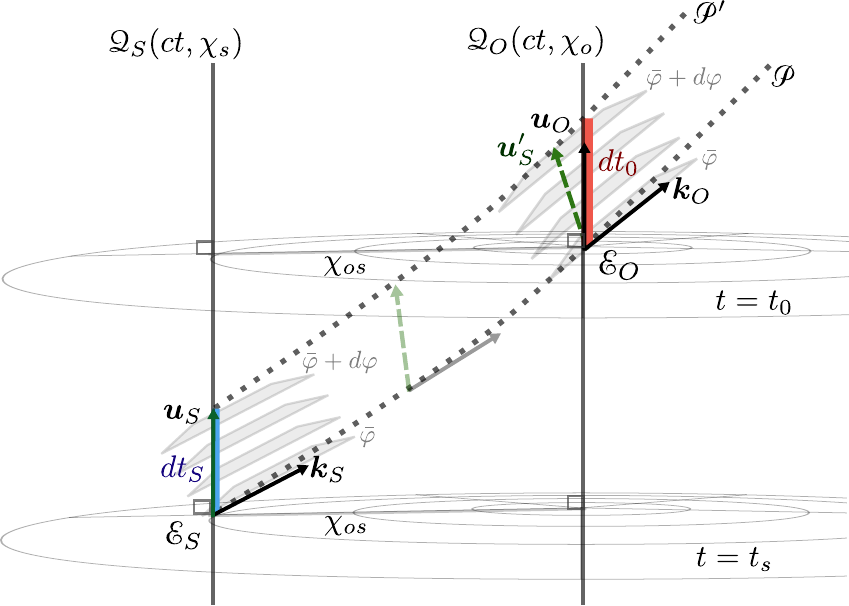}
    \caption{Illustration of a cosmological redshift in an expanding, spatially flat FLRW spacetime (one spatial dimension suppressed). The spacetime can be foliated into slices of constant $t$ which coincide with the proper time of comoving observers when working in comoving coordinates (this does not hold in physical coordinates; see the spacetime diagrams of \cite{Davis2004}). The $\chi$ coordinates on flat space-slices can be drawn as a radial coordinate, but actually is more analogous to a polar angle. \\ Here a distant comoving source travels along a timelike worldline $\mathscr{Q}_S$, and a comoving observer along $\mathscr{Q}_O$; they both have four velocities $u^\mu_S$ and $u^\mu_O$ which are unit vectors in comoving coordinates. The source emits light at event $\mathscr{E}_S$, which then travels along a null geodesic $\mathscr{P}$ until it is seen at event $\mathscr{E}_O$; a short time $dt_s$ later it emits a second signal which travels along a neighbouring null geodesic $\mathscr{P}'$ which the observer receives at a time $t_0 + dt_0$. The time intervals $dt_S$ and $dt_0$ can be interpreted the period of the light wave ($d \uptau = T \frac{d \varphi}{2 \pi}$ setting the change in phase $d \varphi = 2 \pi$) with $dt_0 > dt_s$ due to the expansion $a(t_0) > a(t_s)$. \\
    We could also apply directly that the source measures an emitted photon in their frame to have angular frequency $\omega_S = -c (k_\mu u^\mu)_S$ since $u^\mu$ is the unit vector in the time direction, where $k^\mu = (\frac{\omega}{c}, k^i)$ is the four wave-vector of the photon. A geometric interpretation of this statement comes from illustrating the 1-form $k_\mu$ corresponding to $k^\mu$ as a series of hypersurfaces each with constant phase; i.e. wavefronts. Since the wavevector $k^\mu$ is null, it lies on one of the hypersurfaces of $k_\mu$ (this is non-intuitive due to the Lorentzian geometry; we refer to Figure 2.7 of \cite{Misner1973} for further clarification). The observed frequency $\omega /c$ is the number of times the hypersurfaces of $k_\mu$ is pierced by the unit vector $u^\mu$. \\
    {\hl This means that it is also possible to derive the cosmological redshift using the fact that the phase function $\varphi$, and thus $d \varphi = k_\mu dx^\mu$, is constant along a null geodesic $\mathscr{P}$, where $k_\mu \equiv \partial_\mu \varphi$.} Dividing $k_\mu dx^\mu$ through by proper time or equivalently cosmic time gives $(\omega dt)_S = (\omega dt)_O$. {\hl This is also shown in a more general context in e.g. \cite{Terno2020}.}\\
    Finally, the result can interpreted in terms of special relativistic Doppler shift using ${u'}^\mu_S$, the parallel transport of $u^\mu_S$ along $\mathscr{P}$ to the observer \cite{Narlikar1994, Fleury2015}.}
    \label{fig:flrwredshift}
\end{figure}

Consider two successive light signals emitted from a faraway galaxy $S$ emitted at $\uptau_s$ and $\uptau_s + d \uptau_s$ which are received by an observer $O$ at $\uptau_0$ and $\uptau_0 + d\uptau_0$. If the signals are successive wave crests, then the interval $d \uptau_s$  or $d \uptau_0$ is the period of the light wave measured according to the proper time of the source or observer (more generally, from the definition of angular frequency as $\omega \equiv | \frac{d \varphi}{ d \uptau} |$ we have that $d \uptau = T \frac{d \varphi}{2 \pi}$ where $T$ is the period and $d\varphi$ is the change in the phase over the emission time $d \uptau$). We assume both the source and observer are comoving with the Hubble expansion such that cosmic time coincides with the proper time ($dt = d\uptau$ using $ds^2 = -d(c \uptau)^2$ in comoving coordinates) for both the source and observer. We take for granted that the path of a light ray is given by a \textit{null geodesic} (this is shown from first principles in Section \ref{sec:geometricoptics}), for which $ds^2 = 0$; assuming a light ray radial to the observer at the origin in an FLRW metric gives $\pm d \chi = \frac{c}{a(t)} dt$, where the positive solution corresponds a receding ray (the \textit{future light cone}) and the negative solution to an incoming ray (the \textit{past light cone}). Integrating over the incoming solution separately for both signals, we clearly see that the comoving separation $\chi_{os} = - \int_{\chi_s}^0 d \chi$ between $O$ and $S$ is constant, giving
\begin{equation}
    \int_{t_s}^{t_0} \frac{dt}{a(t)} = \int_{t_s + dt_s}^{t_0+ dt_0} \frac{dt}{a(t)}.
\end{equation}
The scale factor $a(t)$ may be treated as constant over the short time interval $dt$ associated with the period of the light wave, so in terms of the corresponding wavelength $d \lambda = c dt$ we have
\begin{equation}
\boxed{
    1 + z_s \equiv \frac{\lambda_o}{\lambda_s} = \frac{\omega_s}{\omega_o} = \frac{dt_0}{dt_s}  = \frac{a(t_0)}{a(t_s)} = \frac{(k_\mu u^\mu)_s}{(k_\mu u^\mu)_o}} \label{cosmologicalredshift}
\end{equation}
which in an expanding universe such that $t_0 > t_s$ and $a(t_0) > a(t_s)$ is a \textit{cosmological redshift}. A contracting universe would measure a blueshift. The last equality comes directly from writing the frequency of a photon in the form $\frac{\omega}{c}  \equiv | \frac{d \varphi}{ d (c\uptau)} | = |\varphi_{, \mu} u^\mu | = - k_\mu u^\mu$, where $\varphi$ is the phase function, its four-wavevector is $k_\mu \equiv \varphi_{, \mu}$ and the observer four-velocity is $u^\mu \equiv \frac{d x^\mu}{d (c \uptau)}$. We illustrate and discuss additional methods for arriving at Equation \eqref{cosmologicalredshift} in Figure \ref{fig:flrwredshift}. A general formal definition of redshift for an arbitrary spacetime is given by \cite{Perlick2004}.

The cosmological redshift parametrises time since expression \eqref{cosmologicalredshift} shows a one-to-one correspondence between the redshift $z = z_s$ and the cosmic time parameter $t = t_s$ of any distant comoving source. Although $z$ is really a function of two variables $t_0$ and $t_s$, our time of observation or present time $t_0$ is considered a constant value given cosmological time scales. A given $z$ corresponds to a time when our Universe was $1 + z$ times smaller than present day. We can then differentiate the redshift $(1+z) = \frac{a_0}{a(t)}$ with respect to time $t$ to find the expression
\begin{equation}
    \frac{dz}{dt} = -(1+z) H(t). \label{redshiftdrift}
\end{equation}
If we in fact allow for the time of observation to vary, we arrive at the notion of the \textit{redshift drift}. This is the change in redshift for a \textit{single} comoving source over time measured by an observer at present day, found by differentiating $(1+z) = \frac{a(t_0)}{a(t_s)}$ with respect to $t_0$, giving $\frac{dz}{dt_0} = (1+z)H_0 - H(z)$. The redshift drift is expected to be incredibly small, requiring a precision of $\sim 10^{-9}$ to be measured on human time scales \cite{Kim2015}. However, this in principle would provide \textit{direct} measurement of cosmological dynamics, i.e. accelerated or decelerated expansion, if distinguished from changes in the peculiar velocity. A comoving source will exhibit a redshift drift in any FLRW cosmology, except an empty universe where $da/dt = \text{const.}$, including a steady state universe with constant $H$.

Redshift is used as a measure of distance, since the comoving distance \eqref{chi_integral} is parametrised by time $\chi(t)$ and therefore redshift $\chi(z)$ using \eqref{redshiftdrift}
\begin{equation}
    \chi  = - \int_{t_0}^{t_s} \frac{cdt}{a(t)} = \int_0^{z_s} \frac{c dz}{a_0 H(z)}. \label{comovingdistance}
\end{equation}
The expression for $H(z)$ can be found from the Friedmann equations, \eqref{E(z)}. However, the redshift is not an exactly fixed parameter of a comoving object due to the redshift drift effect: it is not precisely interchangeable with a comoving distance. Nonetheless, our measurements e.g. galaxy surveys naturally exist in \textit{redshift space} which are mapped from real space with distortion (e.g. Alcock-Paczy\'nski effect from converting from redshift to real space with the wrong cosmological model, or ``fingers of God'' effect from peculiar velocities).

\paragraph{Interpretation of the Cosmological Redshift}
The interpretation of measured cosmological spectral shifts in terms of a physical cause has been a longstanding source of pedagogical confusion and some professional debate. In classical physics\footnote{The classical Doppler shift formulae are usually demonstrated considering acoustic phenomena and tacitly based on Newtonian approximations, although the kinematics of relativity apply to the propagation of all kinds of signals including sound waves. As velocities must be considered with respect to the medium that transmits the waves, it matters for sound waves whether the source is moving or the observer; whereas for light in Minkowski spacetime, only the relative velocities of observer and source are relevant.}, a \textit{Doppler shift} or change in wavelength is straightforwardly due to the motion of a source relative to an observer. However, the relative velocities of two objects at two spatially separate events is not an \textit{a priori} well-defined concept in general relativity, as we {\hl cannot} subtract vectors at different locations on a curved manifold. This has resulted in attribution of the cosmological redshift being misleadingly {\hl (due to reification)} abstracted to the expansion of space \textit{itself}, or incorrectly -- in the opinions of Synge and Narlikar \cite{Synge1960, Narlikar1994}, although Kaiser, Bunn and Hogg, and Fleury \cite{Kaiser2014, Bunn2009, Fleury2015} do not quite agree -- invoking gravitational redshifting (a cosmological redshift is independent of the spacetime curvature or Riemann tensor; one can transform away first-order derivatives of the metric i.e. Christoffel symbols using appropriate choice of coordinates resulting in a purely kinematic interpretation). 

However, it is standard in differential geometry to compare two vectors on a curved manifold by using \textit{parallel transport} to move one of them to the same point as the other (illustrated in Figure \ref{fig:flrwredshift}). A cosmological redshift in an expanding universe can be equivalently interpreted as a Doppler shift using this modified idea of a relative velocity \cite{Narlikar1994}. This relative velocity is dependent on the path taken for parallel transport; for photons, the path is a null geodesic from the source to the observer.

The idea of a parallel-transported relative velocity may be understood by comparing local velocities using a series of infinitesimal special-relativistic Doppler shifts \cite{Narlikar1994, Peacock1999}. Consider that light redshifted from a \textit{nearby} galaxy $S'$ can be interpreted straightforwardly as a special relativistic Doppler shift due to its recessional velocity, since any spacetime is locally Minkowski. A series of comoving galaxies would each measure light Doppler redshifting purely due to a recessional motion from an adjacent galaxy along the path from $S$ at an arbitrarily large separation from us at $O$. Integrating over infinitesimal Doppler shifts in a general spacetime is not equivalent to a single large Doppler shift in Minkowski spacetime: it possesses path dependence.

\subsubsection{Radar Distance}

We can motivate a distance measure from the non-relativistic expectation that the speed of light is a constant $c = \frac{r}{t}$, where $r$ is an Euclidean distance and $t$ an absolute time. The radar distance $D_\textsc{r}$ is therefore defined in a general spacetime as $c$ times half the duration, as measured by an observer O, of a round trip of a light signal between O and a target
\begin{equation}
\boxed{
    D_\textsc{r} \equiv \frac{c}{2} (\uptau_\textsc{r} - \uptau_\textsc{e})}
\end{equation}
{\hl where $\uptau_\textsc{r}$ and $\uptau_\textsc{e}$ are respectively the proper time of reception and emission.} Its use is mostly limited to short-distance measurements on Earth and within the Solar System where a reflective target exists or may be installed: for example, in interferometry-based experiments (such as gravitational wave detection \cite{LIGO2010}), or the \textit{Lunar Laser Ranging} measurements \cite{Muller2019} of the Earth-Moon distance using retroreflectors set up by lunar missions.

Radar ranging is used to test General Relativity by measuring the gravitational or \textit{Shapiro time delay}. As the path of light is curved in a gravitational field, its travel time as measured by an observer is greater than in flat spacetime. We can deduce this from setting $ds^2 =0$ which corresponds to a photon path and looking at how the coordinate velocity $\frac{dx^i}{d(ct)}$ (i.e. the speed in the global frame) differs in a general metric from the Minkowski metric. We may measure this delay to great precision by timing the echoes of pulsed radar signals sent from Earth to other planets in the Solar system, for example the measured delay for Venus is on the order of 200$\mu$s. The experimental results confirm the predicted time delay calculated using the generalised Schwarzchild metric \cite{Reasenberg1979}.

\subsubsection{Parallax Distance}

Parallax refers to the apparent displacement of an object's position when viewed from different vantage points. We may use parallax to measure distance by observing the displacement of an object against the background when viewed from two different locations. In everyday life, the slight variation in perspective between our two eyes allows for depth perception. We may also use parallax to measure the distance to astronomical light sources such as stars within our galaxy, by comparing their apparent displacement due to our motion against the background of much more distant stars. Such stellar parallax measurements, like those from \textit{Gaia} with microarcsecond accuracies \cite{Gaia2018}, form the first step of the so-called local distance ladder.

In a general spacetime, a light beam from a distant source is distorted by gravitational lensing, meaning we need to account for a varying parallax angle over all observer displacement directions. Therefore, instead of considering a linear displacement, let us consider the observer tracing a circular path on a plane orthogonal to the line of sight. This also conveniently corresponds to the actual experimental technique of using the known diameter of the Earth's orbit around the Sun (referred to as the \textit{baseline distance}) to perform stellar parallax measurements.
\begin{figure}
    \centering
    \includegraphics[width=.8\linewidth]{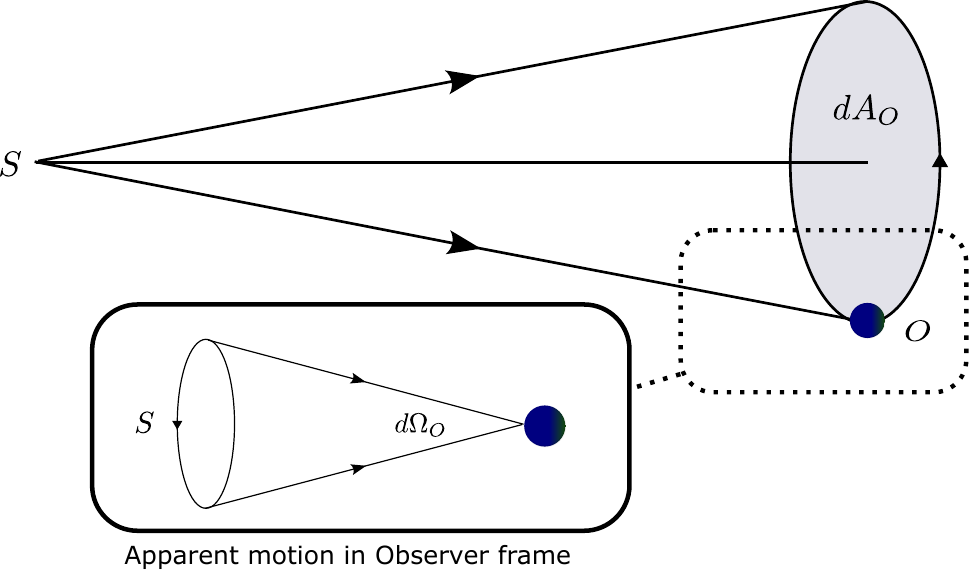}
    \caption{Illustration of a generalised parallax distance, adapted from \cite{Fleury2015}.}
    \label{fig:parallax}
\end{figure}

We may define a generalised parallax distance $D_\textsc{p}$ using the definition of a solid angle $d \Omega$ in an Euclidean spatial geometry, 
\begin{equation}
    d \Omega = \frac{dA}{r^2} \hat{\bm{r}} \cdot \hat{\bm{n}} \label{solidangle}
\end{equation}
where $\bm{r}$ is an Euclidean distance vector, and $\bm{\hat{n}}$ is the unit vector normal to a surface area $dA$. Associating this with the solid angle $d \Omega_o \ll 1$ of a narrow beam of light from the source and $dA_o$ with the proper area encircled by the observer’s trajectory, we define $D_\textsc{p}$ for a general spacetime as 
\begin{equation}
\boxed{
    D_\textsc{p} \equiv \sqrt{\frac{dA_o}{d\Omega_o}}}. \label{genparallax}
\end{equation}
Allowing for line-of-sight motion requires a correction related to a Doppler shift. When there is no shear from gravitational lensing such that the incoming beam has a circular cross-section, then the solid angle is for a cone, $d \Omega_o = 2 \pi ( 1 - \cos d\theta_o) \approx \pi d\theta_o^2$. The corresponding area is $dA_o  = \pi b^2$ where $b$ is half the baseline distance and the expression \eqref{genparallax} reduces to the more common trigonometric parallax definition
\begin{equation}
    D_\textsc{p} = \frac{b}{|d \theta_o|}.
\end{equation}
The absolute value of the \textit{parallax angle} $|d \theta_o|$, which can be related to the divergence of the light beam, ensures $D_\textsc{p} > 0$ in case of convergence from gravitational lensing \cite{Rosquist1988}. It is from this parallax definition that the astronomical unit of the parsec was defined: it is the distance to a source with parallax angle of 1 arcsecond using the baseline distance $2b$ of 2AU (roughly the diameter of the Earth's orbit around the Sun). 

Parallax measurements are generally, or at least until recently, considered to be limited to the distances within our own galaxy, due to the longest possible baseline arising from the Earth's orbit around the sun. Various methods to measure cosmic parallaxes from extragalactic sources, such as quasar or galaxy parallaxes, have long been proposed; for example using the motion of the Earth with respect to the cosmic microwave background \cite{Rasanen2014}. These parallaxes remain below the current detection limit \cite{Gaia2021, GaiaQSO2022}, and observed quasars are instead used to generate an optical reference frame. However, a new technique has enabled preliminary quasar parallax measurements, using reverberation mapping of quasars and spectroastrometry to reduce the required interferometric baseline \cite{Wang2020}. 

\subsubsection{Angular Size Distance}

We may define a different distance measure, also based on the definition \ref{solidangle} of a solid angle in Euclidean space. Consider that an object appears smaller when it is further away. If we know the actual size of this object, then we can use its apparent size to measure its distance from us. The relevant surface area is now the physical area $dA_S$ of the light source, and $d\Omega_o \ll 1$ is its apparent angular size measured by the observer, allowing us to define a surface \textit{area distance} or \textit{angular size distance} $D_{\text{A}}$ for a general spacetime
\begin{equation}
\boxed{
    D_{\text{A}} \equiv \sqrt{\frac{dA_s}{d\Omega_o}}}.
\end{equation}
Once again, we may reduce this expression to a more common one in the absence of shearing from gravitational lensing. This is the \textit{angular diameter distance} 
\begin{equation}
    D_{\text{A}} = \frac{d\ell_s}{d\theta_o}
\end{equation}
where $d\ell_s$ is the proper diameter of the source, and $d\theta_o$ is the angle it subtends on the observer's sky; i.e. this is the Euclidean arc length formula. We will refer to the angular size distance by the more common angular diameter distance interchangeably, although they are not strictly equivalent.

The angular diameter distance therefore shares a kind of symmetry with the parallax distance. Rather than requiring a known distance or baseline at the observer (for parallax), we require the physical size at the \textit{source} be known. The class of objects for which we believe we know the proper size -- or at least can be calibrated by independent experiments -- are called \textit{standard rulers}. One example important for cosmology is the Baryon Acoustic Oscillation (BAO) scale, corresponding to the maximum distance travelled by a sound wave in the early Universe. It is inferred from the Cosmic Microwave Background (CMB) anisotropies and in the distribution of galaxies. The angular diameter distance is the main distance measurement relevant for strong gravitational lensing and time delay experiments.

\begin{figure}
    \centering
    \includegraphics[width=.8\linewidth]{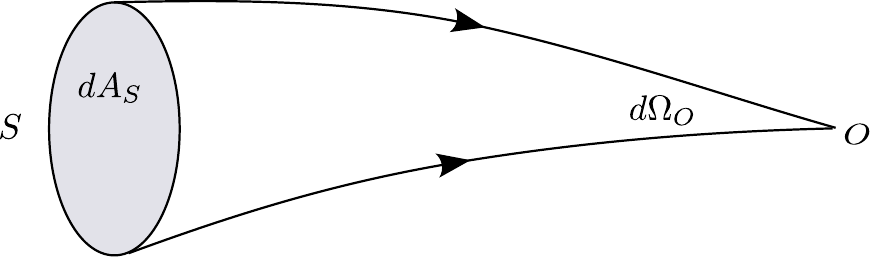}
    \caption{Illustration of an angular \textit{size} distance, adapted from \cite{Fleury2015}.}
    \label{fig:angularsize}
\end{figure}

Assuming an FLRW metric, we deduce from the line element that a proper surface area element $dA_s$ corresponding to a solid angle at the observer $d\Omega_o = \sin \theta_o d \theta_o d \phi_o$ is
\begin{equation}
    dA_s = (a_s f_K (\chi_{os}) d \theta_o) \times ( a_s f_K(\chi_{os}) \sin \theta_o d \phi_o) = a_s^2 f_K^2 (\chi_{os}) d \Omega_o \label{surfacearea}
\end{equation}
where we note the subscripts -- the solid angle at the observer $d\Omega_o$ corresponds to a proper surface area on the celestial sphere $dA_s$. Along with the redshift relation \eqref{cosmologicalredshift}, this gives the angular size distance in terms of the comoving areal radius $f_K( \chi_{os})$ as
\begin{equation}
\boxed{
    D_{os} \equiv D_{\text{A}, o \to s} = (1+z_s)^{-1} a_0 f_K( \chi_{os})}. \label{angulardistanceOS}
\end{equation}
From symmetry, we have a \textit{reciprocity relation} for $D_A$ to an object at $z_d$ and $D_A$ to $z_s$ 
\begin{equation}
D_{od} = D_{do} \left( \frac{1+z_d}{1+z_s} \right).
\end{equation}

There are a few important points to note regarding the angular diameter distance. Firstly, the angular diameter distance measures the distance between us and the object when the light was \textit{emitted}. Secondly, it is not a monotonically increasing function of $z_s$ as one might na\"ively expect of a distance; it reaches a maximum value in most standard cosmologies around $z_s \sim 1$; this is illustrated in Figure \ref{fig:distancemeasures_dodelson}. That is, objects beyond a certain redshift appear larger on the sky. On one hand, an object of a given size looks smaller the further away it is. On the other hand, due to the expansion of the Universe and finite light speed (i.e. the onion shape of the light cone on a spacetime diagram \cite{Lineweaver2005}) very distant objects were closer to us when they emitted the light we see today -- at that time they spanned a larger angle on the sky. The effects are \textit{looking smaller because far away} and \textit{looking larger because closer in the past}. This has the non-intuitive consequence that $D_A$ and the comoving areal radius $f_K( \chi)$ are in general not additive; i.e. $D_{od} + D_{ds} \neq D_{os}$: it is $\chi$ which is additive. Therefore, in flat space, $f_K( \chi)$ is additive but in curved space it is not.

In gravitational lensing, we sometimes need an expression for the angular diameter distance from the observer to the lens $D_{od}$ in terms of the angular diameter distance to the source $D_{os}$ and the angular diameter distance from lens to the source $D_{ds} $, so let us derive one. First, using the alternative notation for the comoving areal distance $f_K (a) = \sin_K (a)$ will aid future transparency when manipulating trigonometric identities generalised with respect to surfaces of constant spatial curvature $K$. We therefore define the following pair of functions
\begin{equation}
    \cos_K (a) \equiv \cos(K^{1/2} a) \quad \text{and} \quad \sin_K (a) \equiv K^{-1/2} \sin(K^{1/2}a) \label{generalisedtrig}
\end{equation}
which are respectively functions of $\cos$ and $\sin$ for $K>0$, $\cosh$ and $\sinh$ for $K <0$. In any given expression we obtain the $K=0$ case by the Taylor series expansions $\cos_K (x) = 1 - \frac{Kx^2}{2} + \mathscr{O}(K^2) + ...$ and $\sin_K (x) = x + \mathscr{O}(K) + ...$, and taking $K \to 0$ by neglecting terms of order $\mathscr{O}(K^2)$ and higher. Using the usual difference of angle formula which holds for $\sin_K (a-b)$ along with $\cos_K (a) = \sqrt{1 - K \sin_K^2 (a)}$, we obtain
\begin{align}
    f_K( \chi_{ds}) &= f_K (\chi_{os} - \chi_{od})
    = \sin_K \chi_{os} \cos_K \chi_{od} - \sin_K \chi_{od} \cos_K \chi_{os}\\
    &= f_K (\chi_{os}) \sqrt{1 - K f_K^2(\chi_{od})} - f_K ( \chi_{od}) \sqrt{1 - K f_K^2(\chi_{os})}
\end{align}
and the angular diameter distance can then simply be written\footnote{The well-cited reference \cite{Hogg1999} is therefore incorrect to state that this formula for the angular diameter distance does not hold for negative spatial curvature; this has also been documented as \href{https://github.com/astropy/astropy/issues/4661}{issue 4661} by the \texttt{astropy} team.}
\begin{equation}
   D_{ds} = \frac{a_0}{(1+z_s)} \left( f_K (\chi_{os}) \sqrt{1 - K f_K^2(\chi_{od})} - f_K ( \chi_{od}) \sqrt{1 - K f_K^2(\chi_{os})} \right).
\end{equation}

\subsubsection{Luminosity Distance} \label{sec:luminositydist}

A light source appears dimmer the further it is away from the observer. We define a distance which matches the non-relativistic, Euclidean inverse-square law for the diminution of light from a point source. 

The \textit{intrinsic luminosity} of the source S is $L_s \equiv \frac{dE_s}{dt_s}$, where $dE_s$ is the energy radiated during during the time interval $dt_s$ as measured at S; and assuming isotropic radiation, the total energy emitted is additionally distributed in space over a spherical shell of surface area $4 \pi r^2$ where $r$ is an Euclidean distance. The observer receives energy per unit time per unit area, or observed \textit{luminosity flux} $F_o = \frac{L_s}{4 \pi r^2}$ which gives the \textit{luminosity distance} for a general spacetime
\begin{equation}
\boxed{
    D_{{L}} \equiv \sqrt{\frac{L_s}{4 \pi F_o}}}.
\end{equation}
Observational astronomers usually measure the brightness of light sources in terms of an apparent magnitude $m$ (the logarithm of the flux observed). The absolute magnitude $M$ of the source is related to the absolute luminosity also as a logarithm. We therefore may find the luminosity distance from the logarithm of its definition, as the \textit{distance modulus} 
\begin{equation}
m - M = 5\log_{10}\left(\frac{D_{L}}{\text{Mpc}}\right) + 25.
\end{equation}
The constants come from the definition of the absolute magnitude defined such that $m - M = 0$ for an object at $10$pc (historically the star Vega).

Much like the angular diameter distance requires knowledge of \textit{standard rulers}, the luminosity distance requires \textit{standard candles} for which the intrinsic luminosity $L_s$ is known. In practice, we do not strictly use standard candles so much as \textit{standardisable} candles. These include Cepheid variable stars which form a crucial step in the local distance ladder, as well as Supernovae Ia (SNe Ia) which provide cosmological distances.

\paragraph{Standard Candles: Cepheids and Supernovae Ia}

Cepheid variable stars are favoured primary distance indicators because they are very luminous and easily identified by their periodicity: they pulse at a rate depending solely on their intrinsic luminosity. Those with longer periods are intrinsically more luminous. This period-luminosity relationship (Leavitt, 1908) allows Cepheids in the Milky Way and nearby galaxies in the Local Group to be used as standard candles once the pulsation period is calibrated, resulting in extremely precise distances ($\sim 3 \%$ per source \cite{Riess2022}). All subsequent rungs in the local distance ladder use Cepheid distances, so although Cepheids are considered to be well understood from a gas physics perspective, their systematics are subject to intense scrutiny \cite{Efstathiou2020}.

The explosion of Type Ia supernovae occurs when the mass of a white dwarf in a binary system exceeds the Chandrasekhar limit, the mass above which electron degeneracy pressure in the star's core is insufficient to balance the star's own gravitational self-attraction, by absorbing gas from its companion star. This means there is always roughly the same amount of energy released in every SN Ia explosion. SNe Ia are considered good standardisable candles due to several properties: the absolute luminosity of SNe Ia is almost constant at maximum brightness (i.e. the dispersion is extremely small at $< 0.3$ mag \cite{Sahni2000}, with peak magnitude $M \sim -19.3$). Furthermore, the supernova light curve is strongly correlated with its intrinsic luminosity: a brighter supernova will have a broader light curve. Standardising the light curves, which involves stretching the time scales of the light curves to fit the norm with the brightness being rescaled by an amount determined by the time stretch, reduces the scatter in the absolute luminosity of type Ia supernovae to $\sim 10\%$ \cite{Perlmutter2003, Sahni2000}. The SNe Ia distance scale then has to be calibrated externally, e.g. with Cepheids. 

\begin{figure}
    \centering
    \includegraphics[width=.8\linewidth]{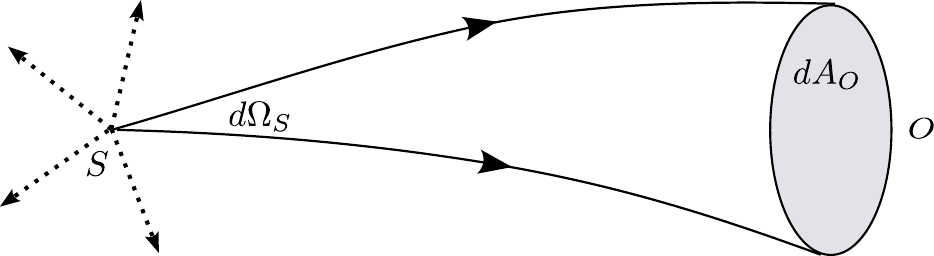}
    \caption{Illustration of a luminosity distance, adapted from \cite{Fleury2015}.}
    \label{fig:luminositydist}
\end{figure}

\paragraph{Surface Brightness Theorem}
The following result can be thought of as an application of  Liouville's Theorem, which states that volume in phase space is conserved. As we will motivate in Section \ref{sec:geometricoptics}, the number of photons $N$ is a conserved quantity, and therefore the number density of the photons is conserved (since the phase-space volume element is an invariant in the absence of absorption, emission, or scattering, e.g. p.290 \cite{Peacock1999}). But when discussing free-streaming photons, we do not usually think in terms of the number density in phase space; we recap a few useful quantities and define one more. The \textit{luminosity} is $L_o \equiv \frac{dE_o}{dt_o}$. The \textit{luminosity flux} is $F_o \equiv \frac{dE_o}{dt_o dA_o} = \frac{L_o}{dA_o}$; i.e. $dA_o$ is the surface area of the detector at O. Note that $dE_o = dE_s$ and $dt_o = dt_s$, so $L_s = L_o$ for the non-relativistic case only. Finally, the \textit{surface brightness} (total over all frequencies) is $I_o \equiv \frac{dE_o}{dt_o dA_o d \Omega_o} = \frac{F_o}{d \Omega_o}$.

First consider non-relativistic, Euclidean space. The surface brightness is the flux per unit solid angle of a spatially extended object
\begin{equation}
    I_o = \frac{F_o}{d \Omega_o} = \frac{L_s}{4 \pi r^2} \frac{r^2}{dA_o} = \text{const.}
\end{equation}
where we used the definition of the luminosity and solid angle respectively. We see that whereas the flux $F_o$ obeys the inverse square law, the surface brightness is constant over a distance $r$ in Euclidean geometry.

Now let us consider FLRW cosmology ({\hl but note that the surface brightness theorem, and hence the distance duality relation, is not limited to Minkowski or FLRW spacetimes, but is valid in any spacetime}). Since the energy of a single photon in the observer frame is $h \nu_o$, where $h$ is the Planck constant, an observer measures a surface brightness
\begin{equation}
    I_o = \frac{h \nu_o dN}{dt_o dA_o d \Omega_o}.
\end{equation}
Assuming that there is no interaction or absorption en route, we relate the surface brightness as measured at one observer O and another at S by equating $dN$
\begin{equation}
    I_o = \frac{\nu_o}{\nu_s} \frac{dt_s}{dt_o} \frac{dA_s}{dA_o} \frac{d \Omega_s}{d \Omega_o} I_s. \label{Ieqn1}
\end{equation}
Now using the surface area element \eqref{surfacearea}, we have that $dA_s = a^2(t_s) f_K^2(\chi_{os}) d \Omega_o$ and $dA_o = a^2(t_0) f_K^2(\chi_{os}) d \Omega_s$. Applying the redshift relation  \eqref{cosmologicalredshift} gives
\begin{equation}
    dA_o d \Omega_o = (1+z_s)^2 d A_s d \Omega_s \label{properarearel}
\end{equation}
and Equation \eqref{Ieqn1} becomes the \textit{surface brightness theorem} or \textit{surface brightness law} \cite{Peebles1993, Peacock1999}
\begin{equation}
\boxed{
        I_o = (1+z_s)^{-4} I_s}. \label{brightnesstheorem}
\end{equation}
We see that there are three separate redshifting effects: one from the redshifting of the frequency (i.e. the energy of the photon), one from the redshifting of the time measurement interval, and one from the proper surface area element. There are a plethora of closely related radiation quantities (which should not be confused) used by astronomers; we refer to Table 9.1 of \cite{Peterson1997} for the transformations between the source- or rest-frame and observer-frame quantities.

\paragraph{Distance duality relation}

\begin{figure}
    \centering
    \includegraphics[scale=0.6]{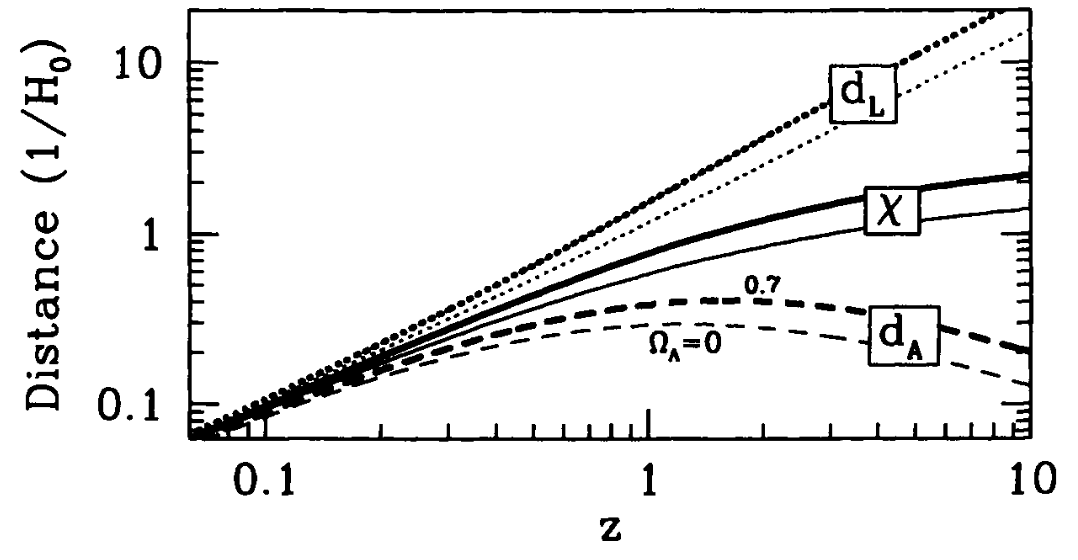}
    \caption{An illustration of the luminosity distance, the comoving distance, and the angular diameter distance (top to bottom) as a function of redshift in a flat expanding universe. The pair of lines in each case correspond to two different cosmologies: the light/lower curve for a flat universe with matter only, i.e. Einstein-de Sitter, and the heavy/upper curve for a $\Lambda$CDM universe. Distances out to fixed redshift are larger in a cosmological constant dominated universe than in a matter-dominated universe.
    Reproduced from \cite{Dodelson2003}.
    }
    \label{fig:distancemeasures_dodelson}
\end{figure}

Whereas the parallax distance contains independent information, the angular diameter distance $D_A$ and the luminosity distance $D_{L}$ are related by the Etherington reciprocity relation. We can use our result \eqref{Ieqn1} and substitute $I_o = \frac{F_o}{d \Omega_o}$ and $I_s = \frac{L_s}{dA_s 4 \pi}$, as the source emits in all directions; but perhaps the following derivation is slightly more lucid. Again $dN$ is the number of photons emitted by the source which radiates in all directions, in a solid angle $d \Omega_S$ and time interval $d t_s$. Again assuming that there is no interaction or absorption en route, we have at the source and the observer's detector with surface area $dA_o$ respectively
\begin{equation}
    dN = ( h \nu_s)^{-1} L_s d t_s \frac{d \Omega_s}{4 \pi} \quad \text{and} \quad dN = (h \nu_o)^{-1} F_o d t_o dA_o.
\label{dN}
\end{equation}
Equating and using the redshift relation \eqref{cosmologicalredshift} gives
\begin{equation}
       D_{{L}} = \sqrt{\frac{L_s}{4 \pi F_o}} = \sqrt{\frac{\nu_s}{\nu_o} \frac{d t_o}{d t_s} \frac{dA_o}{d\Omega_s}} = (1+z_s) \sqrt{\frac{dA_o}{d\Omega_s}}. 
\end{equation}
Using the relation \eqref{properarearel}, we have the distance duality relation
\begin{equation}
\boxed{
    D_{{L}} = (1 + z_s)^2 D_{A}}.
\end{equation}
The luminosity distance can therefore be written through the expression for the angular diameter distance \eqref{angulardistanceOS} as
\begin{equation}
    D_{L} = (1+z_s) a_0 f_K(\chi_{os}) \label{luminositydist_chi}
\end{equation}
which in flat space has a simple expression in terms of the observable redshift using \eqref{comovingdistance} for the comoving distance. Distance measures such as the luminosity distance and the angular size distance are dependent on the expansion of space through the photon path and therefore dependent on the values of cosmological parameters! This will be made explicit in Equation \eqref{luminositydist_z}, and is illustrated in Figure \ref{fig:distancemeasures_dodelson}. 

\section{An Expanding Universe}

\subsection{The Hubble\texorpdfstring{\footnote{The General Assembly of the International Astronomical Union voted in favour of renaming the Hubble Law to the Hubble-Lema\^itre  Law in 2018. There have also been suggestions to include the name of Slipher, who made the earliest measurements in 1917.}}{*} Law}
A spatially isotropic universe has no preferred direction for any of its physical attributes, including its three-velocity field. In three spatial dimensions, this invariance under all possible rotations means that the kinematics can only correspond to expansion or contraction. An easily visualised geometric argument is as follows. Consider drawing a 2-sphere around an observer at the origin and the velocity field leaving or entering that 2-sphere. The hairy ball theorem \cite{Renteln2013} states that given any vector field on an even-dimensional n-sphere, there is at least one point at which the field is purely radial. This means that there is at least one vector that points straight in or out of the sphere so there is necessarily a sense of direction defined by that vector; unless of course all vectors were purely radial.

To find the total inferred three-velocity at an observer we simply differentiate the expression for the proper distance $D^i \equiv a(t) x^i$ given by the FLRW metric with respect to $x^0 = ct$ (see footnote \ref{dagfoot} under Conventions and Notation; this renders velocities a dimensionless fraction of $c$ and is useful in subsequent calculations)
\begin{equation}
    v_{\text{tot}}^i \equiv \frac{dD^i}{d(ct)} = \frac{H(t)}{c} D^i + v^i
\end{equation}
the \textit{Hubble parameter} is $H(t) \equiv \frac{d_t a(t)}{a(t)} = \frac{d}{dt} \ln{a}$, and $d_t$ indicates differentiation with respect to cosmic time $t$ with no extra factor of $c$. We see that the total three-velocity is the sum of an isotropic \textit{Hubble flow} term, or recessional velocity
\begin{equation}
    \bm{v}_H \equiv \frac{H(t)}{c} \bm{D} \label{Hubbleflow}
\end{equation}
which is the only motion inferred by fundamental observers, as well as a (potentially anisotropic) perturbative \textit{peculiar velocity} $v^i \equiv \frac{d x^i}{d(c \eta)}$ arising from gravitational interaction with nearby matter. The peculiar velocity is therefore always a proper quantity and $\ll 1$, i.e. corresponding to physical observables measured by a comoving observer at $x^i$ in their local inertial frame, and does not change if the scale factor is multiplied by a constant \cite{Bertschinger1995}. In contrast, the recessional velocity will become $>1$ (i.e. superluminal) if the proper distance exceeds the Hubble radius: the proper distance is larger than our local inertial frame and therefore loses its physical meaning. In that case the recessional velocity, which is not an invariant, cannot be interpreted physically -- as we noted in Section \ref{sec:distmeasuresincurvedspacetime}, a vector is \textit{fundamentally} a local rather than bi-local concept.

The only possible bulk motion for isotropic models is therefore expansion (or contraction) described by the generalised Hubble law \eqref{Hubbleflow}, where the Hubble velocity $\bm{v}_H$ is proportional to the proper distance to that galaxy. This expansion occurs uniformly everywhere in all directions: it has no centre. The Hubble law is a locally linear relationship (we remind that we defined $v_H$ as fraction of $c$) 
\begin{equation}
    \bm{v}_H \approx \frac{H_0}{c} \bm{D}
\end{equation}
where $H_0 \equiv H(t_0)$ is the \textit{Hubble constant} or present-day Hubble parameter. It was first predicted by Lema\^itre (1927) and observed by Hubble (1929) using the Cepheid distances to nearby galaxies, in terms of the luminosity distance as a function of redshift. The redshift was straightforwardly related to recessional velocity by the Newtonian approximation of the Doppler shift, $v_H \approx z$ \cite{Narlikar1994}. We may also obtain the same result by expanding the scale factor around the present time $t_0$ in a Taylor series
\begin{equation}
    a(t) = a(t_0)( 1 + H_0 (t - t_0) +  \tfrac{1}{2} q_0 H_0^2 (t - t_0)^2 + ...) \label{taylorexpscale}
\end{equation}
where the dimensionless deceleration parameter is $q_0 \equiv - \left( \frac{a'' a}{a'^2} \right) \big|_{t_0} = - \left( \frac{a''}{a H^2} \right) \big|_{t_0}$ (so defined as it was believed in the 1950s that the expansion should be decelerating due to the attractive nature of gravity for matter). Then we take the non-relativistic, Euclidean approximation $t_0 - t_s \approx \frac{D}{c}$ for close light sources giving the observed local Hubble law at first order
\begin{equation}
    z \approx \frac{H_0}{c} D + ...
\end{equation}
for the relation between redshift (and therefore the local relative velocity) and the proper distance to nearby galaxies. It also illustrates directly how redshift may be considered a distance parameter in standard FLRW cosmologies. We see that for small redshifts, the Hubble law is independent of cosmological parameters other than $H_0$, so measuring distances out to higher redshifts is desirable. Very high redshift supernovae have their light shifted to the near infra-red which is strongly absorbed by the Earth's atmosphere; the advent of space-based observatories and telescopes in recent decades have made such measurements possible.

\subsection{Age of the Universe and the Big Bang}

The observed Hubble expansion has the remarkable implication that the age of an FLRW Universe is in fact finite (in contrast with a Minkowski spacetime). Reversing the expansion in time, we see that the distance between us and distant galaxies was smaller than it is now and eventually reaches a singularity. This initial singularity corresponding to the \textit{Big Bang} was a spatially infinite surface at cosmic time $t=0$ (or constant conformal time); i.e. the Big Bang happened everywhere at once. If the Universe has a finite age, then light travels only a finite distance in that time and the volume of space from which we can receive information at a given time is bounded by the \textit{particle horizon}. In other words, it is the width of our past light cone projected on the surface defined by the initial singularity.  We refer to \cite{Mukhanov2005} and \cite{Peacock1999} for further detail on the nuances of defining horizons in cosmology, and \cite{Kinney2009} for details and useful diagrams.

Knowing the value of the Hubble constant, we can obtain a rough estimate for the age of the Universe, since by this linear approximation, all points separated by $D$ today coincided in the past at time $t \approx D c/v_H \approx H_0^{-1}$ (we again reiterate $v_H$ is dimensionless). From the measured value of the Hubble constant, this \textit{Hubble time} $t_H \equiv H_0^{-1}$ is about 15 billion years. The Hubble length or Hubble radius $D_H \equiv c H_0^{-1}$ gives a corresponding approximate scale for the particle horizon. The exact value for the age of the Universe in terms of observables is found by integrating the expression for the change in redshift \eqref{redshiftdrift}
\begin{equation}
    t_0 = \int_0^{\infty} \frac{dz}{H(z) (1+z)} \label{ageofuniverse}
\end{equation}
where the constant of integration is chosen such that $z= \infty$ corresponds to the time of the Big Bang. The expression for $H(z)$ may be written using Equation \eqref{E(z)}, from the Friedmann equation which we will encounter in the following section. Equation \eqref{ageofuniverse} is called the \textit{age-redshift relation}. It is dependent on the composition and curvature of the universe -- for example a universe which is always decelerating is younger than the Hubble time. 

The Big Bang model of cosmology is supported not only by the observed Hubble expansion, but also the light element abundances inferred from absorption lines of stars and galaxies matching predictions by Big Bang nucleosynthesis (BBN); and most strongly the black-body radiation left over from the early Universe, the Cosmic Microwave Background. It was in fact from this second line of evidence (for a review of BBN, see \cite{PDG2022}) that the Big Bang model originated. In the 1940s Gamow, Alpher and Herman suggested that the Universe was once dense and hot enough for nucleosynthesis, cooling with expansion -- thus enabling  a cosmological origin for the abundances of the elements. A prediction of the model was the existence of the cosmic microwave background, a relic background radiation with a temperature $\sim 3$K.

\subsection{The Cosmic Microwave Background in Brief}

The premise of Olber's paradox (1826) is this: if stars are homogeneously distributed in an infinite space, then every line of sight looking out into the night sky should terminate eventually on a star. If space is Euclidean then surface brightness is independent of distance, so the night sky should appear at least as bright as the Sun. Modern cosmology resolves the paradox via the Brightness Theorem \eqref{brightnesstheorem}: the night sky is dark solely as the result of the Universe's expansion. 

With contemporary knowledge, rather than every line of sight terminating on a star, we might expect to see ionised matter with temperature $T \gtrsim 10^3$K in the absence of redshifting from expansion. Baryonic matter was in the form of plasma until $\sim 300,000$ years after the Big Bang, when the Universe had cooled to $\sim 3000$ K \cite{Gawiser2000}.  Until then, the Universe was opaque to electromagnetic radiation due to Compton scattering of the photons by free electrons, and as a result photons and matter were tightly coupled as a \textit{photon-baryon fluid}. The plasma was in a Maxwell-Boltzmann distribution, and the interactions ensured the photons were in a black body distribution at the same temperature.

As the \textit{decoupling}\footnote{See Footnote \ref{footnotedecoupling} regarding terminology.} of matter and radiation occurred the Universe became transparent to light. The Cosmic Background Radiation (CBR) released during decoupling has a mean free path long enough to be considered as free-streaming, i.e. travelling along null geodesics to the present. The expansion along the way reduces the photon energy and thereby cools\footnote{See e.g. \cite{Peacock1999}, \cite{Peebles1993} or \cite{Peter2009} for a full treatment of thermodynamic properties, but since the photon energy is redshifted proportional to $a^{-1}$, the temperature of the black-body radiation scales the same way. In fact, deviations from this relation (which may be detected by measuring the CMB temperature at different redshifts) is a test of standard cosmology.} the temperature of the radiation to $2.73$K, \textit{whilst preserving its black-body distribution function}. We see as this Cosmic Microwave Background (CMB) across the entire sky, as if it were coming from a spherical shell or \textit{surface of last scattering} at a distance of nearly $15$ billion light years \cite{Gawiser2000}. The CMB was first observed by chance (Penzias and Wilson, 1965).

The \textit{Planck} satellite measurement of the CMB is the single dataset with the strongest constraints on cosmology (we discuss this a little further in Section \ref{sec:moderncosmology}). The fact that the observed spectrum is a black-body function with a peak at $\sim 1$mm is the most compelling evidence we have that the Universe was once hot and dense, and has been expanding and cooling since. Furthermore, the initially assumed statistical homogeneity and isotropy of the Universe is strongly justified by the very small magnitude of relative fluctuations (less than 0.01\%) in the CMB, once we subtract the dipole arising from our peculiar motion relative to the CMB rest frame. (In particular, velocities associated with motion within the Solar System are removed for CMB anisotropy studies, whilst peculiar velocities associated with the Local Group are normally removed for cosmological studies \cite{PDG2022}.) 

\section{Einstein's Field Equations in FLRW Cosmology} \label{efeflrw}

A dynamically evolving Universe is an almost inevitable consequence of almost any  cosmological model based on general relativity. It is described for an FLRW metric by the time dependence of the scale factor, determined by the stress-energy content of the Universe through the Einstein field equations applied to FLRW cosmology. These are called the Friedmann equations.

\subsection{The Stress-Energy Tensor}

The mathematical object describing the source of curvature, i.e. gravity, is the stress-energy tensor $T^{\mu \nu}$. Its components $T^{00}$, $T^{0i}$, $T^{i0}$ and $T^{ij}$ in a local coordinate system are (see footnote \ref{ddagfoot} under Conventions and Notation) respectively the energy density, energy flux, momentum density and the spatial stress which arise from matter (including for example, baryons, radiation, dark matter and dark energy) as well as fields (such as electromagnetic fields or neutrino fields). This is a significant departure from Newtonian theory in which only matter possessing mass gravitates.

The threading of spacetime, part of Weyl's Postulate discussed in Section \ref{sec:cosmologicalprinciple}, means that the galaxies can be treated as the fundamental particles of cosmological fluid. A fluid is defined as a dense set of particles treated as a continuum. The equations which describe the state of a fluid may be completely determined by five quantities -- the three components of the fluid velocity $u^i$, and any two thermodynamic quantities of the fluid, such as the fluid density $\rho (\bm{x}, t)$ and the specific entropy $s(\bm{x},t)$. Therefore a complete set of equations should be five also in number: these are the continuity equation, the three components of the Navier-Stokes equation, and the entropy equation. In addition, a general equation of state $P = P(\rho, s)$ relating the pressure to the energy density and the specific entropy must also be defined.

{\hl The cosmological principle, i.e. the conditions of homogeneity and isotropy, restrict the form of the stress-energy tensor to be that of a \textit{perfect fluid} \cite{Baumann2022}.}  A perfect fluid does not possess any anisotropic or shear stress, viscosity or heat conduction. As such, is it described entirely by only two quantities: the energy density $\varepsilon \equiv \rho c^2$ where $\rho$ is the relativistic mass density and pressure $P$ which are related by an equation of state $w = \frac{P}{\varepsilon}$. The pressure includes all kinds -- such as from random motion of stars and galaxies, thermal motion or radiation pressure. The energy density is a total relativistic quantity (it can be thought of as having a contribution from the rest mass density $\varrho$ as well a so-called internal energy component \cite{Schutz2009}). The stress-energy tensor describing a perfect fluid in its rest frame, or equivalently the orthonormal frame of a comoving observer, has the form
\begin{equation}
    T\indices{^{\mu}_{\nu}} = \text{diag}(- \varepsilon, P, P, P). \label{perfectSEtensor}
\end{equation}
The general form for a perfect fluid stress energy tensor, since the only expression linear in $u^\mu u_\nu$ and $g\indices{^\mu_\nu}$ that reduces to \eqref{perfectSEtensor} in the Minkowski limit, is
\begin{equation}
    T\indices{^\mu_\nu} = (\varepsilon + P) u^\mu u_\nu + P g\indices{^\mu_\nu}.
\end{equation}
The energy density and pressure are proper quantities, always taken to be in the fluid's rest frame; and due to homogeneity and isotropy are a function of time $t$ only. The energy density $\varepsilon = \rho c^2 \neq \varrho c^2$ in general: there will be kinetic energy of motion from the random motion of particles even in an average rest frame \cite{Schutz2009}.

In the radiation-dominated early Universe, the cosmological fluid would describe a gas of relativistic particles such as photons; i.e. radiation behaving as a perfect fluid due to interactions with baryons. (After decoupling, the radiation is an ensemble of free photons described by a kinetic equation, i.e. the Boltzmann equation \cite{Mukhanov2005}, and there is no radiation pressure.) In the matter-dominated era, this is ordinary matter (baryons) and Cold Dark Matter (CDM), which can be approximated as \textit{dust}. In cosmology, dust refers to any fluid with pressure much less than the energy density (for the matter-dominated era, this is an observed ratio of about order $10^{-5}$ or $10^{-6}$ \cite{DInverno1992}), such that the pressure may be considered negligible.

In classical general relativity, the stress-energy tensor is expected to satisfy certain \textit{energy conditions}, which correspond to e.g. $\varepsilon \geq 0$, or the dominance of the energy density over pressure. These energy conditions of GR \cite{Poisson2004, Peter2009, Hawking2023} are summarised in Table \ref{tab:energyconditions}.
\begin{table}
    \centering
    \begin{tabular}{ccc}
        & Statement & Conditions \\
        \hline \hline\\
       Weak  & $T_{\mu \nu} u^\mu u^\nu \geq 0$ & $\varepsilon \geq 0$, \; $\varepsilon + P_i >0$ \\
       Null  & $T_{\mu \nu} k^\mu k^\nu \geq 0$ & $\varepsilon + P_i \geq 0$ \\
       Strong & ($T_{\mu \nu} - \tfrac{1}{2}T g_{\mu \nu} )u^\mu u^\nu \geq 0$ & $\varepsilon + \sum_i P_i \geq 0$, \; $\varepsilon + P_i \geq 0$ \\
       Dominant & $-T\indices{^\mu_\nu} u^\nu$ future directed &  $\varepsilon \geq 0$, \; $\varepsilon \geq |P_i|$ \\
    \end{tabular}
    \caption{Adapted from \cite{Poisson2004}. The weak energy condition states that the energy density of any matter distribution $T_{\mu \nu} u^\mu u^\nu$ as measured by any observer with four-velocity $u^\mu$ must be non-negative. The null energy condition replaces the four-velocity in this statement by an arbitrary future-directed null vector $k^\mu$. The fixed component of the pressure is $P_i$; i.e. these statements hold for each component separately. (The perfect fluid stress-energy tensor has $P = P_i$ from isotropy.) In the strong energy condition, the trace of the stress-energy tensor is denoted by $T \equiv T\indices{^\mu_\mu}$. The dominant energy condition corresponds to the idea that matter should flow along only timelike or null worldlines. If $u^\mu$ is an arbitrary, future-directed, timelike vector field, then $-T\indices{^\mu_\nu} u^\nu$ should be a future-directed, timelike or null, vector field.}
    \label{tab:energyconditions}
\end{table}

\subsection{The Einstein Field Equations} \label{efeinbrief}
The Einstein field equations are a set of ten non-linear second-order partial differential equations
\begin{equation}
\boxed{
G\indices{^{\mu}_{\nu}} = \kappa T\indices{^{\mu}_{\nu}}} \label{EFE}
\end{equation}
where $\kappa \equiv \frac{8 \pi G}{c^4}$ is a coupling constant found from correspondence with Newtonian theory (see Section \ref{newtoniancorrespondence}), $G\indices{^{\mu}_{\nu}}$ is the Einstein curvature tensor defined by $G_{\mu \nu} \equiv R_{\mu \nu} - \frac{1}{2}g_{\mu \nu}R$; and in turn $R$ is the Ricci curvature scalar, the contraction of the Ricci curvature tensor $R_{\mu \nu} \equiv R\indices{^\lambda_{\mu \lambda \nu}}$ with the metric. The Riemann curvature tensor $R\indices{^\lambda_{\mu \nu \rho}}$ is a measure of curvature of spacetime defined using the Christoffel symbols 
\begin{equation}
    \Gamma\indices{^{\mu}_{\nu \rho}} \equiv \frac{1}{2} g^{\mu \sigma} ( g_{\sigma \nu , \rho} + g_{\sigma \rho, \nu} - g_{\nu \rho, \sigma}). \label{Christoffeldefn}
\end{equation}
The stress-energy tensor $T\indices{^{\mu}_{\nu}}$ is the source of curvature. As we have seen, it describes the relativistic energy density, momentum density and the spatial stress which arise from all kinds of matter and fields (other than the gravitational field).

The left-hand side of the Einstein field equation was historically constructed from the following assumptions \cite{Misner1973}:
\begin{enumerate}
\item \label{assump1} $G_{\mu \nu} =0$ for flat spacetime.
\item \label{assump2} $G_{\mu \nu}$ is uniquely constructed from only the Riemann curvature tensor and the metric, with the requirements:
\begin{enumerate}
\item It is linear in the Riemann tensor.
\item It must be a symmetric, second-rank tensor.
\item It fulfils the contracted Bianchi identities $G\indices{_{\mu \nu}^{; \nu}} =0$, such that there is a covariant generalisation of energy conservation $T\indices{_{\mu \nu}^{; \nu}} =0$.
\end{enumerate}
\end{enumerate}
A covariant derivative is denoted by $_{;\mu}$ and generalises the idea of a partial derivative to curved spacetime. Since the metric tensor obeys $g\indices{_{\mu \nu}^{; \nu}} =0$, we can modify the Einstein field equations by a term $\Lambda g_{\mu \nu}$ proportional to a \textit{cosmological constant} $\Lambda$ to the left-hand side of \eqref{EFE} and preserve our assumption (\ref{assump2}), although discarding assumption (\ref{assump1}). The cosmological constant was historically introduced by Einstein in 1917 to the field equations in order to achieve a (nonetheless unstable \cite{Bianchi2010}) static matter-dominated Universe. In modern cosmology, it is usually absorbed into the right-hand side as a contribution $T^{\mu\nu}_{\Lambda} = - \varepsilon_{\Lambda} g^{\mu \nu}$, which facilitates the interpretation that $\varepsilon_{\Lambda} \equiv \frac{c^4 \Lambda}{8 \pi G}$ represents the energy density of the vacuum.

The Einstein field equations may be thought of as determining the metric (the gravitational field quantity in general relativity) and its evolution from a given stress-energy tensor, i.e. specifying the matter distribution. One can also try to use the field equations to determine the stress-energy tensor from a given metric; although this often leads to physically unrealistic stress-energy tensors which violate the dominant energy condition. Often, the metric and the stress energy tensor may both be determined in part from physical considerations before being found together by using the field equations as further restrictions on the two sets of partially specified quantities.

\subsection{The Friedmann Equations}
The Einstein field equations give the evolution of the gravitational field quantity, which in general relativity is the metric. In an FLRW spacetime, all the time dependence of the metric is in the scale factor; so the field equations show how $a(t)$ evolves with time. We insert the Einstein tensor for the FLRW metric and the stress-energy tensor for a perfect fluid into \eqref{EFE}, e.g. \cite{Baumann2022}. Given the symmetries, i.e. isotropy and homogeneity of the spacetime, the field equations reduce to just two ordinary differential equations which possess simple analytic solutions. The $00$-component equation is known as the Friedmann equation and the spatial trace or $ii$-component equation as the Raychaudhuri equation; or both together as the Friedmann equations
\begin{gather}
H^2 = \frac{8 \pi G }{3}\rho - \frac{Kc^2}{a^2} \label{friedmann00} \\
2 d_t H + 3H^2 = - \frac{8 \pi G}{c^2} P - \frac{Kc^2}{a^2} \label{Hsquaredeqn}
\end{gather}
where $d_t$ denotes differentiation with respect to cosmic time $t$. The Friedmann equations can be combined to give a second-order differential equation for the dynamics of the scale factor
\begin{equation}
 d_t H + H^2 = \frac{d_t^2a}{a} = - \frac{4 \pi G}{3} \left( \rho + \frac{3}{c^2} P\right). \label{dda/a}
\end{equation}

The Friedmann equation \eqref{friedmann00} may be written in terms of a dimensionless density parameter $\Omega(t) \equiv \frac{\rho(t)}{\rho_{c}(t)}$, where $\rho_c(t) = \frac{3H^2}{8\pi G}$ is the critical density; such that
\begin{equation}
    \Omega(t) -1 = \frac{Kc^2}{(aH)^2}. \label{friedmanndensityparameterisation}
\end{equation}
We can we see that the total energy distribution determines the spatial geometry of our Universe. Since the right hand side of Equation \eqref{friedmanndensityparameterisation} has a fixed sign, then so does the left hand side: $\Omega(t)$ is $\{<1, =1, >1 \}$ for all time if it is so at any given time. The density parameter is a sum of components $\Omega(t_0) = \Omega_{m} + \Omega_{r} + \Omega_{\textsc{de}}$, where $\Omega_{m}$, $\Omega_{r}$, and  either $\Omega_{\textsc{de}}$ or $\Omega_{\Lambda}$ are the \textit{present-day} density parameters for dust, radiation, and either dark energy or the cosmological constant respectively. Some authors also write the spatial curvature $K$ as a present-day curvature ``density''\footnote{It is not, however, a density; curvature does not contribute to the energy content of the Universe and for this reason some authors dislike the notation.} parameter $\Omega_{K} \equiv \frac{-K c^2}{a_0^2 H_0^2}$. Then using the solution to the continuity equation \eqref{continuitysolns}, the Friedmann equation may be written as a sum over the density components
\begin{align}
H^2 &= \frac{8}{3}G \pi \sum_{i} \rho_i = H_0^2 \sum_{i} \Omega_i \left(\frac{a_0}{a}\right)^{3(1+w_i)}  \label{generalisedfriedmann}\\
&= H_0^2 \Omega_{r} (1+z)^4  + H_0^2 \Omega_{m} (1+z)^3  + H_0^2 \Omega_\Lambda - \frac{Kc^2}{a^2}.
\end{align}
It can be convenient to define, for calculating the comoving distance \eqref{comovingdistance} and the age of the Universe \eqref{ageofuniverse}, a function which factors out the Hubble constant
\begin{equation}
E(z)\equiv \frac{H(z)}{H_0}  =\sqrt{ \Omega_{r} (1+z)^4  + \Omega_{m} (1+z)^3  + \Omega_\Lambda - \frac{Kc^2 }{a_0^2 H_0^2}(1+z)^2}.\label{E(z)} 
\end{equation}

\subsection{The Evolution Equations: Fluid Equations}

The Einstein equations via the Bianchi identities require the covariant conservation of the stress-energy tensor
\begin{equation}
T\indices{^{\mu}_{\nu ;\mu}} \equiv T\indices{^{\mu}_{\nu , \mu}} + \Gamma\indices{^{\mu}_{\mu \alpha}}T\indices{^{\alpha}_{\nu}} - \Gamma\indices{^{\alpha}_{\mu \nu}}T\indices{^{\mu}_{\alpha}} =0
\end{equation}
The continuity equation $T\indices{^{\mu}_{0;\mu}} = 0$ is a statement of the local conservation of mass-energy (or the first law of thermodynamics) and $T\indices{^{\mu}_{i;\mu}} = 0$ involves the conservation of momentum. In the case of a cosmological fluid, these are respectively the relativistic versions of the continuity equation and the infamous Navier-Stokes equation of hydrodynamics. The Navier-Stokes equation simplifies to the Euler equation for a perfect fluid.

The continuity equation considering an FLRW spacetime is, using $\Gamma\indices{^{\mu}_{0 \nu}} = 0$ unless $\mu = \nu = i$; then $\Gamma\indices{^i_{0i}} = H$,
\begin{equation}
d_t \varepsilon + 3H\left(\varepsilon  + P \right) =0. \label{continuityeqnflrw}
\end{equation}
The right-hand side is non-zero for $\varepsilon \neq -P$, due to the increasing comoving volume diluting the energy density in an expanding universe. 

\subsection{Single-Component Solutions} \label{sec:singlecomponentsolutions}

Let us assume for the moment that $(\varepsilon + 3 P) \geq 0$, i.e. $w \geq - \frac{1}{3}$, corresponding to the strong energy condition. This is violated in reality by a cosmological constant or any component explaining the observed accelerated expansion $d_t^2 a > 0$, via Equation \eqref{dda/a}: the accelerated expansion cannot be accounted for by Newtonian gravity, as the pressure term disappears in the Newtonian limit.  Under the strong energy condition, however, the curvature ``density'' is analogous to a Newtonian total energy, and the Universe must be expanding or contracting: a closed Universe would collapse in a finite time, whereas a flat or open Universe would expand indefinitely.

 The Friedmann equations \eqref{friedmann00}, \eqref{Hsquaredeqn} with the condition of spatial flatness $K=0$ become a straightforward separable ordinary differential equation $d_t H = - \frac{3}{2}(1+w)H^2$, in terms of the equation of state $w \equiv \frac{P}{\varepsilon}$ for the perfect fluid. When $w \neq -1$ it has the solution 
\begin{equation}
a(t) = a_0 \left(\frac{t}{t_0}\right)^{\frac{2}{3(1+w)}}. \label{asoln1}
\end{equation}
Solving the continuity equation \eqref{continuityeqnflrw} gives
\begin{equation}
    \rho = \rho_0 \left( \frac{a}{a_0} \right)^{-3(1+ w)} = \rho_0 (1+z)^{3(1+w)}. \label{continuitysolns}
\end{equation}

Let us now write a few solutions explicitly for when the cosmological fluid is dominated by a single component, such as the early Universe (radiation-dominated) or for most of the expansion history (matter-dominated). In accordance with the observation that the Universe is entering an epoch of \textit{accelerated expansion} $d_t^2a >0$, we also include a \textit{dark energy} component for which the second-order Friedmann equation \eqref{dda/a} requires the equation of state $w_\textsc{de} < - \frac{1}{3}$. We refer to \cite{Baumann2022} for further details, as well as \cite{DInverno1992} for an illustration categorising the various Friedmann models. 

A gas of radiation or relativistic particles has $w_r= \frac{1}{3}$ (this may be understood due to spatial isotropy and demanding the EM stress energy tensor be traceless) and dust has $w_m=0$, so
\begin{align}
&\text{Radiation-dominated} & a(t) = a_0 \left(\frac{t}{t_0}\right)^{\frac{1}{2}}, && \rho(a) = \rho_0 \left(\frac{a}{a_0}\right)^{-4} \label{radiationdom} \\
&\text{Matter-dominated} & a(t) =a_0 \left(\frac{t}{t_0}\right)^{\frac{2}{3}},  && \rho(a) = \rho_0 \left(\frac{a}{a_0}\right)^{-3} \label{matterdom}
\end{align}
We observe that these solutions correspond to expansion which is \textit{decelerated} since matter is gravitationally self-attractive. The behaviour of $\rho$ with the expansion is easily interpreted: for dust, $\rho$ is the rest mass of the particles (which remains constant) per unit comoving volume which scales as $a^3$. Radiation energy measured by a given fundamental observer, however, redshifts away with an additional $a^{-1}$ due to the expansion, as well as the scaling of the comoving volume.

The dark energy equation of state is required to be $w_{\textsc{de}} < - \frac{1}{3}$, so assuming a positive energy density (stipulated by the weak energy condition), the pressure of dark energy is negative. This negative pressure only alters the effective mass density of dark energy, such that there is a repulsive gravitational effect. Since the Universe is not bounded, and we are considering a homogeneous fluid pervading the entire Universe, there are no pressure gradients and therefore no pressure forces are felt on any local part of the Universe. Furthermore, a hypothetical fluid with a negative pressure in a sealed container would indeed not have any gravitational repulsive effects, as the pressure of the container walls would provide a reaction force \cite{Peebles2003}. 

For general dark energy with unknown equation of state we leave the solution as $\rho_{\textsc{de}} \propto (1+z)^{3(1+w_\textsc{de})}$. However, dark energy corresponding to a vacuum energy or cosmological constant has $\rho_\Lambda = \text{const.}$, so the continuity equation implies that it has $w_{\Lambda}=-1$. This is in fact very close to its measured value. The Hubble parameter is also constant, $H = \sqrt{\tfrac{\Lambda c^2}{3}}$ from the Friedmann equation \eqref{friedmann00}. Then simply integrating the definition of $H$ gives
\begin{align}
\text{Cosmological constant dominated} & \qquad a(t) = a_0 \exp{\left(\sqrt{\tfrac{\Lambda c^2}{3}}(t-t_0)\right)}
\end{align}
which corresponds to the de Sitter model of the Universe. Finally, we mention the case $w_{\textsc{de}} < -1$, which is possible in some scalar field models known as phantom dark energy. Solution \eqref{asoln1} cannot hold since it predicts a contracting universe. However, there is an expanding solution which diverges (implying an infinitely large energy density) at some future time $t_f$. Although a gravitationally bound system is sometimes said to have constant physical size in an expanding universe without caveats (a quantitative discussion can also be found in \cite{Cooperstock1998}), ultimately the condition depends on the higher order derivatives of the scale factor. As the dark energy density increases for this diverging model, it will eventually strip apart all gravitationally bound objects until the Universe ends in a so-called Big Rip singularity at $t_f$ \cite{Caldwell2003}.

\section{Contemporary Cosmology} \label{sec:lambdacdm}

\subsection{Statistical Cosmology} \label{sec:moderncosmology}

During the latter half of the twentieth century, cosmology was focused on understanding the deterministic global dynamics of the Universe described by only a few parameters. In particular, the goal was finding the present-day expansion rate $H_0$ alongside the present-day deceleration parameter $q_0$ in the Taylor expansion of the scale factor \eqref{taylorexpscale} -- encapsulated by the description \cite{Sandage1970} that cosmology was ``a search for two numbers''. However, as cosmological information -- including from different epochs -- has dramatically increased since the 1990s, the focus has shifted from a parametrisation purely based on a finite number of derivatives of the scale factor merely at present day, to parametrisations that use instead the matter and energy content in the Friedmann equations directly. (It is worth noting that the so-called cosmographic approach of expanding the scale factor as a Taylor series around present day is still sometimes used as a model-independent characterisation of cosmology; for a review, see \cite{Dunsby2016}.) If an evolving homogeneous and isotropic universe is an adequate first approximation, then the next step is to account for the observed large-scale\footnote{We mostly restrict analysis to small perturbations away from homogeneity and isotropy. Perturbations in the CMB, or anisotropies since the CMB is defined on a spherical surface of last scattering, have remained small because the photons have freely-streamed to present day. This is why CMB statistics are so good -- (primary) anisotropy calculations are highly accurate within linear perturbation theory \cite{Hu2002}. Small matter perturbations correspond to large length scales (i.e. large-scale structures and galaxy clusters), or the linear to weakly non-linear regime (see Section \ref{sec:perturbationtheory}). On small scales $\lesssim 10$ Mpc cosmological information will have been lost by complicated astrophysical interactions.} inhomogeneities in matter and the observed CMB anisotropies.

The abundance of observational information, in particular measurements of the CMB and extensive galaxy surveys, must be described statistically. Our cosmological models are also inherently statistical, since the inhomogeneities we see in the Universe are presumed according to inflation to have grown (deterministically) due to gravity from initial random fluctuations. To be precise, fractional or relative\footnote{E.g. we don't work with the matter density perturbation field directly, but rather the overdensity field or matter density contrast $\delta_m(\bm{x}) \equiv (\rho_m(\bm{x}) - \bar{\rho}_m)/\bar{\rho}_m$. The CMB temperature field {\hl $T$} is sometimes used directly, or the temperature fluctuation field is defined as $\delta T / T$.} perturbations (in matter density, velocity, CMB temperature...) are treated as classical \textit{random fields} (see e.g. \cite{vanmarcke2010}), defined as functions whose values are stochastic variables. For a given random field $f(\bm{x})$ we may consider a statistical ensemble, a set of functions $f_n(\bm{x})$ each with some probability $Q_n$. The average over this set $\langle f(\bm{x}) \rangle = \sum_{n} Q_n f_n(\bm{x})$ is called the ensemble average and each $f_n(\bm{x})$ is called a realisation of the ensemble.

We can think of our model of the Universe as a stochastic realisation of a statistical ensemble. By postulating that the ensemble is \textit{ergodic}, spatial averaging (in practice, over sufficiently large volumes such that two points are uncorrelated rather than all space) coincides with ensemble averaging and each member of the ensemble has the same statistical properties as the entire ensemble. It is then possible to determine the statistical behaviour of the process from just one member of the ensemble; i.e. just our one Universe.

The quantities that we can determine from a given model of the Universe are therefore statistical: expectation values or moments, rather than exactly the amplitude or values of perturbations. For example, we subtract the mean from the total field to construct the fractional perturbation field $f$, so the most basic statistical quantity we can next construct is called either the covariance or \textit{two-point correlation function}. The two-point correlation function is defined by \begin{equation}
    C(\bm{x} , \bm{x}') \equiv \langle f(\bm{x}) f(\bm{x}') \rangle
\end{equation} and given statistical homogeneity describes the perturbations at different physical scales.

\subsubsection{Matter and CMB Power Spectra}

{\hl It is often useful to perform spectral or harmonic analysis on a general field, decomposing it as a superposition of contributions from different frequencies.} The harmonic decomposition of a field has different forms depending on whether it is defined in Euclidean space, on a sphere or some other space. For example, galaxies are assumed to be distributed in Euclidean space, whereas the CMB is defined on the spherical surface of last scattering. Any field defined in flat space may be represented as a superposition of plane waves, or complex exponentials. This is a summation called a Fourier series when the volume is bounded, or an integral called a Fourier transform when unbounded. Similarly, any field defined on a sphere may be represented as a summation of spherical harmonics. There is no integral version because a sphere is bounded.

The {\hl generalised} Wiener-Khintchine (W-K) theorem states that if $f$ is a statistically homogeneous random field defined over Euclidean space, then the power spectral density or the \textit{power spectrum} $\mathcal{P}$ is the Fourier transform of the autocorrelation function $R(\bm{r}) \equiv \langle f(\bm{x}) f(\bm{x}+\bm{r}) \rangle $, which is \textit{only} dependent on the separation $\bm{r}$. Assuming statistical homogeneity or translation invariance, then the correlation function reduces to the autocorrelation function $C (\bm{x}, \bm{x}') = R(\bm{x}'-\bm{x})$. Therefore, we can find the power spectrum by taking the Fourier transform of the two-point correlation function in Euclidean space. An analogous relation between the two-point correlation function and power spectrum on a sphere exists \cite{Durrer2005, Peebles1993}. Higher-order correlation functions with corresponding $n$-point power spectra can also be constructed.

A power spectrum $\mathcal{P}$ is a function that measures the contribution of each frequency to the field. For example, a periodic signal gives peaks at a fundamental and its harmonics; whereas white noise has equally distributed contributions at all frequencies (i.e. a fractal nature) and therefore has a constant power spectrum. A power spectrum can {\hl usually\footnote{\hl If the field is real, and depending on the choice of harmonic transformation (including taking a Fourier series versus Fourier integral) as well as its normalisation; one may see the generalised W-K theorem as more fundamentally \textit{defining} the power spectrum. Also note that the ``usual'' W-K theorem which gives a similar expression holds for a deterministic rather than stochastic field.} be written in a form resembling}
\begin{equation}
\mathcal{P}(\bm{\xi}) = \langle| \hat{f}(\bm{\xi})|^2 \rangle \label{powerspectrabounded}
\end{equation}
where $\hat{f}(\bm{\xi})$ is the field mapped or ``harmonic transformed'' into a frequency ($\bm{\xi}$) space. Given statistical isotropy, $\hat{f}(\bm{\xi})$ and therefore the power spectrum depends only on $\xi = |\bm{\xi}|$. The frequency $\xi$ is a spatial frequency or the wavenumber $k$ when $f$ is in Euclidean space, or the multipole moment $\ell$ when $f$ is on a spherical surface.

This is therefore a main statistical measure which is applied to the matter and CMB radiation fields as the \textit{matter power spectrum} and the \textit{CMB power spectrum} respectively. (We can relate the observed galaxy distribution to an inferred matter distribution, which also includes dark matter.) This allows us to investigate both the primordial spectrum of fluctuations, and the cosmology-dependent transfer function which converts the initial spectrum to the power spectrum observed today. 

It is often convenient when dealing with observations that the galaxy two-point correlation function is calculated first, and is then Fourier decomposed to give a galaxy-galaxy power spectrum. In turn the galaxy-galaxy power spectrum can be related to the matter power spectrum using a bias parameter \cite{Baumann2022}. For a point process such as the distribution of galaxies, the two point \textit{galaxy} correlation function is defined operationally as the mean excess number of galaxy pairs at separation {\hl $r$}, over that expected for a pure Poisson point process. Theoretical predictions often find the matter power spectrum first. It is specifically is the power spectrum corresponding to the density contrast  $\delta_m(\bm{x}) \equiv (\rho_m(\bm{x}) - \bar{\rho}_m)/\bar{\rho}_m$,
\begin{equation}
\mathcal{P}_{\delta}(k) = \langle |\delta_m(\bm{k})|^2 \rangle.
\end{equation}
{\hl This expression for the matter power spectrum is valid considering that the matter field is defined in a finite box with sides taken to be large compared to wavelengths of interest with periodic boundary conditions; i.e. when considering Fourier series with discrete modes rather than Fourier integrals\footnote{The expression $\langle \delta_m(\bm{k}) \delta^*_m(\bm{k}') \rangle = (2 \pi)^3  \delta_D (\bm{k} - \bm{k}') \mathcal{P}_\delta(k)$ for an infinite Euclidean space can be gained by taking the limit of the volume of the box $V \to \infty$, where the asterisk denotes a complex conjugate and $\delta_D$ is a Dirac delta.} \cite{Peacock1999}.}

We may directly decompose the CMB temperature fluctuation field\footnote{There appear to be a few conventions for defining the CMB temperature power spectrum: $\delta T / T$ \cite{Durrer2005, Hu2002}, $\delta T / \bar{T}$ \cite{Mukhanov2005} or alternatively the full temperature field \cite{Planck2020a, PDG2022}, or even less commonly $\delta T$ \cite{Lineweaver1997}.}  using spherical harmonics
\begin{equation}
\frac{\delta T}{T}(\theta , \phi) = \sum\limits_{\ell =1}^{\infty} \sum\limits_{m= - \ell}^{\ell} a_{\ell m} Y_{\ell m} (\theta , \phi) \label{sphericalharmonics}
\end{equation}
where $T$ is the CMB temperature field, $\delta T$ is a perturbation to the mean temperature $\bar{T} = T - \delta T$, $Y_{\ell m} (\theta , \phi)$ are the spherical harmonic functions and $a_{lm}$ are coefficients. The multipole moment $\ell$ and $m$ are plain (i.e. not tensorial) indices, and heuristically correspond to angular scales and directions on a sphere respectively. The CMB power spectrum is then
\begin{equation}
\mathcal{C}_{\ell} = \langle |a_{lm}|^2 \rangle
\end{equation}
which is independent of $m$ from statistical isotropy. To indicate the approximate contribution to the variance per logarithmic multipole interval, is is convenient to also define an angular power spectrum $\mathcal{D}_\ell = \frac{\ell (\ell +1)}{2 \pi} \mathcal{C}_\ell$ and this is often the power spectrum which is plotted.

\begin{figure}
    \centering
    \includegraphics[width=\textwidth]{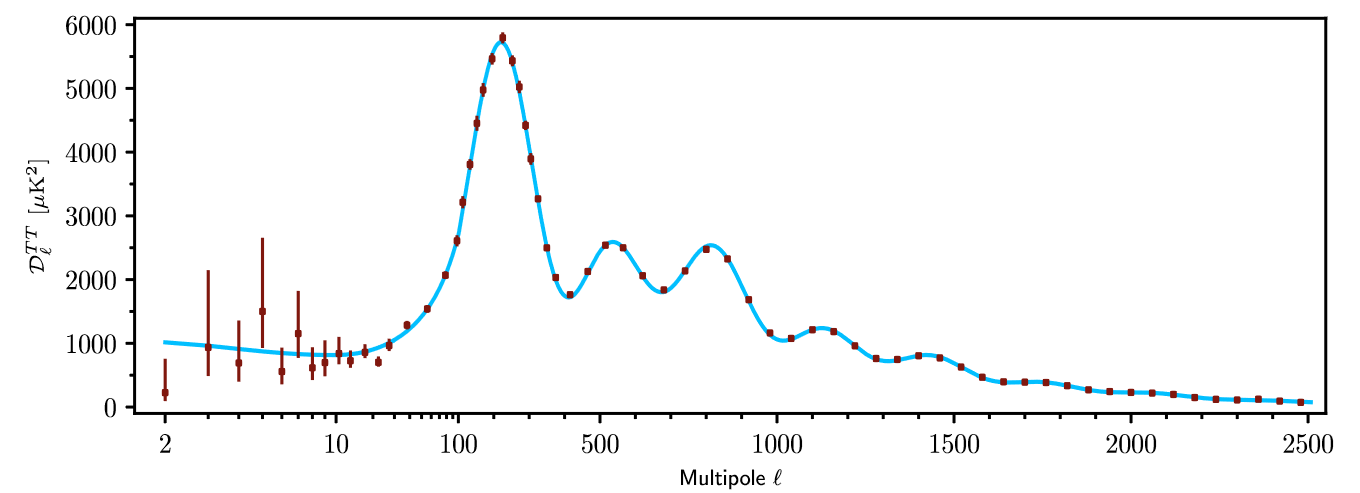}
    \caption{CMB temperature angular power spectrum, taken from \cite{Planck2020a}. The blue line shows the best fit \textit{Planck} 2018 model. }
    \label{fig:planckcmb}
\end{figure}

The CMB power spectrum, as shown in Figure \ref{fig:planckcmb}, displays large errors at low multipoles since the predicted power spectrum is the power in the multipole moment $\ell$ averaged over an ensemble of universes. Our actual measurements are limited to one sky and one Universe: we measure only $2\ell + 1$ numbers for each $\ell$ according to the spherical harmonic decomposition \eqref{sphericalharmonics}. The monopole $\ell=0$ and dipole $\ell =1$ are therefore most affected by this \textit{cosmic variance}.

The angle subtended by the Hubble radius at last scattering corresponds to $\ell \sim 100$ \cite{PDG2022}, so inhomogeneities on larger scales (lower multipoles) which had not entered the horizon at last scattering directly reflect the primordial spectrum. This is observed as an almost constant $\mathcal{D}_\ell$ spectrum (easiest to see on a logarithmic plot) for $10 \lesssim \ell \lesssim 100$, reflecting the assumed nearly scale-invariant primordial spectrum. On the other hand the spectrum at higher multipoles (small scales) have been substantially modified by gravitational instability -- since photons freely stream after last scattering, its temperature from then until it reaches the present day observer can only be affected through gravitational effects.

\subsubsection{Acoustic Oscillations}

\paragraph{CMB Acoustic Peaks}
Perturbations in the gravitational potential caused longitudinal or \textit{acoustic oscillations} in the photon-baryon fluid of the early Universe, giving rise to small time variations in temperature. After last scattering, the temporal phases of the acoustic oscillations were ``frozen in'' to the radiation as a series of harmonic peaks observable in the power spectrum. The length scale associated with the peaks is the sound horizon at last scattering 
\begin{equation}
    r_* \equiv  \int_0^{t_*} \frac{c_s(t)}{a(t)} dt,
\end{equation}
the maximum comoving distance the sound waves with a sound speed $c_s$ in the baryon-radiation fluid could travel from the time of the Big Bang to the time of decoupling\footnote{Often the epoch of the last scattering surface is described as that of \textit{recombination}, i.e. when the first atoms formed and the Universe became neutral, although that is not as precise as simply saying last scattering \cite{PDG2022}. Often the term \textit{decoupling} is avoided, because (p.72,  \cite{Dodelson2003}) although the photons stopped scattering off electrons at $z \sim 1000$, there were many more photons than baryons and electrons must be considered to scatter from photons until much later. \label{footnotedecoupling}} of photons from matter at $t_*$. The sound speed may be approximated as that of a pure photon gas $\sim \frac{c}{\sqrt{3}}$; in reality this should be modified to account for the baryons.

Low multipole moments $\ell$ correspond to large angular scales and high $\ell$ to small scales. There is no one-to-one conversion between a multipole $\ell$ and an angular scale $\theta$ on the sky \cite{PDG2022}, but the spherical harmonics have oscillatory patterns with roughly $\ell$ ``wavelengths'' over a great circle of the sphere. Therefore we can estimate the separation between harmonic peaks (i.e. between minima and maxima, or half-``wavelengths'') as\footnote{The angular scale of the acoustic peaks $\theta_n$ is often \textit{defined} as the angle subtended by a comoving sound horizon by the comoving distance to the last scattering surface. This requires the assumption of spatial flatness, such that the angular diameter distance is simply proportional to the comoving distance.}
\begin{equation}
\ell_{n} \sim \frac{\pi}{\theta_{n}} = \frac{n \pi D_{\textsc{sls}}}{a(t_*) r_*} \approx\frac{n \pi \chi_{\textsc{sls}}}{c_s  \eta_*}\label{lpeak}
\end{equation}
where $D_{\textsc{sls}}=D_A(t_*)$ is the angular diameter distance to the surface of last scattering and approximated as $a(t_*) \chi_{\textsc{sls}}$ assuming spatial flatness, and $n$ is an integer labelling the peaks.  Odd acoustic peaks correspond to matter falling into, and even number peaks is an acoustic wave expanding out of the gravitational potential wells. We will also denote the first peak with $\theta_*$ and $\ell_*$. The CMB data supplies measurements of the angular size of the sound horizon from the angular location $\theta_n$ of the acoustic peaks, and the physical sound horizon serves as a standard ruler. We can give a straightforward rough estimate \cite{Reid2002} of $r_*$ using the Friedmann equation \eqref{friedmanndensityparameterisation} assuming a spatially flat dust-only universe (since $z_* \sim 1100$ according to recombination models, almost independently of the main cosmological parameters \cite{PDG2022}, during the matter-dominated epoch). Hence
\begin{equation}
     r_* \approx c_s\int_0^{t_*} \frac{dt}{a(t)} \approx  - \frac{c}{\sqrt{3}}\int_{\infty}^{z_*} \frac{dz}{H(z)} \approx \frac{2c}{H_0 \sqrt{3  \Omega_m}} (1+z_*)^{-\frac{1}{2}}
\end{equation}
which is about 280 Mpc. The actual value, which requires accounting for baryons giving a reduced sound speed, is $\sim 144$ Mpc \cite{Planck2020}.
On the other hand, the angular diameter distance to the surface of last scattering $D_{\textsc{sls}}$ is dependent on the expansion rate of space since the early Universe, and therefore dependent on cosmological density parameters. Assuming only $\Omega_m$ and $\Omega_\Lambda$ components and spatial flatness (i.e. flat $\Lambda$CDM), we have from the definition of angular diameter distance \eqref{angulardistanceOS}, the comoving distance in terms of redshift \eqref{comovingdistance} and the Friedmann equation \eqref{friedmanndensityparameterisation} we can estimate
\begin{equation}
D_{\textsc{sls}} = \frac{c}{(1 + z_*) H_0} \int_0^{z_*} \frac{dz}{\sqrt{\Omega_m(1+z)^3 + \Omega_{\Lambda}}}.
\end{equation}
This may be estimated either by binomial expansion or numerically, and the predicted first acoustic peak is $\ell_* \sim \frac{\pi \chi_{\textsc{sls}}}{c_s \eta_*}  \sim 200$, or $\theta_* \sim 1 ^\circ$. The measured acoustic peak scale and its angular extent provides a measurement of the angular diameter distance to last scattering, and is therefore a sensitive probe of cosmological parameters which determine the expansion history.

\paragraph{Baryon Acoustic Oscillations: A Statistical Standard Ruler}

The acoustic oscillations of the photon-baryon fluid also imprinted on the distribution of the baryons. This is observed as a density excess in the galaxy and matter power spectra, on a preferred scale $r_d$ -- the sound horizon at the \textit{cosmic drag} epoch (or the time of baryon decoupling) which is only slightly later than the sound horizon $r_*$ at last scattering. We can measure the amount of excess clustering at all redshifts, so the Baryon Acoustic Oscillation (BAO) scale $r_d$ is considered a \textit{statistical standard ruler}. The BAO scale is fixed in \textit{comoving} coordinates \cite{Planck2020a}; as it is so large ($\sim 150$ Mpc at present day), it has been preserved by linear evolution from astrophysical processing. 

\subsection{The Concordance Model: \texorpdfstring{$\Lambda$}{Lambda}CDM} \label{lcdm}

In contemporary cosmology, the notion of a cosmological parameter is therefore very general and there is no unique set. Typical models used for comparison with observations have between five to ten parameters. As we have seen, we use parameters to describe what the Universe is composed of: baryons, photons, neutrinos, dark matter, and dark energy. We also need to describe the nature of perturbations away from the FLRW approximation through e.g. the matter and radiation power spectra (and the {\hl $n$}-point spectra). The amplitude of matter fluctuations, for example, is conventionally represented by the parameter $\sigma_8 = \sigma_8(z=0)$, defined by the root-mean-square fluctuation of the matter overdensity $\rho_m / \bar{\rho}_m$ in spheres of radius $8 h^{-1}$Mpc at redshift $z=0$. Any time we add extra degrees of freedom to the model, more parameters arise. As alluded to in discussions of acoustic oscillations, if we add perturbations to a perfect fluid, then the perturbed fluid equations can be linearised to yield a wave equation \cite{Peacock1999}. This is characterised by a sound speed $c_s^2 \equiv \frac{\delta p}{\delta \rho}$ arising from an extra degree of freedom in a component of the cosmological fluid in addition to the equation of state $w$. It has also been advocated that the background CMB temperature should be included as a cosmological parameter \cite{Yoo2019}. There may also be parameters describing the physical state of the Universe, such as the ionisation fraction as a function of time during the era since recombination. For example, the six parameters of the base cosmological model used by the \textit{Planck} Collaboration \cite{Planck2020a} are the dark matter density $\Omega_c h^2$, the baryon density $\Omega_b h^2$, {\hl an approximation to the observed angular size of the sound horizon at recombination, the initial super-horizon amplitude of curvature perturbations, scalar spectral index, and reionisation optical depth.} Here $h$ is the dimensionless Hubble parameter, or ``little h'', which arises from the historically motivated convention of writing $H_0 \equiv 100 \, h \, \text{km} \, \text{s}^{-1} \text{Mpc}^{-1}$ such that formulae and results are expressed in terms of {\hl $h$} and its powers, so that a different preferred value for $H_0$ could more easily be substituted \cite{Hogg1999, Croton2013}.

Since the turn of the century, a wide range of observational data, theoretical arguments as well as simplifying assumptions have motivated the adoption of the \textit{standard} or \textit{concordance} $\Lambda$CDM model of cosmology. Its main features can be summarised as follows \cite{Planck2020a}
\begin{enumerate}
    \item Laws of physics are the same throughout the observable Universe.
    \item General relativity is an adequate description of gravity.
    \item The Universe is statistically homogeneous and isotropic when averaged on sufficiently large scales.
    \item The Universe has been expanding and cooling since early times from a hot and dense initial state.
    \item There are five main components to the energy and matter of the Universe:
    \begin{enumerate}
        \item dark energy which acts like the vacuum energy density (i.e. the cosmological constant $\Lambda$)
        \item dark matter which is stable, pressureless and interacts with ordinary matter only through gravity (i.e. CDM)
        \item ordinary or baryonic matter
        \item photons observable as the CMB
        \item neutrinos, which are nearly massless.
    \end{enumerate}
    \item The spatial curvature is very small or zero.
    \item Variations in density (allowing for eventual structure formation) formed in the early Universe and, as predicted by inflation, are Gaussian, adiabatic and nearly scale-invariant.
    \item The observable Universe has a \textit{trivial topology}; it is not periodic or multiply connected.
\end{enumerate}

Despite the precision fit to a variety of observations, the $\Lambda$CDM model does not offer any conceptual understanding of its namesake predominant features. The main components of matter and energy in the Universe are cold \textit{dark matter} at 26.1\% and \textit{dark energy} which behaves like a cosmological constant at 68.9\%, with baryons or ordinary matter only making up 4.9\% of the total (the numbers quoted here are from the analysis of \textit{Planck} CMB data \cite{Planck2020}). It is worth noting that the baryon density, measured to an accuracy of a percent, also lacks an underlying theory which predicts its value within orders of magnitude \cite{PDG2022}. Dark matter and dark energy are in a sense misnomers\footnote{This is perhaps easily forgiven; there is historical precedent for matter that was not easily detectable originally. For example, Neptune was inferred from the motion of Uranus (we might say it was \textit{dim matter}...) and neutrinos were predicted from the range of energies in beta decay. On the other hand, there were numerous purported discoveries of a hypothesised planet \textit{Vulcan} nearby causing the unexplained precession in the orbit of Mercury, until the advent of General Relativity.}, as they do not appear to interact with the electromagnetic potential i.e. are invisible --  neither absorbing (typically what we would call \textit{dark}), emitting nor scattering light of any wavelength.

\subsubsection{Dark Matter} \label{sec:darkmatter}

Dark matter constitutes $\sim$ 85\% of total mass, and yet its nature is unknown and it has never been directly detected\footnote{Besides the controversial DAMA \cite{DAMA2013} results which have been unable to be reproduced.}. In order to explain the growth of matter perturbations between the time of matter-radiation equality and the time of recombination, which is inferred from the CMB power spectrum, dark matter must have formed by matter-radiation equality. This eliminates macroscopic objects such stars and planets as candidates as a main component of dark matter. However, one kind of macroscopic object that could have formed early in the Universe are primordial black holes (PBH), as high densities and inhomogeneous conditions could lead sufficiently dense regions to undergo gravitational collapse. Therefore dark matter could be very heavy (up to $10^5 M_\odot $ for PBH) modulo gravitational lensing limits or for fundamental particles, up to Planck mass. At the low mass end there is a limit from requiring the dark matter wavelength $\lambda_{DM} \leq$ size of smallest DM structures $\sim 1 \text{kpc} \sim 3 \times 10^{19} \text{m}$. Non-exhaustive particle dark matter categories include: ultralight bosons, axions, bosonic or fermionic thermal dark matter, or more than one new particle \cite{Young2017}.

The primary evidence for dark matter is from calculations of galactic dynamics. These predict that if dark matter did not exist, that many galaxies would fly apart instead of rotating. Other (non-exhaustive) lines of evidence include observations in gravitational lensing, the CMB and matter power spectra, from the formation and evolution of galaxies, from mass location during galactic collisions, and from the motion of galaxies within galaxy clusters. We refer to \cite{Young2017, PDG2022} for reviews on dark matter; {\hl and to \cite{zavala2019} and \cite{vegetti2023} for reviews on dark matter structure formation and dark matter in the context of strong gravitational lensing respectively.}

Zwicky showed in 1933 that the galaxies of the Coma Cluster were moving too fast for the cluster to be bound together as predicted by the Virial Theorem. The idea of dark matter did not gain traction for another fifty years (when it was re-introduced to account for the large rotational velocities of hydrogen gas within galaxies) since there was no capacity then to measure hot gas in clusters (the intracluster medium or ICM). Estimates roughly suggest that 14\% of a cluster is gas; with 85\% being DM and 1\% being stars \cite{Roos2012}.

\paragraph{Galactic Dynamics: Rotation curves and velocity dispersions}

Spectroscopic radial velocities can be used to determine the rotational velocities of {\hl disk} galaxies which are inclined to the line-of-sight through the Doppler shifted spectra. Typical lines that are used are $H\alpha$, [NII], CO and the 21cm line produced by neutral hydrogen in interstellar space \cite{BovyInPrep}.

The \textit{rotation curve} of a galaxy is the velocity a test particle or star would have if it were in circular orbit as a function of radius, reflecting the distribution of mass in the galaxy. As a first example, we can estimate the mass $M$ inside a certain radius $R$ by equating the Newtonian gravitational acceleration to the centripetal one $-\frac{v^2}{R}= -\frac{GM}{R^2}$ where v is the rotation velocity. In general, the shape of the rotation curve depends on the halo virial mass.

\begin{figure}
    \centering
    \includegraphics{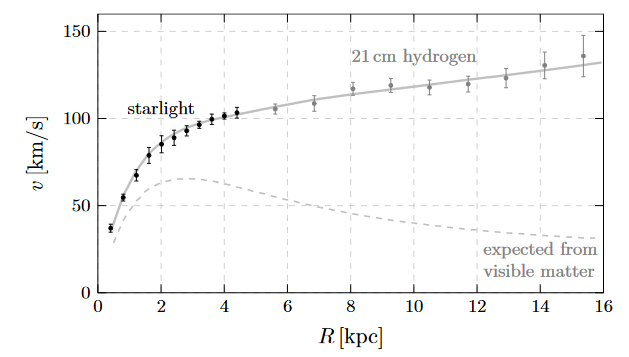}
    \caption{An example galaxy rotation curve (M33), reproduced from \cite{Baumann2022}. The dashed line is the expected rotation curve accounting only for the visible matter in the stellar disk.}
    \label{fig:rotationcurve}
\end{figure}

Baryonic matter in the form of stars and gas dominate the potential in the central region of galaxies. When most of the mass is at the centre, the velocity decreases with the square root of the radius. This behaviour is known as \textit{Keplerian decline}, since this is the case for planets in the Solar System following Kepler's laws. However, this is almost never observed for galaxies. 

Most {\hl disk} galaxies have rotation curves that show solid body rotation (such that the velocity increases linearly with radius) in the very centre, following by a slowly rising or constant velocity rotation in the outer parts; this is illustrated in Figure
\ref{fig:rotationcurve}. A flat rotation curve, where the velocity is constant over some range of radii, implies that the mass is increasing linearly with radius. Rotation curves of most galaxies therefore can be fit with a superposition of components: stellar bulge, stellar and gaseous disks and a dark halo that is usually modelled as a quasi-isothermal sphere. This is strong evidence for dark matter, more so than the Coma cluster as it shows that the dark matter is distributed differently than the luminous matter: galaxies are a \textit{biased} tracer of matter.

Whilst baryons can collapse or accrete into a galaxy or black hole by shedding angular momentum through collisions, and also radiate photons to shed energy, it is harder for dark matter which does not interact as much to do the same. Fewer collisions and the conservation of angular momentum therefore explains why the dark matter halo of a galaxy is at a larger radius with negligible angular velocity relative to the baryons of the galaxy.

\paragraph{The Bullet Cluster}
The Bullet Cluster \cite{Randall2008} is a pair of colliding galaxy clusters at redshift $z \sim 0.3$. It is the first example of a system where the centre of mass and centre of baryons are distinctly separate from each other, showing that dark matter and baryons interact differently during a collision. Most components of this merger can be spatially traced -- galaxy plasma emits X-ray Bremsstrahlung radiation measured using the Chandra space telescope; and stars emit optical and infrared light, measured using Hubble and Magellan. Gravitational lensing of background galaxies caused by the Bullet Cluster spatially place the DM mass within the cluster.

\subsubsection{Accelerated Expansion: Dark Energy} \label{sec:darkenergy}

The accelerated expansion of the Universe, first observed via SNe Ia in the late 1990s \cite{Perlmutter1997, Riess1998}, is another puzzling phenomena in modern physics: even the most conservative explanations require a pervasive new constituent to the stress-energy content of the Universe. The properties of \textit{dark energy} include an unusual equation of state $w_\textsc{de} < - \frac{1}{3}$ which results in repulsive gravity (as discussed in Section \ref{sec:singlecomponentsolutions}), that it is close to spatially homogeneous and only recently relevant (combining low and high redshift data sets shows that the dark energy contribution goes from less than 10\% to nearly 70\% of the total within the last e-fold\footnote{The quantity $N= \ln a$ is a natural scale of the dynamics and called the number of e-folds. It arises in the \textit{e-folding method} \cite{Amendola2010} of solving differential equations with non-constant coefficients such as $\mathscr{H}$ in cosmology, where the substitution of $N$ for the scale factor cancels the coefficients.} of expansion \cite{Planck2020a}).

In the simplest case, dark energy is the cosmological constant without further physical basis. If $G_{\mu \nu} + \Lambda g_{\mu \nu} = \kappa T_{\mu \nu}$ is {\hl the} correct form of the EFEs, we would still need to reproduce Newtonian gravity in the local weak field limit. The Newtonian correspondence which we derive in Section \ref{newtoniancorrespondence} gives $\delta g_{\mu \nu}= -2 \frac{\Phi}{c^2}$ where $\Phi$ is the Newtonian gravitational potential, so the $00$ component of the EFEs is
\begin{equation}
    \nabla^2 \Phi = 4 \pi G \rho - \Lambda c^2,
\end{equation}
so $\Lambda c^2$ must be small in comparison to $4 \pi G \rho$ to reproduce Newtonian gravity. Since $\Lambda$ has dimensions of $\Length^{-2}$, we deduce that the length scale of the cosmological constant must be much larger than the scale of stellar objects where Newtonian gravity is still a very good approximation.

The fact that dark energy is only recently relevant, i.e. cosmic acceleration begins around redshift $z \sim 1$, is often discussed as a coincidence problem \cite{Amendola2010}. We recall that the matter energy density scales as $a^{-3}$ whereas the energy density of $\Lambda$ is constant; hence their ratio has changed by a factor of $\sim 10^{27}$ since BBN. It may appear to be remarkably coincidental the current values that we observe are within the same order of magnitude ($\Omega_m \sim 0.3$, $\Omega_\Lambda \sim 0.7$); although, much like any other fine-tuning arguments (such as the flatness problem), whether this is indeed a problem is often a point of contention in the literature \cite{Bianchi2010}.

On the other hand, the age of the Universe also supports the existence of dark energy: the age of the Universe in the absence of the cosmological constant is in the range 8.2 Gyr $< t_0 < $ 10.2 Gyr, less than estimated ages ($> 11$ Gyr) of globular clusters \cite{Amendola2010}. This was known before the SNe Ia evidence for dark energy; the accelerated expansion makes up for this discrepancy.

\paragraph{Observational Evidence for Accelerated Expansion}

Well-established methods for measuring the accelerated expansion include the luminosity distances and redshifts of Type Ia supernovae, the CMB, BAO and large-scale structure. The first clear evidence for a late-time cosmic acceleration was reported in 1998 from two independent teams, the High-redshift Supernova Search Team (HSST) \cite{Riess1998} and the Supernova Cosmology Project (SCP) \cite{Perlmutter1997}, both from observations of distant SNe Ia.

\begin{figure}
    \centering
        \centering
        \includegraphics[scale=0.8]{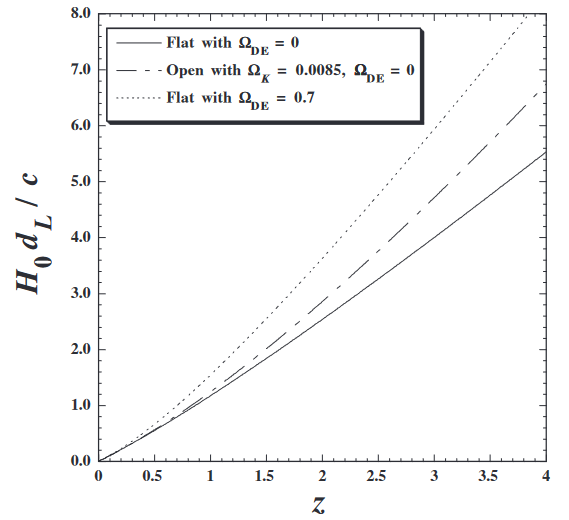}
        \caption{Theoretical Hubble diagrams, or the dependence of luminosity distance on redshift for different cosmologies. In the limit $z \ll 1$, $\frac{H_0}{c} D_{L} \approx z$ which is the original linear Hubble law. Diagram is reproduced from \cite{Amendola2010}.}
        \label{fig:luminositydistredshift}
\end{figure}

\begin{figure}
    \centering
        \includegraphics[scale=0.4]{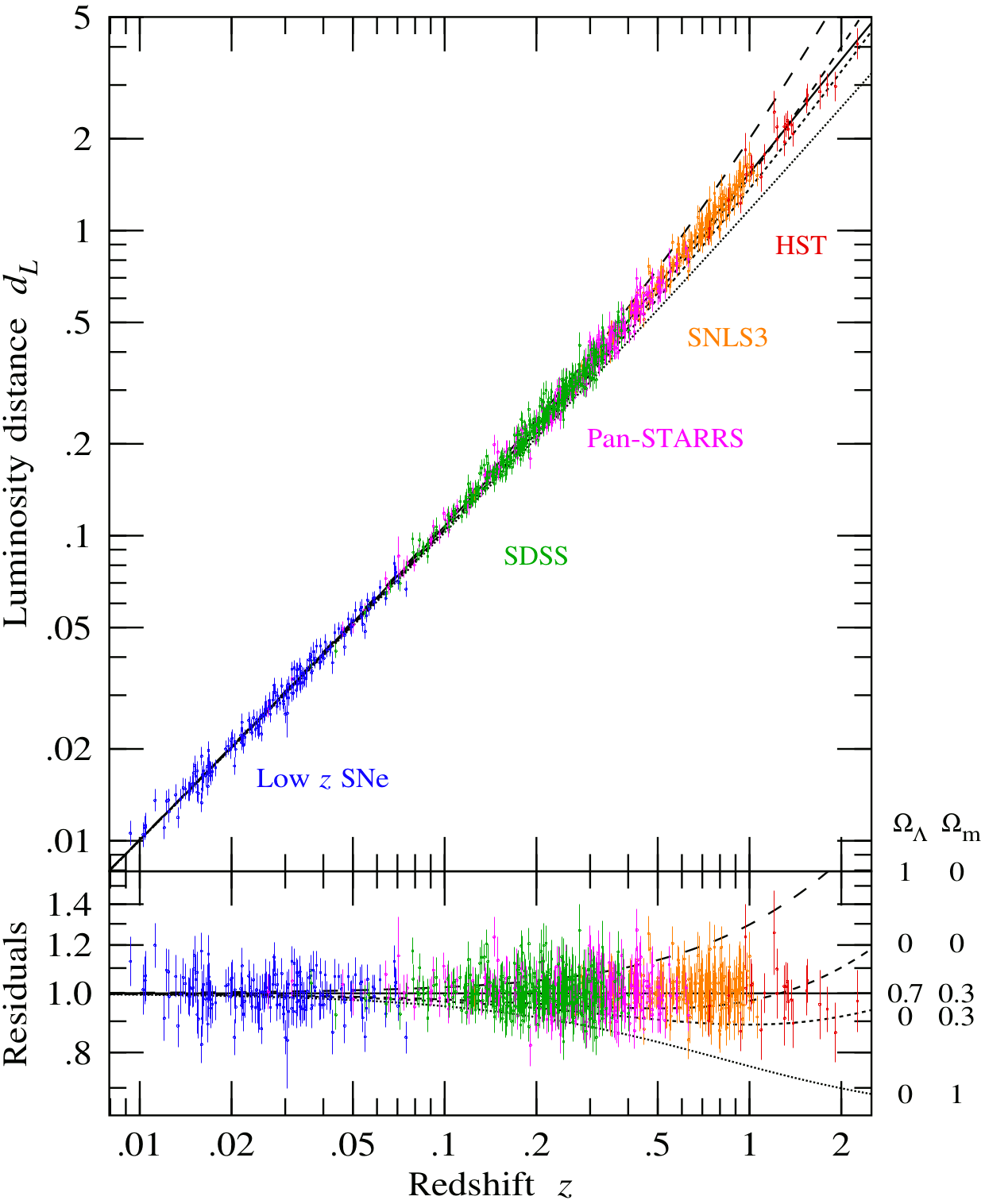}
        \caption{Log luminosity distance (i.e. $\propto$ the distance modulus) versus log redshift, or Hubble diagram, of 1048 Type Ia supernovae from the \textit{Pantheon} compilation \cite{Scolnic2018}, reproduced from \cite{Hamilton2022}. See also the very similar \textit{Pantheon}+ Hubble diagram, Figure 4 of the recent publication \cite{Brout2022}. However for pedagogical clarity we choose to include this version including the dashed lines corresponding to alternative cosmologies.}
    \label{fig:pantheonhubblediagram}
\end{figure}

The question of observational evidence for accelerated expansion is largely a question of measuring distances, since  observational distance measurements are dependent on the expansion history. As mentioned in Section \ref{sec:luminositydist}, SNe Ia are a standardisable candle, providing luminosity distance measurements as a function of redshift; explicitly using the Friedmann equation \eqref{E(z)} in the luminosity distance \eqref{luminositydist_chi} 
\begin{equation}
    D_{L} = \frac{c}{H_0}(1+z_s)\Omega_K^{-\frac{1}{2}} \sinh \left( \Omega_K^{\frac{1}{2}} \int_0^{z_s} \frac{dz}{E(z)} \right). \label{luminositydist_z}
\end{equation}
Note that for local distance ladder measurements, it is usually the combination $H_0 D_{L}$ which is constrained \cite{PDG2022}; that is, the Hubble constant scales the distances inferred from redshifts. This dependence of the luminosity distance on cosmology is illustrated as a Hubble diagram, shown in Figure \ref{fig:luminositydistredshift} and for real data in Figure \ref{fig:pantheonhubblediagram}. For example, the Pantheon+ analysis of 1550 SNe Ia, ranging in redshift from $z = 0.001$ to $2.26$, found for a flat $\Lambda$CDM model $\Omega_m = 0.334 \pm 0.018$ and $\Omega_\Lambda = 0.666 \pm 0.018$ from SNe Ia alone \cite{Brout2022}. Considering distance uncertainties of 5–8\% per well observed supernova, a sample of around 100 SNe already achieves sub-percent statistical precision. The current 1–2\% systematic uncertainties are dominated by uncertainties associated with photometric calibration, dust extinction, observed dependence of luminosity on host galaxy properties; and potentially intrinsic redshift evolution of the supernovae \cite{PDG2022}.


CMB observations are also consistent with the presence of dark energy, although evidence from the CMB alone is limited. Rather, the CMB constrains geometry, matter and radiation content which can be used in conjunction low-redshift observations. The properties of dark energy also affect the CMB through secondary anisotropies, such as the Integrated Sachs-Wolfe (ISW) effect (see Section \ref{sec:perturbedlightrays}) at late times. During matter domination the potential is constant, but as dark energy begins to dominate, the gravitational potential felt by the CMB photons decays. We expect to see the properties of dark energy encoded in the exact nature of the ISW effect in the CMB angular power spectrum.

The measurements of BAO are another independent test for the existence of dark energy, favouring a Universe with dark energy rather than a CDM Universe. The BAO data constrains two ratios; the first $\delta z_s (z) = \frac{r_d }{c}H(z)$ measures fluctuations along the line of sight, whereas analogous to the CMB acoustic peak angle, $\theta_s(z) = \frac{r_d}{(1+z)D_A(z)}$ measures those orthogonal. Since errors are correlated, in practice what is constrained is an effective distance 
\begin{equation}
    D_V(z) \equiv \left((1+z)^2 D_A^2(z) \frac{cz}{H(z)} \right)^{\frac{1}{3}}.
\end{equation}
BAO measurements are complementary to SNe ones, as they provide distance measurements calibrated using the sound horizon from the CMB, with greater precision possible at high redshift (limited by statistical rather than systematic uncertainties even for the largest feasible surveys) \cite{PDG2022}.

\paragraph{The Cosmological Constant as a Vacuum Energy}

It is possible to associate the cosmological constant with a vacuum energy which includes contributions from the zero-point energies of quantum fields. It is worth noting that this association in itself is not necessarily a \textit{fundamental} one \cite{Bianchi2010}. A vacuum according to quantum field theory, which forms the foundation of modern particle physics, is not truly empty but filled with virtual particles; a quantum vacuum contribution to the gravitational stress-energy would be analogous to the Lamb shift in atomic physics when considering electromagnetic effects. A na\"ive calculation in much of the early work including an unpublished piece by Pauli \cite{Rugh2002} involves the following (see also the later calculation by Zeldolvich \cite{Zeldovich1968}, and reviews \cite{Carroll2004, Copeland2006, Milonni1994}). Considering the total zero-point energy of quantum electrodynamics, the energy density of the vacuum is
\begin{equation}
\varepsilon_{\text{vac}} =  \frac{1}{2} \int_0^\infty c \sqrt{(mc)^2 + p^2} \frac{d^3 \bm{p}}{(2 \pi \hbar)^3} = \frac{4 \pi c}{2(2 \pi \hbar)^3} \int_{0}^{\infty} p^2  \sqrt{(mc)^2 + p^2}   dp \propto p^4
\end{equation}
where $m$ is the mass of the field, $\bm{p} = \hbar \bm{k}$ is the three-momentum and $\bm{k}$ the corresponding 3-wavevector. That is, taking the Fourier transform of a free quantum field means that we have an infinite number of harmonic oscillators in momentum space, with each mode contributing to a zero-point energy of $\frac{1}{2}\hbar \omega$ which gives a divergent result. Prior to the discovery of the accelerated expansion it was generally assumed that a correct calculation of the quantum vacuum energy would give a zero value.

The approach was then to introduce a cut-off scale $k_{\text{max}}$ such that the integral is finite, in which case we have $\varepsilon_{\text{vac}} \sim c \hbar k^4_{\text{max}}$. If we expect that quantum field theory is valid up to the Planck scale, then $k_{\text{max}}^{-1} \sim l_{\text{pl}} = \sqrt{\hbar G / c^3}$. We find that 
\begin{equation}
    \varepsilon_{\text{vac}} \sim  10^{76} (\hbar c)^{-3} \text{GeV}^4 \sim 10^{113} \text{J}/ \text{m}^3
\end{equation} 
which corresponds to the heuristic statement that the ``natural'' value for a quantum vacuum density is a Planck mass per Planck length cubed. It is infamously $\sim 120$ orders of magnitude larger than the observed value, $\Omega_\Lambda = \rho_{\Lambda} / \rho_{c} \approx 0.68$ \cite{Planck2020}  where $\rho_{\Lambda} \equiv \varepsilon_{\Lambda} c^{-2}$ corresponding to
\begin{equation}
\varepsilon_{\Lambda} \approx 3.9 \times 10^{-47} (\hbar c)^{-3} \text{GeV}^4 \approx 8 \times 10^{-10} \text{J}/{\text{m}^3}.
\end{equation}
We have already noted this value is very small such that Newtonian gravity is reproduced on small scales, and also drives the accelerated expansion on the scale of the Hubble time $t_H \sim 10^{18}$s rather than Planck time $t_\text{pl} \sim 10^{-43}$s.  

The second-order calculation carried out by Zeldovich obtained a value of $\rho_{\Lambda}$ which still contradicted the observational bound by 9 orders of magnitude, as critiqued in \cite{Padmanabhan2003}. Other avenues of attack have also been considered, such as from supersymmetry, but all differ from the observed value by many orders of magnitude \cite{Copeland2006}. It is possible that the discrepancies may be cancelled by introducing counter terms, which gives rise to what is sometimes described as a fine-tuning problem \cite{Padmanabhan2003, Carroll2004}. Whether any such calculations of the the zero point energy in field theory are realistic is debatable; so many other models of dark energy have been considered, such as being due to the dynamics of a light scalar field \cite{Copeland2006}.

\paragraph{Alternatives}
As we have seen, current observations constrain the equation of state of dark energy close to $w=-1$, consistent with a cosmological constant. However, if the equation of state evolves with time, then scalar fields which arise in particle physics are a natural alternative candidate for dark energy. Scalar field models proposed include quintessence, K-essence and tachyon fields. These have a lower bound on the equation of state as $w \geq -1$. On the contrary, phantom fields have an equation of state $w \leq -1$. Still other solutions have been proposed, that are fundamentally not interpreted as a dark energy fluid, such as modifications to general relativity itself. We refer the interested reader to \cite{Amendola2010, Padmanabhan2003, Copeland2006, PDG2022, Weinberg2013}.

\section{Challenges to the Consensus}

Aside from the nature of dark matter and dark energy, two of the most widely discussed unresolved issues that are \textit{a priori} within the framework of standard cosmology are foremost the \textit{Hubble tension} and secondly the \textit{$\sigma_8$} or \textit{$S_8$ tension} \cite{Abdalla2022}. If the tensions are not accounted for by unknown systematic errors, then physics beyond $\Lambda$CDM is required.

Fundamental assumptions and observations have also been directly questioned or tested. Modifications to general relativity have been suggested at different scales as an alternative to dark matter and dark energy, notably Modified Newtonian Dynamics (MOND) to replace dark matter on galactic scales and $f(R)$ gravity instead of dark energy.

On the empirical side, the evidence for large-scale isotropy and accelerated expansion has also been challenged; in particular, there is discussion around a number of cosmic dipoles and their interpretation. To a lesser extent, there are also discrepancies between $\Lambda$CDM predictions and simulations on small scales in the deeply non-linear regime, although many of these to do with galaxy statistics have recently been declared resolved \cite{Sales2023, Kim2018, Sawala2022, Ostriker2019, Delpopolo2021}.

\subsection{\texorpdfstring{$H_0$}{H0} and \texorpdfstring{$S_8$}{S8} Tensions}
The SH0ES (Supernovae and H0 for the Equation of State of dark energy) program's stated goal is to reach a percent-level measurement of $H_0$, using observations from the Hubble Space Telescope to break the degeneracy among cosmological parameters used for CMB data and the equation of state for dark energy \cite{Riess2022}. In approaching this goal with a current (2022) measurement of $H_0 = (73.04 \pm 1.04) \, \text{km} \, \text{s}^{-1} \, \text{Mpc}^{-1}$, SH0ES further entrenched at $5\sigma$ a decade-long discrepancy with the prediction of $H_0 = (67.4 \pm 0.5)\, \text{km} \, \text{s}^{-1} \, \text{Mpc}^{-1}$ from the \textit{Planck} CMB analysis under $\Lambda$CDM. The disagreement persists in models that allow non-zero curvature and low redshift evolution of the DE equation of state. Both teams are confident in their analyses, and do not believe the tension can be explained by measurement uncertainties. Measurements of $H_0$ using independent techniques are therefore needed to verify or refute the apparent tension.

\begin{figure}
    \centering
    \includegraphics[scale=0.7]{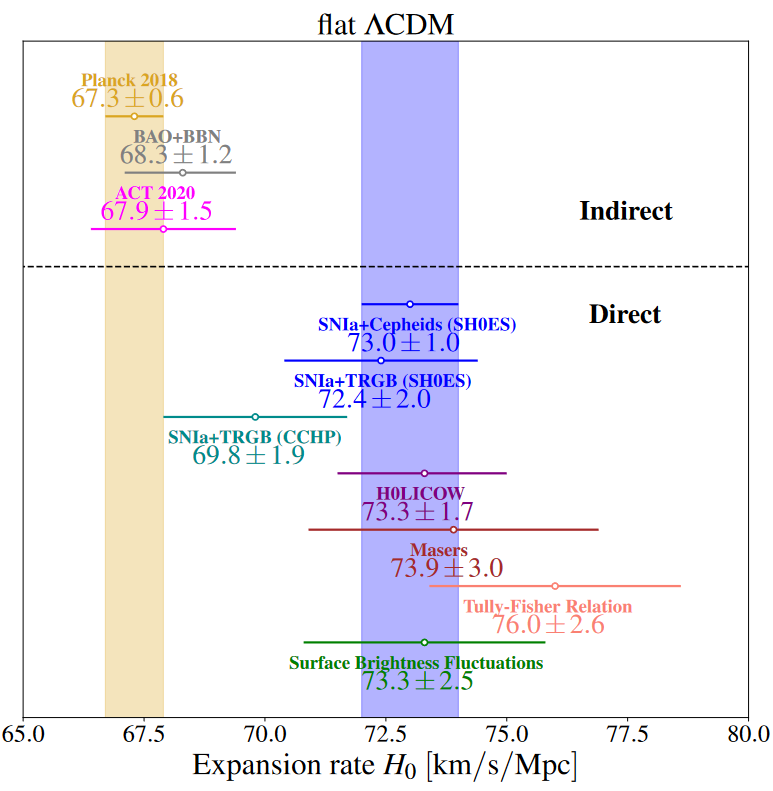}
    \caption{A selection of recently published $H_0$ values under $\Lambda$CDM, reproduced from \cite{Poulin2023}; a more detailed plot is found in \cite{Abdalla2022}.}
    \label{fig:hubbletension}
\end{figure}

The SH0ES baseline measurement is consistent with other primary distance indicators in the hosts of SNe Ia, which lends support to the idea that tension may be intrinsic to low-redshift or late measurements versus high-redshift or early measurements; for example it could be due to new physics in the early Universe that re-scales the CMB sound horizon. In fact there is broadly a consensus in the indirect model dependent estimates at early times, such as CMB and BAO experiments, and likewise in the direct late-time $\Lambda$CDM-independent measurements, such as distance ladders and strong lensing \cite{DiValentino2021}; this is shown in Figure \ref{fig:hubbletension}.

On the other hand, disagreements in the value of $H_0$ have a very long history (since measuring cosmological distances {\hl is} so difficult) and over this history, the estimated error in the Hubble constant has been quite small compared with the subsequent adjustments to the preferred value \cite{Kirshner2004}. This sets a historical precedent for chronically underestimated systemic errors which contextualises the current Hubble tension.

\begin{figure}
    \centering
    \includegraphics[scale=0.35]{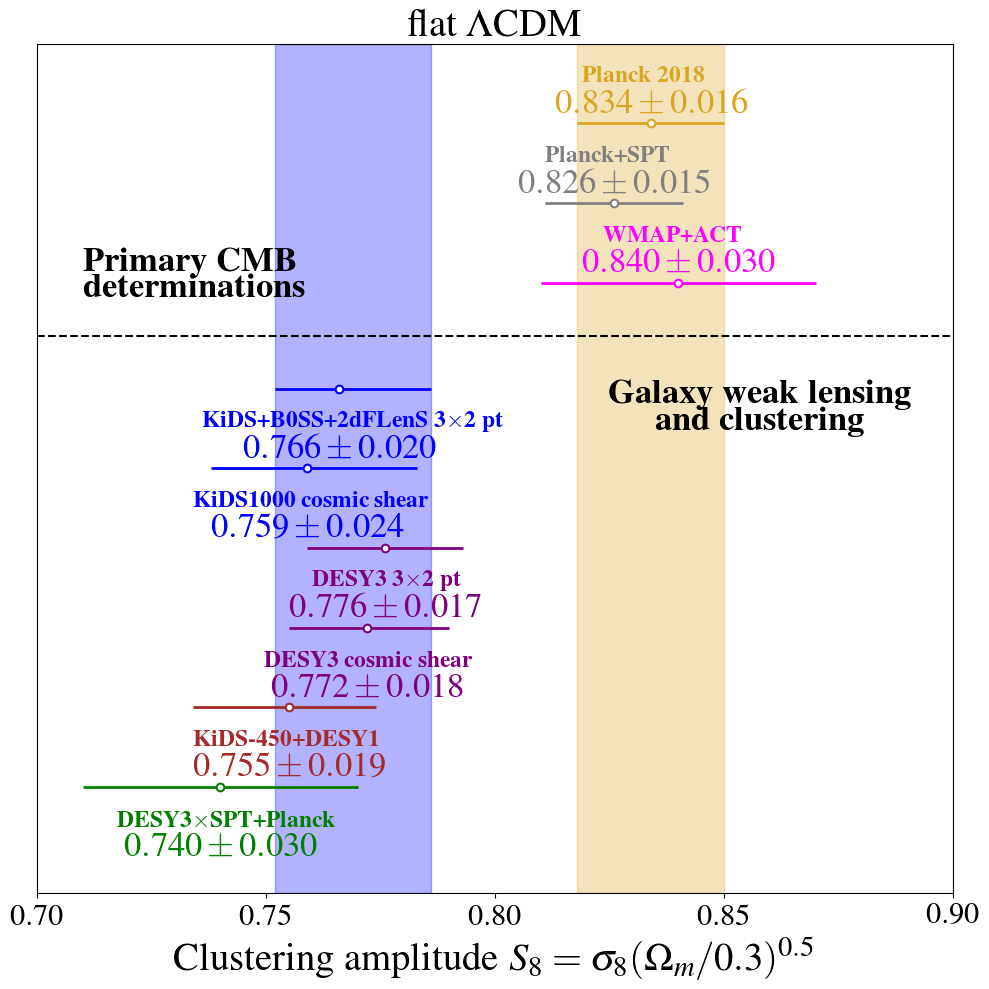}
    \caption{{\hl $S_8$} or $\sigma_8$ tension highlighting the discrepancy between low and high redshift measurements, courtesy of Vivian Poulin. A more detailed plot is found in \cite{Abdalla2022, DESKIDS2023}.}
    \label{fig:s8tension}
\end{figure}

Another tension is currently observed in the parameter describing the amplitude of matter fluctuations. The degree of clustering of matter is often measured by the quantity $S_8 \equiv \sigma_8(\Omega_m/0.3)^{0.5}$ which sets the scale for the amplitude of the weak lensing (cosmic shear, see Section \ref{sec:cosmicshear}) measurements; and $\sigma_8$ is the root-mean-square matter overdensity in a sphere of radius $8h^{-1}$Mpc. Measuring $S_8$ is model dependent; the standard flat $\Lambda$CDM model provides a good fit to the data from all probes. However, similar to the $H_0$ dichotomy, lower redshift probes have a preference for a lower $\sigma_8$ value, corresponding to a lower level of structure formation compared with the high redshift CMB estimates \cite{Abdalla2022}; this is illustrated in Figure \ref{fig:s8tension}. 

\subsection{Modified Gravity}

\subsubsection{Modified Newtonian Dynamics}

Although multiple pieces of evidence at different redshifts and on different scales point to the existence of CDM, no Standard Model particle meets its description. None of theoretically well-motivated dark matter particle candidates have been found by indirect, direct or collider searches either, and the ``natural'' parameter space for these candidates is shrinking. If dark matter does not exist, then the next most likely explanation must be that general relativity, or its Newtonian limit, is incorrect and should be modified.

The most popular modified gravity theory to replace dark matter is Modified Newtonian Dynamics (MOND), and its various relativistic extensions (including TeVeS gravity) \cite{Milgrom2001, Sanders2014, Skordis2021}. MOND posits that standard Newtonian dynamics is an adequate effective gravitational theory only for accelerations that are much larger than an additional constant $a_0$ with units of acceleration. In effect, MOND calculates an effective gravitational force from an observed (baryonic) matter distribution, which works remarkably well on galactic scales in the prediction of disk galaxies' rotation curves. Notable shortfalls for MOND include that it cannot account for the differing measurements of $\Omega_m$ and $\Omega_b$, and the existence of phenomena such as the Bullet Cluster described in Section \ref{sec:darkmatter}, which has an apparent centre of mass significantly offset from the baryonic centre of mass.

\subsubsection{Modifications to General Relativity} \label{sec:modifiedgrav}

Modifications to general relativity have been proposed to explain the accelerated expansion, effectively through extra geometric terms in lieu of dark energy. Whilst dark energy models modify the stress-energy content of the Universe, modified gravity theories alter the left-hand side beyond a simple cosmological constant; for example modifying the Einstein-Hilbert action. For example, it is proposed in $f(R)$ theories 
\cite{Sotirou2006} that a low-energy limit of a quantum gravity theory will manifest itself as higher-order corrections to general relativity, and these corrections may be found by modifying the action. Rather than assuming that the gravitational Lagrangian is only the Ricci scalar, it might be any function of the Ricci scalar.

Modifications to our understanding of gravity may be classified by whether they violate the Equivalence Principle, on which General Relativity is founded, in its Weak, Einstein or Strong forms. The \textit{Weak Equivalence Principle} (WEP) states that inertial mass is the same as gravitational mass, which is equivalent to saying that the gravitational field described by some metric is coupled to all matter. It is a statement of the \textit{universality of free fall}, that all test particles of any composition fall along geodesics of this metric. A test particle is one which is sufficiently small such that tidal forces from inhomogeneities in the gravitational field may be neglected (i.e. it allows the experiment to be taken as local, or confined within an infinitesimally small region).

The \textit{Einstein Equivalence Principle} (EEP) adds the principle of relativity, i.e. local position invariance and local Lorentz invariance to the WEP: the non-gravitational laws of physics are independent of the time, position and velocity of the laboratory in which they are tested. The EEP therefore states that in a local frame, the effects of gravity cannot be distinguished from the effects of a linear acceleration relative to an inertial frame. As a consequence of local position and Lorentz invariance, fundamental constants such as the fine-structure constant should not vary, nor should we observe preferred directions in space \cite{PDG2022}. 

The \textit{Strong Equivalence Principle} (SEP) generalises the EEP to include massive test bodies with self-gravitational interactions, e.g. gravitationally bound objects such as planets, stars or black holes; and all laws of physics (thereby allowing gravitational tests, e.g. Cavendish experiments). The SEP is therefore the strictest version of the Equivalence Principle, and General Relativity appears so far to be the only metric theory of gravity which satisfies it.

Distinguishing between dark energy and modified gravity is not necessarily straightforward, but one possible prescription categorises any theory which obeys the SEP as dark energy, and anything which does not as modified gravity \cite{Joyce2016}. Heuristically the SEP forbids fifth forces; we generically encounter additional degrees of freedom whenever we depart from General Relativity, e.g. scalar charge, which manifests as an additional force beyond that of gravity. The violation of the SEP can also lead to a difference in the inertial and passive gravitational mass of a self-gravitating body dependent on its gravitational binding energy \cite{Bonvin2018}. On the other hand, a violation of the WEP could occur if the metric couples differently to different matter species, for example arising from an interacting dark sector, or coupling between dark and baryonic matter. In some simple models of modified gravity, deviation from GR may be measured as a change in a rate of growth of structure parameter with respect to standard dark energy models; the recently launched \textit{Euclid} satellite aims to test GR in this manner \cite{Euclid2011}.

The main drawback to modified gravity theories is the fact that none have been found that can yet significantly challenge general relativity at all scales, and that gravity is severely constrained by tests of General Relativity within the Solar System (i.e. ``classical'' tests such as gravitational lensing and the Shapiro time delay; contrary to \cite{PDG2022}, the gravitational redshift is a test of the EEP rather than of GR \cite{Will2018}) as well as laboratory tests. This is illustrated in Figure 1.13 of \cite{Peter2009}. To escape these constraints, mechanisms which suppress the effects of additional degrees of freedom such as those which ``screens'' a fifth force are needed. 

\subsection{Cosmic Anisotropies}

In terms of direct empirical challenges for $\Lambda$CDM, one area of interest involves observations of possible large-scale anisotropy, in particular cosmic dipoles. There are two preferred frames we can measure, defined respectively defined by the vanishing of the dipole in CMB temperature and in galaxy number counts. Furthermore, it is crucial that any measured dipole must be correctly separated into components of kinematic origin and intrinsic origin. The CMB and matter dipoles must be aligned and purely kinematic for the cosmological principle to hold. Contrary to early claims that cosmic variance renders distinguishing a cosmological dipole from a kinetic dipole an impossibility \cite{Hu2002} however, there are in fact constraints on an intrinsic dipole from Doppler-like and aberration-like effects in the CMB \cite{Abdalla2022}.

A number of distinct dipole measurements have been reported in the literature, including in the measurement of the fine-structure constant from quasar spectra, in the clustering of galaxies and their bulk flow, in radio galaxies, quasars, and SNe Ia; but there is no immediately obvious correlation in their locations \cite{Abdalla2022}. In particular, the evidence for large-scale isotropy, including the interpretation of the CMB dipole, and the resultant analysis of SNe Ia data indicating accelerated expansion has been disputed by Sarkar and collaborators \cite{Colin2019, Mohayee2021, Secrest2022}.


\chapter{Light, Lens and Action} 
\label{chap:lightlensaction} 
\captionsetup{width=.9\linewidth}
\section{The Action Principle} \label{actionpriniple}

The dynamics of a physical system, such as particles and fields, form a large number of separate differential equations. However, the entire set of tensorial differential equations may be generated from a unifying variational principle\footnote{Not all physical systems may be described by an action principle: in particular, those involving forces that are not derivable from a single scalar quantity known as the work function \cite{Lanczos1986}. This applies to, for example, friction or viscosity.} using just one fundamental scalar quantity. This quantity is a functional called the action $S$, and the \textit{action principle} states that
\begin{displayquote}
The true dynamics of the system is the one where its action is stationary: $\delta S =0$.
\end{displayquote}
It is also commonly known as the principle of least action for historical reasons, although the action is in general only required to be a stationary value rather than an extremal or minimum value (e.g. when considering arbitrary rather than infinitesimal lengths of a path of integration p.247 \cite{Arnold1978}, p.316 \cite{Misner1973}, or \cite{Landau1951}). The mathematical methods for finding the critical points of \textit{functionals} is known as the calculus of variations. To find the equations of motion of a particle, we may assume the field to be given and vary the trajectory of the particle; the true path out of all the possible trial paths is the one where the action is stationary. On the other hand, to find field equations which describe the evolution of the field using the action principle, we assume the motion of the particle to be given and vary the field quantity.

The fact that the action is a scalar automatically ensures invariance, such that the action principle depends on neither the observer nor the choice of coordinates. The action is usually (but not always) an integral of a function known as the Lagrangian for particles, or the Lagrangian density for continua and fields (since the position goes from being a dynamic variable parametrised by time, to being just a label of the field which is now the dynamic variable). 

For example, consider a physical system which consists of a gravitational field and an electromagnetic field, sourced by continuous pressure-less matter (dust) possessing a charge distribution. Then we could write the action as
\begin{equation}
    S = S_g + S_m + S_f + S_q = \int_{V}  \left( \frac{c^3}{16 \pi G} R  -(-\mathscr{p}^\mu \mathscr{p}_\mu)^{\frac{1}{2}} - \frac{1}{16 \pi c}  F_{\mu\nu}F^{\mu \nu} - \frac{1}{c^2} A_\mu j^\mu \right) d V. \label{particularaction}
\end{equation}
The Einstein-Hilbert action $S_g$ gives the gravitational field contribution, and {\hl $R$} is the Ricci scalar. The action for dust $S_m$ with momentum \textit{density} $\mathscr{p}^\mu \equiv \rho u^\mu$, where $\rho = \varrho$ is the mass density for dust and $u^\mu$ is the fluid four-velocity, reduces to $S_p = -mc^2 \int d \uptau$ in the limit of an isolated particle with mass $m$. The Maxwell action $S_f$, given here in Gaussian units, is the contribution from an electromagnetic field in the absence of a source, described by the electromagnetic field strength tensor (or the Faraday tensor) $F_{\mu \nu}$. Finally, $S_q$ is the action for the interaction of the fields with continuous charged matter distribution with a four-current density $j^\mu$. The invariant volume element is $d V \equiv \sqrt{|g|} d^4x$, where $g$ is the determinant of the metric; in the case of Minkowski spacetime {\hl (under the standard choice of coordinates)} then $\sqrt{|g|} = 1$.

If we vary this example action \eqref{particularaction}, we find several sets of simultaneous equations \cite{ Dirac1975}. The variation with respect to the metric gives the Einstein field equations sourced by a stress-energy tensor consisting of two parts, one from the matter stress-energy and one from the electromagnetic field stress-energy tensor. The variation with respect to the electromagnetic four-potential $A_\mu$ gives Maxwell's equations in the presence of charges. Finally, the variation with respect to a displacement gives the geodesic equation (i.e. the equation of motion for a free massive particle, or in this case mass element) modified by an additional term accounting for the Lorentz force. We will examine each of these fundamental equations separately in the following sections of this thesis.

We refer the reader to \cite{Dirac1975, Landau1951} for a full treatise on classical field theory, including action principles in general relativistic field theory and calculations that may be omitted here. \cite{Misner1973}, \cite{poisson2014} and \cite{Harte2019} are recommended for discussions on geometric optics in general relativity. We also recommend \cite{Lanczos1986} for variational principles in classical mechanics.

\section{Electromagnetism in Curved Spacetime}

\subsection{Maxwell's Equations}
As mentioned in Section \ref{actionpriniple}, Maxwell's equations in curved spacetime can be generated from an action varied with respect to the electromagnetic four-potential $A_\mu$. The relevant physical system is that of an electromagnetic field and charged particles within it, but we omit immediately $S_m$ in the total action as it has no dependence on $A_\mu$. The relevant action is then
\begin{equation}
     S_{em} = S_f + S_q = \int_V  \left( - \frac{1}{16 \pi c}  F_{\mu\nu}F^{\mu \nu} - \frac{1}{c^2} A_\mu j^\mu \right) d V. \label{emaction}
\end{equation}
Once again the invariant volume element is $d V \equiv \sqrt{|g|} d^4x$ which ensures that the action is a scalar quantity under coordinate transformations in a general spacetime, and is the only point of difference from the action in Minkowski spacetime under the principle of \textit{minimal coupling}. The principle of minimal coupling states that we should not spuriously add terms explicitly containing the Riemann curvature tensor to the action\footnote{There may be ambiguity as to the inclusion of curvature terms which may arise or not, depending on the order in which covariant derivatives are taken, due to the non-commutation of the covariant derivative. To resolve such ambiguities, the principle of minimal coupling needs to be applied to on the level of physical fields or the action, rather than the level of the vector potential. Otherwise, charge conservation would not be maintained.} when transitioning from special to general relativity. Again we use Gaussian units; the values of the coefficients differ depending on the choice of units.

In relativistic theory, the electromagnetic field is represented by a real, trace-free, anti-symmetric tensor of rank 2 known as the electromagnetic field strength tensor or Faraday tensor $F_{\mu \nu}$. As in flat spacetime, it is the exterior derivative of $A_{\mu}$
\begin{equation}
    F_{\mu \nu} \equiv A_{\nu , \mu} - A_{\mu , \nu} \label{faradaytensor}
\end{equation}
{\hl due to the symmetry of the Christoffel symbols, $A_{\mu; \nu} - A_{\nu; \mu} = A_{\mu, \nu} - \Gamma\indices{^\rho_{\mu \nu}}A_\rho - ( A_{\nu,\mu} - \Gamma\indices{^\rho_{\nu \mu}}A_\rho) = A_{\mu, \nu} - A_{\nu, \mu}$}. Maxwell's equations in a general spacetime are a set of linear, first-order partial differential equations
\begin{empheq}[box=\fbox]{gather}
    F_{[\mu \nu; \lambda]} =0 \label{generalhomogeneousMaxwelleqn}\\
    {F^{\mu \nu}}_{; \nu} = \frac{4 \pi}{c} j^{\mu} \label{generalinhomogeneousMaxwelleqn}
\end{empheq}
where the inhomogeneous Maxwell equations \eqref{generalinhomogeneousMaxwelleqn} arise from the variation of the action \eqref{emaction}, and the homogeneous Maxwell equations \eqref{generalhomogeneousMaxwelleqn} are automatically satisfied in the potential formulation \eqref{faradaytensor} of electrodynamics. Maxwell's equations in curved spacetime look almost identical to those in a flat spacetime, only swapping partial derivatives for covariant derivatives. It satisfies the equivalence principle in the form of this so-called \textit{comma-to-semicolon rule}: there is no change physically or geometrically to the laws of non-gravitational physics with or without gravitation; rather the only change arises from switching from Minkowski to non-Minkowski frames.

To find the wave equation which governs the vector potential, we write the inhomogeneous equations \eqref{generalinhomogeneousMaxwelleqn} in terms of the vector potential
\begin{equation}
  {A^{\nu ;\mu}}_{;\nu} - {A^{\mu ; \nu}}_{;\nu} =  - {A^{\mu ; \nu}}_{;\nu} + R\indices{^{\mu}_{\rho}} A^{\rho} - {{A^{\nu}}_{; \nu}}^{; \mu}  = \frac{4 \pi}{c} j^{\mu}
\end{equation}
where we used the commutation relation ${A^{\rho}}_{;\mu \nu} - {A^{\rho}}_{; \nu \mu} = R\indices{^{\rho}_{\sigma \mu \nu}} A^{\sigma}$, and $R\indices{^{\rho}_{\sigma \mu \nu}}$ is the Riemann curvature tensor whose contraction over the first and third indices is the Ricci curvature tensor $R_{\mu \nu}$. This is an intractable equation as it stands. However, the vector potential $A^{\mu}$ is neither classically observable nor uniquely determined by the field. Any gauge transformation $A_{\mu} \rightarrow A_{\mu} + s_{,\mu}$ where $s$ is a scalar function leaves $F_{\mu \nu}$ unchanged. We can therefore exploit this gauge invariance to simplify the problem, by \textit{fixing} the gauge. The Lorenz gauge condition in curved spacetime is given by
\begin{equation}
    {A^{\mu}}_{;\mu} = 0
\end{equation}
and so we have the wave equation, or d'Alembert equation, for the vector potential
\begin{equation}
      - {A^{\mu ; \nu}}_{;\nu} + R\indices{^{\mu}_{\nu}} A^{\nu} = \frac{4 \pi}{c} j^{\mu}.
\end{equation}
The operator in the first term is the covariant d'Alembertian and may also be written using the notation $\Box \equiv \nabla^{\nu} \nabla_{\nu}$. As our purpose is the analysis of electromagnetic waves (that is, electromagnetic fields occurring in vacuum in the absence of charges), the electromagnetic wave equation henceforth becomes
\begin{equation}
\boxed{
      {A^{\mu ; \nu}}_{;\nu} - R\indices{^{\mu}_{\nu}} A^{\nu} = 0
      } \label{waveeqnvectorpotential}.
\end{equation}

\subsection{The Geometric Optics Approximation} \label{sec:geometricoptics}

Maxwell's equations cannot be solved explicitly in curved spacetime, except in cases of high symmetry. In particular, the simple plane wave solutions (characterised by the property that the direction of propagation of the wave, and its amplitude, are the same everywhere) available in flat spacetime do not exist. However, many arbitrary electromagnetic waves can be considered planar within a small region of space: the amplitude and direction of the wave is practically constant over distances of the order of its wavelength and also the curvature of spacetime. If this condition is satisfied, we can construct the \textit{wave-surface} or \textit{wavefront} as a hypersurface (3-surface) everywhere at which the phase $\varphi$ of the wave is the same; and the wave(-co)vector $k_{\mu} \equiv \varphi_{,\mu}$ defines the local direction of wave propagation. Rays can then be defined as curves whose tangents at each point are given by the direction of propagation of the wave.

Geometric optics provides a foundation for gravitational lensing and its construction connects it to the underlying fields. In the regime of geometric optics, electromagnetic waves can be treated as the propagation of rays, i.e. as classical point particles following well-defined trajectories and without wave properties (such as diffraction or interference). Bits of information making up a signal can be considered as discontinuities in the field determined by shorter wavelengths than the length scales associated with the curvature of either spacetime or the wave-surface \cite{Harte2019}. Thus the signal transmitted obeys geometric optics. Geometric optics is valid when 
\begin{equation}
    \frac{\omega}{c} l \gg 1
\end{equation}
where $\omega$ is the typical angular frequency and $l$ is a relevant length scale associated with
\begin{enumerate}
    \item the typical length over which the amplitude, polarisation and wavelength varies, such as the radii of curvature of the wavefront or the length of a wave packet produced by, for example, a flare from a quasar; and
    \item length scales in the spacetime curvature
\end{enumerate}
all as measured in a typical local frame (defining frequency in relativity requires reference to an observer). Several length scales may be present simultaneously, and different ones can be relevant for different observables \cite{Harte2019}. Only the first condition, known as the \textit{eikonal approximation}, is necessary in classical optics. As an example, the geometric optics approximation breaks down at \textit{caustics}, predicting infinite intensity where light rays intersect and a radius of curvature of the wavefront $l \to 0$ \cite{Landau1951}. There, wavefronts cannot be considered smooth sub-manifolds \cite{Hasse1996}, forming features such as cusps.

We start with a high-frequency, WKB-like ansatz for the geometric optics vector potential $A^{\mu}$ in order to solve the wave equation. Let $A^{\mu}$ be expanded by the parameter $\epsilon$
\begin{align}
    A^{\mu} (\mathbf{x} ; \epsilon ) &= \operatorname{Re} \left\{ \exp{\left(i \epsilon^{-1} \varphi (\mathbf{x})\right)} \sum_{n=0}^{\infty} \epsilon^{n} A_{(n)}^{\mu} (\mathbf{x}) \right\}\\
    &= \operatorname{Re} \left\{ \exp{\left(i \epsilon^{-1} \varphi (\mathbf{x})\right)}  \left(a^{\mu}(\mathbf{x}) + \epsilon A_{(1)}^{\mu} (\mathbf{x}) + ...  \right) \right\} \label{ansatz}
\end{align}
where we denote $a^{\mu} \equiv A_{(0)}^{\mu}$. The dummy expansion parameter $\epsilon$ tracks how rapidly various terms approach $0$ in calculations and can be eventually set to $1$; one could instead factor out $\omega^{-1}$ from the phase as the expansion parameter \cite{Harte2019, dolan2018}. The phase function $\varphi$ or the \textit{eikonal} does not necessarily have the simple form $\varphi = k_{\mu}x^{\mu} + \text{const.}$ corresponding to a plane wave \cite{Landau1951}; however {\hl in the geometric optics limit} it is necessarily large (since $\frac{\omega}{c} \equiv | \frac{d \varphi}{d (c\uptau)}|$). Truncation of the series at the leading order $n = 0$ corresponds to the geometric optics limit whereas higher order terms give post-geometric-optics corrections. Mathematically speaking, the geometric optics expansion allows us to convert PDEs into an infinite hierarchical sequence of ODEs (cf. Section \ref{sec:perturbationtheory}).

The amplitude is orthogonal to the wavevector $k_\mu a^\mu =0$, i.e. the electromagnetic wave is transverse, only at leading order according to the Lorenz gauge condition on the vector potential ansatz. Inserting the expansion \eqref{ansatz} to leading order into the wave equation \eqref{waveeqnvectorpotential}, we find
\begin{equation}
    \Box a^{\mu}- a^{\mu} \epsilon^{-2} k_{\nu} k^{\nu}  - R\indices{^{\mu}_{\nu}} a^{\nu} + i \epsilon^{-1} \left( 2 {a^{\mu}}^{;\nu} k_{\nu} + a^{\mu} k\indices{_{\nu}^{;\nu}} \right) = 0. \label{ansatzinwaveeqn}
\end{equation}
{\hl Taking the real part only implies that the last term in \eqref{ansatzinwaveeqn} is 0; and also that $\Box a^{\mu} - R\indices{^{\mu}_{\nu}} a^{\nu} - a^{\mu} \epsilon^{-2} k_{\nu} k^{\nu} = 0$. In addition, we have from the geometric optics approximation (and assuming $A^{\mu} (\mathbf{x} ; \epsilon ) = a^{\mu} \neq 0$) that $\Box a^{\mu} - R\indices{^{\mu}_{\nu}} a^{\nu} \ll a^{\mu} \epsilon^{-2} k_{\nu} k^{\nu}$.} This gives two equations
\begin{gather}
 k_{\nu} k^{\nu} = \varphi_{, \mu} \varphi^{, \mu} = 0 \label{nullwavevector} \\ 
2{a^{\mu}}^{;\nu} k_{\nu} + a^{\mu} k\indices{_{\nu}^{;\nu}} = 0 \label{evolutionequation}
\end{gather}
the first of which is a constraint equation, and the second of which is the evolution of the amplitude.

The constraint equation \eqref{nullwavevector} is the condition that the wavevector is null, or the eikonal equation when written in terms of the phase function. The eikonal equation is the same as the Hamilton-Jacobi equation \cite{Ehlers2000} generalised from classical mechanics. The evolution of the amplitude contains both photon conservation and the parallel transport of the polarisation state (i.e. the direction of the vector amplitude).

The Faraday tensor in the geometric optics regime may also be found by inserting the expansion \eqref{ansatz} into its definition {\hl \eqref{faradaytensor} and keeping only terms of order $\epsilon^{-1}$. Writing $A^{\mu} = \operatorname{Re} \left\{ a^{\mu} \exp{\left(i \epsilon^{-1} \varphi (\mathbf{x})\right)} \right\}$, then setting $\epsilon =1$ we have}
\begin{align}
    F_{\mu \nu} =  \operatorname{Re} \left\{\exp{\left(i \epsilon^{-1} \varphi \right)} \left(a_{\nu; \mu} - a_{\mu; \nu} + i \epsilon^{-1} (a_\nu k_\mu - a_\mu k_\nu) \right) \right\} = \operatorname{Re} \left\{2 i A_{[\nu}k_{\mu]}\right\}.
\end{align}
Taking the covariant derivative of the null wavevector condition \eqref{nullwavevector}
\begin{equation}
    ( k_{\nu} k^{\nu} )_{; \mu} =  2 k^{\nu} k_{\nu ; \mu} =0
\end{equation}
and using the property that as $k_{\nu}$ is a gradient that the indices of $k_{\nu ; \mu}$ can be swapped due to the symmetry of the Christoffel coefficients $k_{\nu; \mu} = (\varphi_{, \nu})_{; \mu} = (\varphi_{, \nu})_{,\mu} - \Gamma\indices{^{\rho}_{\mu \nu}} \varphi_{, \rho}$  gives
\begin{equation}
    \boxed{
    k^{\nu}k\indices{^{\mu}_{;\nu}} = 0\label{nullgeodesic1}
    }.
\end{equation}
The light rays are the integral curves $x^{\mu}(\sigma)$ of the vector field of $k^{\mu}$ (i.e. the curves to which $k^{\mu}$ are everywhere tangent), implying $\varphi^{,\mu} = k^{\mu} = \frac{d x^{\mu}}{d \sigma}$ where $\sigma$ is an affine parameter. It is also valid in geometric optics to consider light rays as the worldlines of photons with four-momenta $p^\mu = \hbar k^{\mu}$ where $\hbar$ is the reduced Planck constant. In terms of the light rays, and writing out the covariant derivative in terms of the Christoffel symbols, equation  \eqref{nullgeodesic1} becomes
\begin{equation}
    \frac{d^2 x^{\mu}}{d\sigma^2} + \Gamma\indices{^{\mu}_{\nu \rho}} \frac{dx^{\nu}}{d \sigma}\frac{dx^{\rho}}{d \sigma}  = 0 \label{nullgeodesic2}
\end{equation}
which explicitly shows that light rays are null geodesics. These null geodesics form a ``twist-free'' congruence away from caustics, i.e. neighbouring rays do not rotate, since in the geometric optics regime $k^\mu$ can be written as the gradient of a scalar field (i.e. it is orthogonal to the constant phase hypersurface) \cite{Harte2019, Griffiths2009}. They need not be the $45^{\circ}$ lines of Minkowski diagrams (plotted against coordinate time) but are curved due to gravity; and neighbouring geodesics can expand or contract, and shear from one another in a general spacetime. As we will explore in Section \ref{sec:fermat}, the gravitational field can be thought of as acting like a refractive medium, with typically a spatially-varying refractive index, since the \textit{in vacuo} speed of light is a constant $c$ only in flat spacetime (or in a local inertial frame). In fact it is the only Lorentz scalar which is not a scalar in general relativity \cite{Hehl2018}.

As illustrated in Figure \ref{fig:flrwredshift} of this work and also Figure 2.7 and Box 22.3 of \cite{Misner1973}, that since $k^\mu \varphi_{,\mu} =0$ that even though the $k^\mu$ is orthogonal to the hypersurfaces of constant $\varphi$ (wavefronts), $k^\mu$ also lies on one of these wavefronts due to the Lorentzian geometry: $k^\mu$ is self-orthogonal and non-zero. For conformally static spacetimes, such as the FLRW, we are able to foliate spacetime in hypersurfaces of simultaneity, three-dimensional spatial surfaces $\Sigma_t$. The intersection of a 3-surface of simultaneity with a 3-surface of constant phase or wavefront is the \textit{instantaneous wavefront}, a  (generically) two-dimensional surface of constant phase. The instantaneous wavefront is the concept which is more familiar from classical optics: it is perpendicular to the spatial projection of the wavevector $k^i$. 

A \textit{light cone} is the 3-surface formed by the locus of null geodesics, i.e. possible light rays, through a given event \cite{Carroll2004}. The light cone can be considered a special case of a wavefront \cite{Perlick2004} for a small (spherical) source emitting in all directions; caustics are the 2-surfaces formed by the self-intersections of light cones. A general wavefront can be constructed as the envelope of all light cones with vertices on the original wavefront. This is in accordance with Huygens's principle, and is in fact an alternative statement of a generalised Fermat principle \cite{Kovner1990} which we will discuss in Section \ref{sec:fermat}: the worldline of an observer crosses the light cone at one or more observation events; and of all null curves connecting the observer worldline with the emission event only those which lie on the light cone are the true light rays.

We summarise the fundamental laws of geometric optics contained in the solution \eqref{nullwavevector} and \eqref{evolutionequation} to the wave equation
\begin{enumerate}
    \item Electromagnetic fields propagate as \textit{light rays} along null geodesics. (Additionally, the phase is constant along rays. In conformally static spacetimes, this can be visualised as a 2-surface of constant phase advancing in time through three-dimensional space along the rays.)
    \item The polarisation vector is perpendicular to, and parallel transported along the rays.
    \item The number of photons is conserved; or in strictly classical language, intensity obeys an inverse (cross-sectional) area law \cite{Misner1973}.
\end{enumerate}

\section{The Einstein Field Equations in Perturbed FLRW Cosmology}

The Einstein field equations may be also be motivated from an action principle, where the action is the sum of the contributions from the gravitational field (the Einstein-Hilbert action) and any sourcing fields, $S = S_g + S'$ \cite{ Dirac1975, Landau1951}. In general relativity, the metric is the field quantity, rather than the potential of Newtonian gravity; we vary the action with respect to the metric. Doing so, the Einstein-Hilbert contribution gives the Einstein tensor scaled by a coupling constant, i.e. the vacuum Einstein field equations, and the remaining term from $S'$ gives the stress-energy tensor by definition. 

In Section \ref{efeflrw} we explored the Einstein equations for the FLRW cosmology, known as the Friedmann Equations, which gave the evolution of that metric -- i.e. the scale factor. We may make this background FLRW description of the Universe more realistic by adding small inhomogeneities, or perturbations, at linear order and beyond to the background metric and the background stress-energy tensor. Solving the perturbed Einstein field equations allows us to understand the formation and evolution of large-scale structures. We here first find the local Newtonian correspondence, then describe in brief the general perturbations one can construct to the background metric and stress-energy tensor, before taking the Newtonian limit. We refer the interested reader to \cite{Baumann2022, Bertschinger1995, Ma1995} and \cite{Amendola2010} for a full presentation of cosmological perturbation theory (but note \cite{Baumann2022} sometimes uses $x^0 = t$ or $\eta$, rather than our convention of $x^0 = ct$ or $c \eta$ so e.g. the stress energy tensor will appear to carry different factors of $c$, and symbols for the Bardeen potentials are swapped).  It is also standard to work under the assumption of a spatially flat background metric, as this greatly simplifies the calculations \cite{Bertschinger1995}, although they are generalisable to the cases of positive and negative constant curvature. We therefore write all expressions in the following subsections \ref{subsec:perturbedmetric} and \ref{subsec:perturbedstressenergy} unless otherwise stated for the spatially flat case.

\subsection{The Newtonian Correspondence} \label{newtoniancorrespondence}

The equivalence principle, which in its Einstein form can be stated as ``we cannot locally distinguish free fall in a gravitational field from uniform motion in the absence of a gravitational field'', lies at the heart of general relativity. This leads to the local flatness theorem, meaning that at any particular point $\mathscr{E}$, we are able to make a local coordinate transformation from the global coordinates covering the entire spacetime to coordinates which define the local inertial frame or Lorentz frame given by the Minkowski metric: $g_{\mu \nu} (\mathscr{E}) = \eta_{\mu \nu}$ and $\partial_{\rho} g_{\mu \nu} ( \mathscr{E}) = 0$. Geometrically, we could imagine a $1$D curve representing a one-dimensional spacetime, and being able to use a local tangent as a first-order approximation to the curve at that point.

The equations of general relativity must reduce to the Newtonian ones under suitable limiting conditions; this is the intuitive notion that there must be a correspondence between the theory of general relativity and Newtonian theory which adequately describes the everyday (weak-field, low-velocity) world. To find this correspondence, we can expand the metric around a local Minkowski frame $\eta_{\mu \nu}$ by virtue of the local flatness theorem, $g_{\mu \nu} = \eta_{\mu \nu} + \delta g_{\mu \nu}$ where the weak-field limit implies $|\delta g_{\mu \nu}| \ll 1$. We obtain from the line element, using $-c^2 d \uptau^2 = ds^2$ and taking the low-velocity limit where $u^i = \frac{dx^i}{d(ct)}$ also is a perturbative term, $d \tau = \sqrt{(1 - \delta g_{00}) - (u^i)^2} dt$. Inserting this result into the action for a free massive particle and taking a Taylor series expansion gives {\hl $S_p = -mc^2 \int d\uptau \approx -mc^2\int 1 - \frac{1}{2}\left( \delta g_{00} + (u^i)^2 \right) dt$}. The correspondence with the Newtonian action for a massive particle in a  gravitational field demands
\begin{equation}
    \delta g_{00} = -\frac{2\Phi}{c^2}
\end{equation}
where $\Phi$ is the Newtonian gravitational potential. We see that although the components of $\delta g_{\mu \nu}$ other than $\delta g_{00}$ are of the same order of magnitude, they are not first-order in the Lagrangian and cannot be determined by this method: only $\delta g_{00}$ is required for \textit{passive} gravitational effects -- i.e. considering the motion of a test massive particle in the field, rather than considering the particle as source of gravitation.

Likewise the Newtonian correspondence to the Einstein field equations is the Poisson equation. The Einstein field equations can be written in the form $R_{\mu \nu} = \kappa (T_{\mu \nu} - \frac{1}{2} g_{\mu \nu} T\indices{^\mu_\mu})$, by first contracting the indices in the field equations to get $R = - \kappa  T\indices{^\mu_\mu}$. For comparison with the cosmological derivation in the next section where we do not restrict to a local frame, it is desirable to explicitly distinguish between contributions from the Minkowski background and the perturbation, so we write $R\indices{^\mu_\nu} = \bar{R}\indices{^\mu_\nu} + \delta R\indices{^\mu_\nu}$ and $T\indices{^\mu_\nu} = \bar{T}\indices{^\mu_\nu} + \delta T\indices{^{\mu}_{\nu}}$. At zeroth order, we simply have the vacuum equations $\bar{R}_{\mu \nu} =0$. At first order, we introduce the stress-energy tensor of a perfect fluid with negligible pressure (dust) compared with its energy density $|\delta P| \ll \delta \varepsilon \sim \delta \varrho c^2$ as $\delta T\indices{^\mu_\nu} = \delta \varepsilon u^\mu u_\nu$, where $\delta \varrho$ is the \textit{rest} mass density. Since the macroscopic motion must also be non-relativistic, $u^\mu = (1, 0, 0, 0)$ such that the only non-zero component of $\delta T\indices{^\mu_\nu}$ is $\delta T\indices{^0_0} = - \delta \varrho c^2$; and its trace $\delta T\indices{^\mu_\mu}$ has the same value. The Einstein field equations then reduce to $\delta R\indices{^0_0} = - \frac{1}{2} \kappa \delta \varrho c^2$. Calculation of $\delta R\indices{^0_0}$ from the definition simplifies, since derivatives of the Christoffel symbol are all second order in $\Phi$, and time derivatives have extra factors of $c^{-1}$ and are therefore small compared with spatial derivatives. The result is $ \delta R\indices{^0_0} = - \delta R_{00} = \delta \Gamma\indices{^{i}_{00,i}} \approx (-\frac{1}{2} g^{ij} \delta g_{00, j})_{, i} \approx - c^{-2} \bm{\nabla}^2 \Phi$. Thus we regain the Newtonian Poisson equation
\begin{equation}
    \bm{\nabla}^2 \Phi = 4 \pi G \delta \varrho. \label{NewtonianPoisson}
\end{equation}

We summarise the conditions which define the validity of the Newtonian regime as follows
\begin{enumerate}
    \item Distance scales are much smaller than the scale of the curvature\footnote{This is the extrinsic curvature defined for a spatial hypersurface embedded in spacetime rather than the intrinsic curvature of a 3-geometry by itself; see p. 537 of \cite{Misner1973} and also p.40 of \cite{Mukhanov2005}.} of spacetime, which characterises the size of the local inertial frame: in FLRW cosmology, this is given by the Hubble length $cH^{-1}$.
    \item We consider non-relativistic matter, so the pressure is much less than the energy density $|P| \ll \rho c^2$ and we replace the relativistic mass density $\rho$ with the mass density $\varrho$. In Newtonian gravity, the source of gravity is the mass density $\varrho$ rather than the energy density $\varepsilon = \rho c^2$.
    \item The fluid velocity (macroscopic flow) must also be non-relativistic: $|u^i| \ll 1$.
    \item We assume the weak-field limit $|\frac{\Phi}{c^2}| \ll 1$ which allows us to treat the equations perturbatively.
\end{enumerate}

{\hl In this work we are mostly focused on density perturbations forming isolated gravitational lenses, which} occur over scales significantly less than the Hubble length (i.e. we neglect effects from the large scale structure). {\hl Such a gravitational lens} fulfils these conditions and hence should obey the non-relativistic Poisson equation. However, this result needs to be interpreted with a little caution. Our discussion in this section applies to a local coordinate frame; one cannot\footnote{At least, not without borrowing results from modern cosmology and general relativity to correctly address boundary conditions \cite{Bertschinger1995}; doing so leads to a ``pseudo-Newtonian'' approach. However this framework is often inadequate (e.g. it cannot explain accelerated expansion {\hl other than by manually adding a cosmological constant to the Newtonian Poisson equation, but mainly as it does not correctly describe light: it cannot give a radiation-dominated Universe, nor correctly predict gravitational lensing).}} use Newtonian theory to derive cosmological dynamics, for example. As a gravitational lens exists in a cosmological setting, described by a perturbation to a homogeneous background density, we find the first-order perturbed \textit{cosmological} Einstein field equations to make this interpretation explicit.

\subsection{A Note on Perturbation Theory} \label{sec:perturbationtheory}

Cosmology is largely the realm of linear and weakly non-linear processes -- in particular when considering gravity \cite{Amendola2010}. Although a lot of cosmological information in large-scale structure is contained on non-linear scales \cite{Joyce2016}, when equations describing gravitational instability become \textit{highly} non-linear, small-scale astrophysical objects such as galaxies or stars form and the dependence on the global properties of spacetime is swamped by new interactions. Perturbation theory is therefore perhaps one of the most utilised mathematical methods in cosmology. For example, weakly non-linear large-scale structure formation can be studied using a perturbative approach (e.g. the famous Zeldovich approximation based on linear Lagrangian Perturbation Theory), and compared to N-body simulations which can deal with up to $10^9$ particles at a time.

Perturbation theory is a collection of iterative methods for the systematic analysis of the global behaviour of solutions to differential and difference equations \cite{Bender2014}. Generally, we identify a small expansion parameter $\epsilon$ in the equation such that the problem can be solved when $\epsilon = 0$. A perturbative solution is computed by expanding around $\epsilon = 0$ as a series of powers of $\epsilon$; this is local only in the \textit{parameter} $\epsilon$ rather than any variable. If $\epsilon$ is very small, we expect that the true solution is well approximated by only a few terms of the perturbation series. Often we cannot identify a suitable expansion parameter, so it can be introduced artificially before being set to unity in final expressions: informally, we do not need to explicitly include $\epsilon$ when this is the case.

Typically, we approximate the solution of a nonlinear partial differential equation by considering it as a perturbation of a linear equation. This approximate solution is constructed iteratively, with the $n^{{th}}$ order solution being inserted as a source term to determine the ${n+1}^{{st}}$ order solution. In this case, the first order solutions to the (density, velocity, ...) fields are linearly dependent on the initial data. The second order solutions are quadratic in the initial fields, and so on. The physical meaning of the perturbative power series is the description of transition from linear to non-linear dynamics. The variance $\sigma^2$ of the linear fluctuations describes the transition from the linear to the non-linear regime \cite{Scoccimarro1997}. The linear regime is when $\sigma^2 \ll 1$, and the highly non-linear regime is when $\sigma^2 \gg 1$ {\hl (equivalently $\epsilon \gg 1$)}. Perturbation theory provides a framework for calculations (e.g. of fields or statistical functions) in the weakly non-linear regime corresponding to the first few terms in the expansion. For the highly non-linear regime, all terms in the summation are of the same order so perturbation theory is no longer useful.

For example, the complete set of non-linear equations (i.e. the full EFEs plus either the fluid or kinetic equations) is very difficult to solve; this is true even if we linearise gravity but not the hydrodynamic equations -- they become (in Fourier space) coupled integro-differential equations. Perturbation theory allows us to estimate the deviations from the exact linear and background solutions.

\subsection{The Perturbed Metric} \label{subsec:perturbedmetric}

Consider the most general perturbed FLRW metric $g_{\mu \nu} = \bar{g}_{\mu \nu} + \delta g_{\mu \nu}$: $\delta g_{00}$, $\delta g_{0 i}$ and $\delta g_{ij}$ respectively transform as a scalar, vector and tensor under 3-rotations (that is, with respect to the metric of the spatial hypersurfaces). We further decompose $\delta g_{0 i}$ and $\delta g_{ij}$ into (3-)scalar, (3-)vector and (3-)tensor components according to the Helmholtz theorem (or ``SVT'' decomposition). The scalar, vector and tensor perturbations are completely decoupled when considering the Einstein equations at linear order and can therefore be treated separately.

Scalar perturbations to the metric are induced by inhomogeneities in the energy density; that is, perturbations in density and isotropic pressure, as well as in irrotational part of the velocity field and the anisotropic stress. They are the focus of cosmological perturbation theory as they show gravitational instability and are the basis for structure formation in the Universe. Vector perturbations in the metric are related to rotational velocity (vortical) perturbations, for which gravity provides no source. If the vortical perturbations are initially zero they remain zero; if present initially they decay quickly proportional to $a^{-1}$ \cite{Bertschinger1995, Lyth1993}. Tensor perturbations have no analogue in Newtonian theory; they represent gravitational waves, coupled to matter only for anisotropic perturbations and so have no effect on a perfect fluid to first order. As we are only interested in the scalar perturbations decoupled from the vector and tensor perturbations, this leaves four scalar functions $ \hat{\Phi}, \hat{\Psi}, E$ and $B$, defining $D_{i j} B \equiv (\nabla_i \nabla_j - \frac{1}{3} \tilde{\gamma}_{i j} \bm{\nabla}^2)B$, in the metric
\begin{equation}
\delta g_{\mu \nu} = a^2 \begin{bmatrix}
-2\hat{\Phi} & -E_{,i} \\
-E_{,i} & -2\hat{\Psi} \tilde{\gamma}_{i j} + 2D_{i j}B
\end{bmatrix}. \label{perturbedscalarmetric}
\end{equation}

\subsubsection{Gauge Freedom}
As a consequence of the invariance of the spacetime interval $ds^2 = g_{\mu \nu} dx^\mu dx^\nu$ under general coordinate transformations, the metric and matter perturbations must take on different values in different coordinate systems. In other words, there is an ambiguity to splitting a quantity such as the energy density or pressure into a homogeneous background component and a perturbation, since spatial averaging depends on the choice of constant time hypersurfaces. We may therefore think of choosing a \textit{gauge} as a choice of threading and slicing for the perturbed spacetime. This particular gauge freedom is unique to cosmological perturbation theory, additional to the usual gauge freedom of general relativity \cite{Bruni1999}.

Although there are ten degrees of freedom or equations in relating $\delta g_{\mu \nu}$ to the stress-energy tensor, four are redundant, corresponding to \textit{gauge modes}. They arise from the freedom to choose the mapping, i.e. the four functions $\epsilon^\mu (\eta, x^i)$ in an infinitesimal coordinate transformation $x'^\mu = x^\mu + \epsilon^\mu (\eta, x^i)$ under which the spacetime interval remains invariant. This can also be surmised from the fact that the Einstein field relations are also related by the four identities $T\indices{^\mu_{\nu ; \mu}} =0$. Constraining or fixing the allowed transformations $\epsilon^\mu$ corresponds to choosing a gauge: a choice of parametrisation of the physical degrees of freedom in $\delta g_{\mu \nu}$ with convenient properties.

In the Newtonian gauge, we choose to impose the conditions $\delta g_{0 i} = 0$ such that $E=0$, and $B=0$ on the metric. In this gauge, the former condition means the hypersurfaces of constant time are orthogonal to the worldlines of comoving observers, and latter means the spatial hypersurfaces are isotropic. It is a \textit{completely fixed} gauge, as the transformation variables are completely specified: this is not always true, for example with the synchronous gauge. This gives the perturbed FLRW metric as
\begin{equation}
    g_{\mu \nu} dx^{\mu} dx^{\nu}  = a^2(\eta) \left(-\left(1+2\hat{\Phi}\right) d(c \eta)^2 + \left(1 -2 \hat{\Psi} \right) \tilde{\gamma}_{ij} dx^i dx^j \right) \label{perturbedFLRW}
\end{equation}
where the dimensionless $\hat{\Phi}$ and $\hat{\Psi}$ are the gauge-invariant Bardeen potentials. In the non-relativistic approximation, the interpretation of these is that $\Phi = \hat{\Phi} c^2$ is the Newtonian potential and $\Psi = \hat{\Psi} c^2$ is the Newtonian spatial curvature, making the Newtonian gauge useful for the analysis of the formation of large-scale structures and CMB anisotropies. However, as we will see, there are relativistic corrections to the Poisson equation; therefore, it can be argued that neither Bardeen potential is the \textit{same} potential as that of Newtonian gravity. Hence the potential which satisfies the Newtonian Poisson equation is sometimes denoted differently, e.g. \cite{Amendola2010}; we make the distinction by writing the Bardeen potentials in dimensionless form and notated using {\^{}}. The relativistic corrections also make a straightforward implementation of the equations numerically unstable \cite{Hu2004}. There may be many alternate gauges that are chosen for other conveniences. For example, the synchronous gauge, was used in much of the early work \cite{Bardeen1980} as it was favoured for its numerical stability: this corresponds to a threading of the perturbed spacetime or Lagrangian framework of fluid dynamics as opposed to an Eulerian one \cite{Bertschinger1995, Villa2014}.

\subsection{The Perturbed Stress-Energy Tensor} \label{subsec:perturbedstressenergy}

The background stress-energy tensor is that of a perfect fluid in its rest frame, $\bar{T}\indices{^{\mu}_{\nu}} = \text{diag}(- \bar{\varepsilon}, \bar{P}, \bar{P}, \bar{P})$. The perturbed stress energy tensor $T\indices{^{\mu}_{\nu}} = \bar{T}\indices{^{\mu}_{\nu}} + \delta T\indices{^{\mu}_{\nu}}$ is given by
\begin{equation}
T\indices{^{\mu}_{\nu}} = \left(\varepsilon + P\right)u^{\mu}u_{\nu} + P g\indices{^{\mu}_{\nu}} + \pi\indices{^{\mu}_{\nu}}
\end{equation}
where the energy density $\varepsilon \equiv \rho c^2$ and pressure $P$ are always the proper quantities. We have introduced an anisotropic stress term $\pi\indices{^{\mu}_{\nu}}$ which is traceless ($\pi\indices{^{\mu}_{\mu}} =0$) and flow-orthogonal ($\pi\indices{^{\mu}_{\nu}}u^{\nu} =0$); as any trace terms can always be absorbed into the isotropic pressure. In terms of a rest-frame quantity $\Pi\indices{^{\mu}_{\nu}}$, the anisotropic stress becomes a traceless spatial tensor where $ \Pi^{00} = \Pi^{0i} = 0$ and $\Pi\indices{^i_i} =0$.

To find the quantities $\delta T\indices{^{\mu}_{\nu}}$ we must first find the components of both the four velocity and its dual in a general frame of reference. As we are only interested in the scalar perturbations in the Newtonian gauge, we use the metric \eqref{perturbedFLRW} rather than the general perturbed metric to obtain the components of the four velocity through the normalisation condition $u^{\mu} u_{\nu} = -1$, noting the peculiar velocity is $v^i \equiv \frac{dx^i}{d(c\eta)} = \frac{u^i}{u^0}$. Solving $ (u^0)^{-2} = -g_{00} - g_{ii} (v^i)^2$ to first order, we get $u^0 = a^{-1} (1- \hat{\Phi})$; and so $u^i = u^0 v^i= a^{-1} v^i$ to first order. Lowering with the metric to first order gives $u_0 = -a(1 + \hat{\Phi})$ and $u_i = av_i$. Computing each component of the perturbed stress-energy tensor to first order and subtracting the unperturbed stress energy tensor yields
\begin{gather}
    \delta T\indices{^0_0} = -\delta \varepsilon \\
    \delta T\indices{^i_0} = -\left(\bar{\varepsilon} + \bar{P}\right)v^i \\
    \delta T\indices{^0_i} = \left(\bar{\varepsilon} + \bar{P}\right)v_i \\
    \delta T\indices{^i_j} = \delta P \delta\indices{^i_j} + \Pi\indices{^i_j}
\end{gather}

\subsection{Linearised Einstein Field Equations}
To find the Einstein equations for perturbations
\begin{equation}
\delta G\indices{^\mu_\nu} = \kappa \delta T\indices{^\mu_\nu}
\end{equation}
where $\kappa \equiv \frac{8 \pi G}{c^4}$ we must first compute the Christoffel symbols of the metric \eqref{perturbedFLRW}, then from those compute the Ricci tensor, and from that compute the Einstein tensor according to their definitions. These calculations are long but a straightforward application of Ricci calculus. We therefore omit the details which may be found in e.g. \cite{Baumann2022} and write the $00$ (energy equation), $0$i (momentum equation), ii (the spatial trace, or acceleration equation) and ij (anisotropy equation) parts of the linearised Einstein field equations respectively as 
\begin{gather}
 \bm{\nabla}^2 \hat{\Psi} -3 \mathscr{H}(\dot{\hat{\Psi}} + \mathscr{H} \hat{\Phi}) =\tfrac{\kappa}{2} a^2 \delta \varepsilon\\
\dot{\hat{\Psi}} + \mathscr{H}\hat{\Phi} = -\tfrac{\kappa}{2} a^2 \left(\bar{\varepsilon} + \bar{P}\right)v\\
-\tfrac{1}{3}\bm{\nabla}^2(\hat{\Psi} - \hat{\Phi}) + \ddot{\hat{\Psi}} +2 \dot{\mathscr{H}}\hat{\Phi} + \mathscr{H}^2 \hat{\Phi} + 2 \mathscr{H}\dot{\hat{\Psi}} + \mathscr{H}\dot{\hat{\Phi}} = \tfrac{\kappa}{2} a^2 \delta P\\
\hat{\Psi} - \hat{\Phi} = \kappa a^2 \Pi
\end{gather}
where overdots represent partial differentiation with respect to $c \eta$ where $\eta$ is conformal time, $\mathscr{H} \equiv \frac{\dot{a}}{a}$ is the Hubble parameter with respect to $c \eta$, $v$ is the scalar field whose gradient is the irrotational part of the Helmholtz-decomposed peculiar velocity $v^i$, so $v^i = \nabla^i v+ v^{\bot i}$, and similarly $\nabla^i \nabla_j \Pi$ is the scalar component of the anisotropic stress (see e.g. \cite{Baumann2022, Bertschinger2000, Bertschinger1995} or \cite{Amendola2010} for details). The energy and acceleration equations are the perturbed generalisations of the Friedmann equations; whereas the momentum and anisotropy equations are new to the perturbed metric. These four equations have only two linearly independent combinations; we can combine the energy and momentum equations to give a relativistic Poisson equation in a background-FLRW universe, of the same form as the Newtonian Poisson equation
\begin{equation}
    \bm{\nabla}^2 \hat{\Psi} =\frac{4 \pi G}{c^4} a^2 \left(\delta \varepsilon - 3 \mathscr{H} \left(\bar{\varepsilon} + \bar{P} \right)v\right) \label{relativisticPoisson}
\end{equation}
as well as the acceleration and anisotropy equations to give
\begin{equation}
    \ddot{\hat{\Psi}} +2 \dot{\mathscr{H}}\hat{\Phi} + \mathscr{H}^2\hat{\Phi} + 2 \mathscr{H}\dot{\hat{\Psi}} + \mathscr{H}\dot{\hat{\Phi}} = \frac{4 \pi G}{c^4} a^2  (\delta P + \tfrac{2}{3}\bm{\nabla}^2 \Pi).
\end{equation}
The relativistic cosmological Poisson equation \eqref{relativisticPoisson} differs from the Newtonian case \eqref{NewtonianPoisson} by the addition of a relativistic term involving $\mathscr{H}$, which disappears for small scales inside the Hubble radius. In the case that the anisotropic stress $\Pi$ is negligible compared to the energy density perturbation, $\hat{\Psi} = \hat{\Phi}$; this is a good approximation for non-relativistic matter and in general for the matter-dominated cosmological era -- then there is only one scalar field, which may be interpreted as the Newtonian gravitational potential $\frac{\Phi}{c^2} = \hat{\Psi} = \hat{\Phi}$. We here omit presentation of the continuity and Navier-Stokes equations, which may be found in e.g. \cite{Baumann2022, Amendola2010}.

We note that equations of motion in real space are partial differential equations; in particular, the Poisson equation contains a Laplacian. Taking the Fourier transform in the spatial variables in flat space converts PDEs with spatially constant coefficients into ODEs, which are generally easier to solve. Working in Fourier space also helps keep track of the length scales of the perturbations and various terms, and different Fourier modes are decoupled at linear order since the linearised equations of motion are translationally invariant. The physical interpretation is this: we might initially consider dividing the evolving matter distribution into separate volume elements. However, independent evolution would quickly break down due to gravitational interactions between neighbouring volumes, so it is not ideal to view a general perturbation as a sum of spatial components. Instead, if we consider the perturbation as a combination of plane waves, each plane wave will evolve independently in the linear regime.

\section{The Quasi-Newtonian Approximation of Strong Gravitational Lensing} \label{quasinewtonianformalism}

The fundamental equations of gravitational lensing in cosmology are the lens equation and the expression for the deflection angle. A perturbative approach to strong gravitational lensing, rigorously deriving the cosmological lens equation and the deflection angle for curved FLRW spacetimes, is attributed originally to Pyne and Birkinshaw \cite{PyneBirkinshaw1993, Pyne1996}. It has since been incorporated in various forms, mostly working with a local perturbed Minkowski metric in the lens frame, in texts such as \cite{Carroll2004, Dodelson2017, BovyInPrep}. The disadvantage of working locally at the lens is that this loses the generality of the initial work which does not begin with all approximations at once nor treats the path of the light ray differently near to and far from the lens; the advantage is that the derivation is in some ways cleaner. {\hl Other derivations are also found in  the literature e.g. \cite{Bartelmann2001, Bartelmann2010} and p. 356 of \cite{ Schneider2006}, all based in principle on \cite{Seitz1994}; we briefly discuss this method including some subtleties and limitations in Subsection \ref{subsec:perturbedgde}.} A covariant formalism for lensing in completely general spacetimes was recently presented in \cite{Fleury2021}.

It is also possible to invoke an effective refractive index to obtain equations of motion for light propagation for conformally static spacetimes in which the spatial metric is conformal to the Euclidean metric, rather than directly working with the null geodesic equations; this is the approach followed by many textbooks including \cite{Meneghetti2021, Petters2001, Schneider1992}. This is less appealing from an initial pedagogical perspective as firstly the link between the null geodesics and the lens equation may be obscured; and secondly the justification for using an effective refractive index, i.e. formally showing that the general form of Fermat’s principle takes the same form as in ordinary optics in this special class of metrics, is not usually given. However, the general form of Fermat's principle also leads directly to the lens equation and forms an alternative and powerful variational approach which is particularly useful for strong lensing. Therefore, we will follow a perturbative approach in this section before examining Fermat's principle in detail in Section \ref{sec:fermat}. Perlick \cite{Perlick2004} describes the situation {\hl as presented in many sources:}
\begin{displayquote}
The quasi-Newtonian approximation formalism of lensing comes in several variants, and the relation to the exact formalism is not always evident because sometimes plausibility and ad-hoc assumptions are implicitly made[...]
It is not satisfactory if the quasi-Newtonian formalism of lensing is set up with the help of ad-hoc assumptions, even if the latter look plausible. From a methodological point of view, it is more desirable to start from the exact spacetime setting of general relativity and to derive the quasi-Newtonian lens equation by a well-defined approximation procedure.
\end{displayquote}
By following such an approach, the relationship between strong and weak lensing which are often treated separately, additionally becomes clearer. In my following interpretation, I attempt to clearly and methodologically give a well-defined approximation procedure {\hl guided by the framework of \cite{PyneBirkinshaw1993, Pyne1996}}, but otherwise to the best of my knowledge is a unique presentation. The assumptions we will take for the quasi-Newtonian approximation may be summarised as follows 
\begin{enumerate}
    \item Geometrical optics limit -- The lens is transparent, and the wavelength of light is small compared with all other relevant length scales (e.g. the scale of astrophysical objects): this is mostly justified for gravitation lensing. However, in some exceptional cases dealing with e.g. caustic crossing events, a wave optic treatment of lensing \cite{Nakamura1999} is required. For example, the time delay between pairs of images near a fold caustic can become arbitrarily small as they merge \cite{Oguri2019}.
    \item Weak-field, small-angle or flat sky approximation -- The weak-field limit of gravity is valid for almost all relevant astrophysical scenarios \cite{Narayan1997}. It allows for a perturbative approach, leads to assumption that deflection angles are small, and {\hl when considering small sources} also allows us to approximate the celestial sphere as a plane.
    \item Metric -- The background metric is the FLRW metric of standard cosmology, and cosmological distances are angular size distances. Although this may seem like a straightforward assumption, this may not necessarily be the case. The correct notion of cosmological distances in strong lensing corresponding to the choice of the appropriate background metric, with an alternative being the Dyer-Roeder distance, was of great concern in the early literature \cite{Helbig2020}. Alternative metrics are still explored in contemporary research \cite{clarkson2012, fleury2013, Fleury2014, Fleury2015b}.
    \item Born approximation -- Assumes that the potential along the background path is negligibly different to the potential along the true path; it is valid if deflection angles are small. 
    \item  {\hl Conformally static} lens -- The source, gravitational lens and observer are comoving with the Hubble flow, and the mass distribution of the lens is independent of {\hl conformal} time.
    \item Thin lens or sudden deflection approximation -- Light is deflected at a single line-of-sight distance, in the lens plane. The extent of the lens is therefore assumed to be small compared with its distance from both the source and the observer.
\end{enumerate}

Assumptions 1-4 are also valid for weak lensing by large-scale structures. When considering weak lensing of the CMB it is possible to additionally relax the flat-sky approximation \cite{Lewis2006}. One can also avoid (or test) the Born approximation using ray-tracing simulations. Taking additionally Assumptions 5 and 6 restricts us to lensing over short time and length scales -- i.e. consideration of (usually strong) lensing by an isolated lens. It turns out that the \textit{spatial deflection} of light is independent of whether the lens is static, so the lens equation is also applicable to microlensing despite microlensing being in opposition with Assumption 5. We discuss the regimes of strong, weak and micro- lensing in Section \ref{sec:obsregimesgravlensing}.

\subsection{Perturbation of Light Rays} \label{sec:perturbedlightrays}

\subsubsection{Perturbed Geodesic Deviation} \label{subsec:perturbedgde}

{\hl We begin by immediately taking the geometric optics limit and weak field limit of gravity.} As usual in first order perturbation theory or linearised gravity, we may decompose a metric into a background and perturbed part
\begin{equation}
    \bm{g} = \bar{\bm{g}} + \delta \bm{g}.
\end{equation}
Correspondingly, a geodesic in the full metric may be decomposed into background and perturbed parts
\begin{equation}
    x^\mu(\sigma) = \bar{x}^\mu (\sigma) + \delta x ^\mu(\sigma)
\end{equation}
where both the true path $\mathscr{P}$ and the background path $\mathscr{\bar{P}}$ with coordinates $x^\mu (\mathscr{P})$ and $\bar{x}^\mu (\mathscr{\bar{P}})$ of the light ray are affinely parametrised by the affine parameter $\sigma$; this means that the geodesic equation will hold with respect to $\sigma$ for both curves in their respective spacetimes. The corresponding decomposition of the wavevector is $\bm{k} = \bar{\bm{k}} + \delta \bm{k}$, where $k^\mu \equiv \frac{d x^\mu}{d \sigma}$, $\bar{k}^\mu \equiv \frac{d \bar{x}^\mu}{d \sigma}$. Whilst $x^\mu$ and $\bar{x}^\mu$ are the coordinates along paths in general relativity and as such are not vectorial quantities, {\hl the quantity $\delta x ^\mu$ is a vector}.

The geodesic equation for an unperturbed light ray in the background spacetime $\bar{\bm{g}}$ is {\hl the second-order non-linear ordinary differential equation} given by {\hl
\begin{equation}
    \frac{d^2 \bar{x}^\mu}{d \sigma^2} =  - \bar{\Gamma}\indices{^\mu_{\rho \nu}}(\bar{x}^\alpha) \frac{d \bar{x}^\rho}{d \sigma} \frac{d \bar{x}^\nu}{d \sigma}
\end{equation}
where we explicitly include the dependence of the Christoffel symbol on the background path $\bar{x}^\alpha$.}  On the other hand, the geodesic equation for a light ray in the full spacetime $\bm{g}$ may be written as {\hl \begin{equation}
    \frac{d^2 x^\mu}{d \sigma^2} = - \Gamma\indices{^{\mu}_{\rho \nu}}(x^\alpha) \frac{dx^\rho}{d\sigma} \frac{dx^\nu}{d \sigma} 
\end{equation}}
and we expand out both sides with the RHS to first order, to find {\hl
\begin{equation}
    \frac{d^2 \bar{x}^\mu}{d \sigma^2} + \frac{d^2 \delta x^\mu}{d \sigma^2} = - \bar{\Gamma}\indices{^\mu_{\rho \nu}}(x^\alpha) \frac{d\bar{x}^\rho}{d \sigma} \frac{d \bar{x}^\nu}{d \sigma} - 2 \bar{\Gamma}\indices{^\mu_{\rho \nu}}(x^\alpha) \frac{d \bar{x}^\rho}{d \sigma} \frac{d \delta x^\nu}{d \sigma} - \delta \Gamma\indices{^\mu_{\rho \nu}} (x^\alpha) \frac{d \bar{x}^\rho}{d \sigma} \frac{d \bar{x}^\nu}{d \sigma}.
\end{equation}
Note that we cannot immediately subtract the background geodesic equation from the first-order geodesic equation since the connection terms (Christoffel symbols) are evaluated along different paths. Therefore we must perform a Taylor series expansion of the background\footnote{\hl Whether or not we choose to also take the first order Taylor expansion for the perturbation to the Christoffel symbol $\delta \Gamma\indices{^\mu_{\rho \nu}} (x^\alpha) = \delta \Gamma\indices{^\mu_{\rho \nu}} (\bar{x}^\alpha)$ at this point corresponds to whether or not we are already taking the Born approximation mentioned in the previous section; we choose not to at this stage.} Christoffel symbol to first order: $\bar{\Gamma}\indices{^\mu_{\rho \nu}} (x^\alpha) = \bar{\Gamma}\indices{^\mu_{\rho \nu}} (\bar{x}^\alpha) + \bar{\Gamma}\indices{^\mu_{\rho \nu, \gamma}} (\bar{x}^\alpha) \delta x^\gamma$. Substitution and subtracting the background geodesic equation gives the equation for $\delta x^\mu$, the separation between the path in the background spacetime and the path in the perturbed spacetime, at linear order as
\begin{equation}
        \boxed{
        \frac{d^2 \delta x^\mu}{d \sigma^2} + 2\bar{\Gamma}\indices{^\mu_{\rho \nu}}(\bar{x}^\alpha) \bar{k}^\rho \frac{d\delta x^\nu}{d \sigma} + \bar{\Gamma}\indices{^\mu_{\rho \nu, \gamma}}(\bar{x}^\alpha) \bar{k}^\rho \bar{k}^\nu \delta x^\gamma = - \delta \Gamma\indices{^\mu_{\rho \nu}}(x^\alpha) \bar{k}^\rho \bar{k}^\nu.
        }\label{perturbedgde}
\end{equation}
This is a system of second-order linear ordinary differential equations, i.e. a second-order linear ODE with non-constant matrix coefficients with the form
\begin{equation}
    \frac{d^2 \delta x^\mu}{d \sigma^2} + A\indices{^\mu_\nu} \frac{d\delta x^\nu}{d \sigma} + B\indices{^\mu_\nu} \delta x^\nu = f^\mu.
\end{equation}

That this is a second-order linear equation involving a notion of a separation vector is reminiscent of the geodesic deviation equation (GDE) \cite{Synge1960}, also known as the Jacobi equation, describing tidal effects through which one measures spacetime curvature, and thus characterises the effects of gravity. The covariant acceleration of a null background geodesic relative to a neighbouring null background geodesic, to which it is connected by an infinitesimal background separation vector $\bar{n}^\mu$ is given by the background GDE
\begin{equation}
    \frac{D^2 \bar{n}^\mu}{d \sigma^2} - \bar{R}\indices{^\mu_{\rho \nu \gamma}} \bar{k}^\rho \bar{k}^\nu \bar{n}^\gamma = 0
\end{equation}
where $\bar{R}\indices{^\mu_{\rho \nu \gamma}}$ is the Riemann tensor of the background spacetime.

It must be stressed that since $\delta x ^\mu$ represents the separation between rays in different metrics, it is \textit{distinct}\footnote{\hl The distinction is made clear by considering the smallness of $\delta x^\mu$ is governed by the validity of the metric expansion (typically we consider the weak-field expansion) which we could quantify by an arbitrarily small parameter $\zeta$, to first order: \[g_{\mu \nu} (\zeta) = g_{\mu \nu}(\zeta =0) + \frac{\partial g_{\mu \nu}}{\partial \zeta} \zeta \equiv \bar{g}_{\mu \nu} + \delta g_{\mu \nu} \implies x^\mu (\zeta) = x^\mu (\zeta=0) + \frac{\partial x^\mu}{\partial \zeta} \zeta \equiv \bar{x}^\mu + \delta x^\mu.\] On the other hand, the smallness of $\bar{n}^\mu$ between neighbouring geodesics in a single metric -- we choose the background metric $\bar{g}_{\mu \nu}$ to match the discussion in the main text -- is quantified by a different parameter, say $\epsilon$: \[\bar{x}^\mu(\epsilon) = \bar{x}^\mu(\epsilon =0) + \frac{\partial x^\mu}{\partial \epsilon} \epsilon \equiv \bar{x}^\mu(\epsilon =0) + \bar{n}^\mu.\]} from the usual ``separation vector'' or ``deviation vector'' $n^\mu$ between neighbouring geodesics in a single metric which obeys the GDE. Nonetheless, a connection between the equations does exist, as Equation \eqref{perturbedgde} is the \textit{perturbed geodesic deviation equation}: the homogeneous version ($f^\mu=0$) of Equation \eqref{perturbedgde} corresponds to the \textit{geodesic deviation equation belonging to the background spacetime}; and the right hand side is a perturbative driving term $f^\mu$. This is because for \textit{any arbitrary vector} $v^\mu$ we can perform the algebra (Appendix A, \cite{PyneBirkinshaw1993}) to show the equivalence
\begin{equation}
    \frac{D^2 v^\mu}{d \sigma^2} - \bar{R}\indices{^\mu_{\rho \nu \gamma}} \bar{k}^\rho \bar{k}^\nu v^\gamma = \frac{d^2 v^\mu}{d \sigma^2} + 2 \bar{\Gamma}\indices{^\mu_{\rho \nu}} \bar{k}^\rho \frac{d v^\nu}{d \sigma} + \bar{\Gamma}\indices{^\mu_{\rho \nu, \gamma}} \bar{k}^\rho \bar{k}^\nu v^\gamma.
\end{equation}
That is, the LHS of Equation \eqref{perturbedgde} describes the covariant acceleration of a light ray relative to a nearby light ray in the background spacetime -- we can think of the free motion, as defined by the background spacetime, of two photons -- in accordance with the background GDE. The RHS describes the effect of an additional perturbative acceleration to one ray relative to the other, due to the perturbation in spacetime curvature relative to the background.

Notice since the perturbed geodesic deviation equation \eqref{perturbedgde} follows geodesics in the full metric $g_{\mu \nu}$ with respect to some background geodesic, it is describing the geodesics in the full metric with respect to each other at finite separation. If we are concerned with only small distortions of an infinitesimal beam of light (\textit{weak gravitational lensing}), we could use an alternative formalism (the Sachs formalism) which starts from the geodesic deviation equation only \cite{Sachs1961, Seitz1994}, reviewed in detail in \cite{Fleury2015}. It is an attractive way to describe gravitational lensing using only physically observable quantities, but as the GDE does not follow rays at \textit{finite} separation \textit{it is not appropriate for strong gravitational lensing} \cite{Lewis2006, clarkson2016}. Yet the derivation presented in e.g. \cite{Bartelmann2010}, based in concept on \cite{Seitz1994}, appears to start from the GDE and find equations appropriate for strong lensing. This argument is made by exploiting the fact that the lens mapping is \textit{linear for their respective choice of background metric} outside of a single isolated inhomogeneity (which is treated separately in a local perturbed Minkowski frame); this central assumption is pointed out in \cite{Seitz1994} but not in \cite{Bartelmann2010}\footnote{Hence the resultant Equation 49 in \cite{Bartelmann2010} is in principle $d \bm{\beta} = d \bm{\theta} - d \bm{\alpha}$ as angles between neighbouring rays are strictly infinitesimal (cf. the notation in Equation 2.25 of \cite{Lewis2006}; and note that $d \Phi$ was changed to $\Phi$ in Equation 40 to 41 of \cite{Bartelmann2010}). If the background mapping is linear given the choice of metric, and considering a single isolated inhomogeneity with shear assumed to vanish away from the inhomogeneity, the association $d \bm{\beta} = d \bm{\theta} - d \bm{\alpha} \implies \bm{\beta} = \bm{\theta} - \bm{\alpha}$ is valid.}.

Any general second-order linear ODE with non-constant coefficients in the form $\bm{x}'' + A(\sigma) \bm{x}' + B(\sigma) \bm{x} = 0$, where the primes indicate differentiation with respect to $\sigma$, can be cast as a first-order ODE of the form $\bm{y}' = C(\sigma) \bm{y}$, where $C(\sigma) \equiv \big[\begin{smallmatrix} 0 & 1 \\ - B(\sigma) & -A(\sigma) \end{smallmatrix}\big]$, by defining the ``phase space'' or ``state vector'' $\bm{y} \equiv \big(\begin{smallmatrix} \bm{x} \\ \bm{x}' \end{smallmatrix} \big)$. If $C(\sigma)$ and $\bm{x}$ were simply scalar functions, then this would be a separable ODE with an exponential solution. However, since we cannot divide by vectors, nor take the inverse of arbitrary matrices in general, an alternative approach to solving the equation leads to solutions which are called path-ordered exponentials. In this way, the perturbed geodesic deviation equation (as well as the geodesic deviation equation, and also the equation of parallel transport, see Appendix I \cite{Carroll2004}) can be cast as a system of first-order linear higher-dimensional non-constant matrix ODEs and a general solution (for arbitrary perturbed metrics) can be found using such a method \cite{PyneBirkinshaw1993}. Instead of using this more sophisticated formal solution, however, since we are only interested in the FLRW metric we can simply calculate the matrices $A\indices{^\mu_\nu}$ and $B\indices{^\mu_\nu}$ for the FLRW case and perform the summation over the indices: this is the method we will now follow.
}

\paragraph{Conformal Weak Field Perturbed FLRW Metric}

The FLRW metric for first-order scalar perturbations in the Newtonian gauge \eqref{perturbedFLRW} can be written in conformal form $\bm{g} = a^2 \bm{\tilde{g}}$ such that
\begin{equation}
    \tilde{g}_{\mu \nu} dx^{\mu} dx^{\nu} =  -\left(1+ 2\hat{\Phi}\right) d(c \eta)^2 + \left(1 - 2 \hat{\Psi}\right) \tilde{\gamma}_{ij} dx^i dx^j. \label{conformalperturbedmetric}
\end{equation}
As the metric is diagonal, to find the inverse metric we simply take the reciprocals of each element such that to first order
\begin{equation}
    \hl \tilde{g}^{\mu \nu} \partial_{\mu} \partial_{\nu} = - \left(1- 2 \hat{\Phi} \right) \partial_{(c \eta)}^2 + \left(1 + 2 \hat{\Psi} \right) \tilde{\gamma}^{ij} \partial_i \partial_j.
\end{equation}
We work throughout the extended abstract calculations of this section directly with the conformal metric $\tilde{g}_{\mu \nu}$ rather than the true metric $g_{\mu \nu}$ as null geodesics are invariant in conformally related metrics (the affine parameter is transformed according to the conformal factor, see Appendix D of \cite{Wald1984}): we therefore \textbf{drop the tildes in the rest of this section referring to the conformal quantities} to declutter the notation. We refer to Table 5.1 of \cite{Fleury2015} for a useful ``conformal dictionary'' to relate quantities (such as the wavevector) calculated with the conformal metric to the true metric. Note that it is much more elegant to derive the following results by defining an alternative conformal metric \cite{Durrer2005, Lewis2006}, which would allow us to \textit{combine the Bardeen potentials} from the start as the Weyl potential $\Phi_W \equiv \frac{1}{2}( \hat{\Phi} + \hat{\Psi})$, {\hl but this anticipates the results before deriving them}. We calculate the following non-zero Christoffel symbols from their definition \eqref{Christoffeldefn} using the conformal metric \eqref{conformalperturbedmetric} (note partial derivatives with respect to $x^0$ should be understood to mean with respect to $c \eta$)
\begin{align*}
&\begin{aligned}
    &\Gamma\indices{^0_{\rho 0}} = \Gamma\indices{^0_{0 \rho }} = \hat{\Phi}_{,\rho} \qquad \; && \Gamma\indices{^i_{i0}} = \Gamma\indices{^i_{0i}} = - \hat{\Psi}_{,0}\\
    &\Gamma\indices{^0_{ii}} = - \hat{\Phi}_{,0} &&
    \Gamma\indices{^i_{00}} = \gamma^{ij} \hat{\Phi}_{,j}
\end{aligned}\\
    &\Gamma\indices{^i_{jl}} = g^{ii} g\indices{_{ii(,l}} \delta\indices{^i_{j)}} - \frac{1}{2} g^{ii} g\indices{_{jj,i}} \delta\indices{^j_l} = \gamma^{ii} \gamma\indices{_{ii(,l}} \delta\indices{^i_{j)}} - 2 \hat{\Psi}\indices{_{(,l}} \delta\indices{^i_{j)}} + \hat{\Psi}_{,i} \gamma^{ii} \gamma_{jj} \delta\indices{^j_l} 
\end{align*}
where we explicitly used the fact the metric is diagonal and define the symmetric index notation $T_{(\mu \nu)} \equiv \frac{1}{2} (T_{\mu \nu} + T_{\nu\mu})$. 
Thus for $\Gamma\indices{^i_{jl}}$ we need only calculate $\gamma^{ii} \gamma_{ii,l}$ which are only nonzero for 
$\gamma^{\theta \theta} \gamma_{\theta \theta, \chi} = \gamma^{\phi \phi } \gamma_{\phi  \phi , \chi} = 2 \frac{f'_K(\chi)}{f_K(\chi)}$ and $\gamma^{\phi \phi } \gamma_{\phi  \phi , \theta} = 2 \frac{f'_K(\chi)}{f^2_K(\chi)} \cot \theta$. This gives the background spatial coefficients with $l= \chi$ as
\begin{equation}
    \bar{\Gamma}\indices{^i_{\chi j}} = \gamma^{ii} \gamma\indices{_{ii(,\chi}} \delta\indices{^i_{j)}} = \frac{1}{2} \gamma^{ii} ( \gamma\indices{_{ii, \chi}} \delta\indices{^i_j} + \gamma\indices{_{ii,j}} \delta\indices{^i_\chi} ) 
    = \frac{1}{2} \gamma^{ii} \gamma\indices{_{ii, \chi}} \delta\indices{^i_j}
    = \frac{f'_K(\chi)}{f_K(\chi)} ( \delta\indices{^i_j} - \delta\indices{^i_\chi} \delta\indices{^\chi_j} )
\end{equation}
and the first order spatial coefficients
\begin{equation*}
    \delta \Gamma\indices{^i_{jl}} = -2 \hat{\Psi}\indices{_{(,l}} \delta\indices{^i_{j)}} + \hat{\Psi}\indices{_{,i}} \gamma^{ii} \gamma\indices{_{jj}} \delta\indices{^j_l}.
\end{equation*}

\paragraph{Null Wavevector Condition}
The null wavevector condition is
\begin{equation}
    g_{\mu \nu} k^{\mu} k^{\nu} = 0.
\end{equation}
At zeroth order, this gives
\begin{equation}
    |\bar{k}^0|^2 = \gamma_{ij} \bar{k}^i \bar{k}^j
\end{equation}
and at first order,
\begin{align}
        2 \bar{g}_{\mu \nu} \bar{k}^{\mu} \delta k^{\nu} + \delta g_{\mu \nu} \bar{k}^{\mu} \bar{k}^{\nu} &= 0 \label{firstordernullwavevectorcond}\\
        - \bar{k}^0 \delta k^0 + \gamma_{ii} \bar{k}^i \delta k^i - (\hat{\Phi} + \hat{\Psi}) |\bar{k}^0|^2 &= 0 \label{perturbednullwaveveccond}
\end{align}

\paragraph{Background null geodesics}
The background radial null geodesics are given by
\begin{equation}
    \bar{\chi} = \eta_0 - \eta , \; \bar{\theta} = \text{const.} , \; \bar{\phi} = \text{const.}
\end{equation}
where we have chosen the negative sign in the solution which corresponds to a ray moving towards the observer at the origin. The \textit{conformal} wavevector along this path is $\bar{k}^{\mu} = (\bar{k}^0, \bar{k}^{\chi}, 0 , 0) = \text{const.}$, and $\bar{k}^{\chi} = \frac{d \chi}{d \eta} \frac{d \eta}{d \sigma} = - \bar{k}^0$. In the conformal metric, light rays correspond to straight lines (whereas had we used cosmic time $t$ instead, then the path of the light would be curved). We can check this radial solution fulfils the background geodesic equation.

\paragraph{Perturbed Geodesic Deviation}

{\hl We now calculate the coefficient $B\indices{^\mu_\nu} \equiv \bar{\Gamma}\indices{^\mu_{\rho \gamma, \nu}}\bar{k}^\rho \bar{k}^\gamma$ in Equation \eqref{perturbedgde} for the conformal perturbed FLRW metric \eqref{conformalperturbedmetric}. Since $\bar{\Gamma}\indices{^0_{\mu \nu}} =0$, we need only $B\indices{^i_\nu}$. From the background solution $\bar{k}^\nu$ is non-zero only for its temporal and radial component, so the only relevant Christoffel symbols are $\bar{\Gamma}\indices{^i_{00}}$, $\bar{\Gamma}\indices{^i_{0 \chi}}$ and $\bar{\Gamma}\indices{^i_{\chi \chi}}$; all of which were calculated to be $0$. Hence, the perturbed geodesic deviation equation \eqref{perturbedgde} reduces in the case of the conformal perturbed FLRW metric \eqref{conformalperturbedmetric} to\footnote{The derivation presented in \cite{Fleury2015} written in terms of the perturbed wavevector misses the $B^\mu_\nu$ term for a general metric. It also argues that $\bar{k}^{\mu}_{,\nu} \delta k^{\nu} +  \bar{\Gamma}^{\mu}_{\nu \rho} \bar{k}^{\nu} \delta k ^{\rho} = \delta k^{\nu} \nabla_{\nu} \bar{k}^{\mu} = 0$; however, this is not justified, and results in a missing factor of $2$ from $A^\mu_\nu$. In addition, $\frac{d \delta k^\mu}{d \sigma} \neq \bar{k}^\nu \delta k^\mu_{, \nu}$ in general; consider that to first order
$\frac{d \delta k^\mu}{d \sigma} = \frac{d^2 (x^\mu - \bar{x}^\mu)}{d \sigma^2}  = k^\nu k^\mu_{,\nu}  - \frac{d^2 \bar{x}^\mu}{d \sigma^2} = \bar{k}^\nu \bar{k}^\mu_{, \nu} - \frac{d^2 \bar{x}^\mu}{d \sigma^2}  + \bar{k}^\nu \delta k^\mu_{, \nu} + \delta k^\nu \bar{k}^\mu_{,\nu} = \bar{k}^\nu \delta k^\mu_{, \nu} + \delta k^\nu \bar{k}^\mu_{,\nu}.$} 
\begin{equation}
    \boxed{
    \frac{d^2 \delta x^\mu}{d \sigma^2} + A\indices{^\mu_\nu} \frac{d\delta x^\nu}{d \sigma} = f^\mu }\label{flrwperturbedgde}
\end{equation}
where $A\indices{^\mu_\nu} \equiv 2\bar{\Gamma}\indices{^\mu_{\rho \nu}} \bar{k}^\rho$ and $f^\mu \equiv - \delta \Gamma\indices{^\mu_{\rho \nu}} \bar{k}^\rho \bar{k}^\nu$.
}

{\hl 
\subsubsection{Time and Frequency Perturbations}

Setting $\mu =0$ in Equation \eqref{flrwperturbedgde}, we note that $\bar{\Gamma}\indices{^0_{\nu \rho}} = 0$, 
so we have simply that
\begin{align}
\frac{d^2 \delta x^0}{d \sigma^2} &= f^0\\
   &= - \delta \Gamma\indices{^0_{00}} |\bar{k}^0|^2 -2 \delta \Gamma\indices{^0_{0i}} \bar{k}^0 \bar{k}^i - \delta \Gamma\indices{^{0}_{ij}} \bar{k}^i \bar{k}^j\\
   &= -2 \hat{\Phi}\indices{_{,i}} \bar{k}^0 \bar{k}^i + \hat{\Phi}\indices{_{,0}} ( - |\bar{k}^0|^2 + \gamma_{ij}\bar{k} \bar{k}^j)\\
   &= -2 \hat{\Phi}\indices{_{,i}} \bar{k}^0 \bar{k}^i\\
   &= -2 \bar{k}^0 \frac{d \hat{\Phi}}{d \sigma} + 2 |\bar{k}^0|^2 \hat{\Phi}\indices{_{,0}} \label{acceleqn0}
\end{align}
where we made use of the background null wavevector condition and $\frac{d \hat{\Phi}}{d \sigma} = \bar{k}^0 \hat{\Phi}\indices{_{,0}} + \bar{k}^i \hat{\Phi}\indices{_{,i}}$.

Integration over the spacetime path $x^\alpha(\sigma)$ gives the perturbation to temporal component of the deviation vector: that is, the time delay due to the potential
\begin{align}
    \frac{d \delta x^0}{d \sigma} &= \delta k^0|_o - 2 \bar{k}^0 \hat{\Phi} +2 \bar{k}^0 \hat{\Phi}|_o + \int_{\sigma_o}^{\sigma'} 2 | \bar{k}^0|^2 \hat{\Phi}_{,0}(x^\alpha(\sigma)) d \sigma \label{timewavevectorpert}\\
    [\delta x^0]^s_o &= (\delta k^0 + 2 \bar{k}^0 \hat{\Phi})|_o \int_{\mathscr{P}} \frac{d \ell_K}{|\bar{k}^0|} -2 \bar{k}^0 \int_{\sigma_o}^{\sigma_s} \hat{\Phi} (x^\alpha(\sigma))d \sigma +\int_{\sigma_o}^{\sigma_s} \int_{\sigma_o}^{\sigma'} 2 |\bar{k}^0|^2 \hat{\Phi}_{,0}(x^\alpha(\sigma)) d \sigma d \sigma'
\end{align}
where we used $d \ell_K \equiv \sqrt{\gamma_{ij}\frac{dx^i}{d \sigma}\frac{dx^j}{d \sigma}}d \sigma = |\bar{k}^0| d \sigma$ (employing the background null wavevector condition) to write the first term on the RHS (the boundary term) with respect to the spatial path length rather than the affine parameter. The final term can also be written as a single integral via Fubini's theorem; however if we assume that the potential is conformally static then it is $0$.

We now also need to consider the correct initial conditions for the problem: we have integrated from the observer located at the origin towards to the source so we may impose that the background and perturbed rays are spatially aligned at the observer: $\delta x^i|_o =0$ and $\delta k^i|_o =0$ (parallel). However, we cannot in general take the wavevector of the two geodesics to coincide fully at the observer (i.e. $\delta k^\mu|_o \neq 0$) because the geodesics must each be null in their respective, different metrics. Instead we may make use of the null wavevector condition \eqref{perturbednullwaveveccond}, setting $\hat{\Phi} = \hat{\Psi}$  such that $- \bar{k}^0 ( \delta k^0 + \delta k^\chi) - 2 \hat{\Phi} |\bar{k}^0|^2 =0$. Evaluating this constraint at the observer with the assumption $\delta k^\chi|_o =0$ gives $(\delta k^0 + 2 \bar{k}^0\hat{\Phi})|_o=0$, and so the boundary term is $0$. We then have
\begin{equation}
      [\delta x^0]^o_s = -2 \bar{k}^0 \int_{\sigma_s}^{\sigma_o} \hat{\Phi} (x^\alpha(\sigma))d \sigma
\end{equation}
as the time delay at the observer of the perturbed light ray compared with the background light ray due to the potential. However, as we will see in Section \ref{subsubsec:timedelays}, this calculation omits an important contribution to the total time delay due to the geometric difference in spatial path length between the path $x^i(\mathscr{P})$ and the path drawn directly from the observer to the source (which is OS in Figure \ref{fig:lensingsetup}, whereas the background ray $\bar{x}^i(\bar{\mathscr{P}})$ for our choice of initial conditions refers to OI).}

We may also look at the perturbation at the level of the wavevector: Equation \eqref{timewavevectorpert}, and using $d(c \eta) = \bar{k}^0 d \sigma$ gives
\begin{equation}
    [ \delta k^0]^o_s = -2 \bar{k}^0 [\hat{\Phi}]^o_s + 2 \bar{k}^0 \int_{\eta_s}^{\eta_o} \partial_{\eta}\hat{\Phi} (\eta, \bar{x}^i (\eta)) d \eta \label{freqpert}
\end{equation}
{\hl where we also took the Born approximation, evaluating the potential along the background path $\bar{x}^\alpha$ instead of $x^\alpha$.} To physically interpret this result, consider the definition of frequency. In the rest frame of an observer with four velocity $\bm{u}$, the electromagnetic potential in the geometric optics limit is $\bm{A} = \text{Re} \{\bm{a} e^{- i \varphi} \}$ and the angular frequency $\omega$ is defined as the rate of change of phase $\varphi$ with respect to the proper time $\uptau$ of the observer 
\begin{equation}
    \frac{\omega}{c} \equiv \left| \frac{d \varphi}{ d(c\uptau)} \right| = \left| u^{\mu} \varphi_{, \mu} \right| = - u^{\mu} k_{\mu}
\end{equation}
where the negative is due to picking  the wavevector $k^{\mu} \equiv \varphi^{, \mu}$ to be future oriented, such that $k^0 > 0$ in any coordinates.

Expanding out the frequency to first order, noting $\bar{u}^\mu = \delta^{\mu}_{0}$ for the background observer at rest in comoving coordinates (such that $\bar{\omega} = -c \bar{u}^0 \bar{k}_0 = c \bar{k}^0$) and the background wavevector is purely radial, we have 
\begin{align}
     \omega &= \bar{\omega} - c\bar{u}_\mu \delta k^\mu - c\delta u_{\mu} \bar{k}^\mu\\
     &= \bar{\omega} + c\delta k^0 - c\bar{k}^0 (\delta u_0 - \delta u_i \delta^i_\chi)\\
     &= \bar{\omega} + c\delta k^0 + c\bar{\omega} (\hat{\Phi} + v_i \delta^i_\chi)
\end{align}
where we obtained the components of the four velocity through the normalisation condition $g_{\mu \nu} u^{\mu} u ^{\nu} = -1$, and noting that the peculiar velocity is defined by $v^i = \frac{u^i}{u^0}$. Solving $- (u^0)^{-2} = g_{00} + g_{ii} (v^i)^2$ and expanding to first order, we get $u^0 = (1- \hat{\Phi})$; and so $u^i = u^0 v^i = v^i$ to first order. Lowering with the metric to first order gives $u_0 = -(1 + \hat{\Phi})$ and $u_i = v_i$. Inserting Equation \eqref{freqpert} gives the result
\begin{equation}
    \frac{- \delta z}{1 + \bar{z}} = \frac{\delta \omega_o}{\bar{\omega}_o} - \frac{\delta \omega_s}{\bar{\omega}_s} = \left[ - \hat{\Phi} + v_i \delta^i_\chi \right]^o_s + 2 \int_{\eta_s}^{\eta_o} \partial_{\eta}\hat{\Phi} (\eta, \bar{x}^i (\eta)) d\eta \label{eq:frequencyperturb}
\end{equation}
where we used the redshift relation $1 +z = \frac{\omega_s}{\omega_o}$ and expanded using $(\bar{\omega}_o + \delta \omega_o)^{-1} = \bar{\omega}_o^{-1} ( 1+ \frac{\delta \omega_o}{\omega_o})^{-1} = \omega_o^{-1} ( 1- \frac{\delta \omega_o}{\omega_o})$ to first order. As the result involves ratios of frequencies evaluated at both $O$ or both $S$, the expression holds for both the conformal and true metric.

Under the standard assumptions listed in Section \ref{quasinewtonianformalism} of an isolated strong lens, the potential considered is a perturbation to the homogeneous background and isolated at one or multiple intermediate line of sight distances such that the potential may be considered static as the light ray passes it. It is then straightforward that both terms containing the potential are negligible, and the redshift of a source's image is unperturbed.

Frequency perturbations are often classified explicitly as an effect \textit{separate from gravitational lensing} which refers to the \textit{deflection of light} only \cite{Lewis2006}. That is, frequency perturbations arising from a gravitational potential which evolves with time, or equivalently line-of-sight gradients in the potential, are sometimes not considered as ``lensing''; as opposed to spatial perturbations (deflections) arising from transverse gradients in the potential.

However, these redshift perturbations are considered as additional temperature anisotropies in the context of the CMB, as the potential cannot be assumed static over the time taken for CMB photons to traverse large-scale structures. The first term of Equation \eqref{eq:frequencyperturb} $\big[-\hat{\Phi} \big]^o_s$ known as the Sachs-Wolfe effect (or Ordinary Sachs-Wolfe effect) is a gravitational redshift corresponding to the change in the potential evaluated at the observer and source, the second corresponds to the peculiar velocities of the observer and the source; and the third term is the Integrated Sachs-Wolfe (ISW) effect corresponding to the integrated change in the potential along the path of the light ray. In other words, the ISW effect quantifies the difference between the blueshift of a photon entering an evolving potential and the redshift as it exits.

The Sachs-Wolfe term $\big[-\hat{\Phi} \big]^o_s$ is sometimes simplified to $\hat{\Phi}|_s$ since the potential at the observer effects equally the entire source. We refer to e.g. \cite{Amendola2010} and \cite{Durrer2005} for a brief description of this result applied to the CMB. It is observable only if the spatial extent of the source is very large, i.e. on very large scales on the CMB's last scattering surface.

The solutions to the linearised Einstein field equations show that in a dust-only universe, the matter perturbations have a growing mode proportional to the scale factor whilst the gravitational potential is static or constant in time e.g. \cite{Baumann2022}, \cite{Amendola2010}. (This is true even when non-linear for objects approximated as undergoing \textit{spherical collapse}, from Newton's shell theorem or Birkhoff's theorem.) Therefore, even in the general relativistic case, there is no Integrated Sachs-Wolfe in the linearised regime during matter domination. This is different at early times i.e. shortly after recombination, when the radiation content cannot be neglected, and also at very late times when dark energy content becomes important. Large-scale gravitational potentials decay under the late-time accelerated expansion, leading to the late-time ISW effect. For density perturbations which become non-linear, the gravitational potential starts growing, corresponding to a late-time integrated Sachs–Wolfe effect on very small scales (Rees–Sciama effect) \cite{Durrer2005}. The ISW effect can also be related to the Birkinshaw-Gull effect arising from a transverse peculiar velocity of a lens \cite{Birkinshaw1983, PyneBirkinshaw1993, Merkel2013}.

\subsubsection{Spatial Perturbation}
Let us set $\mu =i$ to investigate the spatial perturbation to the wavevector. The LHS of Equation \eqref{flrwperturbedgde} gives
\begin{align}
        \text{LHS} &= \frac{d \delta k^i}{d \sigma} + 2 \bar{\Gamma}\indices{^{i}_{\nu \rho}} \bar{k}^{\nu} \delta k ^{\rho}\\
        \intertext{and since $\bar{\Gamma}\indices{^i_{00}} = \bar{\Gamma}\indices{^i_{i0}}  = \bar{\Gamma}\indices{^i_{0i}} = 0$, and using that the background ray is purely radial}
        &= \frac{d \delta k^i}{d \sigma} + 2 \bar{\Gamma}\indices{^{i}_{l j }} \bar{k}^{\chi} \delta\indices{^l_\chi} \delta k ^{j}\\
        &= \frac{d \delta k^i}{d \sigma} + 2 \frac{f'_K(\chi)}{f_K(\chi)} ( \delta\indices{^i_j} - \delta\indices{^i_\chi} \delta\indices{^\chi_j} ) \bar{k}^{\chi} \delta k ^{j}.
\end{align}
Now, the RHS of Equation \eqref{flrwperturbedgde} gives
\begin{align}
    f^i &= - \delta \Gamma\indices{^{i}_{\nu \rho}} \bar{k}^{\nu} \bar{k}^{\rho}\\
    &= - \delta \Gamma\indices{^i_{00}} |\bar{k}^0|^2 - 2 \delta \Gamma\indices{^i_{0i}} \bar{k}^0 \bar{k}^i - \delta \Gamma\indices{^i_{jl}} \bar{k}^j \bar{k}^l \\
    &= - \hat{\Phi}\indices{_{,i}} \gamma^{ii} |\bar{k}^0|^2 + 2 \hat{\Psi}\indices{_{,0 }} \bar{k}^{0} \bar{k}^i + 2 \hat{\Psi}\indices{_{,j }} \bar{k}^{j} \bar{k}^i - \hat{\Psi}\indices{_{,i}} \gamma^{ii} \gamma_{jj} \bar{k}^j \bar{k}^j \\
    &= -( \hat{\Phi} + \hat{\Psi})\indices{_{,i}} \gamma^{ii} |\bar{k}^0|^2 + 2 \hat{\Psi}\indices{_{,\nu}} \bar{k}^{\nu} \bar{k}^i.
\end{align}
Putting these together, the {\hl perturbed GDE} becomes
\begin{equation}
    \frac{d \delta k^i}{d \sigma} + 2 \frac{f'_K(\chi)}{f_K(\chi)} ( \delta\indices{^i_j} - \delta\indices{^i_\chi} \delta\indices{^\chi_j} ) \bar{k}^{\chi} \delta k ^{j} = -(\hat{\Phi} + \hat{\Psi})\indices{_{,i}} \gamma^{ii} |\bar{k}^0|^2 + 2 \hat{\Psi}\indices{_{,\nu}} \bar{k}^{\nu} \bar{k}^\chi \delta\indices{^i_\chi}.
\end{equation}

\subsection{Lens Equation} \label{sec:lenseqn}
We are only interested in the perturbations to the angular position on the observer's sky, so we restrict $ i \rightarrow A = \{1,2\}$ in the first-order geodesic equation; such that $\theta^A \equiv (\theta, \phi)$. A number of terms vanish, leaving
\begin{equation}
    \frac{d \delta k^A}{d \sigma} + 2 \frac{f'_K(\chi)}{f_K(\chi)}  \bar{k}^{\chi} \delta k ^{A} = - 2 \hat{\Phi}_{,B} \gamma^{AB} |\bar{k}^0|^2
\end{equation}
where we have set, assuming GR and no anisotropic stress as is relevant for the rest of this work, $\hat{\Phi} = \hat{\Psi}$. For more general usage it is better to use the Weyl potential $\Phi_W \equiv \frac{1}{2}( \hat{\Phi} + \hat{\Psi})$, to show with clarity the Newtonian potential and the Newtonian curvature contribute in equal parts to the driving term on the right hand side.

At this point, there should be little confusion over if we are working over four $\mu$, three $i$ or two $A$ indices other than in the wavevector. Hence for convenience we use boldface notation, defining the angular position \textit{tuple} $\bm{\theta} \equiv \theta^A$, $\delta \bm{k_\theta} \equiv \delta k^A$, and the angular gradient of the potential $\bm{\nabla_{{\bm{\theta}}}} \hat{\Phi} \equiv \hat{\Phi}_{,B} \gamma^{AB}$. Multiplying through by $f^2_K( \chi)$ and noting that $\bar{k}^\chi = \frac{d \chi}{d \sigma}$, we get
\begin{align}
    \frac{d}{d \sigma} \left( f^2_K (\chi) \delta \bm{k_\theta} \right) &= - 2 f^2_K(\chi) \bm{\nabla_{{\bm{\theta}}}} \hat{\Phi}({x^\mu}(\sigma)) |\bar{k}^0|^2\\
     f^2_K(\chi) \delta \bm{k_\theta} &= \bm{C} - \int_{\sigma_o}^{\sigma} 2 f^2_K( \chi(\sigma')) \bm{\nabla_{{\bm{\theta}}}} \hat{\Phi}({x^\mu}(\sigma'))|\bar{k}^0|^2 d \sigma' \label{integralwavector}
\end{align}
and since $\delta \bm{k_\theta} =0$ {\hl at the observer}, $\bm{C} =0$. The total displacement in angular position $\delta \bm{\theta}$ is the change in the wavevector (i.e. the tangent vector) along the path by definition, so
\begin{equation}
    \left[\delta \bm{\theta} \right]_o^s = -2 \int_{\sigma_o}^{\sigma_s} \int_{\sigma_o}^{\sigma''} \left( \frac{f_K(\chi(\sigma'))}{f_K(\chi(\sigma''))} \right)^2 \bm{\nabla_{{\bm{\theta}}}} \hat{\Phi}({x^\mu}(\sigma')) |\bar{k}^0|^2 d \sigma' d \sigma''.
\end{equation}

\paragraph{Born Approximation}
The spatial path length can be related to the affine parameter via $dx^\mu = k^\mu d \sigma$. Instead of integrating over the true path, we use the \textit{Born approximation} (a name borrowed from scattering theory) and integrate along the unperturbed background path, i.e. the line-of-sight coordinate $\chi$. This assumes that the potential evaluated along the background path is negligibly different to the potential evaluated along the true path. Corrections to the Born approximation are second order in the potential \cite{Schneider2006}. The definition of the background wavevector gives $d(c\eta) = \bar{k}^0 d \sigma$ and the background solution implies $d (c\eta) = - d \chi$ such that $d \chi = - \bar{k}^0 d \sigma$.

Additionally, we wish to write the double integral as a single integral. Since $\bm{\nabla_{{\bm{\theta}}}} \hat{\Phi} $ is independent of $\chi''$, we change the order of integration using Fubini's theorem so we may perform the integral over $\chi''$ first. Skipping the working (see e.g. \cite{Fleury2015}) for this second step we have
\begin{align}
    \left[\delta \bm{\theta} \right]_o^s  &= - 2 \int_{\chi_o}^{\chi_s} \int_{\chi_o}^{\chi''} \left( \frac{f_K(\chi')}{f_K(\chi'')} \right)^2 \bm{\nabla_{{\bm{\theta}}}} \hat{\Phi}( \eta, \chi', \bm{\theta}) d \chi' d \chi''\\
    &= - 2 \int_{\chi_o}^{\chi_s} \frac{f_K(\chi_s - \chi')}{f_K(\chi')f_K(\chi_s)} \bm{\nabla_{{\bm{\theta}}}} \hat{\Phi} ( \eta, \chi', \bm{\theta}) d \chi'.
\end{align}

The full solution is given by adding the background null solution where $\bar{\bm{\theta}} = \text{constant}$ to the perturbed solution. {\hl For reasons of convention (we typically have $\theta_o > \theta_s$), we reverse the limits} 
\begin{equation}
    {\hl \left[\bm{\theta} \right]^o_s = \left[\bar{\bm{\theta}} + \delta \bm{\theta} \right]^o_s}.
\end{equation}
{\hl The angular coordinates $\bm{\theta}_o$ ``at'' the observer O, who is situated at the coordinate pole, are unambiguously defined by the tangent to the true light path at the observer, and hence are the coordinates of the light source seen by the observer.} We recognise the \textit{lens equation} which relates the {\hl observed coordinates $\bm{\theta}_o$ of the source to the unobservable unlensed coordinates $\bm{\theta}_s$ of the source}
\begin{equation}
\boxed{
    \bm{\beta} = \bm{\theta} - \bm{\alpha}(\bm{\theta}) \label{eq:lensmapping}
    }
\end{equation}
using the standard convention of strong lensing where {\hl $\bm{\theta} \equiv \bm{\theta}_o $ and $\bm{\beta} \equiv \bm{\theta}_s $. In general these coordinates are taken relative to an arbitrary constant ray of the background metric; it is conventionally drawn between the lens and the observer as the optical axis. Note in this work we have picked initial conditions such that $\bar{x}^\mu$ are the coordinates of the background radial ray from the observer towards the image, so additionally $\bm{\theta}_o = \bm{\bar{\theta}}$.} We refer to {\hl Figure \ref{fig:generallensdiagram}} as a visual aid.

{\hl We define $\bm{\alpha}(\bm{\theta})$ as an \textit{outward} displacement in the angular position}
\begin{equation} \hl
    \boxed{\bm{\alpha} (\bm{\theta}) \equiv \left[\delta \bm{\theta} \right]^o_s  = 2 \int_{\chi_o}^{\chi_s} \frac{f_K(\chi_s - \chi')}{f_K(\chi')f_K(\chi_s)} \bm{\nabla_{{\bm{\theta}}}} \hat{\Phi} ( \eta, \chi', \bm{\theta}) d \chi'}.
\label{reduceddeflection}
\end{equation}
 The deflection angle of light is independent of the photon frequency: we can treat the deflection of e.g. X-rays the same as infra-red light. This is often stated as ``lensing is \textit{achromatic}''; but, following our discussion regarding the frequency perturbation, this requires defining lensing to refer only to effects from the transverse gradient of the potential and a further caveat is mentioned in Section \ref{sec:magnification}.

The displacement in the image position on the observer's sky is usually called the \textit{scaled} deflection angle $\bm{\alpha}$ in relation to the physical deflection angle $\hat{\bm{\alpha}}$ defined in the lens frame (rather than at the observer) which we will encounter in the next section. The physical deflection angle is often the appropriate quantity to calculate as lensing phenomena was historically observed within our own solar system using the Schwarzchild metric (see e.g. \cite{DInverno1992}), and in cosmological scenarios may be similarly defined in the local perturbed Minkowski frame of the lens (given the thin lens approximation) before being related to the angle at the observer via the definition of angular diameter distances. We note we may write the expression for $\bm{\alpha}$ in terms of the angular diameter distance instead of $f_{K}$.

 The lens equation defines a mapping of the observed image position $\bm{\theta}$ on the sky to the unperturbed source position $\bm{\beta}$. There are potentially many different values of $\bm{\theta}$ for which the right-hand side is equal to $\bm{\beta}$, so there are potentially many different images of a single source. The definition and properties of this lens mapping, including conditions for multiple imaging are discussed in Section \ref{sec:lensmappingproperties}.

\subsection{Thin Lens Approximation and The Deflection Angle} \label{sec:thinlensapprox}

The thin lens approximation assumes that the deflection occurs at a single line of sight distance; that is the potential can be assumed to act at a length scale much smaller than the cosmological distances between the lens and the source and the observer and the lens. For this reason, the thin lens approximation is also referred to as the sudden-deflection approximation.

\begin{figure}
    \centering
    \includegraphics[scale=1.2]{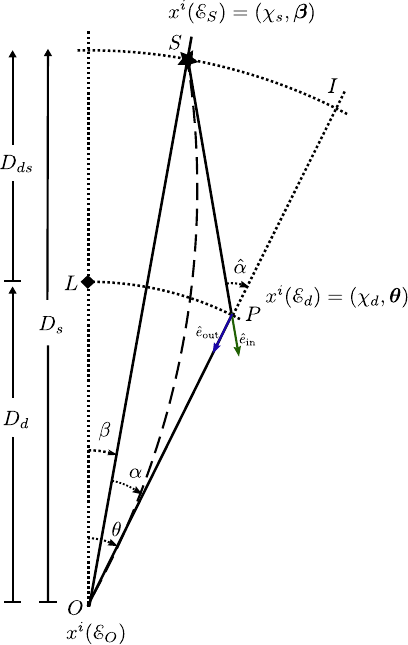}
    \caption[General lensing diagram.]{The general lensing set up: since we work in FLRW spacetime with comoving coordinates, we are free to project the spacetime diagram onto a spatial slice. We depict only two spatial dimensions, so with time and one angular spatial dimension suppressed (Figures 3.2 and 3.3 of \cite{Petters2001} illustrates additionally the additional space dimension and time dimension). A source emits a photon at an emission event $\mathscr{E}_S$ which is seen by the observer at $\mathscr{E}_O$. The lens position is marked by the diamond at point $L$, the source by the star at point $S$ with angular position $\beta$ and the apparent image seen by the observer is at point $I$ with angular position $\theta$.
    Under the flat-sky approximation, the spheres at the source and lens are replaced by their tangent planes (i.e. the arcs LP and SI are replaced by their tangents at L and S respectively). The spatial projection of the true spacetime path {\hl $x^i(\mathscr{P})$} of the light ray is marked by the dashed line and the spatial projection of {\hl the background path $\bar{x}^i(\bar{\mathscr{P}})$ which corresponds to the chosen initial conditions $\delta x^i|_o =0$, $\delta k^i|_o =0$ is the line from O to I.} Under the thin lens approximation, the true spatial path is replaced by the piecewise spatial path SPO constructed from the tangents to the true path at S and O. P therefore represents a single deflection event $\mathscr{E}_d$. The piecewise spacetime path corresponding to SPO through $\mathscr{E}_S$, $\mathscr{E}_d$ and $\mathscr{E}_O$ will be called $\mathscr{P}_{sdo}$. {\hl The angle between SP and SI is the magnitude of the physical deflection $\bm{\hat{\alpha}}$ which is defined by the difference in the normalised tangent vectors $\bm{\hat{e}}_{\text{in}} - \bm{\hat{e}}_{\text{out}}$ at P.} Observationally, we use angular diameter distances $D_A$ instead of the comoving coordinate distance $\chi$: the relevant angular diameter distances from observer to lens ($D_d$), observer to source ($D_s$) and lens to source ($D_{ds}$) are marked.} 
    \label{fig:generallensdiagram}
\end{figure}

In the case of a background Minkowski spacetime, this leads to the definition of the deflection angle $\bm{\hat{\alpha}}$ at the lens as the angle between the asymptotic tangents to the light rays at the observer and the source. In a curved spacetime, this is generalised by approximating the true path of the light ray by two piecewise null \textit{background} geodesics, one from the source and one from the observer respectively, to the celestial sphere of the lens. These null geodesics parallel-propagate their own (tangent) wavevectors from the end points to the lens frame. The vector angle at their intersection is the deflection angle. The set up is illustrated in Figure \ref{fig:generallensdiagram}.

\paragraph{Flat Sky Approximation}
Since all angles involved in strong lensing are very small, we can approximate the celestial sphere at the lens and the source with their tangent planes.  A proper distance on the source plane then replaces the arc length labelled SI on the source sphere in Figure \ref{fig:generallensdiagram} with extent $D_{ds} \bm{\hat{\alpha}} \approx D_s \bm{\alpha}$ by definition of the angular diameter distance. This gives the relation between the scaled deflection angle $\bm{\alpha}$ seen by the observer and the physical deflection angle $\bm{\hat{\alpha}}$ in the lens frame
\begin{equation}
    \bm{{\alpha}} = \frac{D_{ds}}{D_{s}} \bm{\hat{\alpha}}
\end{equation}
where the ratio of the angular size distances $\frac{D_{ds}}{D_s}$ is often referred to as the \textit{lensing efficiency} (e.g. \cite{Golse2002}).

The characterisation of the thin lens approximation as a sudden deflection suggests that the {\hl driving term of the perturbed GDE} can be modelled as an impulse at the lens plane, {\hl that is $f^i \propto \delta$ where $\delta$ is a Dirac delta generalised function.} Furthermore, it means that the potential can be considered to be static, that is, the Newtonian potential varies negligibly in the time the light ray takes to pass the lens. To this end, we accordingly modify Equation \eqref{integralwavector} by changing the integration variable to $\chi$, inserting a Dirac delta function at $\chi_d$ in the integrand and drop the time dependence in the potential (i.e. suppressing writing $\hat{\Phi} (\eta_d, \chi', \bm{\theta})$ in full). {\hl In essence, we are finding a scaled Green's function for the system.} This gives
\begin{align}
    \delta \bm{k_\theta} (\chi, \bm{\theta}) &= 2 f^{-2}_K(\chi) \int_{\chi_o}^{\chi} f^{2}_K( \chi') \bm{\nabla_{{\bm{\theta}}}} \hat{\Phi} (\chi', \bm{\theta}) \delta(\chi' - \chi_d) \bar{k}^0 d \chi'\\
    &= \hl 2 \bar{k}^0 \frac{f^{2}_K( \chi_d) }{f^{2}_K( \chi)}\bm{\nabla_{{\bm{\theta}}}} \hat{\Phi}(\chi_d, \bm{\theta}) H(\chi - \chi_d) \\
    &= \hl \begin{cases}
        0 & \chi_o \leq \chi < \chi_d\\
        2 \bar{k}^0 \frac{f^{2}_K( \chi_d) }{f^{2}_K( \chi)} \bm{\nabla_{{\bm{\theta}}}} \hat{\Phi}(\chi_d, \bm{\theta}) & \chi \geq \chi_d
    \end{cases} \label{heaviside}
\end{align}
{\hl where $H$ is a Heaviside step function defined according to \eqref{heaviside} with restriction $\chi \geq \chi_o$ implicit. If we were to integrate once more for the solution to path of the ray, it would therefore involve a ramp function. This more formally explains how the thin lens assumption allows us to approximate the true light ray by two piecewise null background geodesics. One can think in terms of Newtonian analogy, with an impulsive force acting on a free particle resulting in a change from one constant velocity to another. The physical deflection angle is found by subtracting the outgoing (P $\to$ O) unit tangent vector (normalised spatial wavevector) from the incoming (S $\to$ P) unit tangent vector $\hat{e}_{\text{in}} - \hat{e}_{\text{out}}$ at the deflection event in order to define an \textit{outwards} deflection; we refer to {\hl Figure \ref{fig:generallensdiagram}}. For a single impulse perturbation this is}
\begin{equation}
    \hl \bm{\hat{\alpha}} = - \frac{\Delta \bm{k_\theta}(\chi_d)}{\bar{k}^0}  = \frac{(\bm{k}_{\theta} \text{\scriptsize (S$\to$P)} - \bm{k}_{\theta} \text{\scriptsize (P$\to$O)})|_P }{\bar{k}^0} = 2 \bm{\nabla}_{\bm{\xi}} \hat{\Phi}(\chi_d, \bm{\xi})
\end{equation}
where $\Delta$ is the ``change in'', and the minus sign after the first equality accounts for the observer looking backwards along the light ray. In the final equality we also applied the flat sky approximation, such that the 2D angular tuple $\bm{\theta}$ fixed at the lens with $\chi = \chi_d$ is replaced at the same point by the 2D Cartesian vector $\bm{\xi} = D_d \bm{\theta}$ lying on the lens plane. 
Note under these assumptions the first order perturbation to the wavevector is in fact orthogonal to the background wavevector, so we could also write $\bm{\nabla}_{\perp} = \bm{\nabla}_{\bm{\xi}}$. This impulse lensing angle for a localised\footnote{When considering gravitation or a metric by itself, it is in general only sensible to refer to global quantities (the equivalence principle states that we can transform away all gravitational effects locally); but if we consider material particles as well we may use them to identify points \cite{Rovelli1991}. Here the position of the perturbation is identified by the lens object.} perturbation is {\hl exactly the same as that of a point mass lens in a spatially flat spacetime (see also e.g. Section 16.1.1 of \cite{BovyInPrep}).}

More generally, in the weak-field limit where gravity is linearised, the superposition principle holds and {\hl we may sum contributions from a continuum of point masses in the vicinity of the lens}. The deflection arising {\hl from integrating the continuum of impulse deflections along the proper line-of-sight under the Born approximation\footnote{As a contrasting example, adding deflections in the manner of a random walk (on cosmological scales) gives the recursive lens equation of the multi-plane lensing formalism \cite{Petters2001}, where the Born approximation is not valid although the weak-field regime is valid everywhere.} is}
\begin{equation}
\boxed{ \hl
    \bm{\hat{\alpha}} = 2 \int^{\zeta_s}_{\zeta_o} \bm{\nabla}_{\bm{\xi}} \hat{\Phi}(\zeta, \bm{\xi} = D_d \bm{\theta}) d \zeta
} \label{alphahat}
\end{equation}
in Cartesian proper coordinates ($\bm{\xi}$, $\zeta$) at the lens. Necessarily there are two line-of-sight distance scales involved in this formulation, the one involving the proper line-of-sight distance $\zeta$ in the frame of the lens and the cosmological comoving radial distance $\chi$. They may be related at the lens by the scale factor at the time the light ray reaches the lens, $\zeta = a_d \chi$. {\hl Therefore in our choice of coordinates, $\zeta_o < \zeta_s$; it is \textit{opposite} in convention to e.g. \cite{Petters2001} (p.62, coordinates defined p.56).}

In terms of $\bm{\hat{\alpha}}$, the lens equation becomes
\begin{equation}
\boxed{
    \bm{\beta} = \bm{\theta} - \frac{D_{ds}}{D_s} \bm{\hat{\alpha}}(\bm{\theta}).} \label{lenseqn_alphahat}
\end{equation}
We also note some authors do not use the symbols $\bm{\beta}$ and $\bm{\theta}$ in strong lensing \cite{Petters2001}, to emphasise that under the thin lens and flat sky approximation the tuples $\bm{\beta}$ and $\bm{\theta}$ become vectors in $\mathbb{R}^2$ (the source and lens planes) which are sometimes referred to as ``angular vectors''. We do not make this distinction.

In this form, the only contribution from general relativity to the lens equation, other than the distance measure which corresponds to the choice of the appropriate background cosmology, is from the factor of 2 in the expression for the deflection angle. It differs from the Newtonian prediction due to the addition of the curvature potential; the non-Euclidean nature of both the temporal and spatial parts of the metric play equal parts in the deflection of light. (As we saw in Section \ref{newtoniancorrespondence}, or see also e.g. \cite{Peacock1999} or \cite{Misner1973} p.415, only the $\delta g_{00}$ term is required for calculating the effect on \textit{massive} test particles in nearly Newtonian systems using \textit{passive gravity}.) Hence gravitational lensing strictly is dependent on two potentials $\hat{\Phi}$ and $\hat{\Psi}$, whilst the relativistic Poisson equation \eqref{relativisticPoisson} relates the energy density to only $\hat{\Psi}$; lensing is a test of GR (i.e. the equivalence principle(s), see e.g. \cite{Bonvin2018}) and anisotropic stress. The {\hl ratio of} the two potentials $\hat{\Phi}$ and $\hat{\Psi}$ is sometimes referred to as the gravitational slip. We conclude this section with another quote from Perlick \cite{Perlick2004}
\begin{displayquote}
    In this way, the geometric spacetime setting of general relativity is completely covered behind a curtain of approximations, and one is left simply with a map from a plane to a plane.
\end{displayquote}

\section{Fermat's Principle} \label{sec:fermat}

The motion of a free massive particle, i.e. subject only to gravitation manifesting as curved spacetime, is given by the variation of its action with respect to possible trajectories 
\begin{equation}
    \delta S_p = -mc^2 \; \delta \int_{\uptau_\mathscr{A}}^{\uptau_\mathscr{B}} d \uptau = -mc \;  \delta  \int_{\sigma_\mathscr{A}}^{\sigma_\mathscr{B}} \sqrt{ - g_{\mu \nu} (x^\lambda) \frac{dx^\mu}{d \sigma} \frac{d x^\nu}{d \sigma}} d \sigma = 0 \label{eq:actionfreemassiveparticle}
\end{equation}
where $\sigma$ is a parameter along the path, and the action functional is simply proportional to the proper time functional, i.e. proper time interval $\Delta \uptau$ between two timelike separated events $\mathscr{A}$ and $\mathscr{B}$. Applying the action principle leads via the Euler-Lagrange equation to the geodesic equation, i.e. geometrically defining a timelike geodesic as
\begin{displayquote}
    The worldline of a free test particle between two timelike separated events $\mathscr{A}$ and $\mathscr{B}$ renders stationary the proper time between them.
\end{displayquote}

The analogous variational principle for photons is Fermat's principle. In the classical theory, Fermat derived the laws of refraction in geometric optics by requiring that light travels on the path of least time \cite{Born1980}. Yet in general relativity there is no concept of a global time for an arbitrary spacetime, and light rays correspond to null geodesics\footnote{Photons travel on \textit{affine geodesics} along which its tangent vector is parallel propagated; they may be interpreted as the ``straightest possible'' curves; rather than the \textit{metric geodesics} arising from the action principle where the \textit{proper time} functional of massive particles (or length functional for spacelike geodesics) is stationary \cite{DInverno1992}. Metric and affine geodesics coincide for massive particles in GR.} such that the line element $ds^2 = -c^2 d \uptau^2$ is always zero: loosely speaking, photons do not themselves possess a proper time unlike massive particles. More precisely, the form of the action $S_p$  proportional to the proper time functional is not valid for null geodesics due to the square root. For variations of null curves it is ill-defined: even infinitesimal variations of a null curve will make some sections of the curve spacelike and other sections timelike (see e.g. p.44, \cite{Wald1984}).

In the general relativistic formulation of Fermat's principle, valid for arbitrary gravitational fields and in the geometric optics approximation, it is the \textit{arrival time functional} of the photon as measured along the worldline of a particular observer which plays the role of the action. This was first put forward by Kovner \cite{Kovner1990} with a full proof by Perlick \cite{Perlick1990}:
\begin{displayquote}
    Let $\mathscr{P}$ be a null curve from a fixed (photon) emission event $\mathscr{E}_s$ to a possible observation event  $\mathscr{E}_o$ at its intersection with a timelike observer worldline $\mathscr{Q}$. $\mathscr{P}$ is a null geodesic (light ray) if and only if the functional of its arrival time, measured along $\mathscr{Q}$, is stationary with respect to variations of $\mathscr{P}$ restricted to null curves.
\end{displayquote}

\begin{figure}
    \centering
    \includegraphics[scale=0.7]{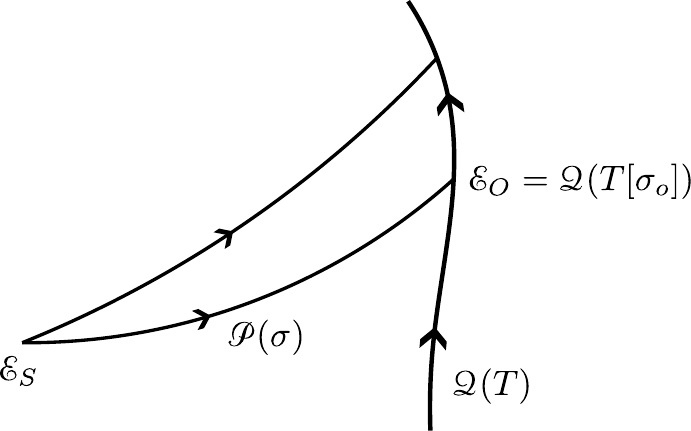}
    \caption{The observer travels in spacetime along a worldline $\mathscr{Q}(T)$, and a trial null path $\mathscr{P}$ connects the emission event $\mathscr{E}_S$ to a possible observation event $\mathscr{E}_O$ along the observer's worldline. The possible photon path $\mathscr{P}$ is parametrised by $\sigma$, so the possible observation events occur at time $T_o = T[\sigma_o]$. Actual paths taken by photons from the emission event to an actual observation are the null paths which make stationary the arrival time functional $T[\sigma_o]$ determined by $\mathscr{P}$.}
    \label{fig:fermatcoordfree}
\end{figure}
 If the timelike curve $\mathscr{Q}(T)$ is parametrised by $T$, let us use the notation $T[\mathscr{P}(\sigma_o)]$, or the shorthand $T[\sigma_o]$, to denote the arrival time functional of the null curve $\mathscr{P}(\sigma)$.\footnote{This is strictly overloading notation by using the same symbol for both the time parameter $T$ and the time functional $T[\sigma_o]$. In other works, the path dependence is sometimes not explicitly written and $T$ is used to refer to both the parameter and functional. However, this is fairly standard (analogous to the also common practice of using the same notation for the proper time functional and the proper time parameter, or proper length functional and length parameter, e.g. p.106 \cite{Carroll2004} and p.84 \cite{DInverno1992}).} We may choose the worldline of the observer to be parametrised by, and so measure the arrival time with the proper time $\uptau$, but any smooth parametrisation also holds. This is analogous to the parametrisation-invariance of the action for a free massive particle \eqref{eq:actionfreemassiveparticle}, where we could choose any parameter $\sigma \rightarrow \sigma'(\sigma)$ to measure the length along the path between $\mathscr{A}$ and $\mathscr{B}$. The restriction to null trial paths ensures that locally, the photon moves at $c$ at every point.

The generalised Fermat's principle enjoys a convenient symmetry between future-oriented and past-oriented light rays: it can be equivalently expressed instead considering a fixed observation event and considering all the past-oriented null curves joining the observation point and the timelike worldline of a continuously emitting source, with the relevant quantity being extremised being the proper time of the source.  One may picture the arrival time functional as an \textit{arrival time surface} over a space of trial paths, where the local minima, maxima\footnote{The general form does not permit local maxima - we can always make any given path longer by adding arbitrary wiggles - but when the trial space is restricted and therefore finite e.g. by the thin lens approximation, then there can be local maxima.} and saddlepoints of the functional pick out the geodesics -- the curves actually followed by light rays.

The generalised Fermat’s principle for light rays can be written in coordinate form and then reduced for three increasingly restricted classes of spacetimes: the conformally stationary, conformally static and conformally spatially Euclidean metrics \cite{Perlick1990, Perlick2004}. Although the generalised formulation was first presented comparatively recently in 1990, the version restricted to static spacetimes was derived by Weyl in 1917. 

\subsection{Standard Variational Principle for Geodesics}

In this subsection, we review the standard variational principle leading to the geodesic equation. The variation of an action with respect to a path with coordinates $x^\mu$ and parameterised by $\sigma$, defining the tangent vector $\dot{x}^\mu \equiv \frac{d x^\mu}{d \sigma}$, is
\begin{align}
    \delta S  &= S[x^\mu(\sigma) + \delta x^\mu (\sigma)] - S[x^\mu (\sigma)] \\
    &= \int_{\sigma_\mathscr{A}}^{\sigma_{\mathscr{B}}} L(x^\mu +  \delta x^\mu, \dot{x}^\mu + \delta \dot{x}^\mu; \sigma ) - L(x^\mu, \dot{x}^\mu ; \sigma) d \sigma\\
    &= \int_{\sigma_\mathscr{A}}^{\sigma_\mathscr{B}} \left( \frac{\partial L}{\partial x^\mu} \delta x^\mu + \frac{\partial L}{\partial \dot{x}^\mu} \delta \dot{x}^\mu \right)  d \sigma\\
    &= \left[  \frac{\partial L}{\partial \dot{x}^\mu} \delta x^\mu \right]^{\sigma_\mathscr{B}}_{\sigma_\mathscr{A}} + \int_{\sigma_\mathscr{A}}^{\sigma_\mathscr{B}} \left(  \frac{ \partial L}{\partial x^\mu} - \frac{d}{d \sigma} \left( \frac{\partial L}{\partial \dot{x}^\mu}\right)\right)  \delta x^\mu d \sigma \label{actionvariation}
\end{align}
using a first order Taylor expansion, and integration by parts since $\delta x^\mu$ and $\delta \dot{x}^\mu$ are not independent. The \textit{standard} action principle involves setting $\delta S =0$ along with fixed endpoints, such that the first term in Equation \eqref{actionvariation} vanishes. Since $\delta x^\mu$ is completely arbitrary, we obtain the usual Euler-Lagrange equations $\frac{ \partial L}{\partial x^\mu} - \frac{d}{d \sigma} \left( \frac{\partial L}{\partial \dot{x}^\mu}\right) =0$. In the derivation\footnote{See also from p.324 of \cite{Spivak1999} for details, as well as p.320 justifying the use of a more transparent, modern notation for variational calculus which we do not follow as it has more ``moving parts'' -- i.e. explicitly considering an entire family of curves rather than a single one.} of equations of motion for a  particle, or more generally any timelike, null or spacelike geodesic, the Euler-Lagrange equations lead to the geodesic equation for a suitable Lagrangian. In contrast to the standard variational principle, the generalised Fermat principle will allow one endpoint to vary and is subject to an additional constraint equation. 

The action or Lagrangian given in Equation \eqref{eq:actionfreemassiveparticle} is not unique; any Lagrangian is valid if it yields the correct equations of motion. One example of an alternative Lagrangian is proportional to the square of the original (the proper time functional). It is therefore sometimes called a ``dynamic'' -- i.e. analogous to a kinetic energy -- Lagrangian or action, as opposed to the ``geometric'' path length or proper time. This new action is valid for null curves, due to removing the square root, in addition to timelike and spacelike curves, but it is no longer parametrisation invariant
\begin{equation}
    L(x^\mu(\sigma), \dot{x}^\mu(\sigma) ; \sigma) = \frac{1}{2} g_{\mu \nu}( x ^\lambda) \dot{x}^\mu \dot{x}^\nu. \label{L2}
\end{equation}
This amounts to having picked a gauge for the parametrisation, and requires the parameter to be affine (e.g. proper time in the case of a timelike curve) in order for the action to be stationary for geodesic curves; i.e. it leads to the affine geodesic equation 
\begin{align}
    0 = \frac{ \partial L}{\partial x^\mu} - \frac{d}{d \sigma} \left( \frac{\partial L}{\partial \dot{x}^\mu}\right) = - g_{\mu \nu}   \left( \frac{d \dot{x}^\nu}{d \sigma}  + \Gamma \indices{^\nu_{\alpha \lambda}} \dot{x}^\alpha \dot{x}^\lambda \right) \equiv - \frac{D \dot{x}_\mu}{d \sigma}
\end{align}
using the definition of the Christoffel symbols $\Gamma\indices{^\mu_{\nu \rho}} \equiv g^{\mu \alpha} \Gamma_{\alpha \nu \rho}$ with $\Gamma_{\alpha \nu \rho} \equiv \frac{1}{2} (g_{\alpha \nu, \rho} + g_{\alpha \rho, \nu} - g_{\nu \rho, \alpha})$ and the definition of a covariant derivative of a vector along a curve (directional covariant derivative). That  $\sigma$ is necessarily affine, i.e. gives a constant tangent vector along the curve \cite{Spivak1999}, can be seen from calculating the change in the length of the tangent vector:
\begin{align}
    \frac{d}{d \sigma} \left( g_{\mu \nu} \dot{x}^\mu \dot{x}^\nu \right) &= g_{\mu \nu, \alpha} \dot{x}^\mu \dot{x}^\nu \dot{x}^\alpha + g_{\mu \nu} ( \ddot{x}^\mu \dot{x}^\nu + \ddot{x}^\nu \dot{x}^\mu)\\
    &= ( \Gamma_{\mu \nu \alpha} + \Gamma_{\nu \mu\alpha}) \dot{x}^\mu \dot{x}^\nu \dot{x}^\alpha + g_{\mu \nu} ( \ddot{x}^\mu \dot{x}^\nu + \ddot{x}^\nu \dot{x}^\mu)\\
    &= ( g_{\mu \nu} \ddot{x}^\mu + \Gamma_{\nu \mu \alpha} \dot{x}^\mu \dot{x}^\alpha) \dot{x}^\nu + (g_{\mu\nu} \ddot{x}^\nu + \Gamma_{\mu \nu \alpha} \dot{x}^\nu \dot{x}^\alpha ) \dot{x}^\mu
\end{align}
and so if the geodesic equation is obtained, both terms in parentheses are $0$.

Alternatively, we can use an action containing an einbein \cite{Green2012, Polchinski1998, Mitsou2020}. The mass-shell constraint (null wavevector constraint) appears directly in the action principle such that the parametrisation invariance of the original action $S_p$ is recovered whilst being unified for both massless and massive particles. Let us modify the Lagrangian function \eqref{L2} to the following
\begin{equation}
    L(x^\mu(\sigma), \dot{x}^\mu(\sigma), e(\sigma); \sigma) = \frac{g_{\mu \nu}\dot{x}^\mu \dot{x}^\nu}{2e} - \frac{m^2 c^2}{2}e. \label{einbeinlagrangian}
\end{equation}
Here, $e(\sigma)$ is an auxiliary field called the \textit{einbein}, an unknown independent function. Since the Lagrangian is independent of $\dot{e}$, the einbein plays the role of a generalised Lagrange multiplier. The geometric interpretation is that the einbein specifies a $1D$ metric $g_{\sigma \sigma} = - e^2$ on the particle worldline; hence it  is analogous to a lapse function \cite{Mitsou2020, Gielen2023} in the threading or foliation formalisms which we will discuss in Section \ref{sec:threading} and determines the parametrisation of the worldline. The etymology is from the German ``one-leg'' (it is the 1D version of what is sometimes called an n-bein or tetrad formalism).

Variation of the action now includes an extra term due to variation with respect to $e$, exactly as we would have for a Lagrange multiplier
\begin{equation}
    \delta S = \left[ \frac{\partial L}{\partial \dot{x}^\mu}  \delta x^\mu \right]^\mathscr{B}_{\mathscr{A}} + \int_{\sigma_\mathscr{A}}^{\sigma_\mathscr{B}} \left( \left( \frac{\partial L}{\partial x^\mu} - \frac{d}{d \sigma}\left( \frac{\partial L}{\partial\dot{x}^\mu}\right) \right) \delta x^\mu + \frac{\partial L}{\partial e} \delta e \right) d \sigma \label{einbeindeltaS}
\end{equation}
We compute $\frac{\partial L}{ \partial x^\mu} = \frac{1}{2}g_{\lambda \nu, \mu} \dot{x}^\lambda \dot{x}^\nu e^{-1}$, the conjugate momenta $p_\mu = \frac{\partial L}{\partial \dot{x}^\mu} = \dot{x}_\mu e^{-1}$, $\frac{d}{d \sigma} \left( \frac{\partial L}{ \partial \dot{x}^\mu} \right) = \left( \ddot{x}^\lambda g_{\mu \lambda} + \tfrac{1}{2}( g_{\mu \lambda, \nu} + g_{\nu \mu, \lambda}) \dot{x}^\lambda \dot{x}^\nu \right) e^{-1} - \dot{x}_\mu e^{-2} \dot{e}$ and $\frac{\partial L}{\partial e} = - \frac{1}{2} \left(\dot{x}_\mu \dot{x}^\mu e^{-2} + m^2c^2\right)$. The action principle $\delta S =0$ using fixed endpoints now gives the mass shell constraint in addition to the Euler-Lagrange equation
\begin{gather}
    \frac{D \dot{x}^\mu}{d \sigma} - \dot{x}^\mu \dot{e} e^{-1} = 0 \label{einbeineom}\\
    p_\mu p^\mu = g_{\mu \nu} \dot{x}^\mu \dot{x}^\nu e^{-2} = -m^2 c^2. \label{einbeinmassshell}
\end{gather}
Solving the constraint for the einbein and taking the positive root only (such that the action corresponds to the standard minimum for infinitesimal paths of integration) gives
\begin{equation}
    e = \frac{(- g_{\mu \nu}\dot{x}^\mu \dot{x}^\nu)^{\tfrac{1}{2}}}{mc}\label{einbeinsoln}
\end{equation}
i.e. the einbein specifies the norm of the tangent vector. The case for massless particles with $m=0$ is very simple: equation \eqref{einbeinmassshell} becomes $\dot{x}_\mu \dot{x}^\mu e^{-2} =0$, and substitution into \eqref{einbeineom} gives the usual geodesic equation $\frac{D \dot{x}^\mu}{d \sigma} =0$. The einbein $e$ therefore remains unspecified (with restriction $e>0$); it is a gauge freedom.

On the other hand, solving the equation of motion \eqref{einbeineom} for $m\neq 0$ requires choosing a gauge. For example, choosing $e= \frac{1}{mc}$ such that $\dot{e} = 0$ means that $\sigma$ is affine (i.e. proper time) and we recover the geodesic equation $\frac{D \dot{x}^\mu}{d \sigma} =0$. Note that we can also recover the original square root Lagrangian for massive particles by substituting \eqref{einbeinsoln} back into the Lagrangian \eqref{einbeinlagrangian}.

\subsection{Proof of Fermat's Principle}

\paragraph{Fermat's Principle}

Kovner's version of Fermat's Principle has two particular points of difference from the usual action principle. First, it \textit{allows for one endpoint to vary} along an observer trajectory, and secondly it requires that \textit{the only allowable trial curves are those which obey the mass shell constraint}. Let the timelike observer curve $\mathscr{Q}$ be given by the coordinates $x^\mu = q^\mu (\uptau)$ parametrised by proper time $\uptau$ (although any other smooth time parametrisation also works) and tangent four velocity $u^\mu = \frac{d q^\mu}{d (c \uptau)}$.  We consider the case of a photon intercepting the observer worldline, but given the previous section, analogous results are also valid for a massive particle. Let a trial photon path $\mathscr{P}$ with coordinates $x^\mu = r^\mu (\sigma)$ and tangent vector $k^\mu = \frac{d r^\mu}{d \sigma}$, with constraint $k_\mu k^\mu =0$ such that the trial path is always null, run from a fixed emission event $\mathscr{E}_S$ to the observer curve. This is illustrated in Figure \ref{fig:fermatcoordfree}, substituting proper time $\uptau$ for the arbitrary $T$. The observation event is allowed to vary as the intersection of the trial null path at $\sigma_o$ with the observer curve
\begin{equation}
     r^\mu (\sigma_o) = q^\mu (\uptau[\sigma_o])
 \end{equation}
Therefore, the boundary term at the observation event no longer automatically vanishes; and we can write the variation of the action \eqref{einbeindeltaS} as
\begin{equation}
 \delta S =   \left[\frac{k_\mu}{e} u^\mu c\right]_{\sigma_o} \delta \uptau[\sigma_o] - \int_{\sigma_s}^{\sigma_o} \left(\left(\frac{D k^\mu}{e d \sigma} - \frac{\dot{e} k^\mu}{e^2} \right) \delta r^\mu + \left( \frac{k_\mu k^\mu}{2e^2} + \frac{m^2c^2}{2} \right) \delta e \right) d \sigma =0. \label{variationfermat}
\end{equation}
where $m=0$. The assumption that any trial curve obeys the mass shell constraint or null wavevector condition means the last term must vanish. Another viewpoint is: we do not need to start from the action principle; computing the variation of the mass shell condition gives the boundary and Euler-Lagrange terms which together are then identically $0$ \cite{Kovner1990, Schneider1992}.

\paragraph{Geodesic $\implies \delta \uptau[\sigma_o] =0$} If the trial null curve is geodesic, then $\frac{D k^\mu}{e d \sigma} - \frac{\dot{e} k^\mu}{e^2} = 0$. Since $k_\mu$ is null and $u^\mu$ is timelike, they cannot be orthogonal to one another, i.e. $k_\mu u^\mu \neq 0$. In fact they are both future oriented  $k_\mu u^\mu >0$ and $e>0$; so $\delta \uptau[\sigma_o] =0$. However, we now need to show sufficiency; i.e. that if $\delta \uptau[\sigma_o] =0$ then the trial curve is geodesic.

\paragraph{$\delta \uptau[\sigma_o] =0 \implies$ geodesic}
 Due to the null wavevector condition for all trial curves, $\delta r^\mu$ can no longer be considered completely arbitrary and the vanishing of the middle term (Euler-Lagrange term) cannot be assumed to automatically guarantee the geodesic equation. To show sufficiency, we follow an argument given by \cite{Fleury2015}. The trick is that although $\delta r^\mu$ is not arbitrary, $\delta r^i$ can be considered arbitrary with $\delta r^0$ then determined by the null condition.

Since we are in a completely general spacetime, we define a local observer rest frame using a tetrad basis \cite{Mitsou2020, Carroll2004}, i.e. a set of four orthonormal vectors $\bm{e}_{\hat{\alpha}}$ where the tetrad indices $\hat{\alpha} \in \{0,1,2,3\}$ label each vector. In terms of the spacetime components ${e_{\hat{\alpha}}}^\mu$, the observer's four-velocity sets ${e_{\hat{0}}}^\mu =u^\mu$, and the spatial frame components have arbitrary orientation, but the orthonormality restriction requires $u_\mu {e_{\hat{i}}}^\mu =0$ and ${e_{\hat{i}}}_\mu {e_{\hat{j}}}^\mu = \delta_{\hat{i} \hat{j}}$ (i.e. $g_{\mu \nu} {e_{\hat{\alpha}}}^\mu {e_{\hat{\beta}}}^\nu = \eta_{\hat{\alpha}\hat{\beta}}$). The matrix ${e_{\hat{\alpha}}}^\mu$ (sometimes called the \textit{vierbein}, ``four-leg'') encodes the changing of frames between the tetrad frame and the coordinate frame.

We construct such a tetrad at the observation event, using the observer's four velocity $u^\mu|_o$, and parallel transport the tetrad back along the null trial curve, generating a tetrad field $\bm{e}_{\hat{\alpha}}(\sigma)$ along the null curve. We can understand this field as a field of the frames of hypothetical observers, each with $\bm{e}_{\hat{0}}$ in the direction of their four-velocity but each of their spatial frames is arbitrarily oriented. We can then write $\delta r^\mu = \delta r^{\hat{0}} {e_{\hat{0}}}^\mu + \delta r^{\hat{i}}{e_{\hat{i}}}^\mu$ for the variation of the trial null curve. The value of $\delta r^{\hat{i}}$ is therefore arbitrary with $\delta r^{\hat{0}}$ then determined by the null condition; i.e. that a particle on the varied path moves at $c$ in each of the hypothetical observer frames. Consider that variation of the null wavevector condition, in tetrad components, requires\footnote{Cf. equation \eqref{firstordernullwavevectorcond} with an unperturbed metric, and dropping the barred notation for the trial path as standard for variational calculus; i.e. the varied curve is $r^\mu + \delta r^\mu$.} that
\begin{equation}
    k_{\hat{\alpha}} \delta k^{\hat{\alpha}} =0.
\end{equation}
Expanding the summation and integration by parts gives
\begin{align}
    [\delta r^{\hat{0}}]_{\sigma_o} &= -\int_{\sigma_s}^{\sigma_o} \frac{k_{\hat{i}}}{k_{\hat{0}}} \delta k^{\hat{i}} d \sigma\\
    &= \left[ \frac{k_{\hat{i}}}{k_{\hat{0}}} \delta r^{\hat{i}} \right]_{\sigma_o}^{\sigma_s} + \int_{\sigma_s}^{\sigma_o} \frac{d}{d \sigma} \left( \frac{k_{\hat{i}}}{k_{\hat{0}}} \right) \delta r^{\hat{i}} d \sigma.
\end{align}
Since the spatial variation $r^{\hat{i}}$ is zero at the end points, the first term  (boundary term) disappears. Then if $c \delta \uptau[\sigma_o] = [\delta r^{\hat{0}}]_{\sigma_o}  = 0$, we have that $\frac{d}{d \sigma} \left( \frac{k_{\hat{i}}}{k_{\hat{0}}} \right)= \frac{d k_{\hat{i}}}{d \sigma} - \frac{k_{\hat{i}}}{k_{\hat{0}}} \frac{d k_{\hat{0}}}{d \sigma} = 0$
since $\delta r^{\hat{i}}$ is arbitrary; and this expression is also valid for the $\hat{\alpha}=0$ component by direct substitution. In terms of spacetime coordinates this gives
\begin{equation}
  \left(\frac{d k_\mu}{d \sigma} - \frac{k_\mu}{k_0} \frac{d k_0
  }{d \sigma}\right) {e_{\hat{\alpha}}}^\mu = 0.\label{notgeodesicyet}
\end{equation}
Since the tetrad basis is parallel transported, the covariant derivative and the total derivative along the curve coincide. This can be seen by using 
\begin{equation}
    \frac{D {e_{\hat{\alpha}}}^\mu}{d \sigma} = \dot{r}^\nu \nabla_\nu {e_{\hat{\alpha}}}^\mu = \dot{r}^\nu {e_{\hat{\alpha}}}^\mu_{, \nu} + \dot{r}^\nu \Gamma\indices{^\mu_{\nu \lambda}} {e_{\hat{\alpha}}}^\lambda = \dot{r}^\nu \Gamma\indices{^\mu_{\nu \lambda}} {e_{\hat{\alpha}}}^\lambda = 0 
\end{equation}
and from $\dot{r}^\nu \Gamma\indices{^\mu_{\nu \lambda}} {e_{\hat{\alpha}}}^\lambda =0$ we obtain 
\begin{equation}
    \frac{D (k_\mu {e_{\hat{\alpha}}}^\mu)}{d \sigma} = k_{\mu; \nu} \dot{r}^\nu {e_{\hat{\alpha}}}^\mu  =  ( k_{\mu, \nu} + \Gamma\indices{^\lambda_{\mu \nu}} k_\lambda ) \dot{r}^\nu  {e_{\hat{\alpha}}}^\mu =  k_{\mu, \nu} \dot{r}^\nu  {e_{\hat{\alpha}}}^\mu = \frac{d (k_\mu  {e_{\hat{\alpha}}}^\mu)}{d \sigma}.
\end{equation}
Though we treated $k_\mu$ temporarily as though it is defined in the full spacetime and not just along the path, the final result does not require $k_\mu$ to be defined other than on the path. Formally one could treat $k_\mu$ defined everywhere along a family of null curves. Applying this result to Equation \eqref{notgeodesicyet}, we have a geodesic equation for a non-affine parameter (e.g. p.75, 101 \cite{DInverno1992}), showing that the tangent vector is parallel-propagated
\begin{equation}
\frac{D k_\mu}{d \sigma} -  \frac{\dot{k_0}}{k_0} k_\mu = 0.
\end{equation}
We can therefore regain the affine null geodesic equation by choosing an affine parameter so $d \sigma ' = k_0 d \sigma$, i.e. $\sigma' = r_0$ (under this transformation the derivative of $r_0$ with respect to itself is 1, and then second order derivative is 0 and hence the second term vanishes; equivalent to setting the einbein $e=\text{const.}$). We see that $\delta \uptau[\sigma_o] =0 \implies$ the trial null curve is a geodesic.

\subsection{Eikonal Form}

{\hl Fermat's principle can also be written in terms of the phase function or eikonal; this is often stated for Euclidean space in textbooks (e.g. \cite{Landau1951}, additionally understood in terms of the classical limit of the path integral formulation of quantum mechanics \cite{Thorne2017}, p.21 \cite{Landau1977a}). Using the Lagrangian \eqref{L2}, we can write
\begin{equation}
   \delta \int_{\mathscr{E}_s}^{\mathscr{E}_o} k_\mu dx^\mu  = \delta \int_{\mathscr{E}_s}^{\mathscr{E}_o} d \varphi =  \delta \varphi[ \mathscr{P}] = 0
\end{equation}
thus the action is the phase \textit{functional} $\varphi[ \mathscr{P}] \equiv \varphi(\mathscr{E}_o) - \varphi(\mathscr{E}_s)$, defined as the difference in the phase \textit{function} evaluated at the end points of the path $\mathscr{P}$; analogous to how the time functionals of the previous section were defined as time intervals. The phase functional is the constant of motion associated with integrating both sides of the null wavevector condition along a geodesic: $\int k_\mu k^\mu d \sigma = \int k_\mu dx^\mu = \varphi[\mathscr{P}] = \text{const.}$ (p. 176, \cite{Hartle2003}); and from the constancy of the phase function (the first law of geometric optics) for a true path $\varphi[ \mathscr{P}] =0$ also.

Fermat's principle is now in a form reminiscent of the principle of Maupertuis, which applies in classical mechanics when the Hamiltonian is associated with the total energy and is constant. In general relativity the energy is a component of a four-vector rather than a scalar, i.e. observer-dependent. Nonetheless it is still possible to treat photon dynamics and gravitational lensing with a Hamiltonian formalism, albeit the Hamiltonian is \textit{not} the energy \cite{Bertschinger1999a, bertschinger1999b, bar-kana1997}.

We may also write $\delta \varphi[\mathscr{P}] = k_\mu \delta x^\mu$ where $x^\mu(\sigma)$ are the coordinates along the path $\mathscr{P}$. The ``conservation of $k_\mu \delta x^\mu$'' along rays can also be proven as a separate theorem \cite{Kermack1932, Sachs1961} and in a different form \cite{Synge1960} p.21, \cite{Poisson2004} p.17.} There is also an alternative formulation of Fermat's principle as a variational principle for instantaneous wavefronts as opposed to rays \cite{Fritelli2002b}, see also \cite{Perlick2004}, using the eikonal equation as the starting point.

\subsection{The Threading of Spacetime} \label{sec:threading}

To aid the application of Fermat's principle in a coordinate form, we consider the \textit{threading} of spacetime into a set of timelike integral curves. This is very similar to, but more general than, Weyl's Postulate discussed in Section \ref{sec:cosmologicalprinciple}: the threading formalism allows for worldlines to intersect \cite{Bertschinger1995} (i.e. the curves do not necessarily form a congruence) and does not require foliating the spacetime. When made explicit in the metric, it is also known as the ``1+3'' formalism \cite{Roy2014, Jantzen, Bertschinger1995, Elst1997} -- not to be confused with the ``3+1'' formalism which corresponds to a \textit{foliation} or slicing of spacetime into spatial hypersurfaces.

The integral curves of the threading formalism can be interpreted as the worldlines of a family of test observers comoving with the fluid sourcing this spacetime or \textit{fundamental observers}, characterised by their four-velocity field $\bm{u}$ which fixes the local time direction at each point of spacetime. It corresponds to a Lagrangian description of the fluid dynamics. The 1+3 formalism is particularly useful in the case of stationary spacetimes which possess a timelike Killing vector field.

We write the metric in coordinates $(x^0, x^i)$ with basis $(\bm{e}_{0}, \bm{e}_i)$ adapted to the congruence; such that $\bm{e}_0$ is tangent along each integral curve $\mathscr{Q}_{x^i}$ labelled by the spatial coordinates, and the temporal coordinate $x^0$ is a parameter along the curves, labelling a family of hypersurfaces $\Sigma_{x^0}$. A general metric may be written in the conformal form
\begin{equation}
\boxed{
    g_{\mu \nu} dx^\mu dx^\nu = - g_{00} ( - (dx^0)^2 +2 \hat{g}_{0i} dx^0 dx^i + \hat{g}_{ij} dx^i dx^j  ) }\label{conformalformgeneralmetric}
\end{equation}
where the conformal metric coefficients are $\hat{g}_{\mu i} \equiv g_{\mu i} (- g_{00})^{-1}$ and the metric signature is preserved since $-g_{00} >0$. Often, $g_{00}$ is written in an exponential form to make this condition explicit within the expression itself.

The presence of the cross term involving $\hat{g}_{0i}$ indicates the global basis vectors $\bm{e}_{0}$ and $\bm{e}_i$ are \textit{not} in general orthogonal. We may locally construct spatial bases orthogonal to the curves, which form the local observer frame, but if there is a non-vanishing fluid vorticity, these do not knit together into global integrable hypersurfaces (see \cite{Wald1984}, \cite{Carroll2004} or \cite{Poisson2004} for a proof using the Frobenius theorem). The $\hat{g}_{0i}$ term can be interpreted as a tilt and/or shearing of the local inertial frame with respect to the globally-defined one. The vanishing of $\hat{g}_{0i}$, the \textit{shift} or \textit{tilt}, implies vanishing vorticity; although the converse is not necessarily true as $\hat{g}_{0i}$ may be separated into a vortical component and the gradient of a scalar function via a Helmholtz decomposition.

\begin{figure}
    \centering
    \includegraphics[scale=0.7]{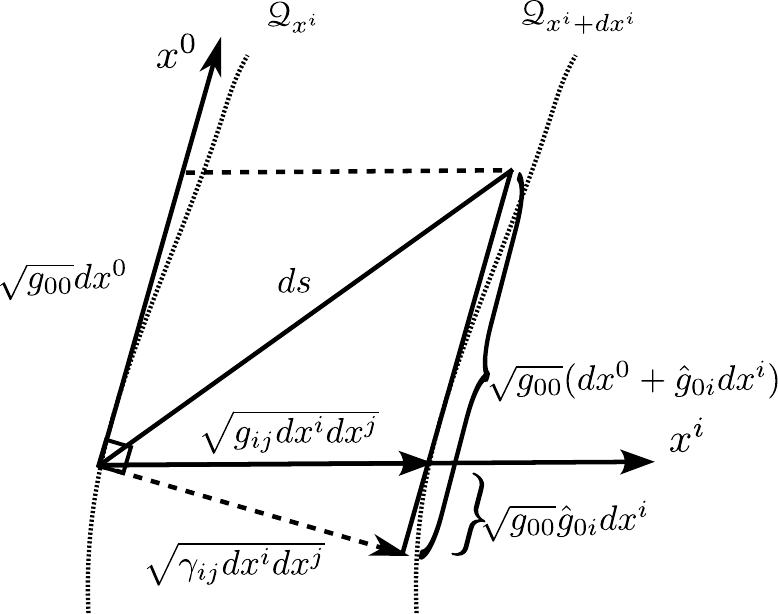}
    \caption[Orthogonalisation of spacetime.]{Two neighbouring time-like curves in a set filling a general spacetime, labelled by $\mathscr{Q}_{x^i}$ and $\mathscr{Q}_{x^i +dx^i}$. They may be interpreted as the worldlines of comoving observers. The tangent vector, or four-velocity, field defines the time direction of the global coordinates $(x^0, x^i)$ at every point in spacetime. The hypersurfaces constant in $x^0$ are not orthogonal to the $x^0$ curves. However, we may locally decompose the basis vectors into orthogonal ones (i.e. a local inertial frame) and thus the metric via a completion of squares. Orthogonality here is illustrated as Euclidean, but should be interpreted in the Lorentzian sense; and we have suppressed two spatial dimensions. The choice of decomposition here keeps the temporal direction, but changes the spatial direction; this is known as the ``1+3'' decomposition corresponding to the threading of spacetime by the worldlines of comoving observers. The reciprocal decomposition which keeps the spatial directions but alters the temporal one is the ``3+1'' decomposition corresponding to the foliation of spacetime. Figure 4.1 of \cite{Fleury2015} may be a helpful complementary (foliation-based) visualisation. Figure 3 of \cite{Jantzen} gives an illustration respecting the true Lorentzian nature of the inner product.}
    \label{fig:time_orthog}
\end{figure}

The general metric \eqref{conformalformgeneralmetric} can be decomposed in the 1+3 formalism via a completion of squares -- keeping the basis vector in the time direction, but orthogonalising the spatial bases. The procedure is illustrated by Figure \ref{fig:time_orthog}, and gives the form
\begin{align}
    g_{\mu \nu} dx^\mu dx^\nu  &= - g_{00} ( (\hat{g}_{ij} + \hat{g}_{0i} \hat{g}_{0j}) dx^i dx^j -( \hat{g}_{0i} \hat{g}_{0j} dx^i dx^j - 2 \hat{g}_{0i}dx^0 dx^i + (dx^0)^2 ))\\
    &= - g_{00} ( \hat{\gamma}_{ij} dx^i dx^j -(- \hat{g}_{0i} dx^i + dx^0)^2) \label{1+3decompmetric_one}
\end{align}
where we defined the conformal orthogonal spatial projection of the metric $\hat{\gamma}_{ij} \equiv \hat{g}_{ij} + \hat{g}_{0i} \hat{g}_{0j}$. If we consider the decomposition with respect to the four velocity field $\bm{u}$ explicitly, we may also write $g_{00} = -u_0^2 = -M^2 $ and $\hat{g}_{0i} = u_i u_0^{-1} = M_i$, in terms of the \textit{threading lapse function} $M(x^0, x^i)$ and the \textit{threading shift one-form} $M_i (x^0, x^i)$. (Alternative names for special cases specific to lensing are also given on p.44 of \cite{Perlick2004}.) They are related to the fluid four-velocity by $u^{\mu} = M^{-1}(1, 0, 0, 0)$ and $u_{\mu} = M(-1, M_i)$; the lapse function specifies the relation between the coordinate time and proper time along the threading, and the shift function may also be interpreted as the relative velocity between the comoving and local inertial frames. To do this, we make the following arguments.

Consider that we may decompose any vector $\bm{v}$ into components parallel to and perpendicular to $\bm{u}$ as
\begin{equation}
    v^\mu = v^{\mu}_{\parallel \bm{u} } + v^{\mu}_{\perp \bm{u}} = - u^\mu u_\nu v^\nu + (\delta\indices{^\mu_\nu} + u^\mu u_\nu ) v^\nu
\end{equation}
and so we see that $-u^\mu u_\nu$ and $\gamma\indices{^\mu_\nu} \equiv \delta\indices{^\mu_\nu} + u^\mu u_\nu$ are respectively the parallel and orthogonal projection operators to $\bm{u}$. The metric acts on vectors $\bm{v}$ and $\bm{w}$ such that
\begin{align}
    \bm{g} ( \bm{v}, \bm{w}) &= \bm{g} ( \bm{v}_{\parallel \bm{u} }, \bm{w}_{\parallel \bm{u} }) + \bm{g} ( \bm{v}_{\perp \bm{u} }, \bm{w}_{\perp \bm{u} }) + \bm{g} ( \bm{v}_{\parallel \bm{u} }, \bm{w}_{\perp \bm{u} }) + \bm{g} ( \bm{v}_{\perp \bm{u} }, \bm{w}_{\parallel \bm{u} })\\
    &= \bm{g} ( \bm{v}_{\parallel \bm{u} }, \bm{w}_{\parallel \bm{u} }) + \bm{g} ( \bm{v}_{\perp \bm{u} }, \bm{w}_{\perp \bm{u} })
\end{align}
where we took the last two terms as $0$ as the input vectors are orthogonal. Now define the part of $\bm{g} \perp \bm{u}$, that is the orthogonal spatial projection of the metric, or the local spatial metric at the observer, as $\bm{\gamma}(\bm{v}, \bm{w}) = \bm{g} ( \bm{v}_{\perp \bm{u} }, \bm{w}_{\perp \bm{u} })$. Using 
\begin{equation}
    \bm{g} ( \bm{v}_{\parallel \bm{u} }, \bm{w}_{\parallel \bm{u} }) = g_{\mu \nu} v^\mu_{\parallel \bm{u} } w^\nu_{\parallel \bm{u} } = g_{\mu \nu} u^\mu u_\rho u^\nu u_\sigma v^\rho w^\sigma = -u_\mu u_\nu v^\mu w^\nu \end{equation}
we can write the metric tensor in coordinate form as
\begin{equation}
     g_{\mu \nu} = \gamma_{\mu \nu} - u_\mu u_\nu
\end{equation}
and we see that the orthogonal spatial metric  $\gamma_{\mu \nu} $ is the index-lowered projection operator. From expanding the corresponding line element $(\gamma_{\mu \nu} - u_\mu u_\nu) dx^\mu dx^\nu = - u_0^2 (dx^0)^2 -2u_0u_i dx^0 dx^i + (\gamma_{ij} - u_i u_j) dx^i dx^j$, we see that $g_{00} = -u_0^2 = -M^2 $ and $\hat{g}_{0i} = u_i u_0^{-1} = M_i$.

Since the proper time along each comoving observer worldline is related to the parameterisation by the coordinate time through $u^{0} = \frac{d x^0}{d(c \uptau)} = M^{-1}$, the threading formalism shows that we are free to take Fermat's principle to apply to the chosen coordinate time instead of proper time itself. In the case that $M(x^0, x^i)$ is independent of the spatial coordinates, the coordinate time may be rescaled to measure the proper time along the curve. For example considering the FLRW metric, the coordinate time measures the proper time (simply read off the metric $ds = -c dt$) along the worldlines of all of the comoving observers.

\subsection{Fermat's Principle for Conformally Stationary and Conformally Static Spacetimes}
A purely spatial variational principle arises for a conformally stationary spacetime considering a comoving observer (in the look-forward formulation; otherwise a comoving source in the look-backward formulation); i.e. the time arrival functional is dependent only on the spatial components of both the conformal metric and the photon path. A spacetime is stationary if it possesses a timelike Killing vector field, which means there exists a global preferred time direction. Thus, though a stationary spacetime can still be evolutionary, it can be written in preferred, explicitly time-dependent coordinates $x^{\mu}$ where $g_{\mu \nu , 0} = 0$ and $g_{00} < 0$.

A conformally stationary metric may be written in the form \eqref{conformalformgeneralmetric}, where $-g_{00} >0$ with the additional condition that the conformal metric coefficients $\hat{g}_{\mu i} \equiv g_{\mu i} (- g_{00})^{-1}$ obey $\hat{g}_{\mu i, 0} = 0$. To find an expression for the time arrival functional for a photon trial curve $\mathscr{P}$ parametrised by $\sigma$ with coordinates $x^\mu (\sigma)$ and tangent vector $k^\mu = \frac{d x^\mu}{d \sigma}$ we simply use the null condition $g_{\mu \nu} k^\mu k^\nu = -(k^0)^2 + 2 \hat{g}_{0i} k^0 k^i + \hat{g}_{ij} k^i k^j =0$ which gives
 \begin{equation}
     \frac{dx^0}{d \sigma} = k^0 = \hat{g}_{0i} k^i \pm \sqrt{(\hat{g}_{0i} \hat{g}_{0j} + \hat{g}_{ij}) k^i k^j}
 \end{equation}
and we choose the future-oriented (positive sign) solution. We recognise $\hat{\gamma}_{ij} = \hat{g}_{0i} \hat{g}_{0j} + \hat{g}_{ij}$ is the conformal orthogonal spatial projection of the metric. The arrival time functional of the null trial path $\mathscr{P}(\sigma)$ can be measured according to the time coordinate $\frac{x^0}{c}$, since the timelike observer curve is parametrised by the time coordinate in the threading formalism. In a conformally stationary spacetime, this is the integral
\begin{equation}
\boxed{
    \frac{x^0}{c}[\sigma_o] = \frac{1}{c} \int_{\sigma_s}^{\sigma_o} \hat{g}_{0i} k^i + \sqrt{(\hat{g}_{0i} \hat{g}_{0j} + \hat{g}_{ij})k^i k^j} d \sigma + C
    } 
\end{equation}
where one end of the null curve varies depending on the path and may be considered to define the arrival time on the comoving observer's worldline, $x^0[\sigma_o]/c$. The integration constant $C = x^0[\sigma_s]/c$ is the end of the null curve at the source fixed at the emission event and is irrelevant for the variational problem.

\begin{figure}
    \centering
    \includegraphics[width=0.7\textwidth]{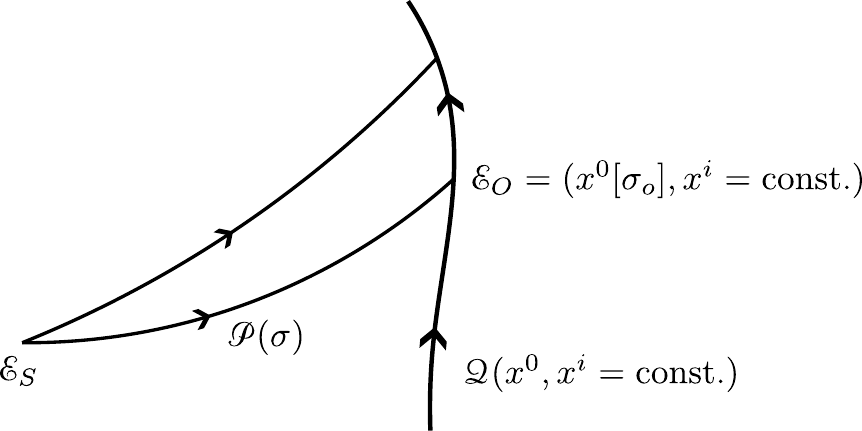}
    \caption[Illustration of Fermat's Principle.]{Illustration of Fermat's Principle in coordinate form, identical to Figure \ref{fig:fermatcoordfree} but with the time parameter along the observer worldline $\mathscr{Q}$ replaced by coordinate time: the observer travels in spacetime along a worldline $\mathscr{Q}(x^0, x^i = \text{const.})$ under the threading approach. A trial null path $\mathscr{P}$ connects the emission event $\mathscr{E}_S$ to a possible observation event $\mathscr{E}_O$ along the observer's worldline. The possible photon path $\mathscr{P}$ is parametrised by $\sigma$, so the possible observation events have coordinates $(x^0[\sigma_o], x^i = \text{const.})$. Actual paths taken by photons from the emission event to an actual observation are the null paths which make stationary the coordinate time (dividing by $c$) arrival functional $\frac{x^0}{c}[\sigma_o]$.}
    \label{fig:fermat}
\end{figure}

In static spacetimes which do not evolve with time, $g_{0i} = 0$ and $x^0$-lines form a timelike congruence of curves which are additionally orthogonal to the slicing of spacetime into global spatial hypersurfaces $x_0 = \text{constant}$; Weyl's Postulate now holds. For conformally static spacetimes which share the causal structure of static spacetimes, we therefore gain the concept of a ``global time'' (simultaneity on the orthogonal spatial hypersurfaces) for the privileged class of comoving observers. Fermat’s principle for conformally static spacetimes is the statement that the light rays are simply the geodesics of the spatial part of the conformal metric, $\hat{g}_{ij}$. The arrival time integral for a null trial path $\mathscr{P}(\sigma)$ in a conformally static spacetime reduces to
\begin{equation}
\boxed{
    \frac{x^0}{c}[\sigma_o] = \frac{1}{c} \int_{\sigma_s}^{\sigma_o} \sqrt{\hat{g}_{ij} k^i k^j} d \sigma + C}.
    \label{eq:statictimearrival}
\end{equation}
This is simply the length functional of the conformal spatial metric $\hat{g}_{ij}$ (modulo a factor of $c$) -- compare with Equation \eqref{eq:actionfreemassiveparticle} -- so the stationary paths correspond to the geodesics of the \textit{Fermat metric} or \textit{optical metric}  $\hat{g}_{ij}$. 

Finally, when the spatial metric is conformally Euclidean, where $\hat{g}_{ij} dx^i dx^j = n^2 \delta_{ij} dx^i dx^j$, Fermat’s principle reduces to the same form as in ordinary optics for a medium with refractive index $n$. Thus we can write 
\begin{equation}
    \frac{x^0}{c}[\sigma_o] = \frac{1}{c} \int_{\ell_s}^{\ell_o} n(x^i) \, d \ell + C \label{eq:arrivaltimeconformalEuclidean}
\end{equation}
where the Euclidean line element is $d \ell = \sqrt{\delta_{ij} \frac{d x^i}{d \sigma} \frac{d x^j}{d \sigma}} d \sigma$. The arrival time is then analogous to the least travel time of classical optics, with the optical path length $n d\ell = c dT$.

Consider that we can write the weak field perturbation of the FLRW metric \eqref{perturbedFLRW} in the following convenient conformal form
\begin{align}
    g_{\mu \nu} dx^{\mu} dx^{\nu} &= -c^2 A(x^i) d t^2 + a^2 (t) B(x^i) d\ell_K^2\\
     &= a^2 (\eta) A(x^i) \left( -c^2 d \eta^2 + \frac{B(x^i)}{A(x^i)} d \ell_K^2 \right) \label{eq:n_conformalFLRW}
\end{align}
We assume the non-relativistic limit such that the potential $\hat{\Phi} = \hat{\Psi} = \frac{\Phi}{c^2}$; and that $\Phi$ is conformally static so $A(x^i) \equiv (1 + 2  \frac{\Phi}{c^2})$, $B(x^i) \equiv (1 - 2 \frac{\Phi}{c^2})$ and the conformal background spatial line element is $d \ell_K^2 = \tilde{\gamma}_{ij} dx^i dx^j$ given by expression \eqref{flrwconformalspatialmetric}.

In the vicinity of the gravitational lens we can set $K=0$ such that $\hat{g}_{ij} dx^i dx^j = \frac{B}{A} d \ell^2$. Therefore, the resulting perturbed Minkowski metric possesses a conformally Euclidean Fermat metric\footnote{In fact, the general FLRW background metric is also spatially conformally Euclidean or flat, but in that case the effective refractive index becomes dependent on the radial coordinate.} and there exists an effective index of refraction at the lens
\begin{equation}
    n = \sqrt{\frac{B}{A}} =  1 - \frac{2 \Phi}{c^2} \label{eq:effecrefracindex}
\end{equation}
to first order in the perturbation. The effective refractive index can be interpreted analogous to classical optics as $n = \frac{c}{c'}$, in terms of a speed of light $c' \equiv \left| \frac{d \ell}{d \eta} \right| = c \sqrt{\frac{A}{B}}$ which is reduced from the value in the absence of gravity $c$. One can derive the thin-lens results found using the geodesic equation in the previous section by applying the variational principle to Equation \eqref{eq:arrivaltimeconformalEuclidean} with the effective refractive index defined in Equation \eqref{eq:effecrefracindex} \cite{Petters2001, Meneghetti2021}. The Euler-Lagrange equations then lead to the deflection angle in the lens frame under the thin-lens approximation.

\subsection{Fermat's Principle for Strong Lensing}

\subsubsection{Scalar Formalism and the Strong Lensing Limit of the Poisson Equation}

Fermat's principle becomes useful for strong lensing in practice when the space of null trial curves can be reduced using the thin lens approximation. Whilst working with differential equations for the lens mapping forms a vectorial approach, the variational approach presents an alternative scalar formalism. This scalar formalism originally emerged from the realisation that since the deflection angle $\bm{\hat{\alpha}}$, or equivalently the scaled deflection $\bm{\alpha}$, is curl-free {\hl (valid only for first-order lensing \cite{Lewis2006})}. Poincar\'e's theorem then implies that it can be written in terms of the gradient of a scalar field \cite{Schneider1985}
\begin{equation}
    \bm{\hat{\alpha}} = \bm{\nabla}_{\bm{\xi}} \hat{\psi}, \quad \bm{\alpha} =\bm{\nabla}_{\bm{\theta}} \psi \label{deflectionpotential}
\end{equation}
which suggested the lens equation can be found via Fermat's principle. To obtain the two-dimensional potentials $\hat{\psi}$ and $\psi$ we only need to note that the integral \eqref{alphahat} for $\bm{\hat{\alpha}}$ is only over the line of sight coordinate $\zeta$ and therefore the transverse gradient may simply be moved outside the integral. This leaves the \textit{dimensionful lensing potential} $\hat{\psi}$ which has dimensions of length, from which we obtain its dimensionless counterpart $\psi$
\begin{equation}
    \hl \hat{\psi} (\bm{\xi} = D_d \bm{\theta}) \equiv \frac{2}{c^2} \int_{\zeta_o}^{\zeta_s} \Phi(\bm{\xi}, \zeta) d \zeta, \quad \psi(\bm{\theta}) = \frac{D_{ds}}{D_d D_s} \hat{\psi}(D_d\bm{\theta}). \label{lensingpot}
\end{equation}

The two-dimensional lensing potential $\hat{\psi}$ is fact the solution to Poisson's equation in the limit of strong lensing. In the non-relativistic limit, the cosmological Poisson equation \eqref{relativisticPoisson} becomes identical to the Newtonian Poisson equation \eqref{NewtonianPoisson} albeit with a specific interpretation. It may be written in comoving coordinates $\bm{x}$ at the gravitational lens with the form and formal solution
\begin{equation}
    \bm{\nabla}^2 \Phi (\bm{x}) = 4 \pi a_d^2 G \delta \varrho(\bm{x}), \quad \Phi (\bm{x}) = a_d^2 G \int \frac{\delta \varrho (\bm{x}')}{|\bm{x} -\bm{x}'|} d \bm{x}'\label{lensingPoisson}
\end{equation}
where the source term $\delta \varrho$ is a perturbation to a homogeneous background \textit{mass density}, $\varrho = \bar{\varrho} + \delta \varrho$; it is assumed to always be positive for strong lensing (see \cite{Petters2001} for a detailed discussion). Throughout this section where we omit the domain of integration it may be assumed to indicate that the integral should formally be taken over the entire two- or three-dimensional space as appropriate, although as derived and in practice the equation applies to the local region of the lens.

The thin lens approximation motivates us to integrate the non-relativistic, cosmological Poisson equation \eqref{lensingPoisson} along the line of sight, reducing it to two spatial dimensions. Assuming that the potential is twice continuously differentiable, which allows the second order partial derivatives to be moved outside the integral, we see that the dimensionful lensing potential $\hat{\psi} (\bm{\xi})$ defined in Equation \eqref{lensingpot} produces a two-dimensional Poisson equation and the solution
\begin{equation}
    \bm{\nabla}^2_{\bm{\xi}} \hat{\psi} (\bm{\xi}) = \frac{8 \pi G}{c^2} \Sigma (\bm{\xi}), \quad \hat{\psi}(\bm{\xi}) = \frac{4 G}{c^2} \int \Sigma(\bm{\xi}') \ln \left| \frac{\bm{\xi} - \bm{\xi}'}{D_d} \right| d \bm{\xi}' \label{2dpoisson}
\end{equation}
where we defined the two dimensional density perturbation $\Sigma (\bm{\xi})$ as the projection of the lens mass density perturbation $\delta \varrho$ on the lens plane as 
\begin{equation} \hl
    \Sigma (\bm{\xi}) \equiv \int_{\zeta_o}^{\zeta_s} \delta \varrho (\bm{\xi}, \zeta) d \zeta
\end{equation}
in proper Cartesian coordinates $(\bm{\xi}, \zeta)$ at the lens {\hl where $\zeta_o < \zeta_s$}.  It is typically known as the surface mass density  of the lens, and fully describes the lensing mass distribution under the thin lens approximation. Finally, in terms of the dimensionless lensing potential, the two-dimensional Poisson equation and its solution become
\begin{equation}
\boxed{
    \bm{\nabla}_{\bm{\theta}}^2 \psi (\bm{\theta}) = 2 \kappa(\bm{\theta}), \quad  \psi (\bm{\theta}) = \frac{1}{\pi} \int \kappa(\bm{\theta}') \ln | \bm{\theta}' - \bm{\theta}| d \bm{\theta}' }\label{dimensionlessPoisson}
\end{equation}
where the convergence or dimensionless surface mass density of the lens is
\begin{equation}
    \kappa (\bm{\theta}) \equiv \frac{\Sigma(D_d \bm{\theta})}{\Sigma_{\text{crt}}}, \quad \Sigma_{\text{crt}} \equiv \frac{c^2 D_s}{4 \pi G D_d D_{ds}}.
\end{equation}
We may now also give the physical and scaled deflection angles for the lens from Equation \eqref{deflectionpotential} respectively as
\begin{equation}
     \bm{\hat{\alpha}}(\bm{\xi}) = \frac{4G}{c^2} \int \Sigma(\bm{\xi}') \frac{\bm{\xi} - \bm{\xi}'}{| \bm{\xi} - \bm{\xi}'|^2} d \bm{\xi}', \quad \bm{\alpha}(\bm{\theta}) = \frac{1}{\pi} \int \kappa(\bm{\theta}') \frac{\bm{\theta} - \bm{\theta}'}{| \bm{\theta} - \bm{\theta}'|} d \bm{\theta}'.
\end{equation}

\subsubsection{Time Delays} \label{subsubsec:timedelays}
In this approach, we do not find the time arrival functional via a full perturbed solution. Rather, we only calculate the arrival times along piecewise null geodesics of the background metric which approximate the trial null curves of the perturbed metric in accordance with the thin lens approximation. 

We can directly apply the null condition $ds^2 =0$ to the weakly perturbed FLRW metric given by Equation \eqref{eq:n_conformalFLRW} with $\hat{\Phi} = \hat{\Psi} = \frac{\Phi}{c^2}$, or equivalently substitute coefficients into Equation \eqref{eq:statictimearrival} {\hl and choose the boundary condition $C = x^0[\sigma_s]/c = 0$}. This gives the conformal arrival time functional which we take along the piecewise approximation $\mathscr{P}_{sdo}$ (see Figure \ref{fig:generallensdiagram}) parametrised by $\sigma$ (although the arrival time now depends only on the spatial projection of the spacetime path)
\begin{equation}
    \eta[\sigma_o] =\frac{1}{c} \int_{\mathscr{P}_{sdo}}  \left(1 - \frac{2 \Phi}{c^2}\right)d \ell_K.
\end{equation}
{\hl We remind that this result is only valid under the assumption that the metric (and therefore $\Phi$) is conformally static.} In addition, for convenience of computation, we compute the time \textit{delay} functional rather than the time arrival, by subtracting the arrival time of the fiducial unlensed ray along $\mathscr{P}_{so}$ parametrised by $\bar{\sigma}$, $\eta[\bar{\sigma}] = \frac{1}{c} \int_{\mathscr{P}_{so}} d \ell_K$ as a constant. We can then define the conformal time delay functional by
\begin{equation}
    \Delta \eta [\sigma_o] \equiv \eta[\sigma_o] - \eta[\bar{\sigma}_o].
\end{equation}
 The time delay functional splits into \textit{geometric} and \textit{potential} contributions,
\begin{equation}
    \Delta \eta [\sigma_o] = \Delta \eta_{\text{geom}}[\sigma_o] + \Delta \eta_{\text{pot}}[\sigma_o]
\end{equation} where we calculate separately
\begin{align}
    \Delta \eta_{\text{geom}} [\sigma_o] &\equiv \frac{1}{c} \left( \int_{\mathscr{P}_{sdo}} d \ell_K - \int_{\mathscr{P}_{so}} d \ell_K\right) \\
    \Delta \eta_{\text{pot}}[\sigma_o] &\equiv - \frac{2}{c^3} \int_{\mathscr{P}_{sdo}} \Phi( x^i (\sigma)) d \ell_K.
\end{align}
We note that the time delay can also be calculated by a separate ``wavefront method'' \cite{Schneider1992} which does not integrate along the rays, but instead considers shifting the position of observer to calculate the difference in time between the deformed sections of wavefronts (corresponding to different images) reaching the observer.

\paragraph{Potential Time Delay}

The potential time delay is just the gravitational or Shapiro time delay. Following a similar procedure to Section \ref{sec:thinlensapprox}, we evaluate the potential time delay in the lens frame at $\sigma_d$ corresponding to the deflection event, by first inserting a Dirac delta at $x^i (\sigma_d)$ to represent a point mass lens
\begin{equation}
    \Delta \eta_{\text{pot}}[\sigma_{d}] = - \frac{2}{c^3} \int_{\mathscr{P}_{sdo}} \Phi (x^i) \delta(x^i(\sigma) - x^i(\sigma_d))  d \ell_K = - \frac{2}{c^3} \Phi(x^i(\sigma_{d})).
\end{equation}
We consider a continuum distribution within the lens frame, dropping the $K$ subscript since the space is locally Euclidean and change from conformal to cosmic time. Using the Born approximation to take the integral over the physical line of sight $\zeta$ we have
\begin{equation}
    \Delta t_{\text{pot}}[\sigma_{d}] = - \frac{2}{c^3} \int_{\mathscr{P}_{sdo}} \Phi ( x^i ) a_{d} d \ell = - \frac{2}{c^3} \int_{\zeta_s}^{\zeta_o} \Phi(\zeta, \bm{\xi}) d \zeta.
\end{equation}
Finally, we use the definition of the lensing potential \eqref{lensingpot} and relate the potential time delay at the lens to the observer through the redshift relation \eqref{cosmologicalredshift}
\begin{equation}
    \Delta t_{\text{pot}}[\sigma_o]= - \frac{(1+z_d)}{c} \hat{\psi}(\bm{\xi} = D_d \bm{\theta}).
\end{equation}

\paragraph{Geometric Time Delay}
The spatial projections of the approximated trial null path $\mathscr{P}_{sdo}$ and the unlensed path $\mathscr{P}_{so}$ form a geodesic triangle of the background spatial metric $\gamma_{ij}$. The geometric time delay may be written
\begin{equation}
    \Delta \eta_{\text{geom}}[\sigma_o] =  \frac{1}{c} \left( \ell_{sdo} - \ell_{s} \right) = \frac{1}{c} \left( \ell_d + \ell_{ds} - \ell_{s} \right)
\end{equation}
where we write the spatial length $\ell_{sdo}$ as a sum of its piecewise components $\ell_{d}$ (the length of the path from $O$ to $P$) and $\ell_{ds}$ (the length of the path from $P$ to $S$).

 Recall from the definitions \eqref{generalisedtrig} the following functions which will allow us to write trigonometric identities with respect to surfaces of constant spatial curvature $K$
\begin{equation}
\cos_K(x) \equiv \cos(K^{1/2} x) \quad \text{and} \quad  \sin_K (x) \equiv K^{-1/2} \sin(K^{1/2}x)
\end{equation}
where the function $\sin_K (x)$ is also the comoving areal distance present in the metric.

Firstly we use the sum-to-product trigonometric identity
\begin{gather}
    K \sin_K \left( \frac{\ell_{sdo} + \ell_s}{2} \right) \sin_K \left( \frac{\ell_{sdo} - \ell_s}{2} \right) = - \frac{\cos_K \ell_{sdo} - \cos_K \ell_s}{2} \\
    \ell_{sdo} - \ell_s \approx  - 2\frac{\cos_K \ell_{sdo} - \cos_K \ell_s}{K \sin_K \ell_s} \label{trigexpression}
\end{gather}
where we use the fact that the lensing angles are small such that $\sin_K \left( \frac{\ell_{sdo} + \ell_s}{2} \right) \approx \sin_K (\ell_s)$ and $\sin_K (\frac{\ell_{sdo} - \ell_s}{2}) \approx \frac{\ell_{sdo} - \ell_s}{2}$.
Secondly, the law of cosines for the background geodesic triangle is
\begin{align}
    \cos_K{\ell_s} &= \cos_K{\ell_d} \cos_K{\ell_{ds}} + K \sin_K{\ell_d} \sin_K{\ell_{ds}} \cos ( \pi - \hat{\alpha})\\
    &= \cos_K( \ell_d + \ell_{ds}) + K\sin_K{\ell_d} \sin_K{\ell_{ds}}( 1- \cos \hat{\alpha}) \\
    &\approx \cos_K \ell_{sdo} + K\sin_K{\ell_d} \sin_K{\ell_{ds}} \frac{\bm{\hat{\alpha}}^2}{2} \label{coslawresult}
\end{align}
where we used the sum of angles formula for cosines in the second equality and the approximation $1- \cos \hat{\alpha} = 2 \sin^2 \frac{\hat{\alpha}}{2} \approx \frac{\bm{\hat{\alpha}}^2}{2}$. Substituting Equation \eqref{coslawresult} into \eqref{trigexpression}, we get 
\begin{align}
    \Delta \eta_{\text{geom}}[\sigma_o]  &\approx  - \frac{\sin_K{\ell_d} \sin_K{\ell_{ds}}}{c \sin_K \ell_s}\frac{\bm{\hat{\alpha}}^2}{2}\\
    &\approx - \frac{1}{2ca_d} \frac{D_d D_{ds}}{D_s}\bm{\hat{\alpha}}^2\\
    &\approx - \frac{1}{2ca_d} D_\tau (\bm{\theta} - \bm{\beta})^2
\end{align}
where by the isotropy of the background FLRW metric, the lengths $\ell \to \chi$ and so we write the ratio of comoving areal distances in terms of the angular diameter distances\footnote{This conversion is eschewed by some authors as a purely historically motivated practice \cite{Schneider2006, HolicowIV}; \textit{comoving} distances are the natural distance measure in lensing. We do not need to include the cumbersome redshift factor $1+z_d$ if the distances are kept as comoving areal distances rather than angular diameter distances. We use angular diameter distances in this work for consistency reasons.} using $D_A(\chi) = a (\chi) \sin_K (\chi)$, and the relationship between $\bm{\hat{\alpha}}$ and $\bm{\alpha}$. The ratio of distances is a quantity useful for cosmology, known as the \textit{time delay distance} 
\begin{equation}
    D_\tau \equiv \frac{D_d D_s}{D_{ds}}.
\end{equation}
In terms of the cosmic time at the observer $t= a_0 \eta$, using the redshift relation $1+z_d = \frac{a_0}{a_d}$, we have
\begin{equation}
     \Delta t_{\text{geom}}[\sigma_o] \approx \frac{(1+z_d)}{2c} D_\tau (\bm{\theta} - \bm{\beta})^2.
\end{equation}

\paragraph{Total time delay and lens equation}

Putting together the results, we have the total time delay as a function of position $\bm{\theta}$ (with the source position $\bm{\beta}$ as a parameter) in the observer frame\footnote{Recall the notation $t[\sigma_o]$ was shorthand for $t[\mathscr{P}_{sdo}(\sigma_o)]$, and the path is fully described by the coordinates $\bm{\theta}$ in the FLRW, thin lens approximation.} $\tau(\bm{\theta}; \bm{\beta}) \equiv \Delta t[\bm{\theta}(\sigma_o)]$ 
\begin{equation}
\boxed{
    \tau (\bm{\theta}; \bm{\beta}) = \frac{(1+z_d)}{c} D_\tau \left( \frac{1}{2} (\bm{\theta} - \bm{\beta})^2 - \psi(\bm{\theta}) \right)} \label{cosmologytimedelay}
\end{equation}
which may also be written as proportional to the \textit{Fermat potential}, defined as
\begin{equation}
    \phi (\bm{\theta}; \bm{\beta}) \equiv \frac{1}{2} (\bm{\theta} - \bm{\beta})^2 - \psi(\bm{\theta}). \label{Fermatpotential}
\end{equation}

Applying Fermat's principle to the time delay function, i.e. setting the gradient of the time delay with respect to the solutions $\bm{\theta}$ as zero
\begin{equation}
    \bm{\nabla}_{\bm{\theta}} \tau (\bm{\theta}) = 0  
\end{equation}
picks out the geodesics from the null curves and therefore gives the lens equation $\bm{\beta} = \bm{\theta} - \bm{\nabla}_{\bm{\theta}} \psi(\bm{\theta})$ which is the same result as the vectorial formalism. The time delay of a single image, $\tau (\bm{\theta})$ is not an observable quantity; however, the time delay between images pairs $\Delta \tau (\bm{\theta})$ of a time-varying source can be measured -- this is discussed in Section \ref{sec:tdcosmography}.

\section{Properties of the Lens Mapping} \label{sec:lensmappingproperties}

In this section we give an overview of the properties of the lens mapping; the reader is referred to e.g. \cite{Petters2001, Mollerach2002, Meneghetti2021} for details and discussions. The lens equation $\bm{\beta} = \bm{\theta} - \bm{\alpha}( \bm{\theta})$ defines a map $\textbf{\varbeta}(\bm{\theta}) \equiv \bm{\theta} - \bm{\alpha}( \bm{\theta})$ from image plane to source plane. The \textit{inverse map} $\textbf{\varbeta}^{-1}$ is from the source plane to image plane ($\mathbb{R}^2 \to \mathbb{R}^2$). The reason to introduce new notation for the map, is that it is more clearly thought of as a different object (a function of $\bm{\theta}$) than the source position or variable $\bm{\beta}$. Care should be taken to general to avoid conflating a function and a variable even when this is a standard practice in the literature (the same care should apply to distinguishing a time variable and the various time functions, to avoid potentially confusing expressions e.g. for a Dirac delta $\delta(t - t(s))$).

In order to investigate the local properties of the lens mapping, we can perform a Taylor series expansion (generalised to apply to a map $\mathbb{R}^n \to \mathbb{R}^m$; for reference, this procedure can be also be found under the analogues of the Taylor polynomials which go under the name \textit{jets} in mathematics). This application of Taylor's theorem gives (using summation notation and recalling that uppercase Latin indices label the vector components and are $\{1,2\}$)
\begin{equation}
    \varbeta^A (\bm{\theta}) = \varbeta^A (\bar{\bm{\theta}}) + J\indices{^A_B}(\bar{\bm{\theta}}) (\theta - \bar{\theta})^B + \frac{1}{2} H\indices{^A_{BC}} (\bar{\bm{\theta}}) (\theta - \bar{\theta})^B (\theta - \bar{\theta})^C + ...
\end{equation}
where $J\indices{^A_B} \equiv \frac{\partial \varbeta^A}{ \partial \theta^B}$ and $H\indices{^A_{BC}} \equiv \frac{\partial^2 \varbeta^A}{ \partial \theta^B \partial \theta^C}$ are respectively the Jacobian and Hessian of the lens mapping.

\subsection{Shear and Convergence}
We notice that neighbouring light rays experience different magnitudes of deflection since the deflection angle depends on the impact parameter. This leads to a prominent feature of gravitational lensing: the cross-sectional area of a bundle of rays gets deformed, i.e. the shape of the lensed image is distorted -- for example, background galaxies can appear as long arcs when lensed around galaxy clusters. The deformation is at first order (i.e. if the source size is small compared to the typical length scale over which the lens deflection field varies significantly, {\hl and if the source is at a non-caustic point \cite{Meneghetti2021, Petters2001}}), described by the Jacobian matrix $\bm{J}$ of the lens mapping, {\hl which is sometimes called the amplification matrix}
\begin{equation}
    J\indices{^A_B} \equiv \frac{\partial \varbeta^A}{ \partial \theta^B} = \left( \delta\indices{^A_B} - \frac{\partial \alpha^A}{\partial \theta^B} \right) = \left( \delta\indices{^A_B} - \frac{\partial^2 \psi }{\partial \theta^B \partial \theta^C} \delta^{AC} \right). 
\end{equation}
As this matrix is symmetric, there is no rotation of the image (this is actually not strictly true for a general spacetime and for multi-plane lensing, but the effect is not usually considered important \cite{Sachs1961, Seitz1994, Fleury2015, Yoshida2005}). We can decompose it into a traceless part and a traceful part that is proportional to the identity $\bm{I}$ (or Kronecker delta $\delta\indices{^A_B}$ in coordinate form). Using the shorthand $\psi_{,AB} \equiv \frac{\partial^2 \psi }{\partial \theta^A \partial \theta^B}$, we can write the isotropic (traceful) contribution as
\begin{equation}
    (\tfrac{1}{2} \text{tr} \bm{J} ) \bm{I} = \tfrac{1}{2} ( 1 -  \psi_{,11} + 1 -\psi_{,22} ) \bm{I} = ( 1 - \tfrac{1}{2} \bm{\nabla}^2_{\bm{\theta}} \psi ) \bm{I} = (1 - \kappa) \bm{I}
\end{equation}
and the anisotropic (trace-free) component defines the shear matrix $\bm{\Gamma}$
\begin{equation}
    \Gamma\indices{^A_B} \equiv J\indices{^A_B} - (\tfrac{1}{2} \text{tr} \bm{J} ) \delta\indices{^A_B} =  - \psi_{,BC} \delta^{AC} + \tfrac{1}{2}(\psi_{,11} + \psi_{,22}) \delta\indices{^A_B} \\
    = \begin{bmatrix}
        \Gamma_1 & \Gamma_2 \\
        \Gamma_2 & -\Gamma_1
    \end{bmatrix}
\end{equation}
where $\Gamma_1 \equiv \frac{1}{2}(\psi_{,11} - \psi_{,22})$ and $\Gamma_2 \equiv \psi_{,12} = \psi_{,21}$. Altogether, the Jacobian matrix of the lens mapping is decomposed into an isotropic transformation described by the the surface mass density or \textit{convergence} $\kappa$ and an anisotropic transformation described by the \textit{shear} $\bm{\Gamma}$
\begin{equation}
    \bm{J}(\bm{\theta}) = ( 1 - \kappa(\bm{\theta})) \bm{I} - \bm{\Gamma}(\bm{\theta}). \label{Jacobianmatrix}
\end{equation}
The Jacobian matrix at a given $\bm{\theta}$ has eigenvalues $\lambda_1 = 1 - \kappa - \Gamma$ and $\lambda_2 = 1 - \kappa + \Gamma$, where the modulus of the shear is $\Gamma \equiv \sqrt{\Gamma_1^2 + \Gamma_2^2}$. {\hl Note that the Hessian of the Newtonian potential is the Newtonian tidal acceleration matrix (e.g. p445-452 \cite{Hartle2003}), and here we are dealing with its two-dimensional projection $\psi_{,AB}$. The Newtonian Poisson equation fully determines the Newtonian potential, but involves only the trace of this Hessian. Convergence is related to this traceful part which is the Newtonian analogue of the Ricci tensor; whereas shearing effects are related to the traceless part of the tidal matrix which is the Newtonian analogue of the Weyl tensor.}

In other words, the local matter density in the lens plane is described the surface mass density $\kappa$, and it causes an isotropic focusing (or divergence) of light rays. It is for this reason that $\kappa$ is also called the convergence (or the expansion, or the Ricci focusing) due to the lens. In the absence of shear, the lensed image changes size but does not change shape.

The shear matrix quantifies the {\hl remaining} effect of (the projection of) the gravitational tidal field; that is, the deflection caused by a mass distribution far away from the light bundle. There is a net effect if the matter distribution outside the light bundle is asymmetric, and this deflection along a particular direction results in the ``shearing'' of the source's intrinsic shape. {\hl The shear defined in gravitational lensing does not actually correspond to the typical mathematical definition of shear \cite{BovyInPrep}; it is a qualitative resemblance.} This anisotropic stretching of the light rays is called astigmatism in the language of optics.

For example, a small circular source with radius $r$ is mapped to an ellipse since the inverse Jacobian is a local linearisation of the inverse lens mapping at first order. The semi-major and semi-minor axes are aligned with the eigenvectors and scaled by the corresponding eigenvalues of the inverse Jacobian (i.e. linearised inverse lens map) since it is symmetric, respectively as $a = |\lambda_1|^{-1}r$ and $b = |\lambda_2|^{-1}r$. The standard way in lensing theory to quantify how this ellipse deviates from a circle is the \textit{ellipticity} defined as $\varepsilon \equiv \frac{a-b}{a+b}$; this quantity is more commonly called the \textit{third flattening} in mathematics. In the absence of shear, the image becomes a circle of radius $|1 - \kappa|^{-1}r$. For details, we refer to \cite{Petters2001, Congdon2018, Meneghetti2021}.

\subsection{Magnification, Caustics and Critical Lines} \label{sec:magnification}
 The geometric optics description of light included a photon conservation law: the total number of photons is a conserved quantity. Gravitational lensing therefore merely redistributes photons by deflection, with an observer seeing possibly more or fewer photons from a given source. This effect is called \textit{magnification} and \textit{de-magnification}. Considering once again a bundle of rays or a beam of light deformed under gravitational lensing, its cross-sectional area is usually different to if the source were unlensed whilst \textit{the surface brightness of the image is conserved under lensing}. This conservation of surface brightness is ultimately a consequence of Liouville's theorem for the phase space distribution \cite{Petters2001}, together with the fact that lensing is independent of frequency or achromatic\footnote{Consideration of the magnification of an extended source can lead to ``colour'' terms in practice, since the surface brightness profile may vary over different wavelengths. These chromatic effects are usually considered in microlensing \cite{Eigenbrod}.} -- cf. the brightness theorem \eqref{brightnesstheorem}. The deformation of the cross-sectional area is described by the magnification matrix $\bm{M}$, which is simply the inverse of the Jacobian of the lens mapping (i.e. the ratio of the differential solid angle element in the image plane to the differential solid angle element in the source plane). The image is magnified if the magnification $\mu$ defined by the determinant is $|\mu| > 1$ and de-magnified if $|\mu| < 1$
\begin{equation}
    \mu (\bm{\theta}) \equiv \det \bm{M} \equiv \frac{1}{\det \bm{J}(\bm{\theta})} = \frac{1}{(1-\kappa)^2 - \Gamma^2}. \label{magnification}
\end{equation}
The eigenvalues of the magnification matrix is simply the inverse of the eigenvalues of the Jacobian matrix. Consequently the sign of the magnification, i.e. of  $\det J$, is related to the \textit{parity} of the image. A negative value of one of the eigenvalues corresponds to negative partial parity, and means that the image is inverted in the corresponding eigenvector direction (i.e. it is mirror-reversed). When both eigenvalues are negative, the total parity of image is positive and it will not be inverted, but appear to be rotated by $\pi \text{rad}$. We refer to \cite{Petters2001, Schneider2006} for further discussion of magnification in a historical context.

A generic strong lens has one or more \textit{critical curves}, i.e. a finite set of closed curves formed by critical points, in the image plane along which $\det \bm{J}(\bm{\theta}) = 0$ and the magnification $\mu$ is therefore formally infinite. They generically consist of smooth sections called folds that meet at zero or more points known as cusps \cite{Congdon2018}. The critical lines map to \textit{caustics} in the source plane, and a source near a caustic is highly magnified and distorted. (Although this mapping is often used to define a caustic in lensing, caustics can also be treated intuitively as the intersection of neighbouring rays \cite{Ehlers2000}, or in terms of wavefronts \cite{Hasse1996, Fritelli2002}.) For example, a circularly symmetric lens has two circular critical curves: the inner one with eigenvalue $\lambda_2 = 0$ called \textit{radial} and the outer one with eigenvalue $\lambda_1 = 0$ called \textit{tangential}, since images near these critical curves are respectively magnified radially and tangentially with respect to the curves (see \cite{Petters2001, Meneghetti2021} for details). As infinite\footnote{For real sources which are extended objects, the overall magnification is given by an averaged value which can be shown to be in fact finite.} magnification is unphysical, the geometric optics approximation formally breaks down for \textit{caustic crossing events} and must be treated using wave optics.

Critical lines and caustics are of additional interest as they demarcate where the number and parity of images changes. For example, starting with a source far from the optical axis to a non-singular lens, there is just one image corresponding to a local minimum of the time delay functional (since the geometric contribution of the time delay surface is a paraboloid at the source position, with the potential contribution deforming the paraboloid -- see \cite{Meneghetti2021} for figures). If this source were to be moved towards the optical axis, then additional images would appear in pairs whenever the source crosses a caustic.

The result, known as either the Odd Number Theorem or Burke's theorem \cite{Burke1981}, is that the total number of images produced by transparent, smooth lenses (such that no images are obscured) must be odd\footnote{This does not directly translate to observations, however. For example, \textit{quadruply-lensed} sources are commonly observed, where the fifth image is close to the centre of the lens and is too dim to be seen. The observational predominance of evenly imaged sources was also taken as a hint by \cite{Gottlieb1994} that the conditions (e.g. Euclidean versus Lorentzian spacetime) under which the Odd Number Theorem was proven \cite{McKenzie1985} are in reality too restrictive. Nonetheless, the literature usually treats the Odd Number Theorem as valid.}. Furthermore, one of the new images always has positive parity and the other has negative parity: critical lines form closed contours which divide the sky into regions where images have a given parity. The Odd Number Theorem can also be understood geometrically by considering the lens mapping as a ``folded sky'' -- we refer to \cite{Mollerach2002}.

Finally, we note that we can rewrite the Jacobian in terms of the time delay functional $\tau$ \eqref{cosmologytimedelay}, or more elegantly the Fermat potential $\phi= \frac{c}{(1+z_d)D_\tau} \tau$ to which the time delay functional is proportional, in the neighbourhood of its critical points
\begin{equation}
    J\indices{^A_B} = \frac{\partial \varbeta^A}{\partial \theta^B} =  \frac{\partial^2 \phi}{ \partial \theta^B \partial \theta^C} \delta^{AC}.
\end{equation}
The Hessian matrix or curvature of the time delay surface at its critical points is thus proportional to the Jacobian matrix and related to the parities of the images; and the Gaussian curvature at a given critical point on the time delay surface is given by $\lambda_1 \lambda_2 = \det \bm{J}$. Since the Hessian of the time delay is inversely related to the magnification, a flat time delay surface implies an infinite magnification, and a large curvature implies a small magnification.

For this reason, it is very interesting that whilst the classification of the stationary points of the action in many branches of physics is not noteworthy -- to the point where the terminology \textit{to make stationary}, \textit{extremise} and \textit{minimise} is often used loosely or interchangeably despite their distinctions, we can classify images according to the type of stationary point of the time delay functional where they arise -- this is tabulated in Table \ref{tab:statpoints}.
\begin{table}
\begin{center}
\begin{tabular}{ c c c c c }
Type & Eigenvalues &   Gaussian curvature & Stationary pt. type &  Point type\\
& of $\bm{J}$  &  of $\tau$ surface & of $\tau$ functional & on $\tau$ surface\\
 &  & ($\det \bm{J}$), equiv. parity & &\\
 \hline \hline
I & $\lambda_1 , \lambda_2 >0$ & $>0$ & local minimum & elliptic\\
II & $\lambda_1 \lambda_2 < 0$& $<0$ & local saddle & hyperbolic \\
III & $\lambda_1, \lambda_2 < 0$ &  $>0$ & local maximum & elliptic \\
IV & $\lambda_1$ or $\lambda_2 \neq 0$ & $0$ & degenerate & parabolic\\
 & $\lambda_1 , \lambda_2 = 0$ & $0$ & degenerate & planar
\end{tabular}
\end{center}
    \caption{Classification of stationary points and image type.}
    \label{tab:statpoints}
\end{table}
We note that we can also interpret the time delay purely as a surface, leading to point classification based on the principal curvatures (dependent on $\lambda_1$, $\lambda_2$) and the Gaussian curvature. This in particular allows description of degenerate stationary points, i.e. those which cannot be classified by the second derivative test, as either a local parabolic point; or a local planar point, in which case the local form of the surface may vary (e.g. a plane or a monkey saddle).

\subsection{Conditions for Multiple Imaging} \label{sec:multipleimaging}
There are two general criteria for multiple imaging \cite{Schneider2006}
\begin{enumerate}
    \item An isolated transparent lens can produce multiple images if and only if there is a point $\bm{\theta}$ with $\det \bm{J}( \bm{\theta}) < 0$. If $\det \bm{J}( \bm{\theta}) >0$ then the lens equation is globally invertible (and so the mapping is bijective or a one-to-one correspondence). If $\det \bm{J}( \bm{\theta}) <0$ at some point $\bm{\theta}$, then this is a local saddle point (recall Table \ref{tab:statpoints}); and the Odd Number Theorem states that there must be at least two additional images.
    \item A \textit{sufficient but not necessary} condition for multiple imaging is that if there exists a point where $\kappa (\bm{\theta}) >1$. A lens that fulfils this condition is dubbed \textit{supercritical}, and an Einstein ring will form if and only if $\kappa (0) >1$ \cite{Congdon2018} for approximately circular lenses.
\end{enumerate}
This second {\hl criterion} is often considered to delineate the regime of strong ($\kappa > 1$) versus weak lensing; although it is not a \textit{necessary}\footnote{\hl For example, a perturbed Plummer lens potential can be everywhere subcritical and still produce multiple images \cite{Petters2010, Petters2001}.} condition for the occurrence of multiple images, the critical surface mass density defines the characteristic scale of the typical features of strong lensing. A lens is also called critical at $\bm{\theta}$ when $\kappa (\bm{\theta}) = 1$ and subcritical when $\kappa (\bm{\theta}) < 1$.

\section{Lens Models} \label{sec:lensmodels}
Having found the general solution to the appropriate Poisson equation, we are now able to solve the lens equation for a variety of lens models -- that is lensing potentials or equivalently their mass density distributions. In practice, this is often done numerically for real lenses, either via parametric or non-parametric modelling \cite{Schneider2006, Meneghetti2017}. Non-parametric modelling (also known as free-form modelling) is a numerical technique where the mass density is divided and allowed to vary over a grid; therefore, assumptions of the form of the density distribution are not necessary. On the other hand, parametric modelling refers to assuming a particular density distribution described by a fixed number of parameters; the most basic parametric models are the analytically tractable ones discussed in this section.

We can divide commonly used lens models as either point or extended mass distributions. Point masses are not used only for simplicity, but are a justified model if the physical angular size of the lens is much smaller than the Einstein angle (the angular scale that characterises the typical angular displacement of an image due to the gravitational lens). This is the case for many astrophysical scenarios where the lens are compact objects such as stars within a galaxy acting on a more distant source (e.g. microlensing). Gravitational lensing by structures like galaxies and galaxy clusters, on the other hand, are modelled by extended mass distributions. In this section we examine the potential and solution to the lens equation for the particular examples (see also \cite{Petters2001}) of the point mass, uniform sheet and some galaxy models, in particular the singular isothermal sphere and the softened elliptical potential.

The above varieties of parametric lens models are usually taken to be special cases of more general power law descriptions of the mass distribution. For example, consider a generic circular lens with a mass density obeying a power law $\varrho = \varrho_0 r^{-n}$. This has a (scaled) deflection angle of magnitude
\begin{equation}
    \alpha ( \theta) = \theta_E \left( \frac{\theta}{\theta_E}\right)^{2-n} = \theta_E^{n-1} \theta^{2-n}
\end{equation}
where the Einstein angle $\theta_E$ sets the scale for deflection angle, and its definition in terms of physical parameters or properties is dependent on the model. The point mass corresponds to the limiting case of $n \to 3$ with $\alpha = \theta_E^2 \theta^{-1}$; the singular isothermal sphere to the case $n=2$ with constant deflection $\alpha = \theta_E$, and a uniform critical sheet to $n \to 1$ with $\alpha = \theta$. The power law given here can be further generalised to account for ellipticity and possibly softening; and that forms a family of potentials usually considered as realistic in contemporary literature -- see \cite{Schneider2006} for details.

\subsection{Simple Models}

\subsubsection{Point Mass}
The point mass potential provides an accurate model for the light curves of most galactic microlensing scenarios; similarly, multiple point mass lenses provide a model of microlensing of quasars by individual stars in a galaxy. The point mass is also a simple first approximation to some two-image systems. Consider a single point mass $M$ at the origin of the lens plane with physical surface mass density $\Sigma(\bm{\xi}) = M \delta (\bm{\xi})$ where $\delta (\bm{\xi})$ is the two dimensional Dirac delta. The dimensionless density or convergence is
\begin{equation}
    \kappa_{pt}(\bm{\theta}) = \pi \theta_E^2 \delta(\bm{\theta}), \quad \theta_E^2 \equiv \frac{M}{\pi D_d^2 \Sigma_{\text{crt}}} = \frac{4GM}{c^2 D_\tau}
\end{equation}
and the angular Einstein radius or Einstein angle $\theta_E$ determines the deflection scale of the lens; and we used $\delta(\bm{\theta}) = D_d^2 \delta(D_d \bm{\theta}) = D_d^2 \delta(\bm{\xi})$. Putting this into Equation \eqref{dimensionlessPoisson}, we obtain the lensing potential for a point mass 
\begin{equation}
    \psi_{\textsc{pt}} = \theta_E^2 \ln{|\bm{\theta}|}.
\end{equation}
The deflection angle is $\bm{\nabla}_{\bm{\theta}} \psi_{\textsc{pt}}  = \theta_E^2 \frac{\bm{\theta}}{\bm{\theta}^2}$ and the lens equation becomes $\bm{\beta} = \left( 1 - \frac{\theta_E^2}{\bm{\theta}^2} \right) \bm{\theta}$. By taking the absolute value of the lens equation we find $|\bm{\theta}|^2 - |\bm{\beta}||\bm{\theta}| - \theta_E^2 = 0$ we may solve for the magnitude $|\bm{\theta}|$ 
\begin{equation}
    |\bm{\theta}|_{\pm} \equiv \frac{1}{2} \left( |\bm{\beta}| \pm \sqrt{4 \theta_E^2 + |\bm{\beta}|^2 } \right) 
\end{equation}
which can be substituted back into the lens equation for $\bm{\theta} = \left( 1 - \frac{\theta_E^2}{\bm{\theta}_{\pm}^2} \right)^{-1} \bm{\beta}$. We therefore get two images for a generic source position, which both lie along the source-lens axis on the sky, on opposite sides of the lens. The angular separation between the two images is given by $\Delta |\bm{\theta}|_{\pm} = 2 \theta_E \sqrt{1 + \frac{\bm{\beta}^2}{4 \theta_E^2}}$; this is approximately $\sim 2 \theta_E$ when the source is close to the optical axis compared with the Einstein angle ($|\bm{\beta}| < \theta_E$). The exception to the presence of two images occurs when $|\bm{\beta}|=0$ such that the source is directly behind the lens. In this case $|\bm{\theta}|_+ = |\bm{\theta}|_- = \pm \theta_E$ and a ring-like image forms, known as an Einstein ring. The potential for the point mass may be generalised to a collection of point masses $M_i$ on the lens plane simply by a summation: $\psi(\bm{\theta}) = \sum_{i} \theta_{E,i}^2 \ln | \bm{\theta} - \bm{\theta}'_{i}|$ with corresponding surface mass density $\kappa(\bm{\theta}) = \pi \sum_{i} \theta_{E,i}^2  \delta(\bm{\theta} - \bm{\theta}'_{i})$. The expressions for the shear and the magnification can be found in e.g. \cite{Petters2001}.

\subsubsection{Uniform Sheet}

A finite disk of uniform density can be used to model the central parts of galaxy clusters; indeed to produce realistic models for the microlensing of quasars, the potential for a sheet of uniform surface mass density (to model continuous matter in a local region near the lens) along with external shear can be added to a point mass model. The potential for a uniform sheet with constant density $\kappa_c$ is
\begin{equation}
    \psi_{\textsc{cts}} (\bm{\theta}) = \frac{\kappa_c}{2} \bm{\theta}^2 \label{uniformsheetpotential}
\end{equation}
which produces no shear, and the corresponding deflection angle is $\bm{\alpha} = \bm{\nabla}_{\bm{\theta}} \psi_{\textsc{cts}} = \kappa_c \bm{\theta}$, and lens equation is
\begin{equation}
    \bm{\beta} = \bm{\theta} ( 1 - \kappa_c).
\end{equation}
This has the single solution $\bm{\theta}_0 = \frac{\bm{\beta}}{1 - \kappa_c}$, which is the image position when $\kappa_c \neq 0$. However in the critical cases when $\kappa_c =1$ and $\bm{\beta} \neq 0$ no image exists; and when $\kappa_c =1$ and $\bm{\beta} =0$, every point of the lens plane is a degenerate lensed image.

\subsection{Galaxy Models}

Galaxies may be considered as a self-gravitating collisionless fluid of stars {\hl and dark matter particles} possessing some phase space distribution function $f(\bm{r}, \bm{v}, t)$, assumed to be in equilibrium, which obeys the collisionless Boltzmann equation coupled with the Newtonian Poisson equation \cite{Binney2011, BovyInPrep}. Since the mass density is related to the gravitational potential through both equations separately, the system is only \textit{self-consistent} if the mass density of the system is proportional to $\int f d \bm{v}$.

There are two general approaches to finding a solution for the system. The direct method of determining mass density profiles and potentials from observations of galaxies can be complicated. Although the distribution of visible matter may be well modelled based on observations of density distributions of stars projected on the sky, the dark matter distribution is mostly only inferred from the dynamics of visible matter. Therefore the total density distribution and the gravitational potential are poorly known, particularly where dark matter dominates far from the centre of the galaxy.

The alternative method is less challenging, as it involves positing a form for a self-consistent distribution function based on either a well-defined or idealised model, from which the density and potential is calculated. In particular, there exist analytically tractable solutions for spherically symmetric density distributions with isothermal velocity distributions (Gaussian velocity distributions with the same dispersion at any radius). There turns out to be an equivalence between a particular functional form for the distribution function describing stellar dynamics, and modelling a galaxy as an isothermal ideal gas. The distribution function for a singular isothermal sphere corresponds to a spherical matter distribution which possesses a flat rotation curve, {\hl matching the observed rotation curves of disk galaxies. It is therefore an important first-approximation galaxy model (or more precisely, for the dark matter halo of galaxies; galaxy lenses are typically elliptical galaxies which typically do not have much rotation). The singular isothermal sphere also adequately reproduces some basic properties of lens systems \cite{Schneider2015}, such as the image separation.}

\subsubsection{Spherical Galaxy Models}

\paragraph{Singular Isothermal Sphere (SIS)}
The singular isothermal sphere (SIS) model treats a galaxy as a spherically-symmetric ideal gas in hydrostatic equilibrium such that the pressure support balances the gravitational potential; a detailed derivation of this model is given in \cite{Binney2011, BovyInPrep}. This model has a three-dimensional mass density which we note is singular at the origin, $\varrho(r) = \frac{\sigma_v^2}{2 \pi Gr^2}$ where $r = \sqrt{\bm{\xi}^2 + \zeta^2}$ is a proper radius and $\sigma_v$ is the velocity dispersion. The rotation curve is flat with rotation velocity $v_c = \sqrt{2} \sigma_v$. The surface mass density $\Sigma ( \bm{\xi} ) = \int_{-\infty}^{\infty} \rho (\sqrt{\bm{\xi}^2 + \zeta^2}) d\zeta= \frac{\sigma_v^2}{2 G \xi}$ gives the dimensionless surface mass density as
\begin{equation}
    \kappa_{\textsc{SIS}} ( |\bm{\theta}| ) = \frac{\theta_E}{2 | \bm{\theta} |}, \quad \theta_E \equiv 4 \pi  \left( \frac{\sigma_v}{c} \right)^2 \frac{D_{ds}}{D_s} \label{siskappa}
\end{equation}
and the corresponding potential is
\begin{equation}
    \psi_{\textsc{SIS}}(\bm{\theta}) = \theta_E | \bm{\theta} | \label{sispotential}
\end{equation}
which gives the reduced deflection angle $\bm{\nabla}_{\bm{\theta}} \psi_{\textsc{SIS}} = \theta_E \frac{\bm{\theta}}{|\bm{\theta}|}$. Once again taking the absolute value of the lens equation to find $|\bm{\theta}|$ case-by-case and substituting back into the lens equation we get
\begin{equation}
    \bm{\beta} = \left( 1 - \frac{\theta_E}{|\bm{\theta}|_{\pm}}\right) \bm{\theta}, \quad |\bm{\theta}|_{\pm} \equiv |\bm{\beta}| \pm \theta_E 
\end{equation}
and solve to find
\begin{equation}
\bm{\theta}_{\pm} = \frac{\bm{\beta}}{|\bm{\beta}|}|\bm{\theta}|_{\pm} \; \;  \text{if } \; 0 < |\bm{\beta}| \leq \theta_E, \quad 
     \bm{\theta}_+ = \frac{\bm{\beta}}{|\bm{\beta}|}|\bm{\theta}|_+  \; \; \text{if }\; |\bm{\beta}| > \theta_E.
\end{equation}

Hence, there are two images if $0 < |\bm{\beta}| < \theta_E$, on opposite sides to the lens centre, with the image separation always $\Delta \theta_\pm = 2 \theta_E$; and if $|\bm{\beta}| > \theta_E$ then there is only one image.  In the case that $|\bm{\beta}| = \theta_E$, then despite the two solutions there is only one image, since the image at $\bm{\theta}_-$ disappears at the origin as its magnification is $0$. Finally, when $|\bm{\beta}| =0$, then we again form a ring-like image since $|\bm{\theta}| = \theta_E$ (in fact, the condition for an Einstein ring to form for any axisymmetric lens in general is that it is ``supercritical'', $\kappa(0) >1$ \cite{Congdon2018}). The magnification turns out to be infinite on this ring; so it also defines the critical curve for this model, and the set of caustics is the point at the origin. The SIS has been commonly used due to its simplicity, but is limited due to being strictly unphysical given the singularity at the origin and the infinite total mass.

\paragraph{Softened (Non-singular) Isothermal Sphere (NIS)}
The singularity at the origin may be avoided by manually adding a central region with finite density to the model, defined by a \textit{core radius} $r_c$ with corresponding $\theta_c = \frac{r_c}{D_d}$, resulting in a \textit{non-singular} or \textit{softened} isothermal sphere model (NIS) with a surface mass density of $\kappa_{\textsc{NIS}} ( |\bm{\theta}| ) = \frac{\theta_E}{2 \sqrt{ \bm{\theta}^2 + \theta_c^2 }}$ and potential
\begin{equation}
    \psi_{\textsc{NIS}}( \bm{\theta}) = \theta_E \sqrt{\bm{\theta}^2 +\theta_c^2}.
\end{equation}
The lens equation becomes
\begin{equation}
    \bm{\beta} = \left(1 - \frac{\theta_E}{\bm{\theta}^2} \left( \sqrt{\bm{\theta}^2 + \theta_c^2} - \theta_c \right) \right) \bm{\theta}
\end{equation}
which can be written as a cubic equation with one or three solutions depending on the source position; see e.g. \cite{Mollerach2002}.

\subsubsection{Ellipsoidal Galaxy Models}

\paragraph{Softened Isothermal Ellipsoid (SIE)}
More realistic models of galaxies must allow for ellipticity, although the lens equation may no longer have analytic solutions. There are several ways this may be achieved, and we refer to \cite{Kassiola1993} for a full discussion. Amongst them, the generalisation from the softened isothermal sphere to a softened isothermal ellipsoid (SIE), i.e. a model with a surface density with elliptic iso-density contours, may be derived (up to a normalisation) by replacing the coordinates $\bm{\theta}$ of the NIS model with coordinates $\bm{\theta}' = \sqrt{{\theta'_x}^2 + q^2 {\theta'_y}^2}$ centred on the lens and aligned with the principle axes of the ellipse, where $q$ is the ratio of the minor axis to the major axis of the projected ellipse \cite{Kormann1994}. This results in the dimensionless surface mass density 
\begin{equation}
    \kappa_{\textsc{SIE}}(\bm{\theta}') = \frac{b_I}{2 \sqrt{\bm{\theta}'^2 + {{\theta'}^2_c}}}
\end{equation}
where $b_I \equiv \frac{\theta_E e}{\arcsin e}$ is dependent on the eccentricity $e$ and the velocity dispersion of the model through the SIS deflection scale $\theta_E$; and ${\theta'}_c \equiv q s_c \equiv q \frac{r_c}{D_d}$ sets the core radius (see \cite{Keeton1998}). In the parameters of \cite{Kormann1994} however, the analytic form of the two dimensional potential is a little complicated. An alternative parametrisation of the same model results in simpler expressions \cite{Keeton1998}; the potential is
\begin{equation}
    \psi_{\textsc{SIE}}(\bm{\theta}') = \theta'_x \alpha_x + \theta'_y \alpha_y - b_I s_c \ln \sqrt{(h(\bm{\theta}')+s_c)^2+(1-q^2)\theta_x'^2)}
\end{equation}
where $h({\bm{\theta}'}) = \bm{\theta}'^2 + {{\theta'}^2_c}$ and the associated scaled deflection angle is
\begin{equation}
    (\alpha_x, \alpha_y) = \frac{b_I}{\sqrt{1-q^2}} \left(\arctan \frac{\theta'_x \sqrt{1-q^2}}{h(\bm{\theta}')+s_c}, \; \text{arctanh} \frac{\theta'_y \sqrt{1-q^2}}{h(\bm{\theta}') +q^2 s_c}\right).
\end{equation}

\paragraph{Elliptical Potentials (EP)}
For analytic simplicity, it can be more convenient to consider a model with elliptical iso-potential contours - i.e. we introduce the ellipticity directly in the potential rather than in the density. A limitation of such elliptical potentials, however, is that they give rise to unphysical features in the mass density distribution such as dumbbell-shaped contours, if the ellipticity is large. For this reason, provided that the ellipticity $\epsilon$ in these models obey $0 < \epsilon \leq 0.2$, they give a reasonable approximation to the isothermal ellipsoidal mass density models. Such an elliptical potential (EP) is given by
\begin{equation}
    \psi_{\textsc{ep}}(\bm{\theta}') = \theta_E \sqrt{s_c^2 + (1- \epsilon) {\theta'}_x^2 + (1+ \epsilon) {\theta'}_y^2}
\end{equation}
where the coordinates $\bm{\theta}'$ are related to the sky coordinates by
\begin{equation}
    (\theta_x', \theta_y') = (\theta_x \cos \varphi  + \theta_y \sin \varphi + \bar{\theta}_{x}, \; - \theta_x \sin \varphi + \theta_y \cos \varphi + \bar{\theta}_{y}).
\end{equation}
When the offset $\bm{\bar{\theta}}$ and the counter-clockwise rotation $\varphi$ vanish, the principle axes of the ellipse are aligned with the sky coordinates such that $\bm{\theta}' = \bm{\theta}$. Once again $\theta_E$, as defined by the SIS model, is the deflection scale of the lens, $s_c$ is the softening or the core radius, and $\epsilon$ is the projected ellipticity related to the projected axis ratio by $q^2 = (1- \epsilon)/(1+ \epsilon)$. We then get the scaled deflection
\begin{equation}
    \bm{\alpha} = \bm{\nabla}_{\bm{\theta}}  \psi_{\textsc{ep}} = \theta_E^2 \psi_{\textsc{ep}}^{-1}(\bm{\theta}') \begin{bmatrix}
    \cos \varphi (1 - \epsilon) \theta_x'  - \sin  \varphi ( 1+ \epsilon) \theta_y' \\
    \sin \varphi (1 - \epsilon) \theta_x' + \cos \varphi (1 + \epsilon) \theta_y' 
    \end{bmatrix}.
\end{equation}
We cannot solve the lens equation for these ellipsoidal or elliptical models analytically; this is done using numerical methods -- for example, one may use a simple optimiser such as \texttt{scipy.optimize.fsolve}, although there exist dedicated software packages such as \href{https://github.com/lenstronomy/lenstronomy}{\texttt{lenstronomy}}. A wide variety of image configurations can be produced for non-axially symmetric lenses.

\subsection{Lensing Degeneracies} \label{sec:lensingdegeneracies}

Most of the observables in gravitational lensing are dimensionless, and this often results in parameter degeneracies when modelling observational data \cite{Saha2000}, i.e. there is usually no unique mass model describing a given system of multiple images. These degeneracies are described by transformations of the lens mapping or equivalently of the time delay surface, which produce unchanged image configurations and magnification ratios. In addition to these global invariance transforms which are applied to the whole lens plane, there may also be local ones, such as the monopole degeneracy \cite{Saha2000, Liesenborgs2012}.

\begin{table}[!htbp]
    \centering
    \begin{tabular}{ c c c c c }
     & Transformation & $ \Delta \tau'$ & $\Delta \bm{\theta}'$ & $\frac{\mu'_i}{\mu'_j}$ \\
     \hline \hline
     Prismatic & $\bm{\alpha}' = \bm{\alpha} + k$ &  inv. & inv. & inv.\\
      & $\bm{\beta}' = \bm{\beta} + k$  & & & \\
     Similarity & $D_A' = s D_A$ & $s \Delta \tau$ & inv. & inv. \\
     Magnification & $\bm{\alpha}' = \lambda \bm{\alpha} - (1 - \lambda) \bm{\theta}$ & $\lambda \Delta \tau$ & inv. & inv.\\
      & $\bm{\beta}' = \lambda \bm{\beta}$ & & &
    \end{tabular}
    \caption[Invariance transformations.]{Invariance transformations and their effect on observables, adapted from \cite{Wagner2018}: the time delay difference $\Delta \tau$, the relative image positions $\Delta \bm{\theta}$ and relative magnifications $\frac{\mu_i}{\mu_j}$. The magnification transformation corresponds to the mass-sheet degeneracy.}
    \label{tab:lensingtransformationstable}
\end{table}

There are three global \textit{invariance transformations} as described by \cite{Gorenstein1988} which can also be combined: these are called the prismatic, similarity, and magnification transformations and are summarised in Table \ref{tab:lensingtransformationstable}. The prismatic transformation (so named as a prism deflects incident light by a constant angle) adds the same constant to both the source position and deflection angle. This is physically interpreted as adding a massive lens at a large transverse distance whilst the source is moved a large distance in the opposite direction, which is not usually considered as an important or likely scenario. The similarity transformation multiplies the lensing distance ratio $D_\tau$ by an arbitrary constant, which leaves the image configuration and magnification ratios invariant but scales the time delay difference. This and related transformations \cite{Saha2000} are usually considered in microlensing, as images are unresolved \cite{Congdon2018}.

\paragraph{Mass-Sheet Degeneracy}

The magnification degeneracy is of particular interest and is usually called the \textit{mass-sheet degeneracy} {\hl \cite{Falco1985}}. The generalised form of the magnification or mass-sheet transform was given by \cite{Liesenborgs2012} and later demonstrated to be a special case of a wider class of transformations, called the source-position transformation \cite{Schneider2014}; but here we will only explore the original form. We find the degeneracy by adding the potential for a uniform sheet \eqref{uniformsheetpotential} to that for an initial potential $\psi_1(\bm{\theta}) = \psi + \frac{1}{2} \kappa_c \bm{\theta}^2$, which leads to the lens equation 
\begin{equation}
    \bm{\beta} = \bm{\theta} - \bm{\nabla} \psi_1 = (1 - \kappa_c) \bm{\theta} - \bm{\nabla} \psi.
\end{equation}
We see that if we additionally multiply both $\bm{\beta}$ and $\bm{\psi}$ by $\lambda \equiv 1- \kappa_c$, we regain the original lens equation. This suggests the magnification transformation or mass-sheet transformation, which for the potential is 
\begin{equation}
    \psi' =  \lambda \psi + \frac{1}{2}(1- \lambda) \bm{\theta}^2. \label{mst_pot}
\end{equation}

The mass-sheet transformation corresponds to rescaling the initial mass distribution $\kappa$ and adding an additional sheet with constant mass density $\kappa_c = (1 - \lambda)$, giving a family of lens models $\kappa' = (1 - \lambda) + \lambda\kappa$ which predicts the same image configuration as the original $\kappa$. This can be interpreted when modelling an observed gravitational lens system as a degeneracy in splitting the mass between a galaxy and the cluster in which it is embedded. The mass-sheet degeneracy is also sometimes called the mass-disk degeneracy, since a circular disk and an infinite sheet are equivalent as long as the disk is larger than the region of images. This also provides the justification that the degeneracy cannot be removed by requiring that $\kappa \to 0$ at large $\theta$.

\begin{figure}
    \centering
    \includegraphics[scale=0.65]{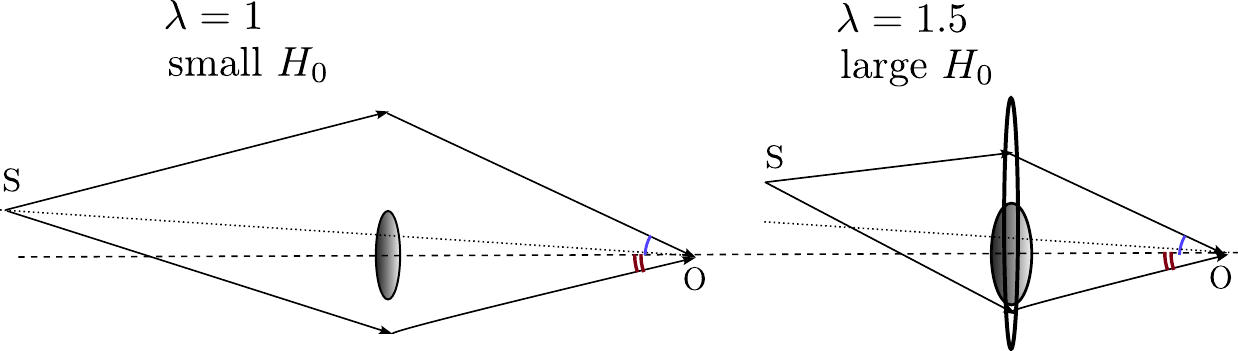}
    \caption{An illustration of the effects of an extreme mass-sheet transformation $\lambda =1.5$ on a lensing configuration, adapted from \cite{Treu2022}. The unobservable source position $\bm{\beta}$ is $\lambda =1.5$ times its original angular position from the optical axis, the lens mass is $\lambda =1.5$ larger, but with a \textit{negative} uniform convergence $\kappa_c = -0.5$ added. The images are observed at the same positions. If the distance ratio $D_\tau \propto H_0^{-1}$ is fixed, then the observed time delay is $\lambda = 1.5$ times larger under a mass sheet transform. Alternatively, the unknown distance ratio $D_\tau$ varies as a free parameter to be inferred, as in this illustration. In this example, an $\lambda = 1.5$ times larger Hubble constant means that all distances, including the ratio $D_\tau$, are reduced by $\lambda^{-1}$ times the original scale. The relation ${D'}_\tau^{-1}\Delta \tau' = \lambda  D_\tau^{-1} \Delta \tau$ therefore predicts the \textit{same} time delay.}
    \label{fig:MST}
\end{figure}

Computing the magnification ratios also shows that they are invariant. To find the transformation of the observable time delay \textit{difference} $\Delta \tau$ between two images, it is convenient to consider the Fermat potential which is proportional to the time delay $\phi = c (1+z_d)^{-1} D_\tau^{-1} \tau$. Recall the Fermat potential is given by \eqref{Fermatpotential}, and so transforms under the mass-sheet transform $\bm{\beta}' = \lambda \bm{\beta}$ and \eqref{mst_pot} as
\begin{equation}
    \phi'(\bm{\theta}) = \frac{1}{2}( \bm{\theta} - \bm{\beta}')^2 - \psi' (\bm{\theta}) =  \lambda  \left(\frac{1}{2}( \bm{\theta} - \bm{\beta})^2 - \psi (\bm{\theta}) \right) + \frac{\lambda (\lambda -1)}{2}  \bm{\beta}^2
\end{equation}
and the difference in the Fermat potential between two images under the mass sheet transform is then
\begin{equation}
    \Delta \phi' = \lambda \Delta \phi.
\end{equation}
Whilst the lens redshift $z_d$ and the speed of light are known or separately measured, the time delay distance $D_\tau$ is usually inferred from the observed time delay. Allowing $D_\tau$ to vary as a free parameter gives
\begin{equation}
    {D'}_\tau^{-1}\Delta \tau' = \lambda  D_\tau^{-1} \Delta \tau
\end{equation}
so a value of ${D'}_\tau^{-1} \propto H_0'$ which is $\lambda$ times the former value ${D}_\tau^{-1} \propto H_0$ gives the same observed time delay: this is illustrated in Figure \ref{fig:MST}. 

On the other hand, if $D_\tau$ is fixed, the observable time delay difference between image pairs transforms as
\begin{equation}
    \Delta \tau' = \lambda \Delta \tau.
\end{equation}
Non-lensing data, such as kinematics of stars in the lens galaxy constraining the mass distribution, or constraints on the absolute size or luminosity of the source, are required to break the degeneracy \cite{Birrer2020, Suyu2014, Schneider2013}, but the mass sheet degeneracy still dominates the uncertainty in strong lensing time delay predictions.
 

\chapter{Cosmology through Gravitational Lensing and Quasars} 
\label{chap:cosmoglqso} 
\captionsetup{width=.9\linewidth}
\section{Observational Regimes of Gravitational Lensing} \label{sec:obsregimesgravlensing}
Gravitational lensing is typically\footnote{This is also not the only way to categorise observations; see p. 144 \cite{Petters2001}.} loosely broken into three observational regimes: strong lensing, weak lensing and microlensing. There is no generally applicable prescriptive definition which divides strong and weak lensing \cite{Bartelmann2001}. \textit{Strong lensing} is our focus in this work and in Chapter \ref{chap:lightlensaction}, typically occurring where $\kappa > 1$ {\hl (the surface mass density of the lens is sufficiently high)} and the source and the lens are closely aligned, with the presence of critical lines resulting in highly magnified and deformed (typically multiple) resolved images of a distant source. As we have seen, it is therefore the regime of rich visual phenomena such as giant luminous arcs and Einstein rings. 

{\hl Strong lensing can be further categorised based on the angular scale of image separations \cite{casadio2021}; unless otherwise specified we are typically concerned with macro-lensing (on the scale of arcseconds or greater) rather than milli-lensing (milli-arcseconds) or micro-lensing (micro-arcseconds).} \textit{Microlensing} is therefore a version of strong lensing where the separations between images are far too small to be resolved, of order micro-arcseconds, far below observational constraints. The total magnification over all the microlensed images can nonetheless be measured for a set-up where the observer, lens and source are in relative motion, through the apparent magnitude as a function of time or \textit{microlensing light curve}. 

Finally, \textit{weak lensing} occurs when $\kappa < 1$  and the lens mapping \eqref{eq:lensmapping} is typically one-to-one, so the source is observed as a single image with small distortions via magnification and shearing far from critical lines. However, recall the criteria for multiple imaging outlined in Section \ref{sec:multipleimaging} -- although the CMB lensing is classified as weak lensing, multiple imaging can still occur for very small deflections if the source is also very small. For example, massive clusters display a CMB Einstein ring of radius $\sim 1$ arcminute which maps onto a point on the last scattering surface \cite{Lewis2006}. Weak lensing is not usually\footnote{Although one famous example is the deflection of starlight at the solar limb during a solar eclipse. Strictly the 1919 Eddington experiment itself measured the deflection statistically as an average over many stars due to difficulty of a single measurement \cite{Petters2001}.} detected in a single object, since its unlensed shape, brightness and/or position would need to be known, but instead is observed statistically.

Of the assumptions we took in Section \ref{quasinewtonianformalism}, we need only drop the assumptions of a static potential and the thin lens approximation (appropriate only for an isolated lens) which led to the quasi-Newtonian strong lensing formalism, to allow for weak lensing by large-scale structure. That is, the lens equation \eqref{eq:lensmapping} is valid for weak lensing. For the weak lensing of sources much smaller than the scale over which the deflection angle varies (e.g. weak lensing of background galaxies, but not the last scattering surface of the CMB \cite{Hanson2010}), the lens mapping can be linearised \cite{Schneider2006, Bartelmann2016} and the Jacobian matrix \eqref{Jacobianmatrix} describing the deformation of an infinitesimal bundle of light rays can be taken to fully describe the mapping and the image deformation. Therefore a thin bundle of light rays is usually followed in weak lensing, rather than a single light ray \cite{Petters2001}.

\section{Gravitational Lensing as a Cosmological Probe}

\paragraph{Observables}
The observables of a gravitational lens system include \cite{Schneider2006, Petters2001}
\begin{enumerate}
    \item \textit{imaging} 
    \begin{enumerate}
        \item \textit{astrometric}: relative positions of the lens and images, number of images
        \item shapes of images (for extended sources)
    \end{enumerate}
    \item \textit{photometric}: apparent brightness and relative fluxes
        \begin{enumerate}
            \item of individual images, or
            \item as a function of time, i.e. \textit{light curves} between images (i.e. time delays of a variable source) or over microlensed images
        \end{enumerate}
    \item \textit{spectroscopic}
    \begin{enumerate}
        \item light spectra (and thereby redshifts) of the lens and sources.
    \end{enumerate}
\end{enumerate}
There is only one image of the source in weak lensing, so in that case the mild deformation of extended sources is the primary observable.

In the case of strong lensing, spectroscopy gives information on the kinematics of the lens, and high resolution imaging allows for modelling the lens environment, i.e. external convergence or line-of-sight effects. Both spectroscopy and high resolution imaging of the lens were limited up until the past decade when new instruments on space-based and ground-based telescopes became available \cite{Treu2022, Eigenbrod}.

The relative image positions form the most important data for the lens system, and the deformation of an extended source (i.e. arcs and rings considering strong lensing) may provide additional constraints. In particular, the occurrence of an Einstein ring in the lensing system in addition to discrete images is very useful as it allows for stronger constraints, e.g. on the Einstein angle.

\paragraph{Applications to Cosmology}
Gravitational lensing finds application in a cosmological context notably through
\begin{enumerate}
    \item \textit{cosmic shear analysis}, which uses weak lensing of background galaxy images by foreground large-scale structures as a technique to probe the growth of structures and the expansion of the Universe
    \item weak lensing by large-scale structure of the CMB
    \item \textit{galaxy cluster strong lensing cosmography}, using strong lensing by clusters of multiple sources
    \item \textit{time delay cosmography}, which infers a dimensionful distance ratio as well as an absolute distance via measurements of time delays \eqref{cosmologytimedelay} between multiple images of strongly lensed, time-variable sources and thereby determining cosmological parameters (mainly $H_0$).
\end{enumerate}
 In addition, gravitational lensing uniquely does not depend on the nature of the matter nor its state (i.e. regardless of dark or baryonic matter or if it is in equilibrium). It can therefore measure the mass of the lens galaxy or cluster, and thereby a dark matter distribution. For example, microlensing is used as a probe of dark matter in the Milky Way halo. As another example, if $H_0$ is taken as a known quantity or only time delay ratios are considered, time delay data could in principle constrain the mass distribution of the lens \cite{Schneider2006}. Table \ref{tab:classificationoflensing} lists some observational classifications of lensing phenomena, and their corresponding applications.

\begin{table}[!htbp]
    \centering
    
    \begin{tabularx}{\textwidth}{ p{2cm} p{2cm} X X}
    Source & Lens & Effects & Applications \\
     \hline \hline
      & & & \\
      & & Microlensing & \\
      \hline
      & & & \\
     Star \newline (Galactic)  & Star \newline (Galactic) & Light curves  &  Extra-solar planets, \newline Milky Way inner structure \\
     Star \newline (Extra-galactic) & Compact object \newline (Galactic) & Light curves & DM in Milky Way \newline halo\\
     & & & \\
     & & Strong lensing & \\
     \hline
     & & & \\
     Quasar & Galaxy & Multiple images, Einstein rings, time delays & Mass of galaxies, substructures, $H_0$, $\Omega_{\Lambda}$\\
     Galaxy & Cluster & Giant arcs, multiple images & Total mass of the cluster, $\Omega_\Lambda$, redshifts\\
     & & & \\
     & & Weak Lensing & \\
     \hline
     & & & \\
     Galaxy & Cluster & Arclets, \newline magnification bias & Cluster mass profile, \newline cluster morphology\\
     & & & \\
     & & Statistical Weak Lensing & \\
     \hline
     & & & \\
     Galaxy & Galaxy & Ellipticity bias, shear-galaxy correlation & Galaxy parameters, halos properties\\
     Galaxy & Large-scale \newline structure & Cosmic shear, shear-shear correlation & Cosmological parameters, power spectra\\
     Quasar & Large-scale \newline structure & Cosmic magnification, \newline quasar-galaxy correlation & Cosmological parameters, power spectra, bias\\
     Last \newline scattering surface & Large-scale \newline structure &  Smoothing of $C_\ell$, \newline $B$ modes & Cosmological parameters, power spectra
    \end{tabularx}
    \caption[Classification of Lensing]{Observational Classification of Gravitational Lensing and Applications, adapted from Table 7.1 of \cite{Peter2009}.
    }
    \label{tab:classificationoflensing}
\end{table}

\subsection{Cosmic Shear} \label{sec:cosmicshear}

\paragraph{Weak Lensing of Small Extended Sources}

The magnification \eqref{magnification} of a small extended source can be approximated to first order by
\begin{equation}
    \mu = \frac{1}{(1 - \kappa)^2 - \Gamma^2} \approx 1 + 2 \kappa
\end{equation}
which shows that in the weak lensing limit, the magnification is determined by the convergence alone. We recall also from Section \ref{sec:magnification} the shape of a small circular source is distorted by weak lensing into an ellipse whose axes are proportional to the inverse of the eigenvalues of the lensing Jacobian. The ellipticity $\varepsilon$ is given by
\begin{equation}
    \varepsilon = \frac{\lambda_1^{-1} - \lambda_2^{-1}}{\lambda_1^{-1} + \lambda_2^{-1}}
 = \frac{\lambda_2 - \lambda_1}{\lambda_2 + \lambda_1} = \frac{\Gamma}{1 - \kappa} \approx \Gamma
\end{equation}
where we took $\kappa \ll 1$ {\hl and $\Gamma \ll 1$} for the final approximation; in this limit which is often taken in weak lensing, the ellipticity is dependent only on the shear.

High-redshift background galaxies are weakly lensed by large-scale matter inhomogeneities, such as the filaments of the cosmic web and foreground galaxy clusters, as well as dark matter along the photon path. Whilst magnification cannot be easily measured, the ellipticity of the lensed images can readily be measured. The lensing induced shear cannot be separated from the intrinsic ellipticity of individual galaxies, but it is correlated for nearby images on the sky (since their light was deflected by the same or nearby massive structures) and this correlation reduces with separation. Assuming that the orientation of intrinsic ellipticities of galaxies is almost entirely random, any systematic alignment between images is caused by lensing. Therefore, the lensing induced shear is observed statistically via averaging over a large number of galaxies. It forms a \textit{cosmic shear} field of order 1\% ellipticity, coherent over scales of around 30 arcmin \cite{PDG2022} from which we can infer the statistical lensing mass distribution \cite{kaiser1992, kaiser1993}. It is also possible to perform \textit{tomography} using extensive surveys, which quantifies how the convergence varies with the redshift bin of the source and gives information on the growth of structure \cite{Congdon2018, Kohlinger2017}.

Statistical weak lensing is therefore considered a very promising probe of cosmology and the dark matter distribution, in particular of the amplitude of the fluctuations and on 10Mpc to 100Mpc scales. At least in principle, a primary challenge is that the signal is generated in part by density fluctuations on small scales, requiring non-linear modelling which distinguishes between baryons and dark matter. The predicted signals scale approximately as $\sigma_8 \Omega_m^\alpha$, where $\alpha \approx 0.3-0.5$ depending on the expansion history and also depend on the distance-redshift relation; to first order this is a sensitivity to $S_8$. SN or BAO measurements are therefore a complementary probe to weak lensing \cite{PDG2022}.

\subsection{Weak Lensing of the CMB}

The path of the CMB photons spans almost the entire observable Universe, along which there are deflections by the gravitational potentials of inhomogeneities, primarily from structures at redshifts of $0.5-10$. These gravitational lensing deflections are described as a \textit{secondary anisotropy} of the CMB, and of order 2' to 3' coherent over patches $2^{\circ}$ to $3^{\circ}$ across \cite{Planck2020a}. Its effects include smoothing out the acoustic oscillations observed in the CMB power spectra by several percent \cite{Lewis2006}, and it also transforms some polarisation E modes into B modes
(which may complicate the detection of primordial gravitational waves). As the deflection of light is achromatic, it preserves the black-body distribution of the CMB.

Whilst the systematics of cosmic shear analyses include shape measurement biases and intrinsic alignments of galaxy orientations, weak lensing of CMB anisotropies does not suffer from much of these observational and astrophysical systematics. Whilst theoretical calculations of the large-scale weak lensing is straightforward, errors for these large-scale signals are dominated by sample variance (i.e. cosmic variance accounting for finite sampling over the sky e.g. removal of portions of the sky covered by the galaxy \cite{Hu2002}). On the other hand, predicting small-scale lensing effects requires modelling complicated astrophysical processes \cite{PDG2022}.

The Planck 2018 measurement of the CMB lensing gave a 3.5\% constraint on $\sigma_8 \Omega_m^{0.25}$, comparable to constraints on $\sigma_8 \Omega_m^{0.5}$ from galaxy lensing with which it is complementary \cite{Plancklensing2018}. Combining with BAO measurements in the galaxy distribution gives a measurement of $\sigma_8$, $\Omega_m$, and $H_0$.

\subsection{Strong Lensing Galaxy Cluster Cosmography} \label{sec:clustercosmography}

Consider a single strongly lensed source, obeying the strong lens equation \eqref{lenseqn_alphahat}
\begin{equation}
\bm{\beta} = \bm{\theta} - \frac{D_{ds}}{D_s} \bm{\nabla}_{\bm{\xi}} \hat{\psi} 
\end{equation}
and we recall from \eqref{lensingpot} that the potential $\hat{\psi}(\bm{\xi})$ is simply the projected Newtonian potential. The dependence on cosmology therefore entirely lies in the lensing efficiency $\frac{D_{ds}}{D_s}$. If we consider only lensing data, e.g. chiefly the position of images which constrains the Einstein radius or the normalisation of the \textit{scaled} lensing potential $\psi$, then the normalisation of the Newtonian lens potential is degenerate with this ratio. Cosmological information cannot be extracted independently unless there is non-lensing data. For example, spectroscopy may separately constrain the kinematics, i.e. the velocity dispersion which usually sets the normalisation of the Newtonian potential (to see this concretely, take the example of the SIS lens \eqref{siskappa}). Although this is possible, this approach can be either observationally expensive, rare, or suffer mass-sheet degeneracies \cite{Caminha2022}.

The above degeneracy can be broken more simply by considering \textit{multiple} sources at different redshifts \cite{Link1998, Golse2002, Acebron2017}. This is also observationally feasible: multiple images can be generated by massive galaxy clusters from a number of sources in a large interval of redshifts. Dividing the image positions (or equivalently the Einstein radii) of one ``family'' of images all arising from one source, with those all from a separate ``family'' cancels the dependence on the velocity dispersion. This leads to a ratio of lensing efficiencies, or the \textit{family ratio}
\begin{equation}
    \Xi = \frac{D_{ds_1}}{D_{s_1}} \frac{D_{s_2}}{D_{ds_2}}.
\end{equation}
which can be used to put constraints on cosmological parameters, in particular $\Omega_m$, $\Omega_k$, and also the equation of state parameter of dark energy. This analysis has been already realised for various galaxy clusters; see \cite{Acebron2017} and \cite{Caminha2022} and references within.

\subsection{Strong Lensing Time Delay Cosmography} \label{sec:tdcosmography}

We reproduce the gravitational time delay function \eqref{cosmologytimedelay} we derived in Chapter \ref{chap:lightlensaction}
\begin{equation}
    \tau (\bm{\theta}; \bm{\beta}) = \frac{(1+z_d)}{c} D_\tau \left( \frac{1}{2} (\bm{\theta} - \bm{\beta})^2 - \psi(\bm{\theta}) \right). \label{eq:timedelayreproduced}
\end{equation}
Although $\tau$ itself is not an observable, the difference in $\tau$ between photometric signals in multiple images of a time variable source is observable, and can be used to constrain cosmology; e.g. defining $\Delta \tau_{BA} \equiv \tau_B - \tau_A$ between images $A$ and $B$:
\begin{equation}
\Delta \tau_{BA} = \frac{D_\tau}{c} (1 +z_d) \Big(\frac{1}{2}  \left(\bm{\theta}^2_B - \bm{\theta}^2_A \right) + (\bm{\theta}_A - \bm{\theta}_B) \cdot \bm{\beta} - \psi (\bm{\theta}_B) + \psi (\bm{\theta}_A)  \Big). 
\end{equation}

The relative image and lens positions may be used to put constraints on the scaled lensing potential $\psi$ and the unobservable unlensed source position $\bm{\beta}$; and the lens redshift may be measured directly. Therefore, the time delay distance $D_\tau$ may be used to constrain cosmological parameters. Since it is a \textit{dimensionful} ratio of angular diameter distances, it is primarily sensitive to the Hubble constant {\hl ($D_\tau \propto \frac{1}{H_0}$)}, but also weakly to several other cosmological parameters such as the dark matter density, spatial curvature and the equation of state parameter for dark energy and its evolution over time. The modelling of the lens potential and the mass-sheet degeneracy (see Section \ref{sec:lensingdegeneracies}, and in particular Figure \ref{fig:MST}) are the primary source of uncertainties. To a lesser extent there are also uncertainties from the limitations of the thin-lens approximation; i.e. the unknown distribution of mass along the line of sight.

In the era of high signal-to-noise spectroscopy, the kinematics of the lens (usually a single galaxy) can also be constrained. Exactly as we discussed for cluster cosmography in Section \ref{sec:clustercosmography}, we are able to tease apart the velocity dispersion which normalises $\hat{\psi}$ from the lensing efficiency $\frac{D_{ds}}{D_s}$. As a result, an absolute angular diameter distance to the lens $D_d$ can be inferred in addition to the ratio $D_\tau$ in time delay cosmography; although the interplay with the mass-sheet degeneracy must be taken into account \cite{Jee2016, Chen2021}. However, the dependence on $H_0$ is more correlated with other cosmological parameters for $D_d$ than $D_\tau$.

Time delay cosmography is an active {\hl area} of contemporary research as it provides a one-step geometric distance independent of the distance ladder. This makes it a promising tool in the resolution of the current Hubble tension. We refer to \cite{Treu2022} for a full and up-to-date review. For example, measurements of time delays between images, high-resolution imaging of the lens system and its environment, together with photometry and spectroscopy, yielded estimates of $D_\tau$ and $D_d$ to 6 lensed quasar systems resulting in a 2.4 per cent precision measurement of {\hl $H_0 = 73.3^{+1.7}_{-1.8} \text{ km } \text{s}^{-1} \text{ Mpc}^{-1}$} in 2020 by the H0LiCOW collaboration \cite{Wong2020}. This was consistent with local distance ladder (SNe Ia) measurements of $H_0$, {\hl and in} the tension with \textit{Planck} CMB measurements.

{\hl However, it was argued that that the lens mass density profiles of either a power-law or stars (constant mass-to-light ratio) plus standard dark matter halos used by H0LiCOW were overconstrained \cite{Kochanek2020, kochanek2021}; and it was the fact that these models were so restrictive which gave the reported precision in $H_0$ rather the spectroscopic data, e.g. the stellar velocity dispersion. Relying on a simply parametrised density profile for the lens corresponds to an ``assertive'' approach to breaking the mass-sheet degeneracy \cite{Treu2022}. The TDCOSMO collaboration, upon relaxation of the form of the mass density profile (and thus with the mass-sheet degeneracy purely constrained by stellar kinematics, a ``conservative'' approach) found instead a $8 \%$ measurement of $H_0 = 74.5^{+5.6}_{-6.1}  \text{ km } \text{s}^{-1} \text{ Mpc}^{-1}$ \cite{Birrer2020}. The TDCOSMO team then used imaging and spectroscopic data from a sample of lenses from the Sloan Lens ACS (SLACS) survey to further constrain the lens mass density profiles. The uncertainty on $H_0$ was reduced to $\sim 5 \%$ with $H_0 = 67.4^{+4.1}_{-3.2}  \text{ km } \text{s}^{-1} \text{ Mpc}^{-1}$ \cite{Birrer2020}, which is closer to the CMB inferred value whilst still statistically consistent with the previous measurements.}

The usage of time delays from strongly lensed \textit{supernovae} to determine $H_0$ was proposed by Refsdal in 1964 \cite{Refsdal1964}, preceding the discovery of the first multiply-imaged quasar by 15 years. Lensed supernovae are much rarer than lensed quasars, and hence quasars are the typical source for current strong lensing time delay measurements. The Hubble Space Telescope provided data of the first resolved and multiply-lensed supernova, dubbed SN Refsdal, relatively recently in 2014. 

{\hl The host galaxy of SN Refsdal is a member of the galaxy cluster MACS J1149.5+2223, and lensed into three macro-images by the cluster. SN Refsdal was observed as four images (S1-S4) in an Einstein cross configuration detected around one of the macro-images of the host galaxy: this macro-image is well aligned with a galaxy belonging to the cluster, thereby adding the effect of galaxy lensing. Remarkably the prediction from lens modelling that another image SX of SN Refsdal would appear in one of the other macro-images} was borne out quite precisely (1.5\% on the time delay) the following year \cite{Kelly2023a}. Interestingly, the estimates of $H_0$ from SN Refsdal \cite{Grillo2018, Kelly2023b} are lower than estimates from quasar strong lensing time delays (see Fig. 5 of \cite{Kelly2023b}); albeit with much larger uncertainties which are attributed mostly to the cluster lens model. Given an exact cluster model, the 1.5\% uncertainty of the SX–S1 time delay measurement would provide an equally precise constraint on $H_0$.

\section{Quasars}

\subsection{Active Galactic Nuclei}

 The first evidence that some galaxies host a strongly emitting (``active'') centre (``nucleus'') or active galactic nucleus\footnote{Again, there is no single strict definition of an AGN. A possible physical definition is that extragalactic object is considered to be an AGN if it contains a massive accreting central black hole. Observationally, defining an AGN is not so clear because of observational limitations and source obscuration; we refer to \cite{Beckmann2012} or \cite{Mo2010} for possibilities.} (AGN) was found by Carl Seyfert in the 1940s, from the optical spectra of six galaxies which showed high-excitation emission lines from the central region, superposed on a continuum spectral distribution \cite{Beckmann2012}. Two initially notable observational characteristics were that the continuum was distinct from an integrated stellar (thermal) continuum characteristic of normal galaxies; and the observed emission lines did not appear to correspond to laboratory wavelengths of known atomic transitions, and some had widths far greater (>1000 km s$^{-1}$ when interpreted as Doppler broadening) than any other known phenomena. These features of the spectra were at the time unexplained.
 
 It was not until the 1960s that the existence of an immensely massive object $\gtrsim 10^{6} M_{\odot}$ at the core of galaxies was put forward, with the process of accretion from a surrounding gaseous disk driving the emission of radiation. The massive central engine was assumed originally to be a hypermassive star: although the theoretical existence of black holes was known at the time as a case of the Schwarzschild metric (1916), this exact solution to the EFEs was not believed to be physical. It became clear by 1939 \cite{Oppenheimer1939a, Oppenheimer1939b} however that general relativity predicts that massive stars collapse into black hole, when internal pressure and stresses cannot balance the gravitational attraction -- i.e. all the matter contained in the star contracts past a (mathematical) event horizon which separates it causally from the rest of the Universe, and in finite time arrives at a singularity.
 
 The Kerr solution to the EFEs for an uncharged, rotating, axially symmetric black hole was discovered in 1963, and it was proposed around the same time that that the central massive object was in fact an astrophysical black hole \cite{Salpeter1964, Zeldovich1964} (the latter article by Zeldovich may be found in \cite{Zeldovich2017}). Astrophysical objects have non-zero angular momentum, so it is assumed that the Kerr metric is appropriate. The strong gravitational potential of the black hole explained not only the large energy output as converted via accretion from gravitational energy, but also the large dynamical (Doppler) broadening of the observed lines, and also the small size of the emitting regions (the observed variability time scales limited its size to dimensions of order light days). The emission lines were recognised as highly redshifted from their rest-frame values over vast cosmological distances. It is now understood that a central supermassive black hole resides at the centre of most galaxies, and an AGN nature is expected whenever enough matter falls within a critical distance. The first direct imaging of a black hole shadow at the centre of a galaxy (M87) was achieved in 2019 \cite{M87} by the Event Horizon Telescope using interferometry.

\begin{figure}
    \centering
    \includegraphics[scale=0.8]{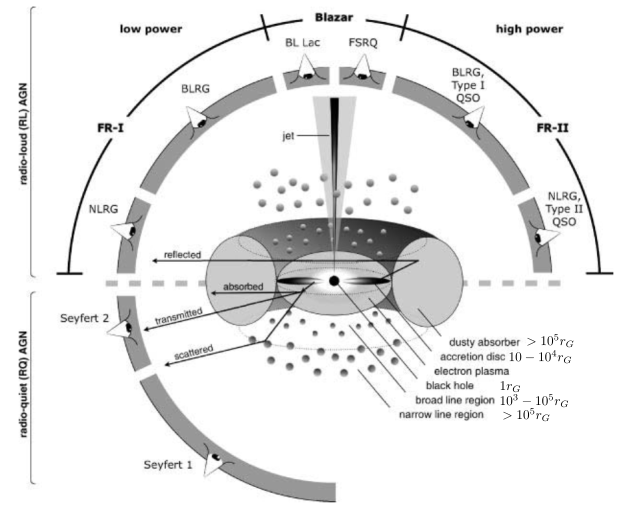}
    \caption{Cartoon, not-to-scale depiction of AGN phenomena, showing a unified model of historically disparate classes of astrophysical objects, reproduced with annotation from \cite{ Beckmann2012}. The observational classification depends on the viewing angle, the amount of jet emission (note the diagram truncates its typically symmetric nature), and the power of the central engine. The broad line region is typically located at $\sim 0.1$ pc whereas the narrow line region is $\sim 100-1000$ pc \cite{Netzer2015}. The host galaxy is up to kiloparsec scale and not displayed. The acronyms used in the figure are: Fanaroff and Riley (FR), broad-line radio galaxy (BLRG), narrow-line radio galaxy (NLRG), quasar or quasi-stellar object (QSO), BL Lacertae object (BL Lac) and flat-spectrum radio quasar (FSRQ).
    }
    \label{fig:AGN}
\end{figure}

Due to their variation in observational features and the history of their discovery, AGN are divided (neither systematically nor uniquely) into a veritable zoo of classes and sub-classes. Their place in a unified AGN model (the details of which are still subject to debate \cite{Netzer2015, Padilla2021}) depends on the viewing angle as well as the luminosity. This model is depicted in Figure \ref{fig:AGN} consisting of the black hole, accretion disk, broad line region (BLR), dusty torus, narrow line region (NLR), and potentially a radio/optical jet.

The detailed description of these AGN classes may be found in e.g. \cite{Beckmann2012, Eigenbrod}. For example, AGN may be divided in radio-loud and radio-quiet, the former exhibiting collimated jets. Radio-quiet AGN include the relatively nearby Seyfert galaxies, subdivided into Seyfert I (possessing broad emission lines or BEL) and Seyfert II galaxies (no BEL). This is explained in the unified model by a ``dusty torus'' which obscures the optical-UV emission at viewing angles corresponding to Seyfert IIs. Radio-loud AGN include blazars (possessing beamed emission from a relativistic jet which is aligned roughly toward the line of sight), and most pertinent to observational cosmology at present, quasars. Quasars were discovered in the early 1960s in the radio domain, and identified later by optical imaging in which they appeared as bright star-like (i.e. point-like as a result of their being extremely distant) sources and therefore dubbed quasi-stellar radio sources (QSRs) or quasi-stellar objects (QSOs). By the 1970s it was known that quasars are actually extended, i.e. residing in galaxies, rather than truly point-like \cite{Beckmann2012}.

We reiterate that classification is neither universal nor unique. For example, a more modern approach followed by \cite{Netzer2013} only refers to ``type-I AGN'' (those objects showing broad, strong UV-optical emission lines in their spectrum) and ``type-II AGN'' (those showing prominent narrow emission lines, very faint, if any, broad emission lines, and a large X-ray-obscuring column). In this scheme, type-I or type-II sources can have low or high ionisation lines, strong or weak radio sources, strong or weak X-ray sources, etc. As another example, SDSS quasars \cite{sdss2001} were defined to mean any extragalactic object with at least one broad emission line and that is dominated by a non-stellar continuum: i.e. Seyfert galaxies and quasars are not considered distinct. Other teams and authors may use an arbitrary convention whereby a quasar is a broad-emission line object is brighter than $M_B = -23$ mag, and if dimmer the object is a Seyfert I \cite{Eigenbrod}.

\begin{figure}
    \centering
    \includegraphics[scale=0.7]{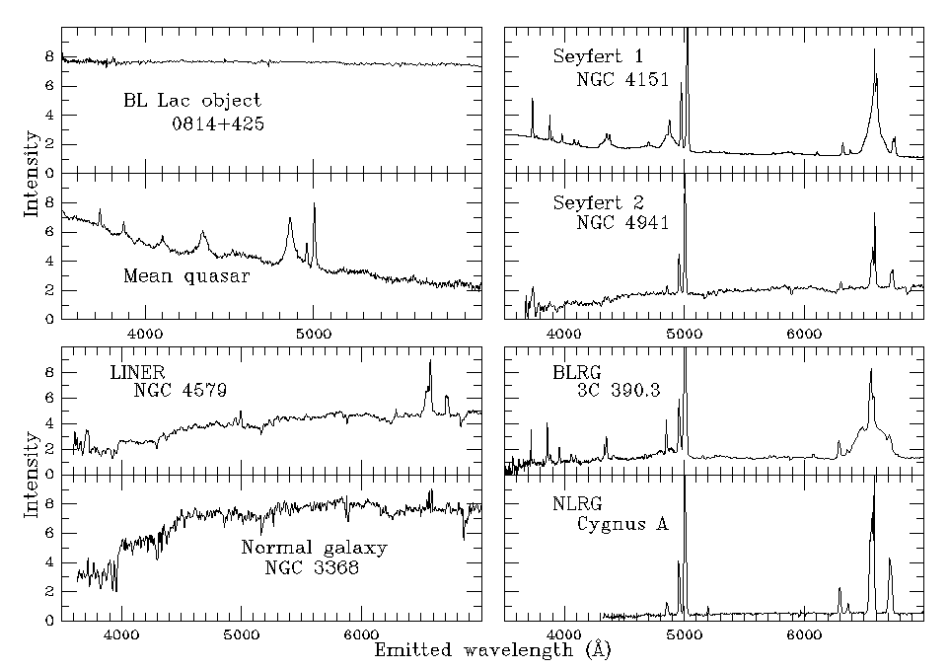}
    \caption{Spectra of AGN (and normal galaxy), showing observational variations which led to their historic classifications. Reproduced from \cite{Oswalt2013}.}
    \label{fig:AGNspectra}
\end{figure}

\subsection{Quasar Spectra} \label{sec:quasarspectra}

Unlike non-active galaxies, quasars and AGN emit electromagnetic radiation over a wide range of energies from radio to gamma-rays, and we refer to e.g. \cite{Schneider2015} for a full description. In this section, we only consider UV-optical-near-IR (rest-frame) wavelengths. In this range, a typical quasar spectrum displays a featureless continuum which is relatively flat or rising towards the UV, produced from the accretion disk together with a series of emission line features, some of which correspond to very high ionisation energy transitions. The emission lines arise from ``clouds'' of matter at larger radii\footnote{There also are emission lines, e.g. the relativistic broad iron line, in the X-ray spectrum arising from the inner radii of accretion disk itself. \cite{Schneider2015}} which is photoionised by radiation (including the continuum emission) from the accretion disk. The clouds are usually assumed to be in photoionisation equilibrium (photoionisation rate equivalent to recombination rate). The relative emission line strengths of quasar spectra and those of ionised gas from e.g. planetary nebulae and H$\textsc{ii}$ regions, are similar: these gases reach photoionisation equilibrium  at similar temperatures of $T \sim 10^4$ K \cite{Peterson2006}.

\begin{figure}
    \centering
    \includegraphics{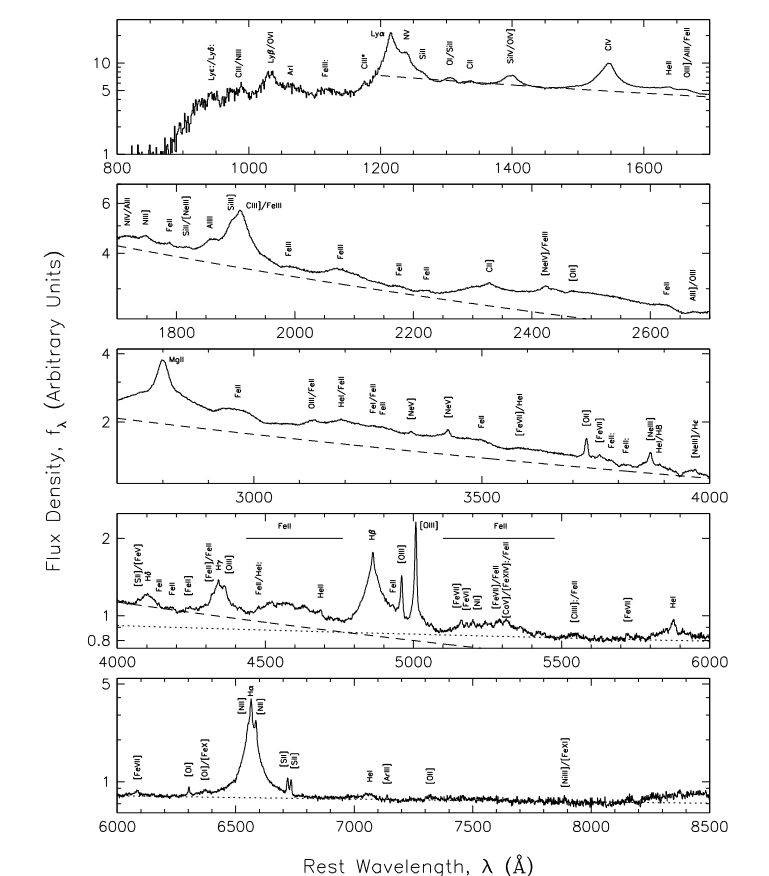}
    \caption{Median quasar composite UV-optical (rest-frame) spectrum on a loglinear scale, showing a series of mostly broad and some narrow emission line features. Emission features are labelled by ion, and those with a colon ( :) are uncertain identifications. Semi-forbidden line transitions are denoted with a single square bracket, e.g. C$\textsc{iii}$], whereas forbidden line transitions are square bracketed, e.g. [O$\textsc{iii}$]. Permitted transitions are unbracketed, such as C$\textsc{iv}$, Mg$\textsc{ii}$ and the Balmer lines. Two power-law continuum fits are shown by dashed and dotted lines. The continuum bluewards of the Lyman-$\alpha$ (Ly-$\alpha$) emission line is heavily absorbed due to the intergalactic medium containing neutral hydrogen H$\textsc{i}$; the dense and uneven series of absorption lines is called the Ly-$\alpha$ forest. The deviation of the spectrum from the continuum fit between 1600 and 3800\AA \, is due to a large number of overlapping iron lines which blend into a quasi-continuum, and Balmer continuum radiation.
    Figure reproduced from \cite{sdss2001}.}
    \label{fig:quasarspectra}
\end{figure}

Quasar emission lines are generally separated into two types: 
\begin{enumerate}
    \item \textit{Broad emission lines} (BELs): permitted and semiforbidden lines are typically Doppler broadened and exhibit variability correlated with that of the continuum emission. In most cases, strong permitted line widths (full width at half maximum or FWHM) exceed $1000 \text{km} \, \text{s}^{-1}$  and can be up to $20 000 \text{km} \, \text{s}^{-1}$ \cite{Netzer2015}.
    \item \textit{Narrow emission lines} (NELs): forbidden lines are narrower with a typical width of $400 \text{km} \, \text{s}^{-1}$ \cite{Schneider2015} and either non-variable or show less variability. These widths are still broader than those of emission lines from non-active galaxies.
\end{enumerate}
They are therefore thought to arise from spatially and kinematically distinct regions, respectively the parsec-scale broad line region (BLR) and the kiloparsec-scale narrow line region (NLR). The cut off between narrow and broad lines is not necessarily unambiguous however (e.g. \cite{Netzer2015} gives the cut off as $1000 \text{km} \, \text{s}^{-1}$ compared to $500 \text{km} \, \text{s}^{-1}$ for the SDSS \cite{sdss2001}). Spectra are very similar from quasar to quasar (compared with stars or galaxies); that is, they are self-similar across a large parameter space (of varying luminosity, black hole mass and redshift). An example composite quasar spectrum is shown in Figure \ref{fig:quasarspectra}.

\begin{table}
    \centering
    \begin{tabular}{cccc}
        \textbf{Region} & \textbf{Length Scale} & \multicolumn{2}{c}{\textbf{Light Crossing Time for $\bm{M_{BH}}$}} \\
         & $[r_G]$ &  $10^7 M_{\odot}$ & $10^9 M_{\odot}$\\
         \hline \\
         Inner Accretion Disk & 10 & 500s & 14h\\
         (X-Ray) & & & \\
         Accretion Disk & $10^2 - 10^4$ & $0.06-6$d & $6-600$d\\
         (UV/Optical Continuum) & & & \\
         BLR &  $10^3 - 10^5$ & $0.6-60$d & $60-6000$d \\
         (UV/Optical BELs) & & & \\
         Dusty Torus & $>10^5$ & $>60$d & $>6000$d \\
         (Near-IR) & & &\\
    \end{tabular}
    \caption{AGN regions, with associated emission. Corresponding length scales in the AGN are given in units of the length scale associated with the black hole, gravitational radius $r_G \equiv \frac{GM_{BH}}{c^2}$; and translated into an associated time scale for the light travel time across the region. Table adapted from \cite{Cackett2021}.}
    \label{tab:AGNstructurescales}
\end{table}

\subsection{Reverberation Mapping}
\label{section:ReverberationMapping}

\begin{figure}
    \centering
    \includegraphics[scale=0.4]{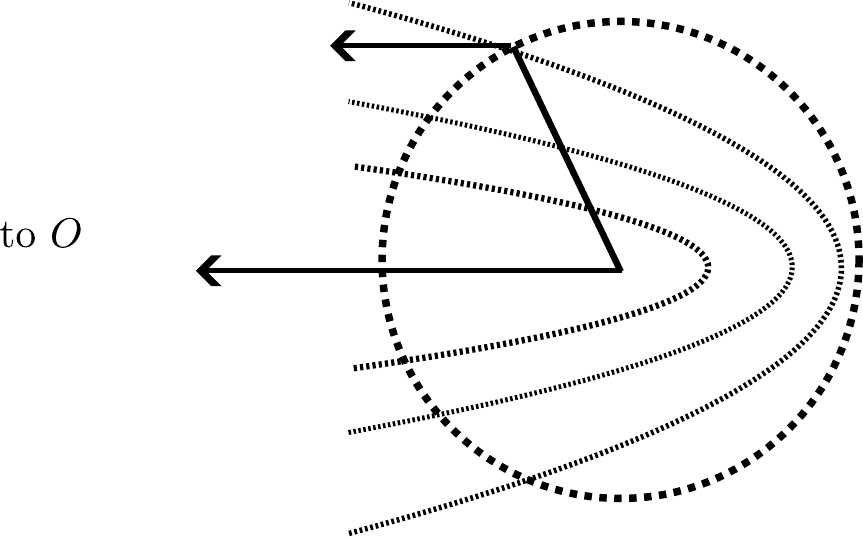}
    \caption{Schematic illustration of reverberation mapping. The observer is to the left at infinity. A geometric time delay arises between the continuum emission (straight light ray) and the line emission (segmented light ray) from the BLR, here depicted as a spherical shell. Material is illuminated at successive times on parabolic surfaces, forming a family of paraboloids of equal time delays, or iso-delay surfaces coaxial with the line-of-sight with a focus at the source. Adapted from \cite{Peterson1993}.}
    \label{fig:RM_isodelay}
\end{figure}

It is currently not possible to spatially resolve the inner regions of AGNs due to their small angular size, with only a few exceptions for some nearby AGN (e.g. M87 as mentioned). For example, the BLR of an AGN at $z \approx 0.012$ ($\sim 50$ Mpc) will have an angular size of $\sim 10^{-5}$ arcseconds \cite{Cackett2021}. Therefore, indirect methods are needed to study the inner regions of AGN. The most established method to do so is reverberation mapping (RM), which uses light ``echos'' to map structure. Light from central regions of the AGN (e.g. continuum emission from the accretion disk) is absorbed and re-emitted by material at a larger radius (e.g. as broad emission lines from the BLR). Since the flux shows variability as an almost defining characteristic of an AGN, the observed light curves (flux as a function of time) from the two regions in the AGN can be cross-correlated to give a time delay or \textit{time lag} -- e.g. emission lines respond to the variability in the continuum light curve with a delay which can be measured. This RM time delay arises from the difference in the light travel time to the observer: this is illustrated in Figure \ref{fig:RM_isodelay}. In effect, reverberation mapping is a technique that swaps observable temporal information and unobservable spatial information: it has been used to constrain the black hole mass and determine of the characteristic size scales of the BLR, and has also been applied to regions associated with X-rays, the UV/optical continuum and dusty torus \cite{Cackett2021}. We limit the following discussion to RM of the BLR, although the same principles apply in general. We refer also to Section 4 of \cite{Ng2023}, included as Chapter \ref{chap:p2} of this thesis, for a review of reverberation mapping; as well as \cite{Blandford1982, Peterson1993, Netzer1990, Netzer2013, Cackett2021}. 

In addition to constraints on the geometry of the BLR through measurements of the time lag from the light curve, velocity-resolved or ``2D'' reverberation mapping gives information on the kinematics of the BLR since the line-of-sight velocity of a BLR cloud is proportional {\hl to} the observed wavelength within the broad emission line (approximating relativistic effects such as the relativistic Doppler effect as negligible). A ``cloud'' refers to a small individual entity in the emission line region. If the velocity of a BLR cloud may be related to its position via a model for the velocity field, then the line-of-sight velocity gives additional information on the position of the BLR cloud as well as the kinematics.

If the continuum emission from the central source occurs at $t=t_i$ (i.e. a Dirac delta signal in time), then the BLR emission occurs at $t=  t_i + t_{\textsc{rm}}$, where the reverberation mapping time delay is
\begin{equation}
t_{\textsc{rm}} (\bm{r}) = \frac{(1+z_s)}{c} \left( r - \bm{r} \cdot \hat{r}_z  \right). \label{tauBLR_ch3}
\end{equation}
Here $\hat{r}_z$ is the unit vector towards the observer and $\bm{r}$ is the spatial position of a given BLR cloud relative to the central source. {\hl It expresses the extra time required for a photon to travel from the central source to a cloud, plus the time accounting for the extra line-of-sight distance of the BLR cloud to the observer.} The cosmological redshift factor $(1+z_s)$ corrects the time delay in the quasar rest frame for the time delay as measured in our observer frame. If we hold $t_{\textsc{rm}}$ fixed, then the ``iso-delay surface'' is a paraboloid of revolution around $\hat{r}_z$. Equation \eqref{tauBLR_ch3} therefore describes a family of nested paraboloids parametrised by $t_{\textsc{rm}}$. The time at which the flux from the BLR is seen is determined by the intersection of the BLR geometry with the iso-delay paraboloids.

Let $\bm{w}$ be the velocity coordinates of a BLR cloud, whereas $v_z$ denotes the line-of-sight velocity variable. The broad emission line spectral flux responds to the continuum flux $C(t - t_{\textsc{rm}})$, which we assume can be given in generality by
\begin{equation}
    L( v_z , t) = \iint j(\bm{r}) C ( t' - \tfrac{r}{c} ) g (\bm{r}, v_z) \delta ( t - (t' - \tfrac{r}{c} + t_{\textsc{rm}}(\bm{r}))) d \bm{r} d t' \label{line}
\end{equation}
where the line-of-sight projected 1D velocity distribution $g(\bm{r}, v_z)$ is defined in terms of the full velocity distribution $f( \bm{r}, \bm{w})$ as
\begin{equation}
    g(\bm{r}, v_z) \equiv \int f( \bm{r}, \bm{w}) \delta ( v_z - \bm{w} \cdot \hat{r}_z ) d \bm{w}.
\end{equation}
The responsivity term $j(\bm{r}) = \frac{\varepsilon(\bm{r})}{4 \pi r^2}$, where $\varepsilon(\bm{r})$ is the reprocessing coefficient of the cloud, is in general dependent on the position of the cloud and contains the physics of the photoionisation of the BLR. The continuum flux received by a gas cloud at time $t'$ and position $\bm{r}$, emitted by the central source at an earlier time $t' - \tfrac{r}{c}$, is $j(\bm{r}) C ( t' - \tfrac{r}{c} )$ and plays the role of the emissivity of the cloud.

The linearity of the response functional \eqref{line} of the BEL flux to the continuum (i.e. the line response scales with the continuum function and superposition of the line response corresponds to input signal) combined with time translation invariance (i.e. a time lag in the continuum corresponds to the same lag in the response) means {\hl that this is a \textit{linear time-invariant system}} and the line response can be written as a convolution of the continuum with {\hl a linear impulse response function (a Green's function)}. {\hl Usually a transfer function is the Laplace or Fourier transform of the impulse response; i.e. it exists in frequency space, but often in the reverberation mapping literature the impulse response function is also called the transfer function. We follow this (conflated) terminology for consistency.} As an aside, convolution can be intuited as a weighted moving average; it is usually described as flipping and shifting one of the two functions before integrating their product (``flip, shift and integrate''). Over an e.g. spatial domain, the flip is often put down to convention, but it is strictly necessary for dealing with time series on account of causality. The shift then allows us the ``average'' around an arbitrary $t$.

The general reverberation mapping problem then involves finding the deconvolution of
\begin{equation}
    L (v_z, t) = \int \Psi(v_z, s) C (t-s) ds \label{convolution}
\end{equation}
since all of the geometrical information is contained within the transfer function $\Psi (v_z, t)$. The transfer function is the line response to a Dirac-delta continuum pulse; replacing $C (t' - \tfrac{r}{c})$ with $\delta (t' - \tfrac{r}{c})$ in Equation \eqref{line} gives
\begin{align}
    \Psi (v_z, t) &= \int j(\bm{r}) g (\bm{r}, v_z) \delta ( t - t_{\textsc{rm}}(\bm{r})) d \bm{r}\\
    &= \int j(\bm{r}) n(\bm{r}) \delta (v_z - u_z(\bm{r})) \delta ( t - t_{\textsc{rm}}(\bm{r})) d \bm{r}. \label{Psi}
\end{align}
The second equality holds when there exists a velocity model $\bm{u}(\bm{r})$ whose corresponding line-of-sight velocity model is $u_z(\bm{r})$, such that we may write the velocity distribution as
$f(\bm{r}, \bm{w}) = n(\bm{r}) f(\bm{w}| \bm{r}) = n(\bm{r}) \delta (\bm{w} - \bm{u}(\bm{r}))$
where $n(\bm{r})$ is the number density of responding clouds. This gives the line-of-sight projected velocity distribution as $g(\bm{r}, v_z) = n(\bm{r}) \delta (v_z - u_z(\bm{r}))$. We can interpret Equation \eqref{Psi} as a probability density function under a change of variables from positions $\bm{r}$ to $(v_z , t)$, written using Dirac delta generalised functions as the dimensionality of the original set of random variables is not the same as the new set of random variables \cite{Au1999}. When the transfer function is visualised using a heat map in $(v_z, t)$-space, it is commonly referred to as a ``velocity-delay map''.

The deconvolution of Equation \eqref{convolution} may be approached using maximum entropy fitting techniques, or regularised linear inversion; and then the outputted velocity-delay map may be compared qualitatively to the theoretical velocity-delay maps produced by simple models. Recovering the transfer function and thereby inferring the physical and kinematic distribution of the BLR using reverberation mapping has been, up until recent improvements of the quality of reverberation data sets, regarded as a technique in a developmental state \cite{Shen2015, Mangham2019, Cackett2021}. Rather than solving this ill-posed inverse problem, reverberation mapping has been often limited to finding only the mean radius for the BLR through measuring the mean time delay for emission lines using cross-correlation analyses, from which it is possible to constrain the mass of the central black hole. Forward modelling of the BLR also offers an alternative Bayesian approach to directly (without explicitly finding the transfer function) providing best-fit parameters of a flexible BLR geometry \cite{Pancoast2011}. 



\subsection{Quasars as a Cosmological Probe}

Quasars are the most luminous persistent sources in the Universe, exceeding the luminosity of normal galaxies by a factor of a thousand \cite{Schneider2015, Moresco2022}. As such, they are amongst the most distant astrophysical objects we observe, with the highest redshifts measured being $z \sim 7$ \cite{Yang2023} -- a substantial fraction of a Hubble time.

Since their discovery, quasars have been an alluring candidate as a cosmological probe. The time delay of strongly lensed quasars as discussed in Section \ref{sec:tdcosmography} is an obvious success (albeit mostly sensitive to $H_0$ and the absolute distance to the lensing galaxy or cluster, rather than the quasar). Also well established is the usage of the Ly-$\alpha$ forest (arising from the line-of-sight absorption of the quasar spectrum -- see Figure \ref{fig:quasarspectra}) to probe the intergalactic medium and large-scale structure, and thereby cosmology. However, most proposed approaches to use quasars directly as a distance indicator, i.e. as a standardisable candle or ruler, have been found to be invalid or ineffective in the end \cite{Abdalla2022} at least in part due to poor understanding of quasar physics \cite{Wang2020}. We nonetheless mention a few methods which either show promise or have led to some initial results.

\subsubsection{Quasars as Potential Standard Candles}

Unlike Sne Ia, the luminosity of quasars covers a large range (over 8 orders of magnitude) \cite{Marziani2014}; and as mentioned in Section \ref{sec:quasarspectra}, their spectra do not appear to show dependence on properties like luminosity or mass. Attempting to use quasars as a standard(isable) candle therefore appears somewhat elusive. 
Some attempts include the Baldwin effect (correlation of the width of certain emission lines with luminosity), or the identification of a sub-population of quasars which radiating near the Eddington limit \cite{Marziani2014}.

\paragraph{RM Time Lag Cosmography}
More notably, there is a relation between the reverberation mapping time delay or lag of BLR emission lines (mostly H$\beta$) and the intrinsic luminosity flux -- i.e. a relation between a radius within the quasar and luminosity -- which can give a luminosity distance \cite{Watson2011, Czerny2023}. The basic idea is that the reverberation mapping time delay should be proportional to the square root of the luminosity of the central source, since the continuum ionising flux incident on the BLR gas drops inversely with the square of the distance from the central engine (cf.  Equation \eqref{line} where $j(\bm{r}) = \frac{\varepsilon(\bm{r})}{4 \pi r^2}$ with $\varepsilon(\bm{r}) = \text{const.}$). The observable ratio of the lag time for a given emission line and the square root of the continuum flux gives a measurement of the luminosity distance \cite{Watson2011}. Reverberation mapping methods (also called ``time lag cosmography'' methods), including earlier attempts using reverberation mapping of the accretion disk \cite{Collier1999} rather than the BLR, are currently limited by various issues -- e.g. by requiring external scaling \cite{Abdalla2022}, and the difficulty of measuring emission-line lags for high-z quasars due to the dilation factor $(1 + z)$ \cite{Wang2020}. Although these methods provide a luminosity distance rather than an angular size distance, the fact that the RM time delays provides the BLR radius means that these methods are sometimes categorised as providing a standardisable ruler (e.g. \cite{Czerny2018}).

\paragraph{UV-X-ray Luminosity Correlation and QSO Hubble Diagram}
What has gained wide attention is the usage of the empirical correlation of quasar luminosites in the UV and X-ray bands, which allows quasars to be used as a standard candle. The UV emission originates from the accretion disk, which is surrounded by the \textit{corona}, a plasma of hot relativistic electrons (see Figure \ref{fig:AGN}) which produces X-rays. The relationship between the UV and X-ray emission determines a luminosity distance and has resulted in the construction of a quasar Hubble diagram \cite{Risaliti2015, Risaliti2019}. The analysis showed a tension with $\Lambda$CDM which only emerges at high redshifts (as the quasar luminosity distances were calibrated using Sne Ia measurements at lower redshifts where data overlapped), and therefore supports the idea of a new model being required (e.g. the increase of the dark energy density with time, or other interpretations such as a negative spatial curvature \cite{Abdalla2022}). Concretely, there is a preference for lower values of luminosity distance at higher redshifts, and whether this result is due to calibration with Sne Ia or otherwise is the subject of some speculation \cite{Abdalla2022}. Time lag cosmography analyses did not such show this tension for instance \cite{Czerny2023}.

\subsubsection{Quasars as Standard Rulers}

\paragraph{SARM Analyses}
Although some structures of nearby AGNs have been resolved\footnote{It is also notable that water megamasers in Seyfert 2 galaxies at low redshift $z \ll 1$ are resolved using Very Long Baseline Interferometry, and have been used to provide angular size distances \cite{Pesce2020, Czerny2018}, but these methods are limited by the rarity of water masers \cite{Wang2020}.}, direct imaging observations cannot resolve the inner structure of quasars (angular size of less than $10^{-4}$ arcseconds) \cite{Sturm2018}. It has not until very recently therefore, that the possibility of using quasars as a standard ruler was demonstrated for the first time. This new development was achieved due to a technique known as \textit{spectroastrometry} (SA), which provides spatial information\footnote{Some authors are particular about the usage of ``spatial resolution'': astrometry determines only the spatial centre of an image; the precision with which this centre is determined is not the same as the angular resolution of an image \cite{Bacciotti2008}.}, e.g. the separation between the photocentres of redshifted and blueshifted emission, on a scale smaller than the angular resolution from the diffraction limit of the telescope.

This technique applied to data from the GRAVITY instrument of the Very Large Telescope Interferometer achieved a spatial resolution $\sim 10^{-5}$arcseconds, thereby allowing the measurements of the characteristic angular sizes of the BLRs of IRAS 09149-6206 at $z = 0.0573$ and quasar 3C 273 at $z = 0.158$ \cite{Sturm2018, Amorim2020}. Since reverberation mapping measures the characteristic linear BLR size, the angular size distances of quasars can then be directly determined. The combination of spectroastrometry and reverberation mapping has been dubbed the SARM approach, and is independent of the calibrations required by distance ladders. 

The first SARM analysis on quasar 3C 273 \cite{Wang2020} simultaneously yielded the angular size distance, the mass of the central supermassive black hole and a value for the Hubble constant with a $\sim$15\% precision. A velocity-resolved RM follow-up was conducted \cite{Li2022}, and the method was also applied to a nearby AGN \cite{Amorim2021}. 

The simulation of mock data predicts \cite{Songsheng2021} that future SARM measurements of 60 AGNs over a larger redshift range ($z <1$ in the mock data set) will be able to achieve a $\sim$2\% precision of $H_0$. Future facilities such as the Extremely Large Telescope and GRAVITY+ will be able to perform SA measurements and detect fainter quasars up to redshifts $z = 2-3$. SARM analyses are expected to provide a robust distance measurement for low-z AGN which can additionally be used to calibrate high-z quasar distances from other methods, e.g. super-Eddington accreting massive black holes \cite{Wang2020}.

\section{Current and Future Cosmology Surveys}

Upcoming observations will bring further precision, and hopefully conceptual clarity or new discoveries, to cosmology. These centre on a number of wide-field imaging and spectroscopic surveys which are already operational, under construction or planned. These are often classified in the literature as ``Stage III'' (current or finishing at time of writing, intermediate-scale) or ``Stage IV'' (longer-term, starting or future) projects, following the Dark Energy Task Force report \cite{Albrecht2006}.

\begin{table}
    \centering
    \begin{tabularx}{\textwidth}{ p{3cm} X p{2cm} p{2cm}}
       Name/ Acronym  &  Data Type  & Base & Status \\
       \hline \hline \\
        & & & Stage III\\
        \hline \\
        SDSS, BOSS & optical imaging, spectroscopic & ground & 2000-\\
        HSC  & optical imaging & ground &  finished\\
        KiDS & optical imaging & ground & 2011-\\
        DES & imaging / photometric & ground & 2013-\\
        &&&\\
        & & & Stage IV\\
        \hline\\
        \textit{Rubin}/LSST & mostly optical imaging & ground & $\sim$2024-2034\\
        DESI & spectroscopic & ground & 2020-$\sim$2026\\
        \textit{Euclid} & optical imaging, \newline near-IR spectro-photometry & space & 2023-\\
        \textit{Roman}/WFIRST & near-IR imaging/spectroscopy & space & planned\\
        SKA & radio & ground & planned\\
    \end{tabularx}
    \caption{Selected examples of current and future surveys (Stage III and IV); a more comprehensive list is found in \cite{Abdalla2022, Snowmass2021}. A more detailed comparison of \textit{Rubin}/LSST, \textit{Euclid} and \textit{Roman}/WFIRST is found in \cite{Chary2020}. Acronyms are as follows: Sloan Digital Sky Survey - SDSS; Baryon Oscillation Spectroscopic Survey - BOSS; HSC - Hyper Suprime Cam; KiDS - Kilo-Degree Survey;  Dark Energy Survey  - DES; LSST - Legacy Survey of Space and Time, formerly the Large Synoptic Survey Telescope; Dark Energy Spectroscopic Instrument - DESI; Wide-Field Infrared Survey Telescope - WFIRST; Square Kilometre Array - SKA.}
    \label{tab:surveys}
\end{table}

The advent of these Stage IV facilities -- most notably \textit{Rubin}/LSST \cite{ LSST2012, LSST2019}, \textit{Euclid} \cite{Euclid2011}, and \textit{Roman} (formerly WFIRST) \cite{Eifler2021a} -- within the next five years therefore marks a major leap in observational capabilities. A comparison of these large optical/NIR surveys is given in \cite{Chary2020}. \textit{Rubin} is a ground-based observatory designed to survey 20000 deg$^2$ of the sky (i.e. the entire Southern hemisphere sky) repeatedly optically imaging each region of sky with deep short exposures, described as a ``wide-fast-deep-repeat'' approach, and will routinely monitor $\sim2$ billion objects for photometric and astrometric changes \cite{LSST2009,  LSST2012, LSST2019}. Euclid is a satellite designed to survey 15000deg$^2$ of the sky, with optical and near-infrared imaging and spectroscopy capabilities \cite{Euclid2018}, mainly designed to investigate dark energy. \textit{Roman} is a multi-purpose near-infrared space telescope surveying much less of the sky at 2200 deg$^2$, with a variety of planned standalone surveys addressing topics from exoplanets, galaxy evolution to supernovae and cosmology. \textit{Euclid} and \textit{Roman} will achieve high angular resolution and extremely stable image quality as space-based observatories.

\textit{Rubin}, \textit{Euclid} and \textit{Roman} data will be able to provide measurements of the spatial density, distribution, and masses of galaxy clusters as a function of redshift; BAO and redshift space distortions as well as weak gravitational lensing/cosmic shear analysis \cite{Abdalla2022, LSST2012, Eifler2021b} -- improving on e.g. $S_8$ constraints from current optical imaging data from the Sloan Digital Sky Survey (SDSS), Dark Energy Survey (DES), Kilo-Degree Survey (KiDS), and HSC Survey \cite{DESKIDS2023}. In the shorter term, the DESI galaxy redshift survey will exceed the size of the SDSS/BOSS surveys by a factor of ten, enabling precision BAO measurements at $z \approx 0.7-1.4$, and at higher redshifts through Ly-$\alpha$ forest maps; and the first percent-level measurements of structure growth through redshift space distortions \cite{PDG2022}.

\textit{Rubin}, and in particular the space-based \textit{Roman} with deep high-resolution NIR capabilities, will also provide distance ladder and next-generation SNe Ia luminosity distance measurements \cite{LSST2012, Abdalla2022, PDG2022}. Over the next few years, improved parallax and new SN Ia data from telescopes such as the Hubble Space Telescope, James Webb Space Telescope and the \textit{Gaia} satellite should also allow further improvements and systematics checks in the Cepheid distance ladder.

\textit{Roman}, \textit{Euclid} and \textit{Rubin} will additionally be able to target strong lensing time delay cosmography. \textit{Rubin}'s extensive survey area will not only lead to the discovery of numerous lenses but high-cadence imaging will also facilitate their identification through variability, and may determine some time delays directly. The high angular resolution and quality of imaging from \textit{Euclid} and \textit{Roman} is more than sufficient to resolve multiple images for typical Einstein radii of order arcsecond, and directly confirm multiply imaged sources without requiring higher-resolution imaging follow-up \cite{Treu2022}.

With the conclusion of the \textit{Planck} CMB observations, future CMB projects such as the Simons Observatory, LiteBIRD and potentially CMB-S4 will target CMB lensing and improved polarisation measurements (in particular the detection of primordial B-mode anisotropies, which would provide evidence for inflation) \cite{PDG2022}.

There are a few notable developing areas of observations we have not yet mentioned, such as gravitational wave and multi-messenger astronomy, {\hl including the increasing interest in the gravitational lensing of gravitational waves}; and the first radio surveys of the redshifted 21-cm line of neutral hydrogen -- which can potentially be measured to redshifts of up to $\sim$20, i.e. currently unexplored epochs of the Universe's history including the cosmic dawn era of the first star-forming galaxies from $z= 14-30$, for which the future Square Kilometre Array (SKA) can provide precision measurements \cite{PDG2022}. There is the interesting potential to use the imprint of acoustic oscillations induced from the relative velocity between dark matter and baryons in the 21-cm power spectrum as a standard ruler at cosmic dawn \cite{Munoz2019}. 
We refer to \cite{Moresco2022}, Table 1 of \cite{Snowmass2021}, as well as Tables 3 and 4 of \cite{Abdalla2022} for a comprehensive list of current and future surveys, experiments and missions.

\chapter{A Geometric Probe of Cosmology -- I. Gravitational Lensing Time Delays and Quasar Reverberation Mapping} 
\label{chap:p1} 
\chaptermark{A Geometric Probe of Cosmology I}


\captionsetup{width=.9\linewidth}
This chapter is published as \fullcite{Ng2020}.
\section{Abstract}
We present a novel, purely geometric probe of cosmology based on measurements of differential time delays between images of strongly lensed quasars due to finite source effects.
Our approach is solely dependent on cosmology via a ratio of angular diameter distances, the image separation, and the source size. It thereby entirely avoids the challenges of lens modelling that conventionally limit time delay cosmography, and instead entails the lensed reverberation mapping of the quasar Broad Line Region.
We demonstrate that differential time delays are measurable with short cadence spectroscopic monitoring of lensed quasars, through the timing of kinematically identified features within the broad emission lines.
This provides a geometric determination of an angular diameter distance ratio complementary to standard probes, and as a result is a potentially powerful new method of constraining cosmology.
%



\section{Introduction}

The time delay between multiple images of strongly gravitationally lensed quasars is a cosmological probe \cite{Refsdal1964, BlandfordNarayan1992, Suyu2014, Treu2013}. Such a time delay is a result of the geometric and gravitational differences in the light paths corresponding to each image. If the lensed source is time variable, such as a quasar, the time delay is detectable via photometric monitoring.

The gravitational lensing time delay is therefore a direct physical measurement of cosmological distances, from which we are able to constrain cosmological parameters. Conventional time delay measurements are most sensitive to the Hubble constant $H_0$, to which it is inversely proportional.
Independent tests of cosmological parameters are especially important given the current tension between determinations of $H_0$ at low redshifts from the cosmic ``distance ladder" and from CMB data. This tension is either due to unknown systematics or new physics \cite{PlanckCollaboration2016, DiValentino2016, Alam2017, Riess2018, Riess2019}. It follows that there is increasing interest in time delay cosmography \cite[e.g.][]{Tewes2013, Courbin2018, Birrer2019a}. One such project is $H_0$ Lenses in COSMOGRAIL's Wellspring \cite[H0LiCOW;][]{Suyu2017}, with a current 3.8 per cent precision measurement of $H_0$ \cite{Bonvin2017} as part of the COSmological MOnitoring of GRAvItational Lenses \cite[COSMOGRAIL;][]{Courbin2004} program.

The main caveat with regards to time delay cosmography is that gravitational lens modelling depends implicitly upon the assumed underlying mass distribution. In particular, there is a systematic problem associated with the ``mass sheet degeneracy" where the degeneracy of ill-constrained lens models leads to uncertainties in estimations of $H_0$ \cite{Falco1985, Saha2000}.

In this paper we present a novel geometric test of cosmology that is \textit{independent} of the lensing potential, by considering \textit{differential} time delays \textit{over} images, originating from spatially-separated photometric signals within a strongly lensed quasar. Measuring these differential time delays, in addition to the standard time delay, will give bounds on cosmological parameters that are essentially based on simple geometry. This is in contrast with conventional methods of determining cosmological parameters, such as steps in the ``distance ladder", which are vulnerable to the details of complicated astrophysics (e.g. supernova explosions and structure formation) and to accumulating systematic errors, as each method is used to calibrate the next \cite{Riess2018}.

The paper is organised as follows: we review conventional gravitational time delay measurements in Section \ref{sec:background} and in Section \ref{sec:timedelaydifferences} we introduce the differential time delays across images. We review reverberation mapping in Section \ref{sec:reverbmapping}, and in Section \ref{sec:lensedrm} combine reverberation mapping with gravitational lensing. Section \ref{sec:spectra} describes a method for measuring differential time delays and we discuss in Section \ref{sec:timescales} the relevant timescales and observational prospects. We outline our conclusions and directions for future work in Section \ref{sec:conclusions}.

\section{Gravitational Lensing} \label{sec:gravlensing}
\subsection{Background} \label{sec:background}
If a gravitational lens produces multiple images of a source, the time required for light to reach an observer will be, in general, different for different paths. The time taken for a given path can be found in the standard cosmological context from the null geodesics of the perturbed Friedmann-Lema\^{i}tre-Robertson-Walker (FLRW) metric.

Consider a photometric signal originating from a point source at a position $\bm{\beta}$ on the sky and at line-of-sight physical distance $l_s$ which, if the source were unlensed, an observer would see at time $t=t_i$. As the source is lensed, an observer sees this signal in a given lensed image X at a position $\bm{\theta}_X$ and at time $t_X = t_i + \tau_X$, where $\tau_X$ is the time delay \cite[e.g.][]{Schneider1992, BlandfordNarayan1992}:
\begin{equation}
    \tau_X \equiv \tau \left(\bm{\theta}_X , \bm{\beta}, l_s\right) = \frac{D}{c} ( 1 + z_d ) \left(\frac{1}{2} ( \bm{\theta}_X - \bm{\beta})^2 - \Psi ( \bm{\theta}_X )\right). \label{tauX}
\end{equation}
Here $D= \frac{D_d D_s}{D_{ds}} \propto \frac{1}{H_0}$ is the ``lensing distance", a ratio of angular diameter distances (subscript $d$, $s$, $ds$ denoting the angular diameter distance to the lens, source and between the lens and source, respectively). The lensing distance is thus the factor containing all the cosmological information. We denote the speed of light by $c$, and the redshift of the lensing mass by $z_d$. The two-dimensional vector positions of image X in the lens plane, and the source in the source plane, are given by $\bm{\theta}_X$ and $\bm{\beta}$ respectively; scaled such that their magnitudes are the observed angular positions relative to the observer-lens axis. The dimensionless ``lensing potential" is denoted by $\Psi$.

The first term in Equation \eqref{tauX} is a geometric component, arising from the difference in path length of a lensed versus unlensed photon; and the second term is a potential term accounting for the gravitational time dilation caused by the lensing mass. We note that the time delay $\tau$ is not an observable quantity, since it is the delay relative to an unlensed photon. We can, however, observe time delays between images, e.g. between images A and B:
\begin{align}
\Delta \tau_{BA} \equiv& \, \tau_B - \tau_A\\
=& \, \frac{D}{c} (1 +z_d) \Big(\frac{1}{2}  \left(\bm{\theta}^2_B - \bm{\theta}^2_A \right) + (\bm{\theta}_A - \bm{\theta}_B) \cdot \bm{\beta} - \Psi(\bm{\theta}_B) + \Psi (\bm{\theta}_A)  \Big). \label{tauBA}
\end{align}
Measuring time delays between images of strongly lensed quasars is a conventional method employed to test cosmology. However, since Equation \eqref{tauBA} is dependent on the  dimensionless lensing potential $\Psi$, time delay measurements are limited by the assumptions and accuracy of the lens model. Simple gravitational lens systems are rare, and observational constraints on the lens model are limited due to the existence, typically, of only two or four images per system \cite{Schneider2013, Birrer2019a}.

Furthermore, \cite{Falco1985} showed that lens models are in fact degenerate. All observables (such as relative image positions and magnification ratios) are invariant, \textit{except for $H_0 \Delta \tau$}, under a family of transformations of the mass profile of the lens along with a translation of the unobservable source position. Any given set of measurements of image positions and fluxes in a lens system therefore can be consistent with a number of different lens models, and therefore a number of different values of $H_0$ whilst preserving the observed time delay $\Delta \tau$. Uncertainties in the estimated values of cosmological parameters from different gravitational lensing systems, and even in separate analyses of the same system can therefore be quite large due to poorly-constrained assumptions made on the mass distribution \cite[for reviews, see e.g.][]{Jackson2015, DeGrijs2011, Schneider2013}.

\subsection{Differential Time Delays} \label{sec:timedelaydifferences}

\begin{figure}
    \centering
    \includegraphics[width=0.6\linewidth]{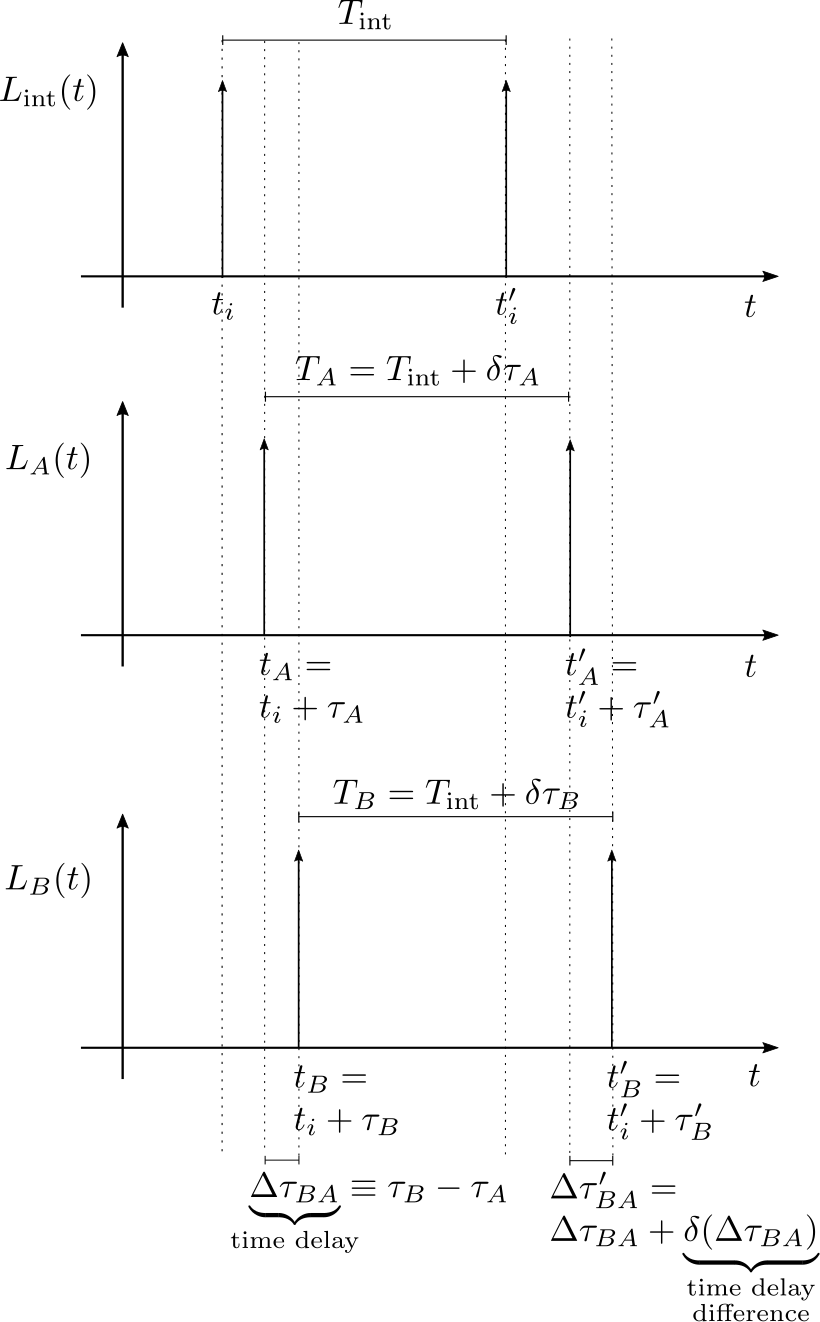}
    \caption{The luminosity of an unlensed source $L_{\mathrm{int}}(t)$ as seen by an observer, and of the corresponding lensed images A and B ($L_A (t)$ and $L_B (t)$ respectively), as a function of time. The signals corresponding to the unprimed times arise from a reference point in the source at $\bm{\beta}$, and the signals corresponding to primed times arise from a spatially perturbed location $\bm{\beta} + \delta \bm{\beta}$. The temporal intervals corresponding to the time delay and the time delay difference are marked.}
    \label{fig:timegraph}
\end{figure}

\begin{figure}
    \centering
    \includegraphics[width=0.65\linewidth]{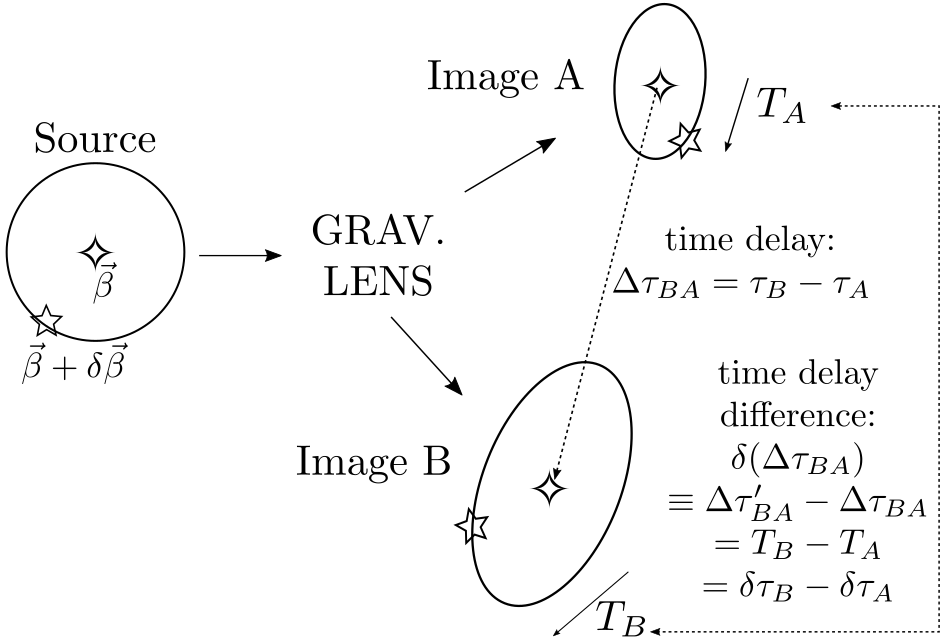}
    \caption{Schematic illustration of two flares, one at a reference point $\bm{\beta}$ in the source, and the other at a spatially perturbed location $\bm{\beta} + \delta \bm{\beta}$ in the source; and their corresponding signals in two lensed images \protect\cite[see][]{Yonehara1999}. The arrows corresponding to $T_A$ and $T_B$ are guides to show $T_X \equiv t_X' - t_X$, as opposed to $t_X - t_X'$.}
    \label{fig:spatialgraph}
\end{figure}

We present a method using differences in differential time delay measurements to probe cosmology. In the following, we show that this approach avoids entirely the systematic lens-modelling problem which usually dominates the error budget.

Let us model a quasar as a collection of point sources, surrounding the central engine. It is convenient to set the central engine as the reference point source at position and line-of-sight physical distance $(\bm{\beta}, l_s)$; we then consider any point source at $(\bm{\beta} + \delta \bm{\beta}, l_s + \delta l_s)$, spatially offset from the central engine. Should an observer measure a hypothetical unlensed signal from the spatially offset point at a time $t=t_i'$, then the lensed signal observed in a given image X would appear at time $t_X' = t_i' + \tau_X'$ and at a position $\bm{\theta}_X + \delta \bm{\theta}_X$. We use the following shorthand notation for the perturbed time delay:
\begin{align}
    {\tau}'_X \equiv& \,\tau(\bm{\theta}_X + \delta \bm{\theta}_X , \bm{\beta} + \delta \bm{\beta}, l_s + \delta l_s)\\
    =& \, \frac{D_d F(l_s + \delta l_s)}{c} ( 1 + z_d) \Big( \frac{1}{2} \left((\bm{\theta}_X + \delta \bm{\theta}_X) - (\bm{\beta} + \delta \bm{\beta})\right)^2 - \Psi(\bm{\theta}_X + \delta \bm{\theta}_X) \Big).
\end{align}
where $F \equiv \frac{D_s}{D_{ds}}$.
We can similarly define a time delay between lensed images for the signal originating from the perturbed source,
\begin{equation}
    \Delta {\tau }'_{BA} \equiv {\tau }'_B - {\tau }'_A.
\end{equation}
The situation is illustrated schematically in Figures \ref{fig:timegraph} and \ref{fig:spatialgraph}. We have further defined $T_{\mathrm{int}} \equiv t_i' - t_i$ as the interval between the hypothetical unlensed signals from the perturbed and reference point sources. From Figure \ref{fig:timegraph} we can see that the interval $T_X$ between signals in a given image X is given by
\begin{equation}
    T_X \equiv t_X' - t_X = {t}'_i - t_i + {\tau }'_X - \tau_X = T_{\mathrm{int}} + \delta \tau_X \label{T_X}
\end{equation}
where we have defined the difference in time delay between the perturbed and reference points \textit{within image X}, the differential time delay, as
\begin{equation}
    \delta \tau_X \equiv {\tau }'_X - \tau_X.
\end{equation}
If the signals are concurrent, and at the same line-of-sight distance within the source, then $T_{\mathrm{int}} =0$ and so $T_X = \delta \tau_X$, i.e. the interval between signals in image X is the differential time delay within image X.

Computing $\delta \tau_X$ to first order in $\delta \bm{\beta}$, $\delta l_s$ and $\delta \bm{\theta}_X$, as shown by \cite{Tie2018}, gives to 1\textsuperscript{st} order
\begin{align}
    \delta \tau_X = &\frac{D}{c}(1+z_d) \Bigg((\bm{\beta} - \bm{\theta}_X)\cdot \delta \bm{\beta} + \frac{\delta l_s}{F} \frac{dF}{dl_s} \left(\frac{1}{2} ( \bm{\theta}_X - \bm{\beta})^2 - \Psi(\bm{\theta_X})\right)\Bigg).
\end{align}
However, the second term corresponding to a displacement of the source in the line of sight of the observer is very small; we can therefore disregard $\delta l_s$ for \textit{lensing} time delays:
\begin{align}
    \delta \tau_X &= \frac{D}{c}(1+z_d)(\bm{\beta} - \bm{\theta}_X)\cdot \delta \bm{\beta} \qquad \text{to 1\textsuperscript{st} order}. \label{DeltaTX}
\end{align}

We furthermore define the \textit{time delay difference} between images A and B as
\begin{align}
    \delta (\Delta \tau_{BA}) &\equiv \Delta {\tau}'_{BA} - \Delta \tau_{BA} = T_B - T_A = \delta \tau_B - \delta \tau_A. \label{timedelaydifferenceequationexact}
\end{align}
We see that it is the difference in the differential time delays of an image pair. Importantly, the time delay difference is independent of $\bm{\beta}$ and $\Psi(\bm{\theta}_X)$ to first order in $\delta \bm{\beta}$ and $\delta \bm{\theta}_X$ \cite{Yonehara1999,Goicoechea2002,Yonehara2003}:
\begin{align}
    \delta (\Delta \tau_{BA}) &= \frac{D}{c}(1+z_d) \left( \bm{\theta}_A - \bm{\theta}_B \right) \cdot \delta \bm{\beta} \qquad \text{to 1\textsuperscript{st} order}. \label{timedelaydifferenceequation}
\end{align}
The time delay difference $\delta (\Delta \tau_{BA})$ is solely determined by cosmology through the lensing distance $D$, the geometry of the lensing configuration and the spatial separation within the source $\delta \bm{\beta}$ in the plane of the sky. The lens redshift and image positions can be measured directly, with the only remaining unknown being $\delta \bm{\beta}$. Remarkably, this expression removes much of the uncertainties associated with lens modelling in using time delay measurements to constrain cosmology. Time delay difference measurements hence appear to be an ideal avenue for testing cosmological models if $\delta \bm{\beta}$ can be determined: we turn to this in the next section.

\section{Broad Line Region Reverberation Mapping} \label{sec:reverbmapping}

The lensing time delay $\tau_X$ measures the time delay between a lensed light ray and its hypothetical unlensed counterpart from the same point source. As we wish to compare arrival times for signals from spatially separated sources, we must include an extra geometric delay arising from the difference in path lengths for the unlensed rays. We also need to know the relative emission times of the signals from the different source positions. The combination of these factors gives the time interval $T_{\mathrm{int}}$ between the signals in the previous section. Furthermore, given Equation \eqref{timedelaydifferenceequation} for the time delay difference, we wish to determine the displacement $\delta \bm{\beta}$ within the source. Although quasars remain spatially unresolved even with the best telescopes and at the lowest redshifts, we are still able to determine the quasar structure.

The spectra of quasars have characteristic broad emission lines corresponding to gas clouds, known as the Broad Line Region (BLR), surrounding the central emitting accretion disk at some distance. Photons travelling outwards from the central source are absorbed and re-emitted by the BLR gas. The broad emission lines therefore respond to variations in the continuum luminosity of the central source with a time delay determined by the BLR geometry.

Using this measured time delay to deduce the geometry of the BLR is an established technique known as reverberation mapping \cite{Blandford1982, Peterson1993, Shen2015, Dexter2019}. Since the BLR response time is set by the speed of light, the BLR reverberation sets an absolute distance scale, ideal for determining $\delta \bm{\beta}$. As the distance scale of the BLR is much greater than the accretion disk, the resultant differential time delay effects are substantially larger than those considered in previous literature \cite[e.g.][]{Yonehara1999, Goicoechea2002, Tie2018}.

For a particular BLR cloud located at a spatial position $\bm{r}$ relative to the central source, the associated time delay relative to the central source is determined by the reverberation mapping constraint equation:
\begin{equation}
\tau_{\mathrm{BLR}} \equiv (1+z_s) \tilde{\tau}_{\mathrm{BLR}} = \frac{(1+z_s)}{c} \left( |\bm{r}| - \bm{r} \cdot \hat{n}  \right) \label{tauBLR}
\end{equation}
where $\hat{n}$ is the unit vector towards the observer. The first term is the time taken for a photon to travel from the central source to the BLR cloud, i.e. the difference in emission times. The second term is the extra time for a photon to travel from a given BLR particle to the observer, as compared with a photon from the central source to the observer. The cosmological redshift factor $(1+z_s)$ corrects the time delay in the quasar rest frame, $\tilde{\tau}_{\mathrm{BLR}}$, for the time delay as measured in our observer frame, $\tau_{\mathrm{BLR}}$. The time interval between an unlensed signal from a point source in the BLR at $(\bm{\beta} + \delta \bm{\beta}, l_s + \delta l_s)$ compared with an unlensed signal from the central engine at $(\bm{\beta}, l_s)$ is therefore $T_{\mathrm{int}} = t_i' - t_i =\tau_{\mathrm{BLR}}$.

The BLR may be approximated by a flat geometry in Keplerian orbit around a central black hole, e.g. \cite{Grier2017}. As a toy example of reverberation mapping, we therefore consider an infinitesimally thin BLR ring, at the same redshift as the central source, facing the observer plane with a constant linear number density. The observed luminosity of the BLR, proportional to the number of responding particles $N$ at a given time, is
\begin{equation}
    L_{\mathrm{ring}}(\tau_{\mathrm{BLR}}) \propto \frac{dN}{d \tilde{\tau}_{\mathrm{BLR}}} \propto  \delta \left(\tilde{\tau}_{\mathrm{BLR}} - \frac{R}{c} \right),  \label{Psithinring}
\end{equation}
i.e. a Dirac delta function at a single time delay value $\tilde{\tau}_{\mathrm{BLR}} = \frac{R}{c}$, where $R$ is the radial distance of the thin ring from the central engine. We recognise intuitively this time delay as the light travel time between the central source and any cloud on the BLR ring, in accordance with the constraint Equation \eqref{tauBLR}. In this simplest form of reverberation mapping, as is often used for estimating black hole masses, BLR time delay measurements directly give an estimate for an average BLR radius \cite[e.g.][]{Bentz2015, Kaspi2017, Shapovalova2017}. This is in contrast with more sophisticated geometric and kinematic models of the BLR, \cite[e.g.][]{Sturm2018, Mangham2019}, which we will explore in future work.

\section{Gravitationally Lensed Broad Line Region Reverberation Mapping} \label{sec:lensedrm}

We continue to consider the toy model for the BLR outlined in the previous section. For a lensed quasar, we observe  multiple images where the BLR ring is distorted into ellipses (see Figure~\ref{fig:lensingsetup}). However, due to the effect of the differential time delay, only part of the ellipse emits the flux visible in a given moment of time. Each point within the lensed quasar images responds to the continuum emission in the source with a total time delay equal to the time delay from the BLR geometry $\tau_{\mathrm{BLR}}$ plus a time delay $\tau_X$ from lensing.

Recall that for an unlensed quasar, there is a Dirac delta signal from the central source at $t=t_i$ across all wavelengths (i.e. the continuum emission). This is followed by a Dirac delta signal from the entire BLR ring at $t= t_i' = t_i + \tau_{\mathrm{BLR}}$ in the broadened emission lines. The time that we see the continuum signal in image X is given by
\begin{equation}
    t_X = t_i + \tau_X. \label{timeXundashed}
\end{equation}
The time that we see a signal originating from a perturbed source position $(\bm{\beta} + \delta \beta, l_s + \delta l_s)$ in image X is similarly given by
\begin{equation}
    t_X' = t_i' + \tau_X' = t_i + \tau_{\mathrm{BLR}} + \tau_X + \delta \tau_X \label{timeXdashed}
\end{equation}
using $\tau_X' = \tau_X + \delta \tau_X$, as we note $\tau_X$ is constant in $\delta \bm{\beta}$ (as opposed to $\tau_X'$).

\begin{figure}
    \centering
    \includegraphics[width=0.7\linewidth]{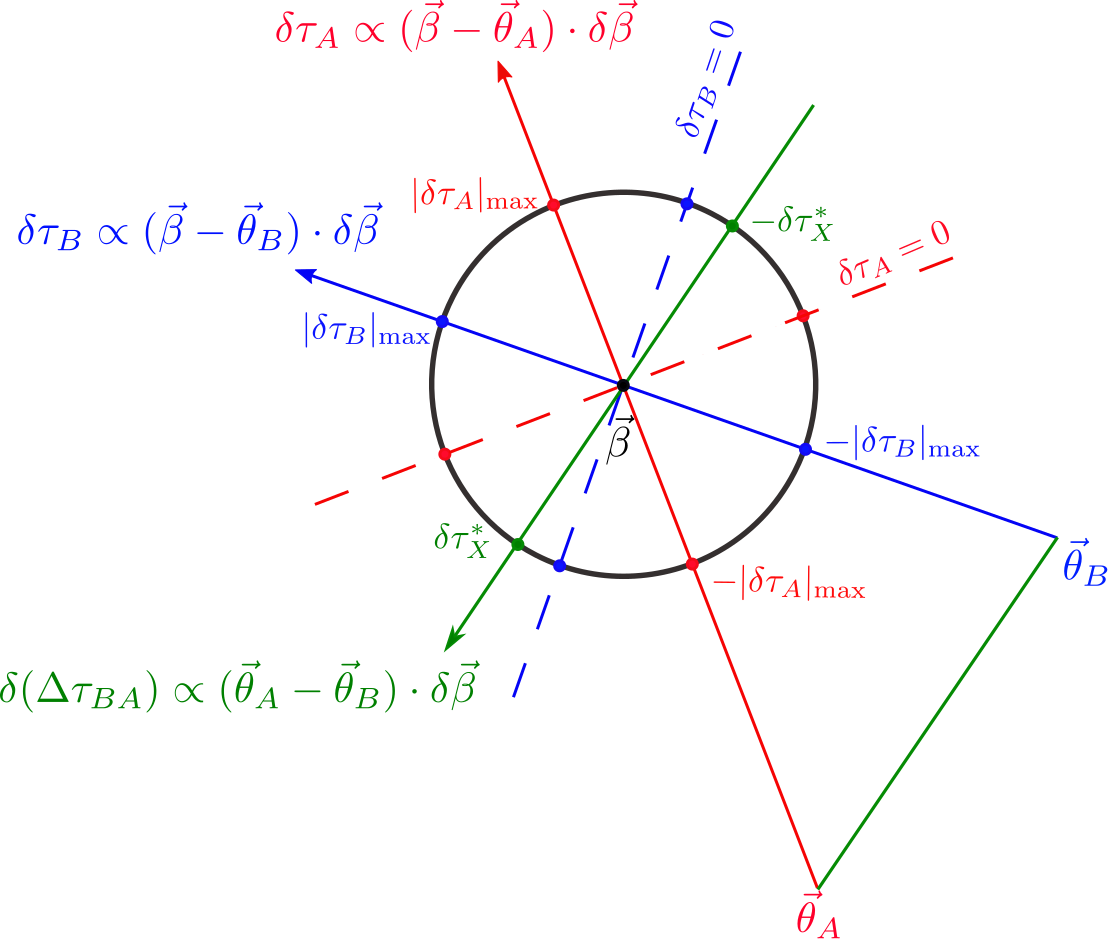}
    \caption{A representation of the source centred at $\bm{\beta}$ surrounded by the thin ring BLR (black circle), not to scale with the labelled image positions $\bm{\theta}_A$ and $\bm{\theta}_B$. Lines in the $\bm{\beta} - \bm{\theta}_A$ direction (red), $\bm{\beta} - \bm{\theta}_B$ direction (blue), and $\bm{\theta}_A - \bm{\theta}_B$ direction (green) through the source are labelled. The points in the source that lie furthest in the $\pm (\bm{\beta} - \bm{\theta}_X)$ direction map to points in image X with the maximal and minimal time delay respectively, $\delta \tau_X = \pm |\delta \tau_X |_\mathrm{max}$, compared to the source centre. The total differential time delay across image X is therefore $2|\delta \tau_X|_\mathrm{max}$. The linear dependence of $\delta \tau_X$ upon the component of the source position in the $\bm{\beta} - \bm{\theta}_X$ direction results in the arcsine distribution of the flux. Similarly, the points in the source that lie furthest in the $\pm (\bm{\theta}_A - \bm{\theta}_B)$ direction give the maximal and minimal time delay differences, $\pm |\delta(\Delta \tau_{BA})|_{\mathrm{max}}$, relative to the source centre. These points, when mapped to image X, have a differential time delay of $\pm \delta \tau_X^*$ respectively, as given by Equation \protect\eqref{deltatauasterisk}. The actual maximal measurable time delay difference is therefore $2|\delta(\Delta \tau_{BA})|_{\mathrm{max}} = 2 (\delta \tau_B^* - \delta \tau_A^*)$, arising from the difference in differential time delays from across the entire BLR diameter.}
    \label{fig:source}
\end{figure}

The variability in the BLR emission lines for a thin face-on ring geometry is a function of only $\delta \tau_X$ since $t_i$ and $\tau_X$ are fixed; and $\tau_{\mathrm{BLR}}$ is constant. We choose coordinates (for a particular image X) such that $\bm{\beta} - \bm{\theta}_X$ points in the $\hat{y}$ direction. From Equation \eqref{DeltaTX}, we have that $\delta \tau_X$ depends linearly on the $y$ component of $\delta \bm{\beta}$. As a result of this spatial dependence, only part of the BLR image emits the flux visible in a given moment of time; see Figures \ref{fig:source} and \ref{fig:lensingsetup}. 

To find the luminosity $L_X(t)$ in image X at time $t$ for a signal originating from an arbitrary location on the BLR ring, we need only find $L_X(\delta \tau_X)$. Let $\delta \bm{\beta} = (x,y)$ and using $\frac{dN}{dy} = \frac{2N}{2\pi|\delta \bm{\beta}|} \frac{|\delta \bm{\beta}|}{\sqrt{|\delta \bm{\beta}|^2 - y^2}}$ we have then that
\begin{align}
\begin{split}
    L_X(t) \propto \frac{dN}{d (\delta \tau_X)} &= \frac{1}{\frac{D}{c}(1+z_d) | \bm{\beta} - \bm{\theta}_X|} \frac{dN}{dy}\\
    &= \frac{N}{\pi \sqrt{|\delta \tau_X|_{\mathrm{max}}^2 - |\delta \tau_X|^2}} , \label{lensingthinringfunction}
\end{split}
\end{align}
where $|\delta \tau_X|_{\mathrm{max}} = \frac{D}{c}(1+z_d) | \bm{\beta} - \bm{\theta}_X| |\delta \bm{\beta}| $. \textit{Lensed BLR signals are therefore widened temporally into an arcsine distribution at each instant.}

\begin{figure}
    \centering
    \includegraphics[width=0.6\linewidth]{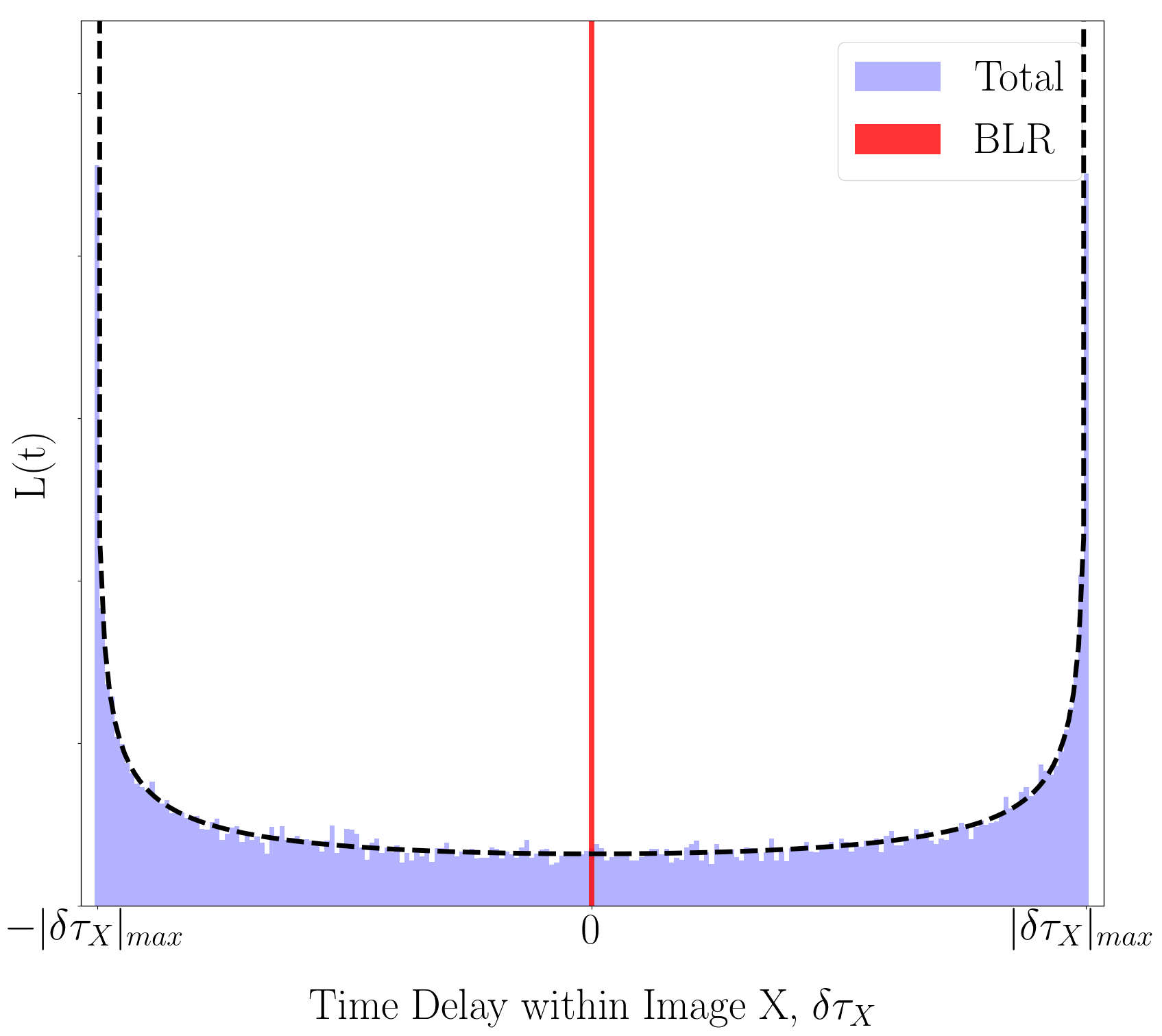}
    \caption{The effect of the differential time delay over an image on the reverberation mapping signal for the toy model considered in this paper. The unlensed view of the flux (red) is a Dirac delta function, as the entire BLR is seen to reverberate at the same time; this is distorted by differential lensing into the extended signal (blue). 
    The width of this signal corresponds to $2|\delta \tau_X|_{\mathrm{max}}$. The dashed line shows the prediction from Equation \protect\eqref{lensingthinringfunction}, an arcsine distribution.
    }
    \label{fig:luminosity_im0_sis}
\end{figure}

The situation is illustrated in Figure \ref{fig:luminosity_im0_sis}. In an image X, we see a Dirac delta function lensed continuum signal at time $t_X = t_i + \tau_X$, whereas the BLR signal in the emission lines is temporally widened into an arcsine distribution due to the spatially-dependent lensing time delay. The point on the ellipse that responds soonest does so at time $t_i + \tau_{\mathrm{BLR}} + \tau_X - |\delta \tau_X |_{\mathrm{max}}$ and the point on the ellipse that responds with the greatest time delay does so at $t_i + \tau_{\mathrm{BLR}} + \tau_X + |\delta \tau_X |_{\mathrm{max}}$. The arcsine distribution is therefore temporally centred at $\bar{t}_X' = t_i + \tau_{\mathrm{BLR}} + \tau_X$, with a width of $2|\delta \tau_X |_{\mathrm{max}}$.

We can measure $\tau_{\mathrm{BLR}}$, and hence $R$, by subtracting the time we see the lensed continuum signal in that image from the time at which the lensed BLR signal is centred: 
\begin{equation}
    |\bar{t}_X' - t_X |= |\tau_{\mathrm{BLR}}| = (1+z_s)\frac{R}{c}.
\end{equation}
We may thereby use reverberation mapping time delays to determine the source size
\begin{equation}
    |\delta \bm{\beta}|=\frac{R}{D_s} = \frac{c |\tau_{\mathrm{BLR}}|}{(1+z_s)D_s}
\end{equation}
of lensed quasars, and we obtain a separate measurement of $R$ for each image. Recalling Equation \eqref{timedelaydifferenceequation}, we have
\begin{equation}
   |\delta(\Delta \tau_{BA})|_{\mathrm{max}} = \frac{D_d}{D_{ds}} \frac{(1+z_d)}{(1+z_s)} |\bm{\theta}_A - \bm{\theta}_B | |\tau_{\mathrm{BLR}}|. \label{finaltimedelaydifferenceequation}
\end{equation}
The redshifts, the image separation $|\bm{\theta}_A - \bm{\theta}_B|$, and the BLR time delay $\tau_{\mathrm{BLR}}$ are measured quantities. Once the time delay difference $|\delta(\Delta \tau_{BA})|_{\mathrm{max}}$ is measured, the ratio of angular diameter distances $\frac{D_d}{D_{ds}}$ may be constrained.

We note that na{\"i}vely subtracting the widths $2|\delta \tau_X |_{\mathrm{max}}$ of lensed signals from two different images does not give a measurement of the time delay difference, including not of the maximum time delay difference: $|\delta(\Delta \tau_{BA})|_{\mathrm{max}} \neq |\delta \tau_B |_{\mathrm{max}} - |\delta \tau_A |_{\mathrm{max}}$. Rather, $|\delta (\Delta \tau_{BA})|_{\mathrm{max}} = \delta \tau_B^* - \delta \tau_A^*$ where $\delta \tau_X^*$ is given by
\begin{equation}
    \delta \tau_X^* = \frac{D}{c} (1+ z_d) \frac{(\bm{\beta} - \bm{\theta}_X) \cdot (\bm{\theta}_A - \bm{\theta}_B)}{|\bm{\theta}_A - \bm{\theta}_B|} |\delta \bm{\beta}| \label{deltatauasterisk}
\end{equation}
as illustrated in Figure \ref{fig:source}.

For more complicated BLR geometries we will need to distinguish the lensing effects from the response function of reverberation mapping. In general, the lensed luminosity function for an arbitrary geometry will be the convolution of the unlensed luminosity function for that geometry with the arcsine response corresponding to a face-on thin ring. Finding the appropriate timescale from the observed luminosity function corresponds to performing an inversion of this convolution, which may be a nontrivial task.

\begin{figure}
    \centering
    \includegraphics[width=0.7\linewidth]{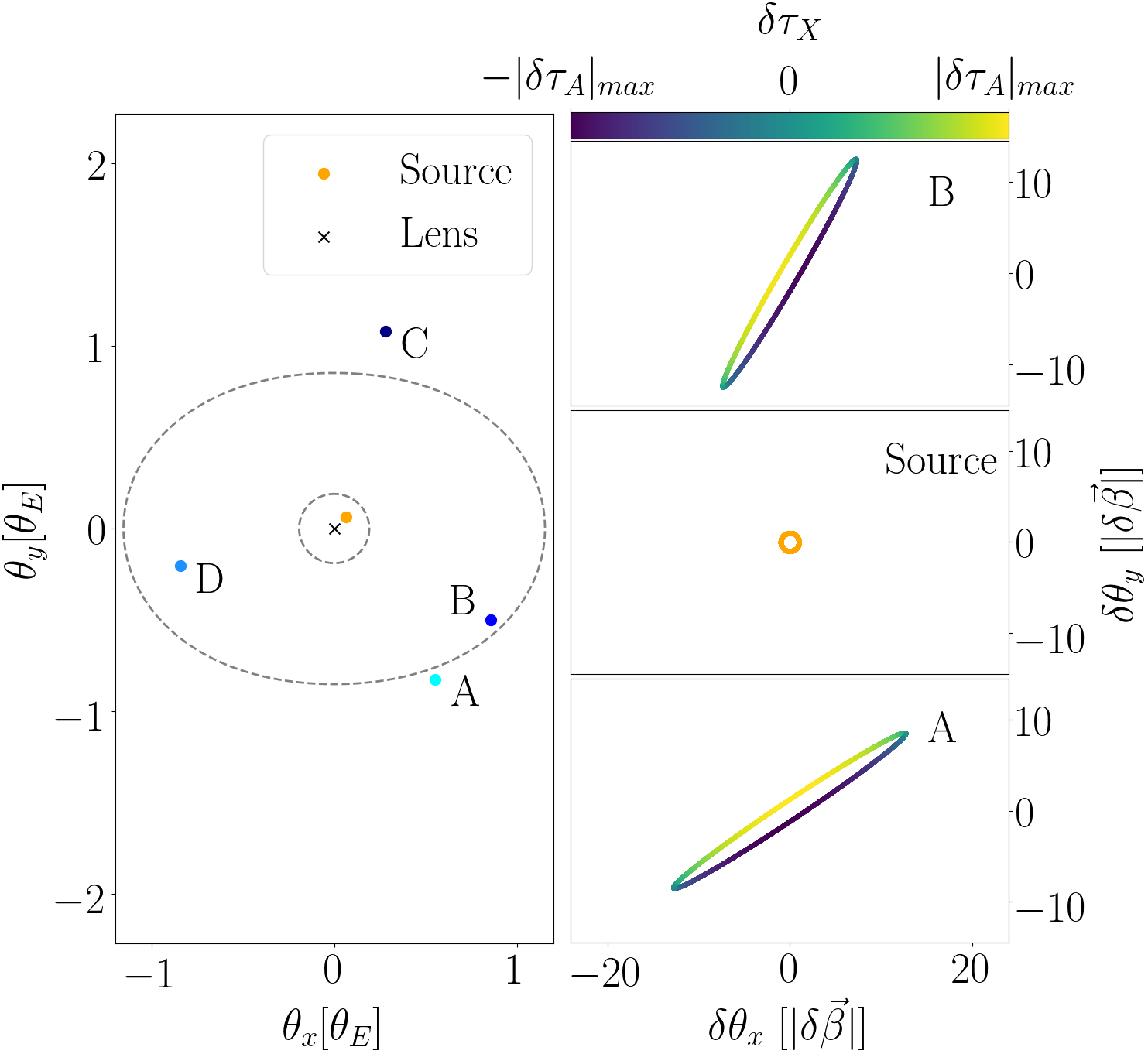}
    \caption{A typical configuration for a lensed quasar, showing the thin ring BLR structure of the quasar and its four images. The critical lines are marked by dashed grey lines. Image positions were calculated using a softened elliptical lens model with a core radius of $0.1 \theta_E$ and ellipticity of $0.1$, setting source at $\bm{\beta} = (0.065, 0.065) \theta_E$, where $\theta_E$ denotes the deflection scale of the lens. The differential time delays within images A and B (relative to the source centre) are shown; their respective maximum values are similar, since $|\bm{\beta} - \bm{\theta}_A| = 1.02 \theta_E \sim |\bm{\beta} - \bm{\theta}_B| = 0.97 \theta_E$.}\label{fig:lensingsetup}
\end{figure}

\section{A Measurement Method Using Lensed Quasar Spectra} \label{sec:spectra}

The spectra of lensed quasars contain more information than a measurement of the flux, as different spatial regions in the quasar are distinguishable through their velocities and hence the measured wavelengths. This additional information enables us to directly measure the differential time delays required to provide cosmological constraints. We leave the full kinematic signature, comprised of multiple spatial and velocity components at each instant, for a future contribution.

As a simplified example, we consider ``cloud 1'' at $\bm{r}_1$ and ``cloud 2'' at $\bm{r}_2$ within an arbitrary BLR geometry, possessing different velocities $\bm{v}_1$ and $\bm{v}_2$ relative to the central source. For a particular Doppler broadened emission line, we see an increased luminosity in the spectra of image A, at a wavelength corresponding to cloud 1, at time $t_{A,1}'$. Similarly, at time $t_{B,1}'$ we see a brightening in the spectra of image B at the same wavelength corresponding to cloud 1. At a time $t_{A,2}'$ we see an increased luminosity in the spectra of image A at a wavelength corresponding to cloud 2; this occurs in image B at a time $t_{B,2}'$.

Recalling $t_X'$ as given by Equation \eqref{timeXdashed}, we have
\begin{equation}
\begin{split}
    t_{X,1}' &= t_i + \tau_{\mathrm{BLR},1} + \tau_X + \delta \tau_{X,1}\\
    t_{X,2}' &= t_i + \tau_{\mathrm{BLR},2} + \tau_X + \delta \tau_{X,2}.
    \end{split}
\end{equation}
Since we are able to measure $t_{X,1}'$ and  $t_{X,2}'$ directly from the lensed spectra of each image, we are able to measure the differential time delays within each image
\begin{equation}
    t_{X,2}' - t_{X,1}' = (\tau_{\mathrm{BLR},2} - \tau_{\mathrm{BLR},1}) + (\delta \tau_{X,2} - \delta \tau_{X,1})
\end{equation}
and take the difference in the differential time delays between the two images
\begin{equation}
    (t_{B,2}' - t_{B,1}') - (t_{A,2}' - t_{A,1}') = (\delta \tau_{B,2} - \delta \tau_{B,1}) - (\delta \tau_{A,2} - \delta \tau_{A,1}).
\end{equation}

This is the time delay difference between clouds 1 and 2, and it will be measured to be a maximal value for a particular wavelength pair. These wavelengths correspond to clouds 1 and 2 located on opposite sides of the BLR in the direction of the image-image axis on the sky (see Figure~\ref{fig:source}), such that:
\begin{equation}
\begin{split}
    (\delta \tau_{B,2} - \delta \tau_{B,1}) - (\delta \tau_{A,2} - \delta \tau_{A,1}) & \underset{\mathrm{max}}{=} 2 \delta \tau_B^* - 2 \delta \tau_A^*\\
    &= 2 |\delta ( \Delta \tau_{BA} )|_{\mathrm{max}}
\end{split}
\end{equation}
where $\delta \tau_X^*$ is given by Equation \eqref{deltatauasterisk}.

Having obtained a measurement of the time delay difference, we have direct observational measurements of all components of Equation \eqref{finaltimedelaydifferenceequation} with the exception of the ratio of the angular diameter distances $\frac{D_d}{D_{ds}}$. This is a determination of cosmological distance ratios that is independent of astrophysical modelling.

\begin{figure}
    \centering
    \includegraphics[width=0.7\linewidth]{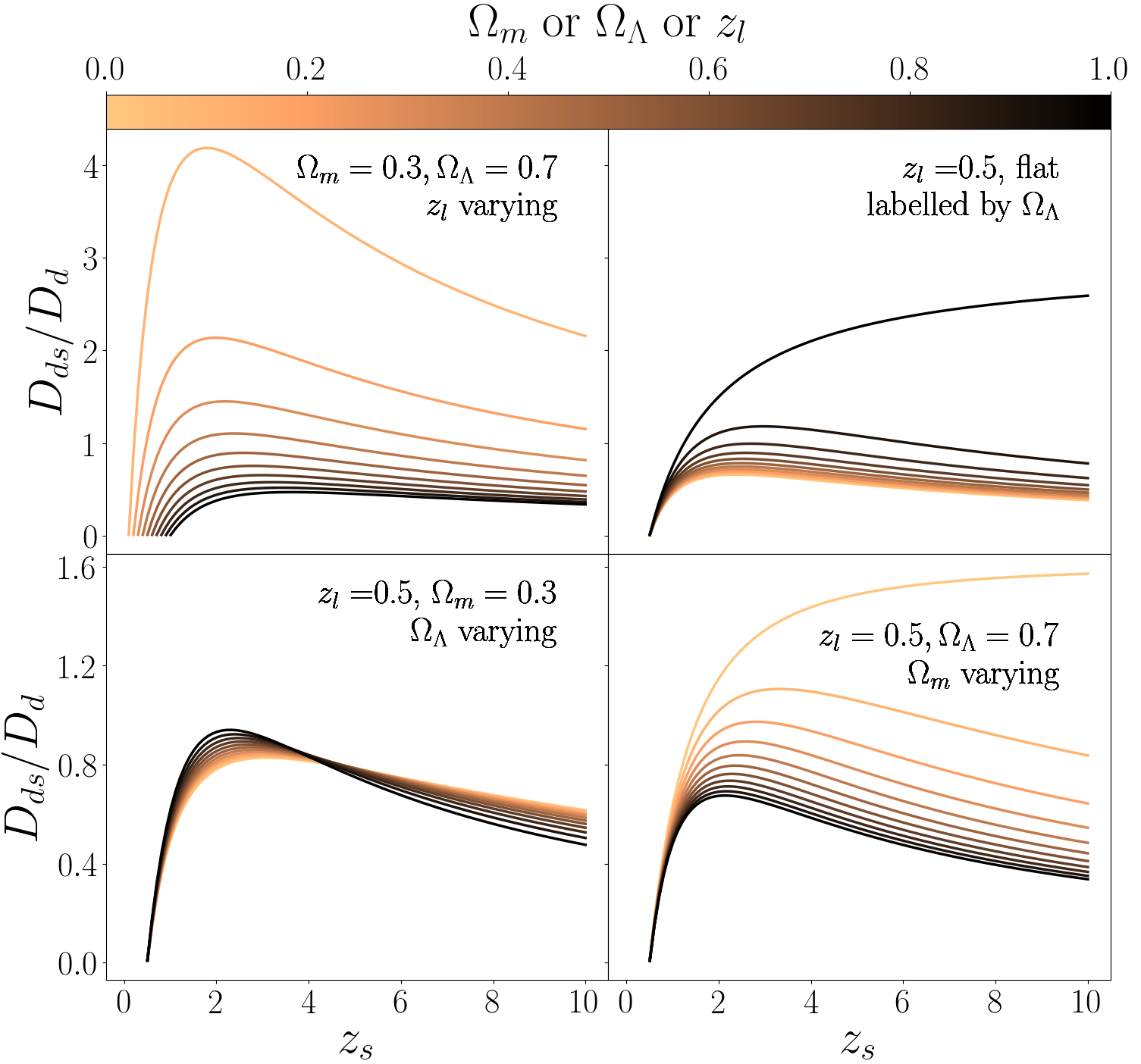}
    \caption{An illustration of the dependence of the ratio of the angular diameter distances $\frac{D_{ds}}{D_{d}}$ on the dark energy density and matter density parameters $\Omega_{\Lambda}$ and $\Omega_m$ assuming $\Lambda$CDM cosmologies. In top left plot we vary $z_l$ in equal intervals from $0.1$ to $1$ for a standard $\Lambda$CDM cosmology; in the top right plot we show the dependence of $\frac{D_{ds}}{D_{d}}$ on the dark energy density $\Omega_{\Lambda}$ (equivalently $\Omega_m$) under assumption of spatial flatness, $\Omega_m + \Omega_{\Lambda} =1$. In the bottom row, we allow curvature to vary from flat. For the top right plot and bottom row, lines are plotted in equal intervals from 0 to 1 in  $\Omega_M$ and $\Omega_{\Lambda}$.}
    \label{fig:paramdependence}
\end{figure}

The measurement of the differential time delay across images is a measure of a dimensionless ratio of angular diameter distances and thus insensitive to $H_0$; it is therefore complementary to probes of cosmology such as the standard time delay \cite[see][for a discussion on the dependencies of other distance ratios and the resulting degeneracies]{Linder2004, Linder2011}. In Figure~\ref{fig:paramdependence} we explore the sensitivity of this dimensionless ratio on various cosmological parameters, revealing that $\frac{D_{ds}}{D_{d}}$ is more sensitive to $\Omega_m$ than to $\Omega_{\Lambda}$, with a strong cosmological lever based upon the source redshift.

We note that by measuring both the conventional time delay between images $\Delta \tau_{BA} \propto \frac{D_d D_s}{D_{ds}}$ and the time delay difference $\delta( \Delta \tau_{BA} ) \propto \frac{D_d}{D_{ds}}$, we may obtain $\frac{\Delta \tau_{BA}}{\delta( \Delta \tau_{BA} )} \propto D_s$, i.e. the angular diameter distance to the source.

\section{Timescales \& Observability} \label{sec:timescales}
Typical BLR radii range from a few light days to a few hundred light days, as estimated from reverberation mapping \cite{Chelouche2014, Bentz2015, Kaspi2017, Shapovalova2017, Lira2018, Sturm2018}, corresponding to an angular extent on the sky on the order of \SIrange[range-units = single]{e-11}{e-10}{\radian}. For a $\Lambda$CDM cosmology considering typical lens redshifts of $0.3-1$ and source redshifts $1-4$ (and potentially up to $z\sim7$), we expect that $\frac{D}{c}(1+z_d)$ is on the order of \SIrange[range-units = single]{e17}{e18}{\second}. Recalling Equation \eqref{DeltaTX} and
Equation \eqref{timedelaydifferenceequation}, we have
\begin{align}
    |\delta \tau_X|_{\mathrm{max}} &\sim |(\bm{\beta} - \bm{\theta}_X)| \times (\num{e6} - \num{e8}) \si{\second}\\
    |\delta (\Delta \tau_{BA})|_{\mathrm{max}} &\sim  |( \bm{\theta}_A - \bm{\theta}_B)| \times (\num{e6} - \num{e8}) \si{\second} \label{orderofmagtimescales}
\end{align}
which sets the timescale for the differential time delay within the image and the timescale for the time delay difference between images. The minimum resolution at which we need to sample for the differential time delay is set by $2\delta \tau_X^*$ which will be some fraction of $2|\delta \tau_X|_{\mathrm{max}}$.

Considering cluster-scale lensing, with an angular deflection scale $\theta_E \sim 10 - 100''$, the largest image separations are twice the deflection scale, $\sim 20 - 200'' \sim$ \num{e-4}$ - $\SI{e-3}{\radian}. This gives a differential time delay within the images $\delta \tau$, as well as a time delay difference $\delta (\Delta \tau)$, on the order of hours to days. On a galaxy scale, the image separations are $\sim 2'' \sim$ \SI{e-5}{\radian}, giving a differential time delay and time delay difference on the order of minutes.

As an example, consider a high redshift quasar at $z_s = 3$ with a $200$ light day BLR radius, lensed by a mass at $z_l \sim 1$, with two images C and D separated by $\sim 1.7 \theta_E$ as illustrated in Figure~\ref{fig:lensingsetup}. For a galaxy scale lens, the differential time delay across the image is $2 |\delta \tau_X|_{\mathrm{max}} \sim  2 \delta \tau_X^* \sim$ \SI{12}{\min}, with a time delay $\Delta \tau_{DC}$ between images on the order of a month. For small clusters, we have $2|\delta \tau_X|_{\mathrm{max}} \sim 2\delta \tau_X^* \sim$ \SI{2}{\hour}; whilst the time delay $\Delta \tau_{DC}$ is on the order of a decade. For a very large cluster lens, the differential time delay within each image is $2|\delta \tau_X|_{\mathrm{max}} \sim  2 \delta \tau_X^* \sim$ \SI{20}{\hour}. However, the time delay $\Delta \tau_{DC}$ is on the order of hundreds to thousands of years, rendering measurements unattainable.

The most promising scenarios for measuring the standard time delays between images $\Delta \tau$ as well as differential time delays $\delta \tau$ are those for which the $\delta \tau$ timescale is still significant whilst reducing the $\Delta \tau$ timescale. In addition to using quasars with emission lines corresponding to clouds at larger radii, we may look for elliptical lenses where image separations can approach zero as the source crosses caustics. Such configurations can result in reasonable timescales for both differential time delays and time delays between images. For example, the quintuply cluster-lensed quasar SDSS J1004+4112 has a close image pair separated by only $3.8''$, such that the time delay between the two images is on the order of 40 days \cite{Fohlmeister2007, Fohlmeister2008}, and the same order of magnitude time delays between close image pairs occur in the sextuply cluster-lensed quasar SDSS J2222+2745 \cite{Dahle2015, Sharon2017}.

To demonstrate using the same example configuration, consider images A and B, with a much smaller separation $\sim 0.4 \theta_E$. For a galaxy scale lens, the differential time delay is $2|\delta \tau_X|_{\mathrm{max}} \sim$ \SI{20}{\min}, with $2 \delta \tau_A^* \sim$ \SI{4}{\min}, $2 \delta \tau_B^* \sim$ \SI{2}{\min}, and the time delay between images is $\Delta \tau_{BA} \sim$ \SI{10}{\hour}. For a small cluster, we have $2|\delta \tau_X|_{\mathrm{max}} \sim$ \SI{2}{\hour}, $2 \delta \tau_A^* \sim$ \SI{40}{\min}, $2 \delta \tau_B^* \sim$ \SI{15}{\min}, and $\Delta \tau_{BA}$ on the order of a month. On very large cluster scales we see a differential time delay on the order of $2|\delta \tau_X|_{\mathrm{max}} \sim$ \SI{20}{\hour}, with $2 \delta \tau_A^* \sim$ \SI{7}{\hour}, $2 \delta \tau_B^* \sim$ \SI{3}{\hour}; whereas the time delay is on the scale of a decade.

The measurement of these time delays within images and time delay differences between images are therefore reliant on the availability of light curves and spectra sampled with a frequency on the order of minutes. Long-term optical variability surveys such as The Optical Gravitational Lensing Experiment \cite[OGLE;][]{Udalski2015} and the Panoramic Survey Telescope and Rapid Response System \cite[Pan-STARRS;][]{Chambers2016} are ideal for measuring time delays in already-monitored lensed quasars, in addition to the dedicated COSMOGRAIL program \cite{Courbin2004}. Recent discoveries of lensed quasars using these variable sky surveys, such as \cite{Kostrzewa-Rutkowska2018} and in particular multiply imaged quasars such as \cite{Berghea2017} are very promising for time delay measurement endeavours. Spectroscopic data needed for time delay difference measurements is obtainable through the Time-Domain Spectroscopic Survey \cite[TDSS;][]{Ruan2016} for photometrically variable targets, including quasars, selected from e.g. Pan-STARRS and archival Sloan Digital Sky Survey imaging. 

The limiting factor of variable sky surveys is the relatively poor time sampling currently available. Quasars behind the Magellanic Clouds detected with OGLE are likely to remain amongst the most densely sampled (\num{1}$-$\SI{3}{\day}) long-term light curves available until the advent of the Large Synoptic Survey Telescope \cite{Kozowski2013}. The main survey of the LSST \cite{LSST2019} will monitor the transient optical sky by tiling the sky with images of approximately ten-square-degrees, with two ``visits" (a pair of 15 second exposures per visit) on a given night, separated by 15-60 minutes \cite{LSST2009}. \cite{Oguri2010} discusses prospects for detecting strongly lensed time-variable sources in current and future time-domain optical imaging surveys; the number is estimated to be $\sim 3000$ for the LSST. The LSST project will furthermore include one hour of intensive observation per night of a set of ``Deep Drilling Fields" (DDFs) which will provide more frequent temporal sampling than the main survey. 50 consecutive 15-second exposures could be obtained in each of four filters in an hour, providing light curves of objects on hour-long timescales, which would be ideal for measuring the difference in time delay within images of lensed quasars; exactly how the LSST observations will be taken and the details of these intensive observations are not yet finalised \cite{LSST2017}.

\section{Conclusions and Future Work} \label{sec:conclusions}

Measurements of differential time delays across the gravitationally lensed images of reverberating BLR quasars offers a new and purely geometric test of cosmology. Such tests are highly relevant in an era when independent determinations of cosmological parameters are crucial to resolving tensions and bypassing systematic challenges associated with conventional methods. The method we have presented of measuring the time delay difference averts the degeneracies and difficulties inherent in lens modelling that typically affect time delay cosmography.

The time delay difference is determined by cosmology via a ratio of angular diameter distances, the image separation on the sky, and the spatial separation within the source. We have shown that gravitationally lensed reverberation-mapped quasars may be used as a means of constraining the source size. The BLR radius of the quasar may be measured, in the simplified case of our toy BLR model, directly from the difference in observed time between the lensed continuum signal and the centre of lensed BLR signal.

Furthermore, the spectra of the lensed quasar may be used to measure the time delay difference, as it carries additional information, the velocity and hence the spatial origin of the BLR cloud, via the wavelength of each signal. The result is that the ratio of angular diameter distances, $\frac{D_d}{D_{ds}}$, may be determined. This is a dimensionless ratio that depends most strongly on the matter density parameter, as well as the dark energy density parameter, and its determination will provide constraints on cosmology complementary to those from standard time delay measurements.

A thorough analysis of the impact of microlensing upon the signatures of the differential time delay will be conducted in a future paper. In particular, we expect that microlensing would impact observations of the continuum and line emissions differently, since the central source is smaller than the BLR. However, microlensing effects may be distinguished from the intrinsic quasar variability since variations from microlensing will be uncorrelated in different quasar images, whereas intrinsic variations appear in both images at different times. Furthermore, the chromatic effects of microlensing will be reduced by considering sources with larger radii (compared with the scale of the Einstein radius of typical microlenses in the source plane). Finally, we may choose lensing systems where the optical depth to microlensing is small, such as systems where images appear in the outer regions of the lensing galaxy.

We will also investigate in future work the sensitivity of the ratio $\frac{D_d}{D_{ds}}$ to cosmological parameters. We will include image configurations from realistic lenses and consider the measurement of time delay differences from highly-magnified images resulting from the source passing through a caustic. We will also consider realistic BLR models possessing a velocity function and include reverberation signals passing through geometric features in the BLR; and consider the effect of the amplitude of quasar variability on the relevant time scales.

\section*{Acknowledgements}
A.L.H.N. acknowledges funding from the Australian Government and The University of Sydney through a Research Training Program scholarship and a Hunstead Merit Award respectively.



 

\chapter{A Geometric Probe of Cosmology -- II. Gravitational Lensing Time Delays and Quasar Reverberation Mapping Revisited} 
\label{chap:p2} 
\chaptermark{A Geometric Probe of Cosmology II}

\captionsetup{width=.9\linewidth}
\newcommand{\appropto}{\mathrel{\vcenter{
  \offinterlineskip\halign{\hfil$##$\cr
    \propto\cr\noalign{\kern2pt}\sim\cr\noalign{\kern-2pt}}}}}

This chapter is published as \fullcite{Ng2023}.

\section{Abstract}
The time delay between images of strongly gravitationally lensed quasars is an established cosmological probe. Its limitations, however, include uncertainties in the assumed mass distribution of the lens. We re-examine the methodology of a prior work presenting a geometric probe of cosmology independent of the lensing potential which considers differential time delays over images, originating from spatially-separated photometric signals within a strongly lensed quasar. We give an analytic description of the effect of the differential lensing on the emission line spectral flux for axisymmetric Broad Line Region geometries, with the inclined ring or disk, spherical shell, and double cone as examples. The proposed method is unable to recover cosmological information as the observed time delay and inferred line-of-sight velocity do not uniquely map to the three-dimensional position within the source.

\section{Introduction}

 The time delay due to geometric and gravitational differences in the light paths between multiple images of strongly gravitationally lensed quasars is a direct measurement of cosmological distances, and therefore of cosmological parameters \cite{Refsdal1964, BlandfordNarayan1992}. Time delay cosmography is now an established independent method by which cosmological information can be determined, e.g. \cite{Tewes2013, Courbin2018, Birrer2019a}, and \cite[H0LiCOW;][]{Suyu2017} with a current 2.4 per cent precision measurement of $H_0$ \cite{Wong2020}. However, gravitational lens modelling depends implicitly upon the assumed underlying mass distribution. In particular, the ``mass sheet degeneracy'' involving the degeneracy of ill-constrained lens models leads to systematic uncertainties in estimations of $H_0$ \cite{Falco1985, Saha2000, Kochanek2020, Birrer2020, Chen2021}.

In a first paper \cite{Ng2020} we presented a geometric probe of cosmology independent of the lensing potential, by considering differential time delays over images, originating from spatially-separated photometric signals within a strongly lensed quasar. In the original work, we presented the predicted signal integrated across all wavelengths as a function of time, from considering a special case of a thin face-on ring geometry of the Broad Line Region (BLR) of the quasar. In this paper we re-examine the methodology and demonstrate some significant difficulties, most notably under-determination. We also consider analytically the picture in the line-of-sight velocity (i.e. spectral) and time delay space, referred to as a ``velocity-delay map''; as well as the signals obtained at any spectral or temporal slice for axisymmetric BLR geometries with the inclined ring or disk, spherical shell, and double cone as examples.

The paper is organised as follows: we briefly review the definitions and the background in Section \ref{sec:p2background} and give an overview of the methodology and difficulties in Section \ref{sec:methodanddifficulties}. We look at the effect of the differential lensing on the velocity-delay maps of different BLR geometries in Section \ref{sec:velocitydelaymaps} and demonstrate the insensitivity of the required BLR parameter estimation to the differential lensing in Section \ref{sec:parameterestimation}.

\section{Background} \label{sec:p2background}

An observer sees a photometric signal from an unlensed point source at line-of-sight physical distance $\xi_z$ and position $\bm{\beta}$ on the sky, at time $t=t_i$. When the source is lensed, the observer sees this signal in a given lensed image X at a position $\bm{\theta}_X$ on the sky and at time $t = t_i + \tau_X$, where $\tau_X$ is the time delay \cite[e.g.][]{Schneider1992, Blandford1986, BlandfordNarayan1992}:
\begin{equation}
    \tau_X \equiv \tau \left(\bm{\theta}_X , \bm{\beta}, \xi_z \right) = \frac{D}{c} ( 1 + z_d ) \left(\frac{1}{2} ( \bm{\theta}_X - \bm{\beta})^2 - \psi ( \bm{\theta}_X )\right). \label{p2_tauX}
\end{equation}
Here $D= \frac{D_d D_s}{D_{ds}} \propto H_0^{-1}$ is the ``lensing distance'' or more commonly the ``time-delay distance'', a ratio of angular diameter distances (subscript $d$, $s$, $ds$ denoting the angular diameter distance from the observer to the lens, to the source and from the lens to the source, respectively). The lensing distance is thus the factor containing all the cosmological information. We denote the speed of light by $c$, and the redshift of the lensing mass by $z_d$. The two-dimensional vector positions of image X in the lens plane, and the source in the source plane, are given by $\bm{\theta}_X$ and $\bm{\beta}$ respectively; scaled such that their magnitudes are the observed angular positions relative to the observer-lens axis. The dimensionless lensing potential is denoted by $\psi$.

Although the time delay $\tau$ is not an observable quantity, the time delays between images of strongly lensed quasars may be measured and is a conventional method employed to test cosmology. The observed time delay between two images A and B are given by
\begin{align}
\Delta \tau_{BA} \equiv& \, \tau_B - \tau_A\\
=& \, \frac{D}{c} (1 +z_d) \Big(\frac{1}{2}  \left(\bm{\theta}^2_B - \bm{\theta}^2_A \right) + (\bm{\theta}_A - \bm{\theta}_B) \cdot \bm{\beta} - \psi (\bm{\theta}_B) + \psi (\bm{\theta}_A)  \Big). \label{p2_tauBA}
\end{align}

Since Equation \eqref{p2_tauBA} is dependent on the  dimensionless lensing potential $\psi $ however, time delay measurements are limited by the assumptions and accuracy of the lens model. Simple gravitational lens systems are rare, and observational constraints on the lens model are limited due to the existence, typically, of only two or four images per system \cite{Schneider2013, Birrer2019a}.

In Paper I \cite{Ng2020}, we instead consider lensed quasars as a finite rather than point source in combination with reverberation mapping \cite{Blandford1982, Peterson1993, Cackett2021} to determine quasar structure. Broad emission lines are a feature of quasar spectra and correspond to gas clouds, known as the Broad Line Region (BLR), surrounding the central emitting accretion disk at some distance. Photons travelling outwards from the central source are absorbed and re-emitted by the BLR gas. The broad emission lines therefore respond to variations in the continuum luminosity of the central source with a time delay determined by the BLR geometry.

For an unlensed quasar, if the continuum emission (across all wavelengths) from the central source occurs at $t=t_i$ (i.e. a Dirac delta signal in time), then the BLR emission occurs at $t=  t_i' =  t_i + \tau_{\textsc{rm}}$, where the reverberation mapping time delay is
\begin{equation}
\tau_{\textsc{rm}} (\bm{r}) = \frac{(1+z_s)}{c} \left( r - \bm{r} \cdot \hat{r}_z  \right). \label{p2_tauBLR}
\end{equation}
Here $\hat{r}_z$ is the unit vector towards the observer and $\bm{r} =(r_x,r_y,r_z)^T = (D_s \delta \bm{\beta}, \delta \xi_z)^T$ is the spatial position of a given BLR cloud relative to the central source. The cosmological redshift factor $(1+z_s)$ corrects the time delay in the quasar rest frame for the time delay as measured in our observer frame. If we hold $\tau_{\textsc{rm}}$ fixed, then the ``iso-delay surface'' is a paraboloid of revolution around $\hat{r}_z$. Equation \eqref{p2_tauBLR} therefore describes a family of nested paraboloids parametrised by $\tau_{\textsc{rm}}$. For the unlensed quasar, the time at which the flux from the BLR is seen is determined by the intersection of the BLR geometry with the iso-delay paraboloids.

\begin{figure}
    \centering
    \includegraphics[width=0.6\linewidth]{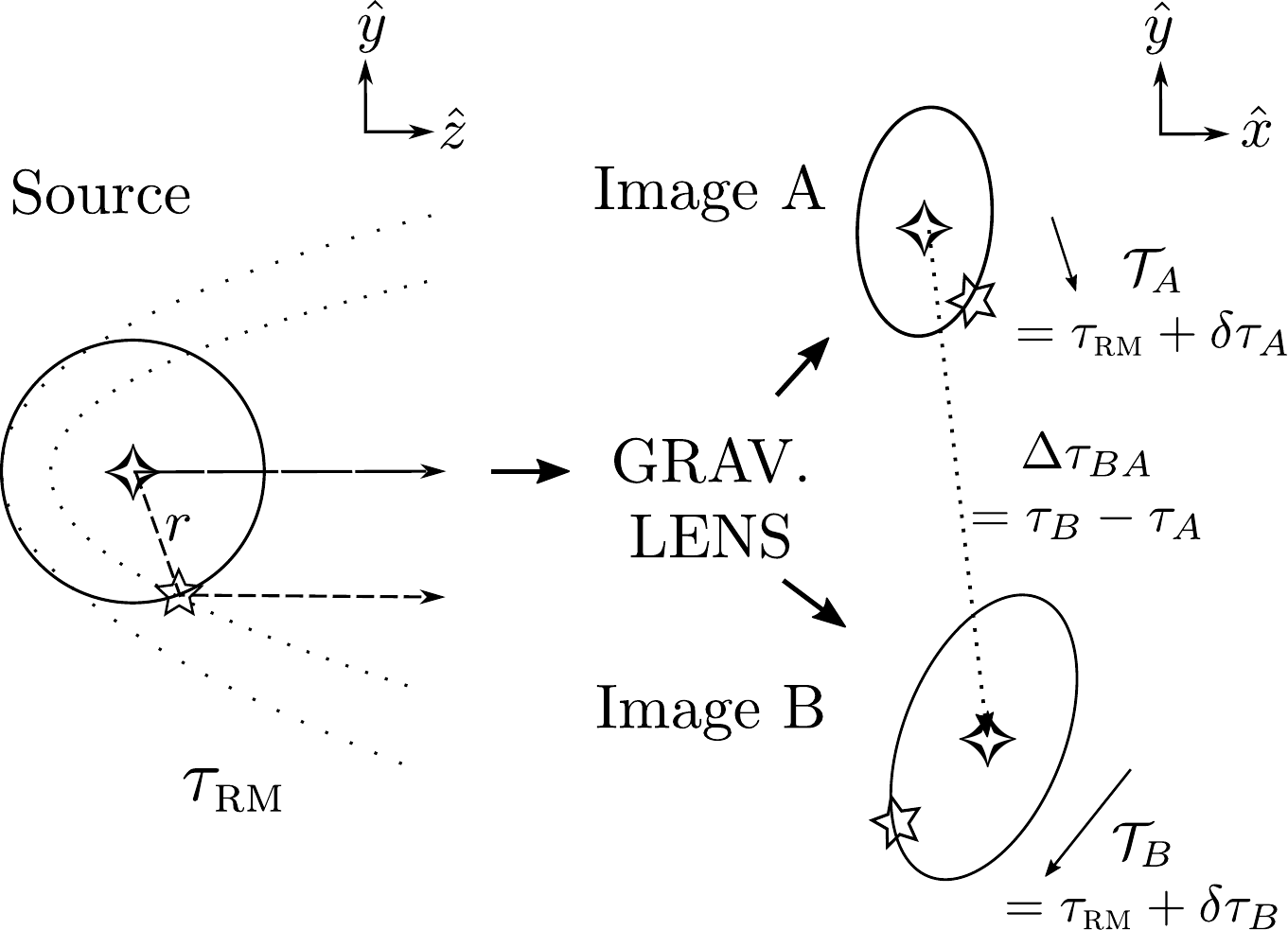}
    \caption{Schematic diagram. The source is shown projected in the y-z plane, where $z$ is in the observer direction, and the lensed images are projected in the x-y (sky) plane. The images are not spatially resolved. The central source is marked by a diamond and a chosen point on the BLR is marked by a star; their corresponding locations within the lensed images are marked similarly. There is a geometric time delay $\tau_{\textsc{rm}}$ between photons from the centre of the source and those reabsorbed and re-emitted by the BLR (paths marked by the dashed arrows; representative slices of iso-delay paraboloids marked by dotted lines); and a further differential time delay arises across each image from gravitational lensing ($\delta \tau_A$ and $\delta \tau_B$).}
    \label{fig:schematicdiagram}
\end{figure}

Considering the effect of the lensing time delay, the time that we see the continuum signal, originating from the centre of the source at $\bm{\beta}$, in image X is given by
\begin{equation}
    t_X = t_i + \tau_X. \label{p2_timeXundashed}
\end{equation}
The time that we see a signal in image X originating from a BLR cloud at position $\bm{r}$, using the shorthand notation ${\tau}'_X \equiv \,\tau(\bm{\theta}_X + \delta \bm{\theta}_X , \bm{\beta} + \delta \bm{\beta}, \xi_z + \delta \xi_z) = \tau_X + \delta \tau_X$, is similarly given by
\begin{equation}
    t_X' = t_i' + \tau_X' = t_i + \tau_{\textsc{rm}} ( \bm{r})+ \tau_X + \delta \tau_X( \bm{r}). \label{p2_timeXdashed}
\end{equation}
The differential lensing time delay within image X, may be found via a Taylor expansion to first order in $\delta \bm{\theta}_X$ and $\delta \bm{\beta}$ (the term corresponding to a line of sight displacement $\delta \xi_z$ may be omitted as it is many orders of magnitude less than that corresponding to a shift in the source plane; see \cite{Ng2020, Tie2018}), using the lens equation $\bm{\nabla}_{\bm{\theta}} \psi = \bm{\alpha}_X  = \bm{\theta}_X - \bm{\beta}$  where $\bm{\alpha}_X \equiv \bm{\alpha} (\bm{\theta}_X)$ is the scaled deflection angle, as
\begin{equation}
    \delta \tau_X (\bm{r}) = \frac{D_d}{D_{ds}c}(1+z_d)(\bm{\beta} - \bm{\theta}_X)\cdot (r_x, r_y)^T. \label{p2_DeltaTX}
\end{equation}

The interval $\mathcal{T}_X(\bm{r})$ between the signal from a BLR cloud at $\bm{r}$ and the continuum from within a given image X is given by
\begin{equation}
\mathcal{T}_X (\bm{r}) \equiv t_X' - t_X = \tau_{\textsc{rm}} (\bm{r}) + \delta \tau_X (\bm{r}) = \frac{(1+z_s)}{c} \left( r - \bm{r} \cdot \bm{e}_X \right) \label{quadricsurface}
\end{equation}
where $\bm{e}_X \equiv \left(\frac{(1+z_d)}{(1+z_s)} \hat{\alpha}_{X,x}, \frac{(1+z_d)}{(1+z_s)} \hat{\alpha}_{X,y}, 1 \right)^T$ and the physical deflection angle is $\hat{\bm{\alpha}}_X \equiv \hat{\bm{\alpha}} ( D_d \bm{\theta}_X ) = \tfrac{D_d}{D_{ds}} \bm{\alpha}(\bm{\theta}_X )$. The magnitude of $\bm{e}_X$ is the eccentricity $e_X$ of the conic section which generates the quadric surface of revolution described by Equation \eqref{quadricsurface}. We regain the unlensed case, i.e. $\tau_{\textsc{rm}}$, simply by setting the deflection angle to $0$. The effect of differential lensing is therefore to deform each iso-delay paraboloid into a one of the sheets of a two-sheeted hyperboloid of revolution about the $\hat{e}_X$ axis, since $e_X >1$ when the quasar is lensed and $e_X=1$ when the quasar is unlensed. For the lensed quasar, the time at which the flux for the BLR is seen is determined by the intersection of the BLR geometry with a family of iso-delay hyperboloids.

The difference in the differential lensing time delays of an image pair is independent of $\bm{\beta}$ and $\psi(\bm{\theta}_X)$ to first order in $\delta \bm{\beta}$ and $\delta \bm{\theta}_X$:
\begin{equation}
\mathcal{T}_{BA} (\bm{r}) = \delta \tau_{BA} (\bm{r}) 
 =  \frac{D_d}{D_{ds}c}(1+z_d) \left( \bm{\theta}_A - \bm{\theta}_B \right) \cdot (r_x, r_y)^T. \label{p2timedelaydifferenceequation}
\end{equation}
In general, we will write $\mathcal{T}_{BA} (\bm{r}) \equiv \mathcal{T}_B (\bm{r})  - \mathcal{T}_A (\bm{r})$ or $ \delta \tau_{BA} (\bm{r}) \equiv \delta \tau_B (\bm{r}) - \delta \tau_A (\bm{r})$ explicitly as a difference to emphasise that the time delay difference involves two separate measurements which are both functions of position. \cite{Yonehara1999, Yonehara2003} proposed and \cite{Goicoechea2002} investigated constraining the locations of spatially separated flares within gravitationally lensed quasars via measurements of this time delay difference.

\section{Proposed Method and Difficulties} \label{sec:methodanddifficulties}

The time delay difference between corresponding photometric signals in images A and B is solely determined by cosmology through the lensing distance, the geometry of the lensing configuration and the spatial separation within the source in the plane of the sky. This unusual property inspired the key proposal of paper I: to extract cosmological information via the ratio of angular diameter distances
\begin{equation}
    \frac{D_d}{D_{ds}} = \frac{c\left( \mathcal{T}_B (\bm{r})  - \mathcal{T}_A (\bm{r}) \right)}{ (1+z_d) \left( \bm{\theta}_A - \bm{\theta}_B \right) \cdot (r_x, r_y)^T} \label{cosmologyeqn}
\end{equation}
whilst alleviating the systematic uncertainties associated with lens modelling in using time delay measurements. The lens redshift and image positions can be measured directly. The time delay within an image $\mathcal{T}_X$ is found by subtracting the time $t_X$ we see a particular signal in the continuum flux, from the time $t'_X$ at which we see the corresponding signal in the flux of a broad emission line originating from a responding BLR cloud at $\bm{r}$.

 The technique as presented, centred on the usage of reverberation mapping to determine the position of the BLR cloud $\bm{r}$ responding at a given measured time delay in each image $A$ and $B$, is ill-motivated rather than one with limitations which could be directly mitigated by future improvements in observations. We summarise the interlinked difficulties as follows
\begin{enumerate}
    \item Reverberation mapping only gives general constraints on the geometric structure of the BLR. It does not uniquely map the location of individual points in the BLR, meaning that Equation \eqref{cosmologyeqn} is underdetermined.
    
    \item The projection of the BLR cloud position on the image-image axis can not be determined from reverberation mapping, but requires e.g. spatially resolving each image.
    
    \item In the general case $\tau_{\textsc{rm}} \gg | \delta \tau_{AB} |$; the flux from each image is largely insensitive to differential lensing.

    \item The determination of the geometric parameters is not a priori independent of the source position.
\end{enumerate}

A main challenge of exploiting Equation \eqref{cosmologyeqn} lies in the task of pinpointing the position of the BLR cloud responding at a given measured time delay in each image. This position also enters separately in the dot product in the denominator of the same equation, i.e. requiring the projection of the BLR cloud position on the image $A$ - image $B$ axis on the sky, which assuming a spatially unresolved BLR is not directly measurable. For example, \cite{Goicoechea2002} determined the position of a flare only along the image-image direction for a given image pair.

A key assumption of \cite{Yonehara2003} was that these flares are discrete, spatially separated and not physically correlated. If there is no physical correlation between discrete flares, then the shape of the flux arising from each of the flares is unique. Together with either lack of or minimal superposition of the flux, this allows each flare to be identified between images.

This assumption is violated when considering reverberation mapping of the BLR, which does not pinpoint the location from discrete flares. Rather, reverberation mapping utilises the response of an \textit{extended region} to the \textit{same} continuum signal to give general constraints on the BLR structure and kinematics in terms of geometric parameters. Determining the time at which we see the response from a particular position in a given image, as required by Equation \eqref{cosmologyeqn}, is in general an underdetermined problem as the response is a superposition of spectral flux from over the entire BLR.

The second consequence of the assumption that flares are not physically correlated is that the time between the discrete flares, denoted by $\Delta t_{\mathrm{burst}}$ in \cite{Yonehara1999}, in general can be arbitrarily small whilst their spatial separation is arbitrarily large.  This allows the fulfilment of the condition $\Delta t_{\mathrm{burst}} \lesssim |\delta \tau_{AB}|$ required for the time delay difference to have an appreciable effect. Choosing to identify $\Delta t_{\mathrm{burst}}$ as a characteristic difference in $\tau_{\textsc{rm}}$ between points on the BLR, this condition was fulfilled in Paper I in the special case of a face-on thin ring BLR geometry which responds simultaneously, but no longer holds for a general configuration in which the difference in $\tau_{\textsc{rm}}$ is on the scale of the light crossing time over the BLR structure. In the general case $\tau_{\textsc{rm}} \gg | \delta \tau_{AB} |$ and so the dependence of the flux from each image on differential lensing is very small.

To see this clearly, in the time delay within the image $\mathcal{T}_X$ given by Equation \eqref{quadricsurface} the terms $\frac{D_d}{D_{ds}}$, $(1+z_s)$ and $(1+z_d)$ are all on the order of unity, whereas $\alpha_{X}$ is on the order of at a minimum $10^{-6}$ to a maximum of $10^{-3}$ radians considering galaxy scale to cluster scale lensing. That is, in general the time scale for the differential lensing (minutes to days, depending on the lens mass) is very small compared to the time scale for reverberation mapping (tens to hundreds of days): $\delta \tau_X \ll \tau_{\textsc{rm}}$ implying $\mathcal{T}_X \sim \tau_{\textsc{rm}}$.

Finally, the determination of the geometric parameters using reverberation mapping is also not a priori independent of the distance ratio $\frac{D_d}{D_{ds}}$ nor the source position $\bm{\beta}$. We show in Section \ref{sec:parameterestimation} of the present paper however that geometric parameters may be inferred accurately without either further data (e.g. image of the Einstein ring) or lens modelling.

As an additional note, \cite{Tie2018} showed in the context of delayed emission across an accretion disk,  whilst the differential lensing time delay given by \eqref{p2_DeltaTX} is negligible, the terms equivalent to the reverberation mapping terms in this work no longer cancel between images when considering the effect of microlensing. Rather, differential magnification over the source is included as an extra weighting factor to these terms which is image and time dependent (as the source moves relative to the stars causing the microlensing). Microlensing thereby adds time delays on the scale of the light crossing time of the relevant quasar structure to the standard time delay \eqref{p2_tauBA}, potentially causing systematic problems for time delay cosmography; although no evidence for this effect has been observed from current data \cite[e.g.][]{Wong2020}. We do not explore this microlensing effect as it does not impact the above conclusions centring on the potential-free expression for the differential lensing time delay.

\subsection{The Degeneracy Problem Exemplified}

We now illustrate the problem of underdetermination with an example. The spectral flux from axisymmetric BLR geometries including the thick ring or disk, the spherically symmetric and the biconical geometry may be considered as the superposition of the contribution of the spectral flux from (possibly inclined) thin rings. We therefore write the Cartesian BLR body coordinates $\bm{p} = r (\sin \vartheta \cos \varphi, \sin \vartheta \sin \varphi, \cos \vartheta)^T$ in terms of spherical coordinates where $\vartheta$ is the zenith angle and $\varphi$ is the azimuthal angle, which is related to the coordinates $\bm{r}$ (which we defined such that $\hat{r}_z$ is in the observer, or line-of-sight, direction) by a rotation through an inclination angle $i \in [ 0 , 2 \pi)$ about the $\hat{r}_x$ axis such that $\bm{r} = \text{Rot}_x(i) \bm{p}$. For example, a ring or disk co-planar with the central ionising source has fixed $\vartheta = \frac{\pi}{2}$, and when its inclination angle is $i=0$ it is face-on and when $i = \frac{\pi}{2}$ it is edge-on with respect to the observer. A spherical geometry may have $i$ set to be $0$.

To demonstrate the degeneracy issue, we restrict Equation \eqref{cosmologyeqn} to the least-degenerate case of an infinitesimally thin inclined ring such that the position within the BLR geometry is parametrised by $\varphi$ alone. 
\begin{equation}
    \frac{D_d}{D_{ds}} = \frac{c \left(\mathcal{T}_B (R, \tfrac{\pi}{2}, \varphi)  - \mathcal{T}_A (R, \tfrac{\pi}{2}, \varphi) \right)}{ R (1+z_d) \left(  \theta_{AB,x} \cos \varphi + \theta_{AB, y} \cos i \sin \varphi \right)} \label{cosmologyeqn_thinring}
\end{equation}
which assuming circular Keplerian orbital motion can be related to the line-of-sight velocity variable $v_z$ (and thereby the observed wavelength) via a line-of-sight velocity model $u_z(\bm{r})$:
\begin{equation}
    u_z (r, \varphi)= u(r) \sin i \cos \varphi
\end{equation}
where $u(r) = \sqrt{\tfrac{GM}{r}}$ and $M$ is the central black hole mass. So for each value of the line of sight velocity, there are two corresponding positions $\varphi^+$ and $\varphi^-$ which are defined by
\begin{equation}
    \sin \varphi^\pm \equiv \pm \sqrt{1 - \left( \tfrac{v_z}{u(R) \sin i} \right)^2}
\end{equation}
such that $0 \leq \varphi^+ \leq \pi$ and $\pi \leq \varphi^- \leq 2 \pi$. One can immediately see that the value of the denominator of Equation \eqref{cosmologyeqn_thinring} is dependent on the choice of $\varphi^{\pm}$ and $\mathcal{T}_{BA} ( \varphi^+ ) \neq \mathcal{T}_{BA} ( \varphi^-)$: even in the simplest case of an infinitesimally thin ring, the positions are degenerate. In Paper I we disregarded this position-wavelength degeneracy (all positions correspond to a single wavelength for a face-on ring) and further exploited the symmetry of the projection of a thin face-on ring in the $\textit{x-y}$ plane: the dependence on position $\varphi$ within the BLR geometry disappears in the denominator when taking the maximum of the time delay for the special case of a face-on ring.

In more realistic models, such as a radially-thick inclined ring, we no longer have a fixed $R$ but rather a range of values of $r$ (and also of $\cos \varphi^\pm$). Each $\mathcal{T}_X(r, \tfrac{\pi}{2}, \varphi^\pm)$ term as well as $\sin \varphi^\pm$ now corresponds to \textit{two} ranges of values; the system is underdetermined and we can not find constraints on $\mathcal{T}_B (\bm{r}) - \mathcal{T}_A (\bm{r})$ for a given inferred value of $v_z$.

\section{Combining Differential Lensing and Reverberation Mapping} \label{sec:velocitydelaymaps}

Reverberation mapping provides constraints on the overall geometry of the BLR since measurements of $T_X$ provides information on $\bm{r}$; and in addition, the observed wavelength within the broad emission line is proportional (assuming relativistic effects are negligible\footnote{A full analysis would consider the impact of relativistic effects which are significant for small BLR radii, such as gravitational redshifting, the relativistic Doppler effect and relativistic beaming, e.g. \cite{Corbin1997, Goad2012}. A consequence of the relativistic Doppler effect is that in general the observed wavelength is not simply proportional to line-of-sight velocity.})
to the line-of-sight velocity of the BLR cloud. If the velocity of a BLR cloud may be related to its position  via a model for the velocity field, then the line-of-sight velocity gives additional information on the position of the BLR cloud as well as the kinematics. A  ``cloud'' refers to small individual entities in the emission line region in reverberation mapping literature; we interchangeably refer to either a cloud or a point particle of the BLR distribution.

We first review the unlensed case \cite{Blandford1982, Peterson1993, Netzer1990, Netzer2013}. Let $\bm{w}$ be the velocity coordinates of a BLR cloud, whereas $v_z$ denotes the line-of-sight velocity variable. The broad emission line spectral flux responds to the continuum flux $C(t - \tau_{\textsc{rm}})$, as given in generality by
\begin{equation}
    L( v_z , t) = \iint j(\bm{r}) C ( t' - \tfrac{r}{c} ) g (\bm{r}, v_z) \delta ( t - (t' - \tfrac{r}{c} + \tau_{\textsc{rm}}(\bm{r}))) d \bm{r} d t'
\end{equation}
where the line-of-sight projected 1D velocity distribution $g(\bm{r}, v_z)$ is defined in terms of the full velocity distribution $f( \bm{r}, \bm{w})$ as
\begin{equation}
    g(\bm{r}, v_z) \equiv \int f( \bm{r}, \bm{w}) \delta ( v_z - \bm{w} \cdot \hat{r}_z ) d \bm{w}.
\end{equation}
The responsivity term $j(\bm{r})$ is in general dependent on the position (or if isotropic, simply the radius) of the cloud and contains the physics of the photoionisation of the BLR. The continuum flux received by a gas cloud at time $t'$ and position $\bm{r}$, emitted by the central source at an earlier time $t' - \tfrac{r}{c}$, is $j(\bm{r}) C ( t' - \tfrac{r}{c} ) = \frac{\varepsilon(\bm{r})}{4 \pi r^2} C ( t' - \tfrac{r}{c} )$ and plays the role of the emissivity of the cloud where $\varepsilon(\bm{r})$ is the reprocessing coefficient of the cloud.

The general reverberation mapping problem involves finding the deconvolution of
\begin{equation}
    L (v_z, t) = \int \Psi(v_z, s) C (t-s) ds
\end{equation}
since all of the geometrical information is contained within the transfer function $\Psi (v_z, t)$. When the transfer function is visualised using a heat map in $(v_z, t)$-space, it is commonly referred to as a ``velocity-delay map''. This deconvolution problem may be approached using maximum entropy fitting techniques, or regularised linear inversion; and then the outputted velocity-delay map may be compared qualitatively to the theoretical velocity-delay maps produced by simple models. Recovering the transfer function and thereby inferring the physical and kinematic distribution of the BLR using reverberation mapping has been, up until recent improvements of the quality of reverberation data sets, regarded as a technique in a developmental state \cite{Shen2015, Mangham2019, Cackett2021}. Rather than solving this ill-posed inverse problem, reverberation mapping has been often limited to finding only the mean radius for the BLR through measuring the mean time delay for emission lines using cross-correlation analyses, from which it is possible to constrain the mass of the central black hole. Forward modelling of the BLR also offers an alternative Bayesian approach to directly (without explicitly finding the transfer function) providing best-fit parameters of a flexible BLR geometry \cite{Pancoast2011}. 

The transfer function, or the Green's function for the system, is the line response to a Dirac-delta continuum pulse; replacing $C (t' - \tfrac{r}{c})$ with $\delta (t' - \tfrac{r}{c})$ gives
\begin{align}
    \Psi (v_z, t) &= \int j(\bm{r}) g (\bm{r}, v_z) \delta ( t - \tau_{\textsc{rm}}(\bm{r})) d \bm{r}\\
    &= \int j(\bm{r}) n(\bm{r}) \delta (v_z - u_z(\bm{r})) \delta ( t - \tau_{\textsc{rm}}(\bm{r})) d \bm{r}.
\end{align}
The second equality holds when there exists a velocity model $\bm{u}(\bm{r})$ whose corresponding line-of-sight velocity model is $u_z(\bm{r})$, such that we may write the velocity distribution as
$f(\bm{r}, \bm{w}) = n(\bm{r}) f(\bm{w}| \bm{r}) = n(\bm{r}) \delta (\bm{w} - \bm{u}(\bm{r}))$
where $n(\bm{r})$ is the number density of responding clouds. This gives the line-of-sight projected velocity distribution as $g(\bm{r}, v_z) = n(\bm{r}) \delta (v_z - u_z(\bm{r}))$. We can interpret the general problem as one of finding a probability density function under a change of variables from positions $\bm{r}$ to $(v_z , T_X)$. As the dimensionality of the original set of random variables is not the same as the new set of random variables, the probability density functions are written using Dirac delta generalised functions.

Including the effect of differential lensing on the transfer function simply involves substituting the reverberation mapping time delay function $\tau_{\textsc{rm}}(\bm{r})$ with the time delay function $\mathcal{T}_X(\bm{r})$ and denoting the time variable with the subscripted $T_X$ to emphasise that it is measured with respect to the lensed continuum signal in each image $X$. It follows that the transfer function and the observed broad emission line spectral flux of each image are respectively given by
\begin{equation}
    \Psi_X (v_z, T_X) = \int j(\bm{r}) n(\bm{r}) \delta (v_z - u_z(\bm{r})) \delta ( T_X - \mathcal{T}_X(\bm{r})) d \bm{r} \label{eq:spectralflux}
\end{equation}
and
\begin{equation}
    L_X (v_z, T_X) = \int \Psi_X(v_z, s) C (T_X -s) ds. \label{lensedlineresponse}
\end{equation}

Although overall deductions about the BLR geometry and kinematics made be made from finding the transfer function or else by direct forward modelling, determining the time at which we see the response from a particular position in a given image, as is required by Equation \eqref{cosmologyeqn}, is in general an underdetermined problem. For the simple BLR models usually considered, e.g. where the (emissivity-weighted) number density distribution is modelled as uniform over the line, surface or volume to which the particles are confined and does not add an extra parameter, the full transfer function does not give information further than its support. For inhomogeneous BLR models which feature over- or under-densities of clouds in particular spatial regions however, the form of the transfer function may further constrain the number density distribution over that region. It can still be somewhat useful to find the theoretical transfer functions for idealised models for the visualisation of the method. 

We therefore present the form of the transfer function in the case of a uniform thin inclined co-planar ring in Section \ref{subsec:inclinedringdisk} below, and leave the details as well as calculations for a disk and spherical shell to Appendix \ref{appendixa}. Throughout these calculations, we assume $j(\bm{r})$ to be a constant \cite[e.g.][]{Pancoast2011}, i.e. that the radiation received and re-emitted by all BLR clouds is constant. A more physically accurate assumption of the dependence of $j(\bm{r})$ on, for example, radius does not impact the overall findings discussed in Section \ref{sec:methodanddifficulties} of this paper. As light is simply redistributed in time due to the differential lensing, the overall effect on the spectral flux tends to be a small change in magnitude of the transfer function whilst spanning a correspondingly larger or smaller domain in time compared to reverberation mapping without lensing.

\subsection{General Effects of Differential Lensing on The Shape of The Velocity Delay Map} 

We can distinguish the time at which the signal from some subset of the clouds, in image A from the time which it is seen in image B, by comparing the support of the transfer function in $(v_z,T_X)$-space from different images; that is, analysing the effect of differential lensing on the shape of the velocity-delay map. This is dictated purely by the first two of the available constraints, the time delay $\mathcal{T}_X(\bm{r})$ and the velocity model $u_z (\bm{r})$.

In this work, we consider axisymmetric broad line region geometries, with three classes as explicit examples: the ring or disk, the spherically symmetric, and the biconical geometry as examples. For non-axisymmetric geometries, the overall principles should still apply, but the analysis may not be as straightforward. It is useful to variously express the time delay within an image $\mathcal{T}_X$, Equation \eqref{quadricsurface}, in spherical coordinates as
\begin{subequations}
\begin{align}
\begin{split}
    \mathcal{T}_X (\bm{r}) 
    &= \tfrac{r}{c} \left(1+z_s + J_X \cos \vartheta +  A_X \sin \vartheta  \sin \varphi + B_X \sin \vartheta\cos \varphi\right) \label{timedelayequation_zero}
\end{split}\\
    &= \tfrac{r}{c} \left( 1+z_s + J_X \cos \vartheta + F_X \sin \vartheta \sin ( \varphi + \phi_X) \right) \label{timedelayequation_one} \\
    &= \tfrac{r}{c} \left( 1+z_s + J_X \cos \vartheta + F_X \sin \vartheta \sin ( \mathrm{sgn}(A_X) \varphi + \phi_X^+) \right) \label{timedelayequation_two}
\end{align}
\end{subequations}
by defining the parameters:
\begin{subequations}
\begin{align}
   J_X &\equiv - (1+z_d) \hat{\alpha}_{X,y} \sin i - (1+z_s) \cos i \\
   A_X &\equiv (1+z_s) \sin i - (1+z_d) \hat{\alpha}_{X, y} \cos i\\
   B_X &\equiv  -(1+z_d) \hat{\alpha}_{X,x}\\
   F_X &\equiv \sqrt{A_X^2 + B_X^2}\\
   \phi_X &\equiv \begin{cases}
    \phi_X^+ \equiv  \arcsin{\frac{B_X}{F_X}} &  A_X \geq 0\\
   \pi - \phi_X^+ &  A_X < 0,\; B_X \geq 0\\
   -\pi - \phi_X^+ &  A_X < 0,\; B_X <0\\ 
   \end{cases}
\end{align}
\end{subequations}
where $\phi_X \in [-\pi, \pi)$ solves $\cos \phi_X = \frac{A_X}{F_X}$ and $\sin \phi_X = \frac{B_X}{F_X}$ simultaneously, and we set $\mathrm{sgn}(0) =1$. The functional forms of $\tau_{\textsc{rm}} (\bm{r})$ and $T_X (\bm{r})$ are identical; we can regain $\tau_{\textsc{rm}}$ by setting the deflection angle $\hat{\alpha}_{X}$ to $0$ such that $J_X = -(1+z_s) \cos i$, $F_X = (1+z_s) \sin i$ and $\phi_X^+ = 0$.

We consider rotational and radial velocity fields whose line-of-sight components are given respectively by
\begin{align}
    u_{\mathrm{Rot}, z } (\bm{r}) &= u(r) \sin i_v \sin \vartheta \cos \varphi \label{rotationalmotioneqn}\\
    u_{\mathrm{Rad}, z} (\bm{r}) &= u(r) (\cos i \cos \vartheta - \sin i \sin \vartheta \sin \varphi ) \label{radialmotioneqn}
    \end{align}
where the rotation axis is in the direction given by $\text{Rot}_x(-i_v) \hat{r}_z$, and $i_v = i$ for disk and biconical geometries and $i_v \in [0, 2 \pi)$ is chosen independent of fixing $i = 0$ for spherical geometries. In the case of ring or disk geometries, we consider circular Keplerian orbits (Equation \eqref{rotationalmotioneqn} with $u(r) = \sqrt{\tfrac{GM}{r}}$ and $\vartheta = \frac{\pi}{2}$) whereas in the case of spherically symmetric geometries we consider solid body rotation (Equation \eqref{rotationalmotioneqn} with $u(r) \propto r$). Combining the time-delay constraint \eqref{timedelayequation_one} with either velocity model \eqref{rotationalmotioneqn} or \eqref{radialmotioneqn} gives a pair of parametric equations for the support of the velocity-delay map. 

\begin{figure}
    \centering
    \includegraphics[width=0.6\linewidth]{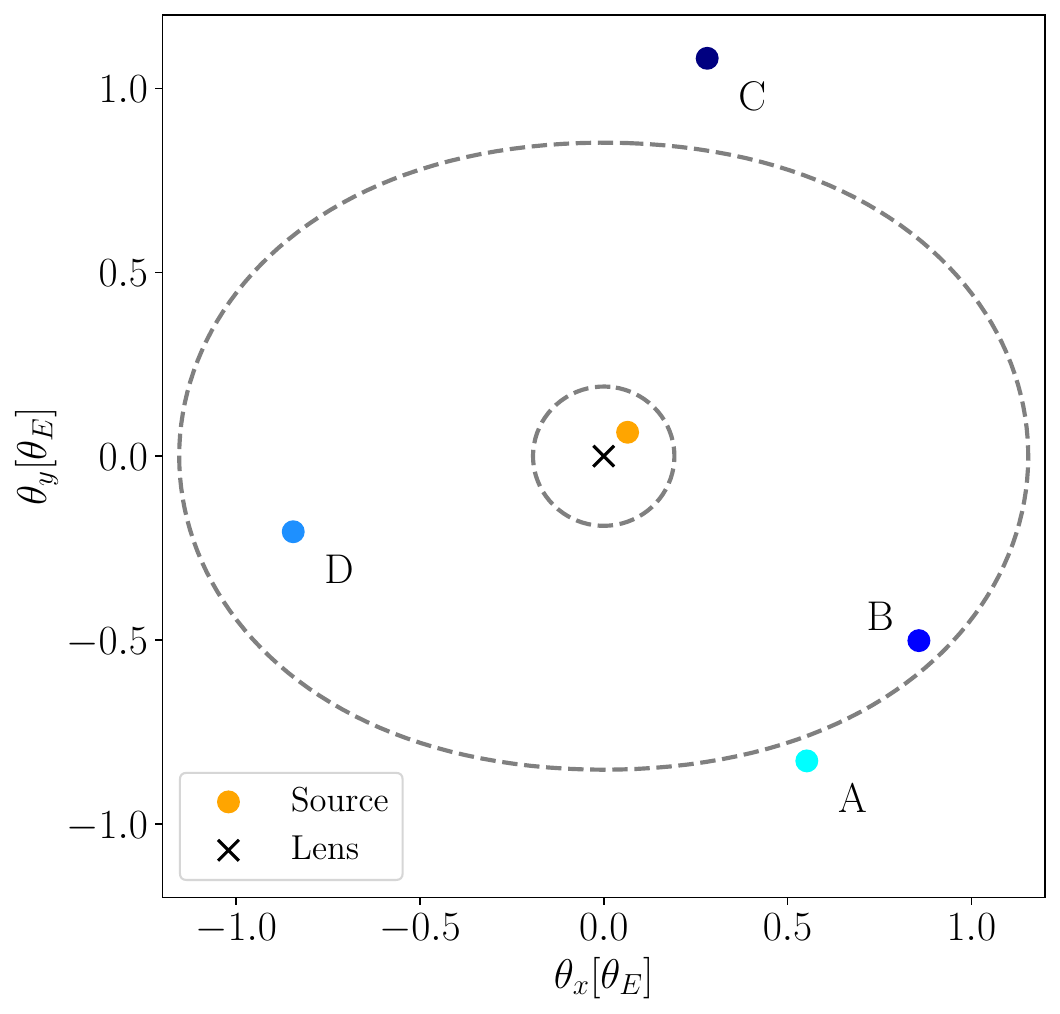}
    \caption{We reproduce the lensing configuration shown in Figure 3 of Paper I, which is based on a softened elliptical lens model with a core radius of $0.1 \theta_E$ and ellipticity of $0.1$, setting the source at $\bm{\beta} = (0.065, 0.065) \theta_E$, where $\theta_E$ denotes the deflection scale of the lens. The critical lines are marked by dashed grey lines. The calculated image positions are used for the illustrative examples throughout this work.}
    \label{fig:sieconfig}
\end{figure}

\begin{figure}
    \centering
    \includegraphics[width=0.7\linewidth]{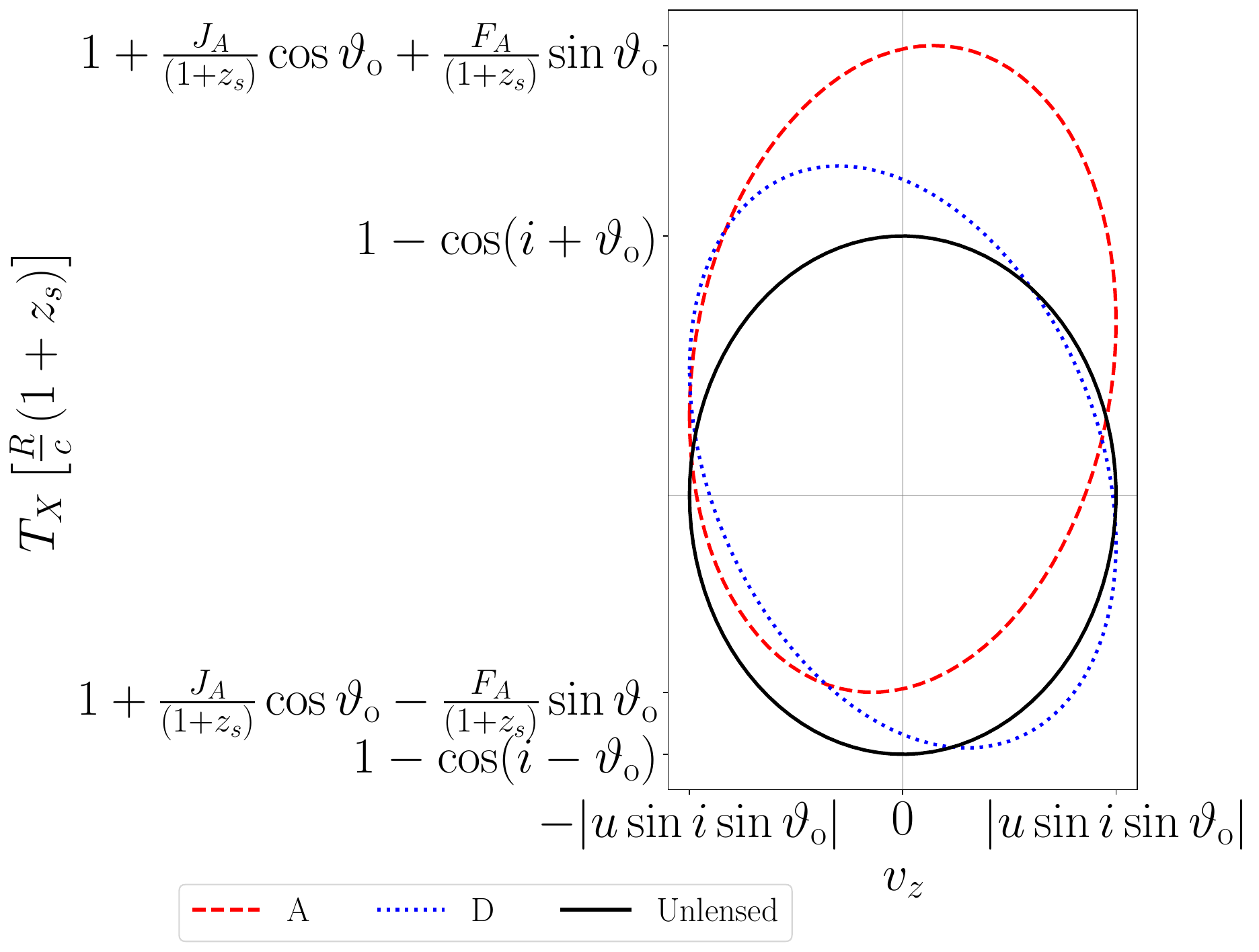}
    \caption{The general effect of differential lensing on the velocity-delay map of a single thin inclined ring component of a BLR geometry in solid body rotation, where the radius of the ring is $r = R$, at fixed polar coordinate $\vartheta_\mathrm{o}$ (i.e. not co-planar with the central source) and with inclination angle $i$. The effect of the lensing is exaggerated by choosing the scaled deflection $\alpha_X$ to be on the order of $10^{-2}$ radians, such that we may distinguish between the ellipses over the entire velocity-delay map for the purpose of this illustration. The observed ellipse is shifted and re-scaled in the time dimension, and rotated by a magnitude which differs between observed images (A and D of Figure \ref{fig:sieconfig} shown here). The axes of the unlensed ellipse are marked by solid lines.}
    \label{fig:singleringdelaymap}
\end{figure}

Recall that the spectral flux from any of these BLR geometries may be considered as the superposition of the contribution of the spectral flux from thin rings. The velocity-delay maps of these geometries may therefore be analysed by decomposing the structures into thin rings, i.e. slicing by holding $r$ and $\vartheta$ constant or alternatively (corresponding to halves of thin rings) $r$ and $\varphi$ constant. In either case, we see that a single thin ring component of a general geometry sketches out an ellipse in $(v_z, T_X)$-space. In either the lensed or unlensed case, the ellipse may be degenerate (i.e. forming a point or an interval, dependent on the parameters).

The general effect of differential lensing is to alter the eccentricity, rotate and scale the axes of, and shift in the time-delay dimension this $(v_z,T_X)$-space ellipse as illustrated in Figure \ref{fig:singleringdelaymap}. The parameter $J_X$ controls the shift in time and $F_X$ the scaling in time or the half-height of the ellipse, whereas the phase-shift $\phi^{+}_X$ controls the rotation and eccentricity. For typical values of the differential lensing time delay, there is negligible rotation of the velocity-delay ellipse for all values of the inclination angle away from the face-on case where $i=0$. For all values of $i$ away from $i=0$, the scale of the velocity-delay ellipse on the time axis is determined by the radius of the thin ring and on the scale of tens to hundreds of days whereas the effect of the lensing is on the scale of minutes. We discuss the consequences in detail for each class of BLR geometry.

\subsection{Inclined Ring or Disk} \label{subsec:inclinedringdisk}

\begin{figure}
    \centering
    \includegraphics[width=0.6\linewidth]{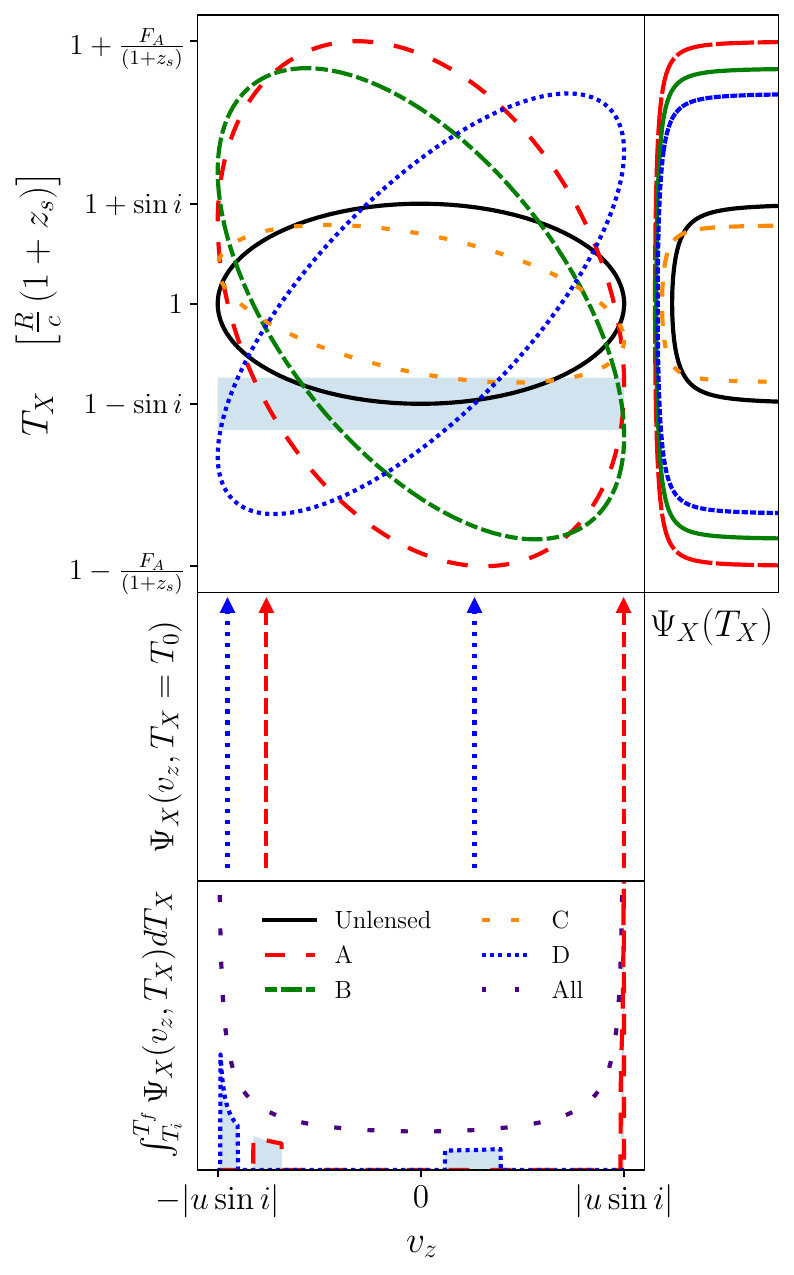}
    \caption{Velocity-delay map given by Equations \eqref{Txthinring} and \eqref{vzthinring}, transfer function integrated over all wavelengths $\Psi_X(T_X)$ given by Equation \eqref{fluxthinring} and transfer function at and over a given time $\big( \Psi_X(v_z, T_X= T_0)$ and $\int_{T_i}^{T_f} \Psi_X (v_z, T_X) d T_X \big)$ from image X $= \{$A, B, C, D$\}$ (see Figure \ref{fig:sieconfig}) of a lensed inclined thin ring BLR in Keplerian orbit. We here have chosen galaxy-scale lensing with a very small inclination angle $i= 10^{-6} \pi \, \mathrm{rad}$ for the purpose of illustration, which leads to the difference in height between the ellipses being a relatively large fraction of their overall heights, and also the relatively dramatic rotations between them. This allows the ellipses of all images and of the unlensed case to be easily distinguished in the illustration. The difference in height between ellipses as a fraction of their overall height becomes very small very quickly away from $i=0$ or $i=\pi$, as does the magnitude of their relative rotations. We show the spectral flux from images A and D at a time delay $T_0$ within images A and D are Dirac delta distributions as given by Equation \eqref{spectralfluxthinring} which when integrated over a time from $T_i = T_0 - \Delta T$ to $T_f = T_0 + \Delta T$ (shaded blue region) form segments of an arcsine distribution, as given by Equation \eqref{spectrafunc_thinring}. 
    The relatively large offsets and asymmetry in the $v_z$ locations of the spectral flux spikes at this arbitrary value of $T_0$ is due to the choice of the small inclination angle. The spectral flux integrated over all time $L(v_z)$ is a complete arcsine distribution given by the dotted purple line, independent of lensing and thus identical for all images. The relative difference in the magnitude of the spectral flux when integrated over the short time interval versus all time is due to the differing contributions from the multiple roots in Equation \eqref{spectralfluxthinring}.}
    \label{fig:thinringveldelmap}
\end{figure}

Consider as the simplest model for a BLR geometry a thin ring co-planar with the continuum source with fixed $\vartheta = \frac{\pi}{2}$ and radius $r=R$, inclined at an angle $i \in [ 0 , 2 \pi)$; in Keplerian orbit with orbital speed $u$. Equations \eqref{timedelayequation_two} and \eqref{rotationalmotioneqn} are then parametrised by the azimuthal angle $\varphi \in [0, 2\pi)$, with the direction of parametrisation determined by $\mathrm{sgn}(A_X)$:
\begin{subequations}
\begin{align}
       \mathcal{T}_X (\varphi) &= \tfrac{R}{c} \left(1 +z_s  +  F_X \sin{\left(\mathrm{sgn}(A_X) \varphi + \phi_X^+  \right)} \right) \label{Txthinring}\\
        u_{z}(\varphi) &= u \sin i \cos \varphi \label{vzthinring}
\end{align}
\end{subequations}
tracing an ellipse in ($v_z, T_X$)-space as illustrated in Figure \ref{fig:thinringveldelmap}. As in the general case, lensing alters the values of $F_X$ and $\phi_X^+$ where $ \frac{R}{c}F_X$ determines the half-height of the ellipse, and the phase shift angle $\phi_X^+$ determines the rotation of the ellipse axes from the $(v_{z}, T_X)$-axes in addition to modifying the eccentricity. However, for the co-planar thin ring, there is no shift in time. 

The velocity-delay map for a radially-thick inclined ring or disk can be generated as a superposition of the velocity-delay maps for the single inclined thin ring over a range of radii: see Figure \ref{fig:alldelaymaps_disk} subplot 2a. This results in a characteristic bell shaped envelope that will be slightly deformed by the lensing. The bell shape is elongated in the time domain in accordance with the maximum and minimum radius orbits, since the top and bottom points of this bell shape corresponds to the maximum and minimum radius orbits respectively.

For the thin inclined ring BLR with uniform linear number density $\mu$, the number density distribution of the BLR model is
$n(\bm{r}) = \mu \delta(r - R) r^{-1} \delta(\vartheta - \frac{\pi}{2})$. The one-dimensional transfer function, as illustrated in Figure \ref{fig:thinringveldelmap}, is given by an arcsine distribution
\begin{equation}
    \Psi_X (T_X) \propto \frac{
    2 \mu c}{\sqrt{F_X^2 - \left(T_X c R^{-1} - (1 +z_s) \right)^2}} \label{fluxthinring}
\end{equation}
when $\left|T_X c R^{-1} - (1 +z_s) \right| \leq F_X$ and is $0$ otherwise. The two-dimensional transfer function is 
\begin{equation}
\Psi_X (v_z, T_X) \propto \mu c R \! \sum\limits_{\pm} \frac{\delta\left(T_X - W^{\pm}_X(R, v_z) \right)}{V^{+}(R, v_z)} \label{spectralfluxthinring}
\end{equation}
where 
\begin{align}
    V^{\pm}(r, v_z) &\equiv u(r) \sin i \sin \varphi^\pm = \pm \sqrt{(u(r) \sin i)^2 - v_z^2}\\
\begin{split}
    W^{\pm}_X(r, v_z) &\equiv \mathcal{T}_X (r, \tfrac{\pi}{2}, \varphi^\pm)\\
    &= \frac{r}{c} \left( 1+ z_s + A_X \frac{V^{\pm}(r, v_z)}{u(r) \sin i}  + B_X \frac{v_z}{u(r) \sin i }\right).
\end{split}
\end{align}
The summation implies that we have a Dirac delta function where $T_X = W^{+}_X(R, v_z)$ and another where $T_X = W^{-}_X(R, v_z)$, as illustrated by Figure \ref{fig:thinringveldelmap}. We note that the form of the transfer function is independent of the image position. When the two-dimensional transfer function is integrated over an arbitrary amount of time, we see that the resulting spectra is given by segments of an arcsine distribution in $v_z$, where the widths and positions of these segments are determined by the integration time,
\begin{equation}
\int\limits_{T_{X,i}}^{T_{X,f}} \Psi_X (v_z, T_X) \, dT_X \propto
 \frac{\mu c R}{V^{+}(R, v_z)} \label{spectrafunc_thinring}
\end{equation}
if $T_{X,i} \leq W^{\pm}_X(R, v_z) \leq T_{X,f}$ and is $0$ otherwise, as shown in Figure \ref{fig:thinringveldelmap}.

\subsection{Spherical Shell}
We can straightforwardly build up the velocity-delay map of a thin spherical shell geometry, which has fixed radius $r=R$, $\vartheta \in [0, \pi]$, $\varphi \in [0, 2\pi)$ and we may set the inclination angle $i=0$. If the spherical shell is in solid body rotation about a rotation axis, which is in the direction given by $\text{Rot}_x(-i_v) \hat{r}_z$ where $i_v \in [0, 2 \pi)$, the constraints become
\begin{subequations}
\begin{align}
    \mathcal{T}_X(\vartheta, \varphi) &= \tfrac{R}{c} \left( (1+z_s) (1- \cos \vartheta ) + F_X \sin \vartheta \sin (\varphi + \phi_X) \right) \label{thinsphericalshelltd}\\
    u_z(\vartheta, \varphi) &= u(R) \sin i_v \sin \vartheta \cos \varphi \label{solidbodyrotation}
\end{align}
\end{subequations}
where here $F_X$ reduces to simply $(1+z_d) \hat{\alpha}_X$.

A thin spherical shell may be considered as being composed of thin face-on rings parametrised by $\vartheta \in [0, \pi]$. Without lensing, we see that each thin ring traces out a horizontal line in observed ($v_z, T_X$)-space and the envelope of these horizontal lines is an ellipse. Including lensing, we have that each horizontal line for each thin ring (fixed $\vartheta$) is distorted into an ellipse in ($v_z, T_X$)-space, illustrated in Figure \ref{fig:alldelaymaps_sphericalandbicone}, subplot 5b. These ellipses have the largest height (largest deviation from a straight horizontal line) when $\vartheta = \frac{\pi}{2}$, the middle of the velocity-delay map.

The distortion we describe over the entire velocity-delay map for the thin spherical shell is also very small. For example, we can check how the duration of the time delay of the overall signal is modified. Recognising that the maximum and minimum of the time delay occurs when $\sin (\varphi + \phi_X) =\pm 1$ respectively, and repeating this argument after combining the remaining $\vartheta$-dependent terms into a single sine term, we have
\begin{equation}
        \tfrac{R}{c} (1+z_s)( 1 - K_X )\leq T_X \leq \tfrac{R}{c} (1+z_s)( 1 + K_X ) \label{eq:domainTxsphericalshell}
\end{equation}
where $K_X \equiv \sqrt{1+F_X^2(1+z_s)^{-2}}$ and $F_X^2(1+z_s)^{-2} \sim 10^{-12}$.

\begin{subfigures}
\begin{figure}
    \centering
    \makebox[\textwidth][c]{
    \includegraphics[width=1.1\textwidth]{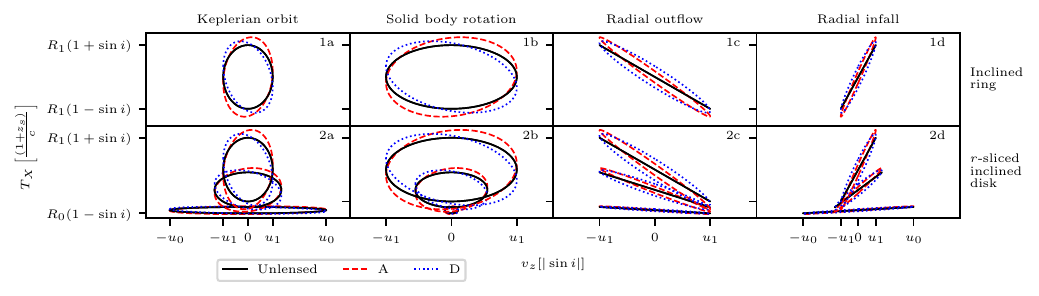}}
    \caption{An illustration showing showing the effect of lensing on velocity-delay maps (the red or blue compared to the solid black lines), greatly exaggerated by choosing the scaled deflection $\alpha_X$ to be on the order of $10^{-2}$ radians. Row 1 corresponds to a thin inclined ring (co-planar with the continuum source) and row 2 to three representative radial slices at $R_0 \equiv R_{\mathrm{min}}$, $R_1 \equiv R_{\mathrm{max}}$ and $ (R_0 + R_1)/2$ of an inclined disk (also co-planar with the continuum source); both with inclination angle $i= \frac{\pi}{4}$. Column a corresponds to Keplerian orbits, column b to solid body rotation, column c to constant radial outflow and column d to radial infall, where $u_0 \equiv |u(R_\mathrm{min})|$ and $u_1 \equiv |u(R_\mathrm{max})|$.}
    \label{fig:alldelaymaps_disk}
\end{figure}
\begin{figure}
    \centering
    \makebox[\textwidth][c]{\includegraphics[width=1.1\textwidth]{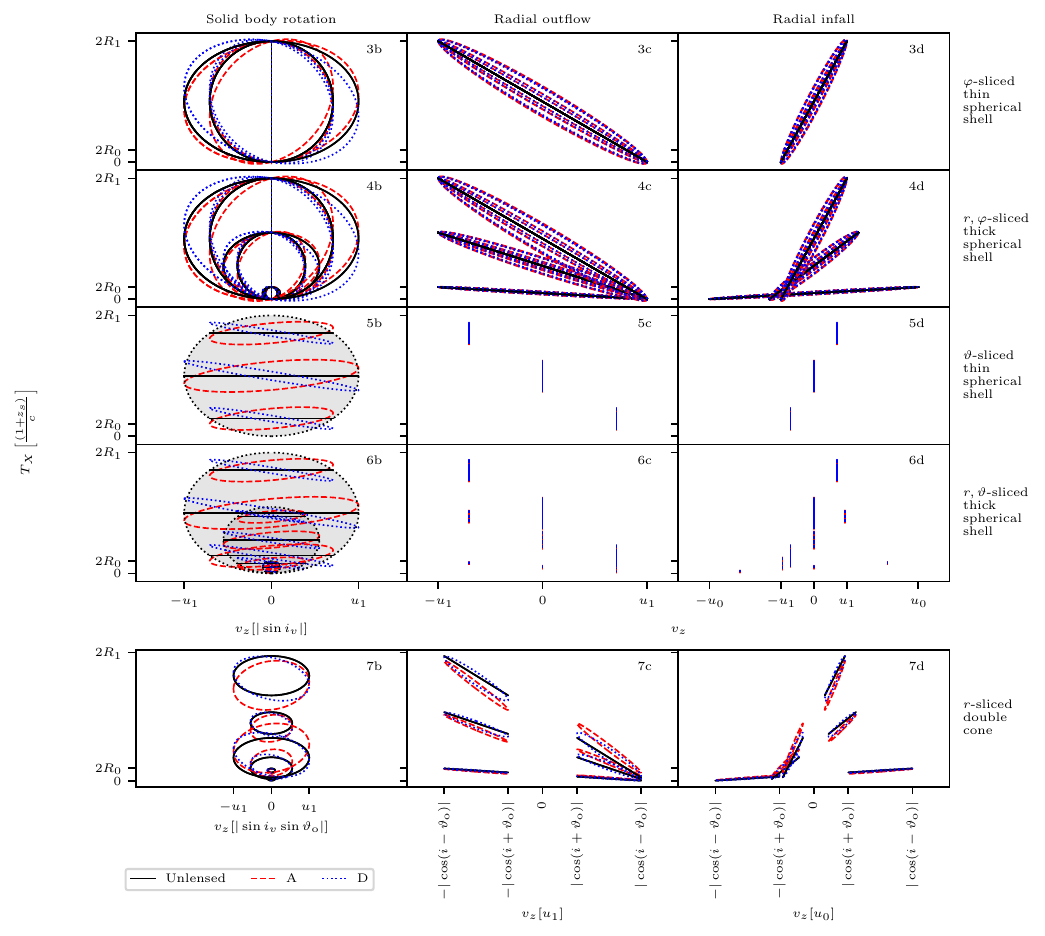}}
    \caption{An illustration showing showing the effect of lensing on velocity-delay maps (the red or blue compared to the solid black lines), greatly exaggerated by choosing the scaled deflection $\alpha_X$ to be on the order of $10^{-2}$ radians. The BLR geometry is varied by row and the velocity field is varied by column as in, and labelled following on from, Figure \ref{fig:alldelaymaps_disk}. Column b corresponds to solid body rotation, column c to constant radial outflow and column d to radial infall, where $u_0 \equiv |u(R_\mathrm{min})|$ and $u_1 \equiv |u(R_\mathrm{max})|$. Rows 3 and 5 correspond respectively to representative $\varphi$-slices and $\vartheta$-slices of a thin spherical shell; and rows 4 and 6 respectively correspond to representative radial and $\varphi$-slices, and radial and $\vartheta$-slices of a thick spherical shell. Row 7 corresponds to representative radial slices of a hollow bicone with half-opening-angle $\vartheta_\mathrm{o} = \frac{\pi}{7}$. All radial slices are taken at $R_0 \equiv R_{\mathrm{min}}$, $R_1 \equiv R_{\mathrm{max}}$ and $ (R_0 + R_1)/2$; five $\vartheta$-slices are taken at equal intervals between $0$ and $\pi$ and nine $\varphi$-slices taken at equal intervals between $0$ and $2\pi$. (Ctd.)}
    \end{figure}
\begin{figure}[t!]
    \ContinuedFloat
    \caption{(Ctd.) In 3b and 4b, whilst the ellipses corresponding to $\varphi = (\pi +) \frac{\pi}{4}$ and $(\pi +) \frac{3\pi}{4}$ are degenerate in the unlensed case, they are not when lensed. Considering instead $\vartheta$-slices, 5b and 6b illustrate that when unlensed, each slice draws out an interval in ($v_z, T_X$)-space (black solid lines), such that the complete unlensed velocity-delay map is a filled ellipse, the envelope of which is shown by the black dotted line. Each of these intervals are distorted into ellipses (red and blue dashed lines) as shown, with the largest distortion in time at $\vartheta = \frac{\pi}{2}$. When the overall structure is considered, this lensing effect is obscured by the contributions from each thin slice. For radial motion, slicing by $\vartheta$ gives vertical intervals at each $\vartheta$ value: showing more $\vartheta$ values would fill out into the ellipses apparent in rows 3 and 4. For the biconical geometry in solid body rotation in 7b, each radial slice of the biconical structure gives a thin ring at $\vartheta = \vartheta_\mathrm{o}$ and another at $\pi - \vartheta_\mathrm{o}$; hence we see three pairs of ellipses. Considering instead radial motion and slicing by $\varphi$ (3c, 3d, 4c and 4d), each thin spherical shell when unlensed maps to diagonal intervals (black) where spherical outflow corresponds to a negative slope and positive for infall (as the observer is in the $+ \hat{z}$ direction). When lensed, these intervals distort into the blue and red filled ellipses. The bicone, as a subset of a thick spherical shell, has its velocity-delay map as a subset of that of the thick spherical shell: i.e 7c is a subset of 4c and 7d of 4d, dependent on the inclination angle $i$ and half-opening angle $\vartheta_\mathrm{o}$ of the bicone.}
    \label{fig:alldelaymaps_sphericalandbicone}
\end{figure}
\end{subfigures}

If we instead consider radial inflow or outflow (Equation \eqref{radialmotioneqn} with $u(r) <0 $ and $u(r) >0$ respectively), we have Equation \eqref{thinsphericalshelltd} combined with
\begin{equation}
    u_z (\vartheta ) = u(R) \cos \vartheta.
\end{equation}
Since here $u_z$ is a function of $\vartheta$ independent of $\varphi$, it is convenient for interpretation to hold $\varphi$ constant instead of $\vartheta$. Without lensing, the velocity-delay map is an interval. By choosing the observer to be in the positive $z$ direction (such that increasing $v_z$ corresponds with increasing wavelength), spherical outflow and inflow corresponds to a negative and positive slope respectively. When the quasar is lensed, i.e. $F_X >0$, the interval deforms into a filled ellipse, since each fixed value of $\varphi$ corresponds to an arc of an ellipse parametrised by $\vartheta$, the envelope of which is given by $\varphi$ corresponding to $\sin( \varphi + \phi_X) =1$. The largest change in the time delay between lensed images corresponds to where $\vartheta = \frac{\pi}{2}$, i.e. $v_z = 0$, and the magnitude of this change is simply 
\begin{equation}
    \mathcal{T}_B ( \varphi) - \mathcal{T}_A ( \varphi) = \frac{R}{c} \sin \vartheta (F_B \sin ( \varphi + \phi_B) - F_A \sin( \varphi + \phi_A) ).
\end{equation}
However, since $\phi_B \neq \phi_A$, the time delay difference cannot be found from the envelopes of the velocity-delay maps which correspond to different $\varphi$. Again, one can easily build up the expected velocity-delay map for a thick spherical shell or ball model of the BLR, by letting $R$ vary over $R_{\mathrm{min}}$ to $R_{\mathrm{max}}$; this is shown illustrated in Figure \ref{fig:alldelaymaps_sphericalandbicone}.

\subsection{Biconical Geometries}

A biconical BLR geometry is a solid or hollow double cone with a given radial interval $r \in [R_{\mathrm{min}}, R_{\mathrm{max}}]$ and a half-opening angle $\vartheta_\mathrm{o} \in \left(0, \frac{\pi}{2}\right]$; aligned along an axis at an inclination angle $i$ to the line-of-sight direction $\hat{r}_z$. In the hollow case, the BLR material is restricted to $\vartheta = \vartheta_\mathrm{o}, \pi - \vartheta_\mathrm{o}$, and in the solid case it is restricted to $\vartheta \in [0, \vartheta_\mathrm{o}] \cup [\pi - \vartheta_\mathrm{o}, \pi]$. A general biconical geometry is therefore a subset of a thick spherical shell geometry: a solid double cone forms a thick spherical shell when $\vartheta_\mathrm{o} = \frac{\pi}{2}$.

A solid double cone can be constructed from an infinite number of hollow double cones of progressively smaller half-opening angles, and a hollow double cone can be composed from an infinite number of thin rings with radial coordinates between $R_{\mathrm{min}}$ and $R_{\mathrm{max}}$ at $\vartheta = \vartheta_\mathrm{o}$ and $\pi - \vartheta_\mathrm{o}$: the parametric equations for the velocity-delay map for the thin rings at radius $R$ and $\vartheta = \vartheta_\mathrm{o}$ and $\pi - \vartheta_\mathrm{o}$ in radial motion are
\begin{subequations}
\begin{align}
\mathcal{T}_X(\varphi) &= \tfrac{R}{c} \left( 1+z_s \pm J_X \cos \vartheta_\mathrm{o} + F_X \sin \vartheta_\mathrm{o} \sin ( \varphi + \phi_X) \right)\\
u_z(\varphi) &= u(R) ( \pm \cos i \cos \vartheta_\mathrm{o} - \sin i \sin \vartheta_\mathrm{o} \sin \varphi).
\end{align}
\end{subequations}
Without lensing, as $\phi_X \to 0$, each thin ring maps to an interval on the $(v_z, T_X)$-plane; whereas with lensing this interval deforms into an ellipse. The full velocity-delay map from the entire structure is then given by the superposition of such intervals or ellipses; and this velocity-delay map is a subset of the region mapped by a spherical geometry with the same velocity field, dependent on the values of $i$ and $\vartheta_\mathrm{o}$. This is illustrated in Figure \ref{fig:alldelaymaps_sphericalandbicone}, subplots 7c and 7d.

We see that for all of the geometries besides a single thin inclined ring, that the signal from a particular position in the BLR structure is typically superimposed by the signal from many other positions in the structure: the time delay and the line-of-sight velocity alone are insufficient in constraining the BLR position.

\section{Parameter Estimation} \label{sec:parameterestimation}

\begin{figure}
    \centering
    \makebox[\textwidth][c]{
    \includegraphics[width=1.1\textwidth]{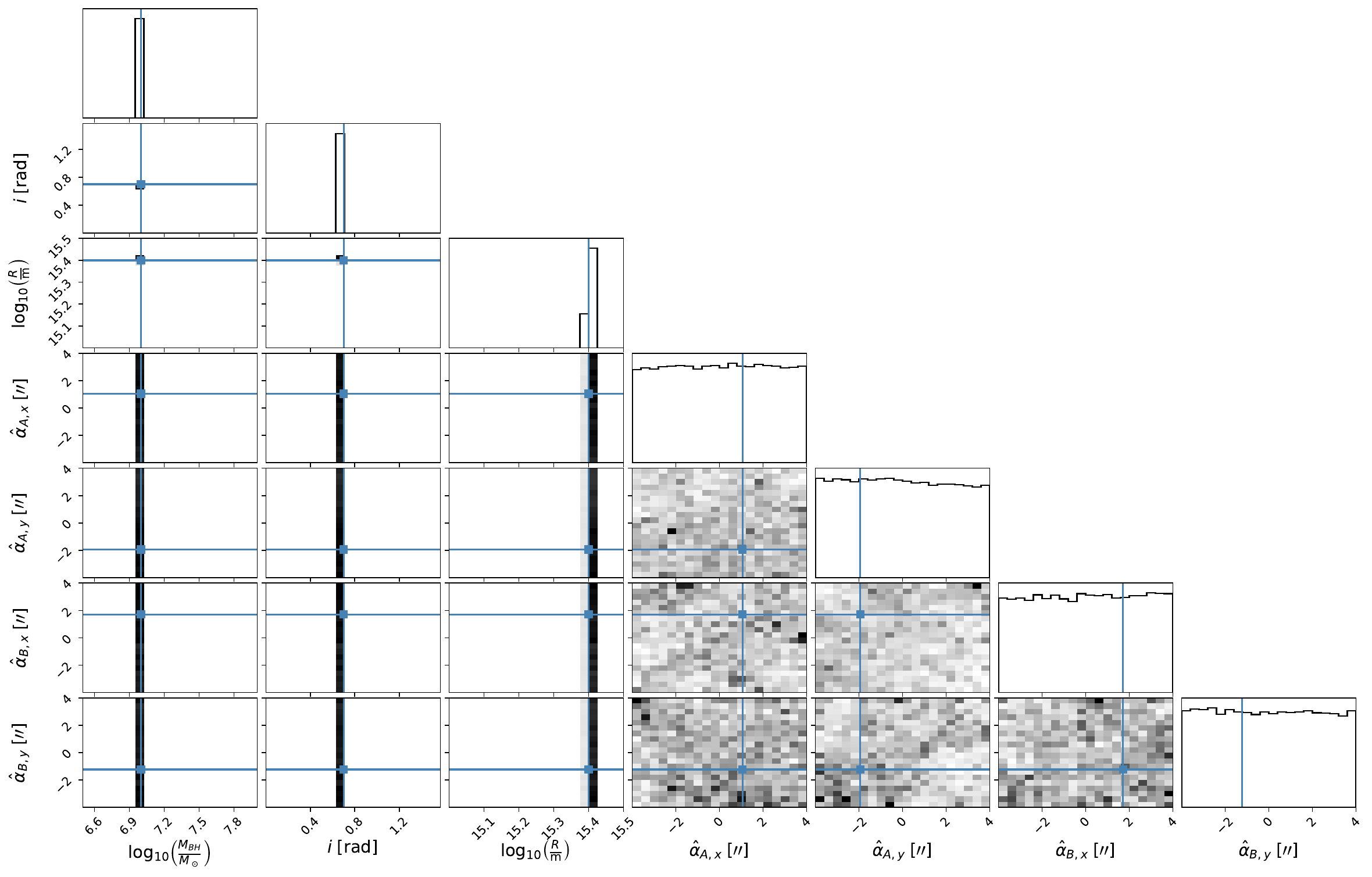}}
    \caption{Corner plot showing the one and two dimensional projections of the posterior probability distributions of the parameters $\log_{10}\left(\tfrac{M}{M_{\odot}}\right)$, $i$, $\log_{10}\left(\tfrac{R}{\textrm{m}}\right)$, $\hat{\alpha}_{A,x}$, $\hat{\alpha}_{A,y}$, $\hat{\alpha}_{B,x}$, $\hat{\alpha}_{B,y}$ corresponding to the black hole mass, inclination angle, BLR radius and deflection angles of each image with their true values (7, 0.7 rad, 15.4, 1.05", -1.94", 1.72", -1.23") marked by the blue squares. The range of each subplot corresponds to the bounds of the priors. We assume an idealised scenario where the lensed transfer functions from two images A and B have been recovered (although one is sufficient) and indicate a thin inclined ring in Keplerian orbit. Using the MCMC algorithm {\sffamily emcee} we illustrate that it is possible to recover the desired BLR geometric parameters even if the lensing parameters (i.e. the deflection angles) are not constrained. That the BLR parameters are extremely well constrained is not a comment on reverberation mapping by itself, but attributed to the idealistic model used to illustrate the insensitivity to lensing.}
    \label{fig:cornerplot}
\end{figure}

Since the available data comes from observations of each unresolved lensed image, the most probable BLR geometric parameters need to be estimated together with lensing parameters. Consequently, the estimates of BLR parameters which are to then be used for estimation of the distance ratio $\frac{D_d}{D_{ds}}$ according to the proposed method are, through the physical deflection angles, implicitly dependent on the source position $\bm{\beta}$. However, we find that it is possible to recover the desired BLR geometric parameters even if the lensing parameters (i.e. the deflection angles) are not constrained, as may be expected from the prior discussions on the smallness of the effect of the lensing for all physically reasonable scales.

As an illustration of this component of the method, we demonstrate finding these parameters using Bayesian parameter estimation, assuming that the transfer function has already been recovered, of the least degenerate, idealistic scenario of contriving the BLR model as restricted to a thin inclined ring. The parameters for this model which we wish to input into the expression for the distance ratio, Equation \eqref{cosmologyeqn_thinring}, are the black hole mass $M$, the inclination angle $i$ and BLR radius $R$.

In Bayesian data analysis, the posterior distribution $\mathcal{P}(\bm{\Theta}_M | \bm{\mathcal{D}}, \mathcal{I})$ is the probability distribution of a given set of parameters $\bm{\Theta}_M$ belonging to a model $M$ given some data $\bm{\mathcal{D}}$ and prior information $\mathcal{I}$. The posterior distribution may be determined through Bayes Rule:
\begin{equation}
    \mathcal{P}(\bm{\Theta}_M | \bm{\mathcal{D}}, \mathcal{I}) = \frac{\mathcal{P}(\bm{\mathcal{D}} | \bm{\Theta}_M , \mathcal{I} ) \mathcal{P} (\bm{\Theta}_M | \mathcal{I})}{\mathcal{P}(\bm{\mathcal{D}}| \mathcal{I})}
\end{equation}
where $\mathcal{P}(\bm{\mathcal{D}}| \bm{\Theta}_M, \mathcal{I})$ is the likelihood, $\mathcal{P}(\bm{\Theta}_M |\mathcal{I})$ is the prior, and the evidence or marginal likelihood is $\mathcal{P}(\bm{\mathcal{D}} | \mathcal{I}) = \int_{\Omega_{\bm{\Theta}_M}} \mathcal{P}(\bm{\mathcal{D}}, \bm{\Theta}_M | \mathcal{I} ) \mathcal{P}(\bm{\Theta}_M | \mathcal{I}) d \bm{\Theta}_M $. The integral is over the domain of the parameter space $\Omega_{\bm{\Theta}_M}$ and as we are only considering a single model, we drop the subscript $M$ for the rest of this work. The posterior may be estimated via numerical methods, usually involving an algorithm that approximates the posterior by generating and weighting a set of discrete samples. Since evaluating the  multi-dimensional integral to compute the evidence is a key difficulty, common numerical methods such as Markov Chain Monte Carlo (MCMC) rely on drawing these samples from a distribution proportional to the posterior distribution. In this work we use the {\sffamily emcee} \cite{Foreman-Mackey2013} implementation of the MCMC algorithm. We assume non-informative priors, such that the posterior is directly proportional to the likelihood
\begin{equation}
    \mathcal{P}(\bm{\Theta} | \bm{\mathcal{D}}, \mathcal{I}) \propto \mathcal{P}(\bm{\mathcal{D}} | \bm{\Theta} , \mathcal{I})
\end{equation}
and the maximum a posteriori estimate of $\bm{\Theta}$ is equivalent to finding the maximum likelihood estimate.

We define our likelihood function, first specifying our model. Let us assume that the original continuum function $C_{\text{original}} (t_X') = \delta( t_X' - t_i)$ is a Dirac Delta function which is then convolved or blurred in the time domain by a Gaussian function such that:
\begin{equation}
    C( t_X') \propto \exp{\left[ - \frac{(t_X' - t_i)^2}{2 \epsilon^2}\right]}
\end{equation}
where $\epsilon$ is the width or standard deviation of the blurred continuum, which allows for numerical sampling of the posterior. Since for a thin inclined ring, the transfer function $\Psi_X (v_z, T_X)$ is given by a sum of Dirac-delta signals  \eqref{spectralfluxthinring}, the line response with $t_i = 0$ becomes
\begin{equation}
\begin{split}
    L_X(v_z, t_X') &\propto \frac{R}{V^+(R, v_z)} \left( \exp{\left[ - \frac{(t_X' -  W_X^+(R, v_z))^2}{2 \epsilon^2}\right]} \right.\\
    &+\left.\exp{\left[ - \frac{(t_X' - W_X^-(R, v_z))^2}{2 \epsilon^2}\right]}\right) \label{modelLX}
\end{split}
\end{equation}
where 
\begin{align}
    V^{+}(r, v_z) &\equiv u'(r) \sin \varphi^+ = \sqrt{u'^2(r) - v_z^2}\\
\begin{split}
    W^{\pm}_X(r, v_z) &\equiv \mathcal{T}_X (r, \tfrac{\pi}{2}, \varphi^\pm)\\
    &= \frac{r}{c} \left( 1+ z_s + A_X \frac{V^{\pm}(r, v_z)}{u'(r)}  + B_X \frac{v_z}{u'(r) }\right)
\end{split}
\end{align}
and
\begin{align}
    u'(r) &\equiv \sqrt{\tfrac{GM}{r}} \sin i \\
    A_X &\equiv (1+z_s) \sin i - (1+z_d) \hat{\alpha}_{X, y} \cos i\\
    B_X &\equiv  -(1+z_d) \hat{\alpha}_{X,x}
\end{align}
and so, fixing $\epsilon$, the parameters associated with this model are
\begin{equation*}
    \bm{\Theta} = \left(\log_{10}\left(\tfrac{M}{M_{\odot}}\right), i, \log_{10}\left(\tfrac{R}{\textrm{m}}\right), \{ \hat{\alpha}_{X,x}, \hat{\alpha}_{X,y} \} \right)
\end{equation*}
where $\{ \hat{\alpha}_{X,x}, \hat{\alpha}_{X,y} \}$ denotes the set of all the physical deflection angles over all the available images (through which there is an implicit dependence on $\frac{D_d}{D_{ds}}$ and $\bm{\beta}$). For the purpose of parameter estimation, the model is considered a function of the parameters $\bm{\Theta}$ instead of the variables $(v_z, t_X')$, which are evaluated at values fixed by each data point and then binned\footnote{Binning is essential: see p.10 \cite{Lindsey1999}.}: $L_x(v_z, t_X') \to L_{X,jk} (\bm{\Theta})$, where $j$ and $k$ respectively denote the time and line-of-sight velocity bins of the data.

Rather than work with the analytic expression \eqref{modelLX} directly, which due to the ${V^+(R, v_z)}^{-1}$ term is singular when $v_z^2 = u'^2(r)$, we instead obtain $L_{X,jk} (\bm{\Theta})$ numerically, utilising $L_x(v_z, t_X') \propto \frac{\partial^2 N}{\partial u_z \partial \mathcal{T}_X} ( v_z, t_X')$. For each BLR cloud we may draw a position sample from the thin ring number density distribution function. To each sample we then assign a time and a velocity from the time delay and velocity models \eqref{Txthinring} and \eqref{vzthinring}, before performing a 2D Gaussian blur. The resultant time delay and line-of-sight velocity values were binned with bin widths of roughly 1d and 50 km s$^{-1}$ respectively.

We generate mock data by fixing the model parameters at a chosen set of true values $\bm{\Theta}_{\text{true}}$ and adding measurement noise. As photon statistics follow a Poisson distribution, the scaled Poisson distribution corresponding to the flux data has a mean at the model value $L_{X,jk} (\bm{\Theta_{\text{true}}})$ and a variance given by $\sigma_{jk}^2 =  a L_{X,jk} (\bm{\Theta_{\text{true}}}) + \sigma^2_{\textrm{abs}}$, where $a$ is a constant and there is an additional background contribution $\sigma^2_{\textrm{abs}}$ to the noise. The parameters for the noise are chosen to correspond to a peak signal-to-noise ratio of approximately 10 with $a \approx 0.05 \, \text{max}(L_{X,jk} (\bm{\Theta_{\text{true}}}))$. As the photon count is large, the distribution is approximated as Gaussian 
\begin{equation}
    \mathcal{D}_{X,jk} = \mathcal{N}(0, \sigma_{jk}^2) + L_{X,jk} (\bm{\Theta}_{\text{true}})
\end{equation}
and the probability density of an individual datum $\mathcal{D}_{X,jk}$ is
\begin{equation}
    \mathcal{P}(\mathcal{D}_{X,jk} | \bm{\Theta} , \mathcal{I}) = \frac{1}{\sqrt{2 \pi} \sigma_{jk} } \exp{\left[ - \frac{(\mathcal{D}_{X,jk} - L_{X,jk} (\bm{\Theta}))^2}{2 \sigma_{jk}^2} \right]}.
\end{equation}

The likelihood function for the line response from a single image $X$, i.e. the joint probability density function $\mathcal{P}(\bm{\mathcal{D}_X} | \bm{\Theta}, \mathcal{I})$, is given (assuming independent data) by the product of the probability densities of the individual measurements $\mathcal{D}_{X,jk}$:
\begin{equation}
    \mathcal{P}(\bm{\mathcal{D}_X} | \bm{\Theta}, \mathcal{I}) = \prod_{j,k} \mathcal{P}(\mathcal{D}_{X,jk} | \bm{\Theta}, \mathcal{I}).
\end{equation}
The log-likelihood for the line response for a single image $X$ is thus
\begin{align}
    &\ln{\mathcal{P}(\bm{\mathcal{D}_X} | \bm{\Theta}, \mathcal{I} )} = \sum_{j,k} \left( \ln{ \left(\frac{1}{\sqrt{2 \pi} \sigma_{jk}} \right)} - \frac{\left( \mathcal{D}_{X,jk} -  L_{X,jk} (\bm{\Theta}) \right)^2}{2 \sigma_{jk}^2} \right).
\end{align}
To find the posterior taking into account the data from all images (assuming independence), we can either iterate over the data from each subsequent image; or perform the analysis in one step, where the full likelihood function is the product of the individual ones
\begin{equation}
    \mathcal{P} ( \{ \bm{\mathcal{D}}_X \} |  \bm{\Theta} , \mathcal{I} ) = \prod_{X} p( \bm{\mathcal{D}}_X | \bm{\Theta} , \mathcal{I})
\end{equation}
where $\{ \bm{\mathcal{D}}_X \}$ is the set of data from all images.

We ran {\sffamily emcee} until the Markov chain was longer than 100 times the integrated autocorrelation time estimate for the chain for each parameter, with the estimated autocorrelation time changing by less than 1\%. Due to the highly restrictive assumptions of this BLR geometric model, the geometric parameter values are very well determined; whereas the deflection angles are unconstrained with the posterior distributions equivalent to the uniform priors. This result for realistic lensing scales could be surmised from the data by eye; although even if the scale of lensing is increased to unrealistically large magnitudes, the geometric parameter values remain very well determined for this model. The corner plot of the one and two dimensional projections of the posterior probability distributions in the case of galaxy-scale lensing is shown in Figure \ref{fig:cornerplot}.

\section{Conclusions} \label{sec:p2conclusions}

In summary, \cite{Ng2020} proposed a method of determining $\frac{D_d}{D_{ds}}$ without lens modelling, via reverberation mapping and measuring times delays of differentially lensed, unresolved quasars. This method suffers from the following:

\begin{enumerate}

    \item Reverberation mapping does not uniquely map the location from discrete flares as is required; rather it only gives general constraints on the geometric structure of the BLR. This is a problem of underdetermination.

    \item The projection of the BLR cloud position on the image-image axis is needed, which requires further data (i.e. spatially resolving each image).
    
    \item The time scale for differential lensing is on the order of minutes to days compared to tens to hundreds of days for the reverberation mapping time delay; $ \tau_{\textsc{rm}} \gg | \delta \tau_{AB} |$ for a general BLR geometry.

    \item Finally, the determination of the geometric parameters is not a priori independent of $\bm{\beta}$. However, the effect of differential lensing is sufficiently small ($\tau_{\textsc{rm}} \gg | \delta  \tau_{AB}|$), implying the geometric parameters may be inferred accurately without additional data or lens modelling.
    
\end{enumerate}

Although imaging the BLR structure of an unlensed quasar requires a resolution of $1-10 \mu$ arcseconds, which is still unfeasible with current interferometry, a high angular resolution between the photocentres of redshifted and blueshifted emission may be achieved using the technique of spectroastrometry \cite{Sturm2018}. This has allowed for initial parallax distance measurements using reverberation mapped quasars \cite{Wang2020}. Spectroastrometry would not be as straightforward (and perhaps superfluous) to apply to lensed quasars and obtaining the source positions from the observed image positions would require lens modelling; although the possibility of image magnification could be an advantage. It is possible that differential lensing time delays may be used to extract cosmological information, although obtaining the source positions from image positions may remain a limitation, if distinct, time-variable features within the quasar structure between images can be identified.

\section*{Acknowledgements}
AN thanks Ed McDonald, C\'{e}line B{\oe}hm and Mark Wardle for useful discussions and comments; Mat Varidel and Josh Speagle for discussions on Bayesian analysis; and Oz Brent for discussions regarding numerical computation. AN also thanks the anonymous reviewer for helpful comments which improved this manuscript. This research was funded in part by an Australian Government Research Training Program Scholarship and University of Sydney Hunstead Merit Award.

\section{Appendix: Calculations of Transfer Functions} \label{appendixa}

\subsection{Thin Inclined Ring}
The number density distribution which describes a thin inclined ring BLR of radius $R$ with uniform linear number density $\mu$, an inclination angle $i$ is
$n(\bm{r}) = \mu \delta(r - R)r^{-1} \delta(\vartheta - \frac{\pi}{2})$. 
\subsubsection{One-Dimensional Transfer Function}
The one-dimensional transfer function is proportional to
\begin{align}
    &\frac{\partial N}{\partial \mathcal{T}_X} (T_X) = \iiint n(\bm{r}) \delta (\mathcal{T}_X(\bm{r}) - T_X) \, d^3 \bm{r}\\
    &= \iiint \mu \delta (r - R) \delta (\vartheta - \tfrac{\pi}{2}) \delta ( \mathcal{T}_X(r, \vartheta , \varphi) -T_X)r \sin \vartheta \, dr d\vartheta d \varphi \\
    &= \int \mu R \delta (\mathcal{T}_X(R, \tfrac{\pi}{2}, \varphi) - T_X) d \varphi \\
    &= \int \mu R \sum\limits_{\tilde{\varphi}} \delta (\tilde{\varphi} - \varphi) \left| \frac{\partial \varphi}{\partial \mathcal{T}_X} (R, \tfrac{\pi}{2}, \varphi) \right| \, d\varphi\\
    &= \sum\limits_{\tilde{\varphi}} \frac{\mu c}{\left| F_X \cos ( \tilde{\varphi} + \phi_X) \right|}
\end{align}
where $\tilde{\varphi}$ are the roots of $\mathcal{T}_X(R, \tfrac{\pi}{2}, \varphi)$.
This gives the result
\begin{equation}
    \frac{\partial N}{ \partial \mathcal{T}_X} (T_X) \propto \frac{
    2 \mu c}{\sqrt{F_X^2 - \left(T_X c R^{-1} - (1 +z_s) \right)^2}} \label{fluxthinringAppendix}
\end{equation}
when $\left|T_X c R^{-1} - (1 +z_s) \right| \leq F_X$ and $0$ otherwise.
\subsubsection{Two-Dimensional Transfer Function}
Although computing the two-dimensional transfer function \eqref{eq:spectralflux} for a general $n(\bm{r})$ is not an easy task, the presence of Dirac delta distributions in $n(\bm{r})$ means we can reduce the dimensionality of the integral. The rule for the composition of a Dirac delta distribution function gives a more general version of the usual change of variables rule (permitting the new and old random variables to not be one-to-one). Assuming the ring is in Keplerian orbit with orbital speed $u$, we obtain
\begin{align}
\begin{split}
    &\frac{\partial^2 N}{\partial u_z \partial \mathcal{T}_X} ( v_z, T_X) \\
    &= \mu \! \iint \! \delta(r-R) r \delta^2( u_z(r,\tfrac{\pi}{2}, \varphi) - v_z, \mathcal{T}_X(r,\tfrac{\pi}{2}, \varphi) - T_X)) \, dr d \varphi
    \end{split}
    \\
    &= \mu \! \sum\limits_{\tilde{r}, \tilde{\varphi}} \iint \! \delta (r - R) r \delta^2(r - \tilde{r}, \varphi - \tilde{\varphi}) \left|\frac{\partial(r, \varphi)}{\partial(u_z,\mathcal{T}_X)}(r, \tfrac{\pi}{2}, \varphi) \right| dr d \varphi\\
    &= \mu R \sum\limits_{\tilde{r}, \tilde{\varphi}} \delta (\tilde{r} - R)  \left|\frac{\partial(r, \varphi)}{\partial(u_z,\mathcal{T}_X)}(R, \tfrac{\pi}{2}, \tilde{\varphi}) \right|
\end{align}
where $\tilde{r}$ and $\tilde{\varphi}$ solve $T_X = \mathcal{T}_X(\tilde{r}, \tfrac{\pi}{2}, \tilde{\varphi})$ and $v_z = u_z(\tilde{r}, \tfrac{\pi}{2}, \tilde{\varphi})$. 

The Jacobian determinant $\left|\frac{\partial (r, \varphi)}{\partial (u_z, \mathcal{T}_X)} (r, \tfrac{\pi}{2}, \varphi) \right|= \left|\frac{\partial u_z}{\partial r} \frac{\partial \mathcal{T}_X}{\partial \varphi} - \frac{\partial \mathcal{T}_X}{\partial r} \frac{\partial u_z}{ \partial \varphi} \right|^{-1}$ when evaluated at the two roots $\varphi^\pm \equiv \tilde{\varphi}$ is
\begin{equation}
\begin{split}
    &\left|\frac{\partial (r, \varphi)}{\partial (u_z, \mathcal{T}_X)} (r, \tfrac{\pi}{2}, \varphi^\pm) \right| = \left| \frac{\partial u(r)}{dr} \frac{r}{c} \frac{v_z}{u^2(r) \sin i }  \right. \\
    & \times \left( A_X v_z  - B_X V^\pm (r, v_z) \right) +\left. V^\pm (r, v_z) W_X^\pm (r, v_z) r^{-1} \right|^{-1}
    \end{split} \label{diskdeterminant}
\end{equation}
where we defined
\begin{align}
    V^{\pm}(r, v_z) &\equiv u(r) \sin i \sin \varphi^\pm = \pm \sqrt{(u(r) \sin i)^2 - v_z^2}\\
\begin{split}
    W^{\pm}_X(r, v_z) &\equiv \mathcal{T}_X (r, \tfrac{\pi}{2}, \varphi^\pm)\\
    &= \frac{r}{c} \left( 1+ z_s + A_X \frac{V^{\pm}(r, v_z)}{u(r) \sin i}  + B_X \frac{v_z}{u(r) \sin i }\right).
\end{split}
\end{align}
In the case of a thin ring the radius is fixed: the first term in Equation \eqref{diskdeterminant} vanishes and we find
$\tilde{r} = \frac{R}{W^\pm (R, v_z)} T_X$, giving
\begin{equation}
    \frac{\partial^2 N}{\partial u_z \partial \mathcal{T}_X} ( v_z, T_X) = \mu c R \! \sum\limits_{\pm} \frac{\delta\left(T_X - W^{\pm}_X(R, v_z) \right)}{V^{+}(R, v_z)}
\end{equation}
\subsection{Inclined Disk}
The number density describing a disk of maximum radius $R$ and uniform surface density $\sigma$ can expressed using the Heaviside step function $H$:
\begin{align}
    n(\bm{r}) &= \sigma  H(R - r) r^{-1} \delta (\vartheta - \tfrac{\pi}{2} )\\
    &= \sigma \int_{-\infty}^{0} \delta( r' + R -r ) r^{-1} \delta (\vartheta - \tfrac{\pi}{2})  \, dr'.
\end{align}
\subsubsection{One-Dimensional Transfer Function}
By recalling the expression for the one-dimensional transfer function for a thin inclined ring, Equation \eqref{fluxthinringAppendix}, we recognise
\begin{equation}
 \frac{\partial N}{\partial \mathcal{T}_X} (T_X) \propto 2 \sigma c \int_{r'_{\mathrm{min}}}^{r'_{\mathrm{max}}} \frac{dr'}{\sqrt{F_X^2 - \left(\frac{cT_X}{r' + R} - (1+z_s) \right)^2}} 
\end{equation}
where $r'_{\mathrm{min }} = \mathrm{max} \left[ - \infty, T_X^+  - R\right] = T_X^+ -R$ and $r'_{\mathrm{max}} = \mathrm{min} \left[0, T_X^- -R \right]$ with $T_X^{\pm} \equiv \frac{c T_X}{(1+z_s) \pm F_X}$. Under a change of variables from $r'$ to $y = r' + R$, such that $y_{\mathrm{min }} =  T_X^+$ and $y_{\mathrm{max}} = \mathrm{min} [R,  T_X^-]$, we have that when $F_X < (1+z_s)$,
\begin{align}
& \frac{\partial N}{\partial \mathcal{T}_X} (T_X) \propto \int_{y_{\mathrm{min}}}^{y_{\mathrm{max}}} \frac{2 \sigma c y \; dy}{\sqrt{(1+z_s)^2 - F_X^2} \sqrt{(y - T_X^+)(T_X^- - y)}} \\
\begin{split}
&\propto \frac{2 \sigma c }{\sqrt{(1+z_s)^2 - F_X^2}} \Bigg[- \sqrt{(T_X^- - y)(y- T_X^+)}\\
&\quad \qquad \left. - \frac{(T_X^- + T_X^+)}{2} \arcsin \left(\frac{2y - (T_X^+ + T_X^-)}{(T_X^+ - T_X^-)} \right) \right]^{\mathrm{min}[R, T_X^-]}_{T_X^+}.
\end{split}
\end{align}
When $y_{\mathrm{max}} = T_X^-$, then $T_X < \frac{R}{c}\left( (1+z_s) - F_X \right)$ and the one-dimensional transfer function is a linear function of $T_X$,
\begin{equation}
    \frac{\partial N}{\partial \mathcal{T}_X} (T_X) \propto 2 \pi \sigma (1+z_s) c^2 \left((1+z_s)^2 - F_X^2\right)^{-\tfrac{3}{2}}T_X
\end{equation}
whereas when $y_{\mathrm{max}} = R$, i.e. $\frac{R}{c} \left( (1+z_s) - F_X \right) \leq T_X \leq \frac{R}{c} \left( (1+z_s) + F_X \right)$, the expression does not simplify as nicely as the former regime.

Similarly, in the case that $F_X > (1 + z_s)$,
\begin{align}
    &\frac{\partial N}{\partial \mathcal{T}_X} (T_X) \propto \int_{y_{\mathrm{min}}}^{y_{\mathrm{max}}} \frac{2 \sigma c y \; dy}{\sqrt{F_X^2 - (1+z_s)^2} \sqrt{(y - T_X^+)(y - T_X^-)}}  \\
\begin{split}
    &\propto \frac{2 \sigma c}{\sqrt{F_X^2 -(1+z_s)^2}} \Bigg[ \sqrt{(y - T_X^+)(y - T_X^-)} \\
    &\qquad \qquad + \left. \frac{(T_X^+ + T_X^-)}{2} \mathrm{arccosh} \left( \frac{2y}{T_X^+ + T_X^-} -1 \right) \right]^{ \mathrm{min}[R, T_X^+]}_{T_X^-}
\end{split}
\end{align}
and finally, in the case that $F_X = (1+z_s)$,
\begin{equation}
   \frac{\partial N}{\partial \mathcal{T}_X} (T_X) \propto \frac{2 \sigma \sqrt{-cT_X (c T_X - 2(1+z_s) R)} ((1+z_s)R + cT_X)}{3(1+z_s)^2 T_X}.
\end{equation}

Since the number density for a radially-thick inclined ring can be found by subtracting the number density of a smaller disk from a larger disk, the one-dimensional transfer function for such a geometry is proportional to
\begin{equation}
    \Psi_{X, \text{thick ring}} (T_X) \propto \left. \Psi_{X, \text{disk}} \right|_{R_{\mathrm{max}}} (T_X) - \left. \Psi_{X, \text{disk}} \right|_{R_{\mathrm{min}}} (T_X).
\end{equation}

\subsubsection{Two-Dimensional Transfer Function}
The two-dimensional transfer function is given by
\begin{align}
\begin{split}
    &\frac{\partial^2 N}{\partial u_z \partial \mathcal{T}_X} (v_z, T_X)\\
    &= \iint \sigma H(R -r) r \delta^2 (u_z(r, \tfrac{\pi}{2}, \varphi) - v_z, \mathcal{T}_X(r, \tfrac{\pi}{2}, \varphi) - T_X) \, dr d \varphi
\end{split}\\
    &= \sum\limits_{\tilde{r}, \tilde{\varphi}} \iint \sigma H(R-r) r \delta^2(r - \tilde{r}, \varphi - \tilde{\varphi}) \left| \frac{\partial (r, \varphi)}{\partial (u_z, \mathcal{T}_X)}( \tilde{r}, \tfrac{\pi}{2}, \tilde{\varphi}) \right| \, dr d \varphi.
\end{align}
Using the Jacobian  determinant \eqref{diskdeterminant}, along with $u(r) \propto r^{-1/2}$ we have the result
\begin{equation}
\begin{split}
&\frac{\partial^2 N}{\partial u_z \partial \mathcal{T}_X} (v_z, T_X) = \sum\limits_{\tilde{r}, \pm} c \sigma H(R- \tilde{r}) \tilde{r} \, \bigg| (1+z_s) V^{\pm}(\tilde{r}, v_z)\\
&+  \left( (V^{\pm}(\tilde{r}, v_z))^2 - \tfrac{v_z^2}{2} \right) \frac{A_X}{u \sin i} + \tfrac{3}{2} V^{\pm}(\tilde{r}, v_z) \frac{v_z}{u \sin i} B_X  \bigg|^{-1}.
\end{split}
\end{equation}
As $T_X = \mathcal{T}_X(\tilde{r}, \tfrac{\pi}{2}, \varphi^\pm )$ and $v_z = u_z(\tilde{r}, \tfrac{\pi}{2}, \varphi^\pm)$ together form a polynomial of 6\textsuperscript{th} degree in $\tilde{r}$, there are 6 (real or complex) roots for which there are no algebraic expressions.

\subsection{Thin Spherical Shell}

A thin spherical shell of radius $r=R$ with a uniform surface density $\sigma$ in solid body rotation, which has a number density given by $n(\bm{r}) = \sigma \delta (r - R)$.

\subsubsection{One-Dimensional Transfer Function}
Calculating the one-dimensional transfer function in the case of a thin spherical shell geometry using the direct method is not straightforward due to the nature of the integral. However, the thin spherical shell may be decomposed into thin face-on rings where there is a unique value of $\tau_{\textsc{rm}}$ for each ring. We happen to know that the conditional probability density function of $\delta \tau_X$ for each given $\tau_{\textsc{rm}}$: as we showed in Paper I, the effect of lensing on a thin face-on ring on its flux is to distort each Dirac delta spike into an arcsine distribution whose width is then dependent on $\tau_{\textsc{rm}}$. In this special case, we may integrate over the contributions from each ring at each time $\tau_{\textsc{rm}}$. Formally,  since $\mathcal{T}_X$ is the sum of $\tau_{\textsc{rm}}$ and $\delta \tau_X$ which may be considered as random variables, $\frac{dN}{d\mathcal{T}_X}$ may be written using the definition of the conditional probability as
\begin{align}
    \frac{dN}{d \mathcal{T}_X} (T_X) &= \int\limits_{-\infty}^{\infty} \frac{\partial^2 N}{\partial \tau_{\textsc{rm}} \partial \delta \tau_X} ( T_X - t, t) \, d t\\
    &= \int\limits_{\tau_{\textsc{rm}}^{ \textsc{min}}}^{\tau_{\textsc{rm}}^{\textsc{max}}} f(t) g(T_X - t,  t) \, d t
\end{align}
when $\tau_{\textsc{rm}}^{\textsc{min}}\leq t \leq \tau_{\textsc{rm}}^{\textsc{max}}$ and $\frac{dN}{d \mathcal{T}_X} (T_X) = 0$ otherwise. Here
\begin{equation}
    f(t) \equiv \frac{\partial N}{\partial \tau_{\textsc{rm}}} (t) = \frac{ 2 \pi \sigma R c }{(1+z_s)} 
\end{equation}
when $0 \leq t \leq \frac{2R}{c}(1+z_s)$ and $f(t) =0$ otherwise. The conditional number density at $\delta \tau_X=s$ given $\tau_{\textsc{rm}}=t$ is  $g(s, t)$, 
where
\begin{equation}
    g(s, t) = 
\frac{1 }{ \pi \sqrt{\left(\tfrac{R}{c} F_X \right)^2 \left( 1- t'^2\right) - s^2}} 
\end{equation}
where $t' \equiv 1- \tfrac{c t}{R(1+z_s)}$
when the condition $\left(\tfrac{R}{c} F_X \right)^2 \left( 1- t'^2\right) \geq s^2$ holds and $0$ otherwise. 
This condition on $t$ is equivalent to $\tau_{\textsc{rm}}^-\leq t\leq \tau_{\textsc{rm}}^+,$
where the bounds $\tau_{\textsc{rm}}^{\pm}$ depend on $s.$
We therefore find $\tau_{\textsc{rm}}^{\textsc{min}}  = \mathrm{Max} [ 0 , \tau_{\textsc{rm}}^{-}] = \tau_{\textsc{rm}}^{-} $ and $\tau_{\textsc{rm}}^{\textsc{max}} = \mathrm{Min} [ \tfrac{2R}{c}(1+z_s), \tau_{\textsc{rm}}^{+}]$; and write
\begin{align}
    \frac{\partial N}{\partial \mathcal{T}_X} (T_X) &= \frac{2 \sigma cR}{(1+z_s) K_X} \int\limits_{\tau_{\textsc{rm}}^{ \textsc{min}}}^{\tau_{\textsc{rm}}^{\textsc{max}}} \frac{d t}{ \sqrt{(t - \tau_{\textsc{rm}}^{ -})(\tau_{\textsc{rm}}^{+} - t)}}\\
    &= \frac{4 \sigma Rc}{(1+z_s) K_X} \arcsin{\left( \sqrt{\frac{\tau_{\textsc{rm}}^{\textsc{max}} - \tau_{\textsc{rm}}^{-} }{\tau_{\textsc{rm}}^{+} - \tau_{\textsc{rm}}^{-}}} \right)}
\end{align}
recalling $K_X \equiv \sqrt{1+F_X^2(1+z_s)^{-2}}$, which gives the result
\begin{equation}
    \frac{\partial N}{\partial \mathcal{T}_X} (T_X) = \frac{2 \pi \sigma R c}{(1+z_s) K_X} 
\end{equation}
when $\tfrac{R}{c}(1+z_s)\left( 1- K_X \right) \leq T_X \leq \tfrac{2R}{c}(1+z_s)$ and
\begin{equation}
    \frac{\partial N}{\partial \mathcal{T}_X} (T_X) = \frac{4 \sigma Rc}{(1+z_s) K_X} \arcsin{\left( \sqrt{\frac{\frac{2R}{c}(1+z_s) -\tau_{\textsc{rm}}^{-}(T_X)}{\tau_{\textsc{rm}}^{+}(T_X) - \tau_{\textsc{rm}}^{-}(T_X)}}\right)}
\end{equation}
when $\tfrac{2R}{c} (1+z_s) \leq T_X \leq \tfrac{R}{c} (1+z_s) \left( 1+ K_X \right)$, and we emphasise the dependence of $\tau_{\textsc{rm}}^{\pm}$ on $T_X$. This gives the domain of $T_X$ in agreement with the expression \eqref{eq:domainTxsphericalshell}. We again note that since $K_X \sim 1$, the lensing effect is extremely small for all images.

\subsubsection{Two-Dimensional Transfer Function}

The two-dimensional transfer function is proportional to
\begin{align}
\begin{split}
    &\frac{\partial ^2 N}{\partial u_z \partial \mathcal{T}_X} (v_z, T_X)\\
    &= \sigma R^2 \iint  \delta^2 (u_z(R, \vartheta, \varphi) - v_z, \mathcal{T}_X(R, \vartheta, \varphi) - T_X) \sin \vartheta  \, d \vartheta d \varphi
\end{split}
    \\
    &= \sigma R^2 \sum\limits_{\tilde{\vartheta}, \tilde{\varphi}}\iint \delta^2(\vartheta - \tilde{\vartheta}, \varphi - \tilde{\varphi}) \left| \frac{\partial (\vartheta, \varphi)}{\partial (u_z , \mathcal{T}_X)} (R, \tilde{\vartheta}, \tilde{\varphi}) \right| \sin \vartheta  \, d \vartheta d \varphi\\
    &= \sigma R^2  \sum\limits_{\tilde{\vartheta}, \tilde{\varphi}} \left| \frac{\partial (\vartheta, \varphi)}{\partial (u_z, \mathcal{T}_X)} (R, \tilde{\vartheta}, \tilde{\varphi}) \right| \sin \tilde{\vartheta}\\
    &= \sum\limits_{\tilde{\vartheta}, \tilde{\varphi}} \frac{\sigma Rc}{\left|u \sin i_v \left( F_X \cos \tilde{\vartheta} \cos \phi_X + (1+z_s) \sin \tilde{\vartheta} \sin \tilde{\varphi} \right) \right|}\\
    &= \sum\limits_{\tilde{\varphi}} \frac{\sigma Rc}{\left|F_X \sqrt{(u \sin i_v)^2 - \frac{u^2}{\cos^2  \tilde{\varphi}}} \cos \phi_X  + (1+z_s) u \tan \tilde{\varphi} \right|}
\end{align}
where $\tilde{\vartheta}$ and $\tilde{\varphi}$ solve $T_X = \mathcal{T}_X(R, \tilde{\vartheta}, \tilde{\varphi})$ and $v_z = u_z(R,\tilde{\vartheta}, \tilde{\varphi})$. Finding the roots requires solving \begin{equation}
\begin{split}
T_X' = &- \frac{F_X v_z'}{(1+z_s)} \left( \sqrt{ \cos^{-2} \tilde{\varphi}- \cos^2 \tilde{\varphi}} \cos \phi_X + \sin \phi_X \right)\\
&\pm \sqrt{1 - \frac{v_z'^2}{\cos^2 \tilde{\varphi}}},
\end{split}
\end{equation} 
where $T_X' \equiv 1 - \frac{c T_X}{R(1+z_s)}$ and $v_z' \equiv \frac{v_z}{u \sin i_v}$, for $\tilde{\varphi}$ in terms of $v_z$ and $T_X$.

Solving for $\tilde{\varphi}$ in terms of $v_z$ and $T_X$ is not straightforward. However, since $\frac{F_X}{(1+z_s)}$ and $\phi_X^+$ are very small compared to $T_X'$ we can take a first order approximation of the solution to $\tilde{\varphi}$ such that $\cos^2 \tilde{\varphi} \approx v_z'^2 (1-T_X'^2)^{-1}$, giving
\begin{equation}
\Psi_X (v_z, T_X) \appropto \frac{2 \sigma Rc |v \sin i_v|^{-1} }{\left| F_X T_X' \cos \phi_X + (1+z_s) \sqrt{1 - T_X'^2 - v_z'^2} \right|}.
\end{equation}
When $F_X =(1+z_d)\hat{\alpha}_X =0$, this is the exact expression for the unlensed transfer function, and is an arcsine distribution in both $v_z$ and $T_X$. From this we see that the effect of an, e.g., larger value for $F_X$ is to depress the approximately arcsine distribution; this effect is extremely small around the centre of the distribution but removes the singularity in the distribution at $v_z' = \pm  \sqrt{1-T_X'^2}$.


\chapter{Conclusions} 
\label{chap:Conclusions} 

In this thesis we reviewed standard cosmology in Chapter \ref{chap:standardcosmo}, and in Chapter \ref{chap:lightlensaction} covered lensing theory with some elements and an approach which differs from many textbooks and reviews. In Chapter \ref{chap:cosmoglqso} we reviewed aspects of lensing and AGNs with a focus on application to cosmology. Chapter \ref{chap:p1} was a publication  (``Paper I'') proposing a new usage of lensing and quasars applied to cosmology; and Chapter \ref{chap:p2} was a detailed and critical follow-up of this idea (``Paper II'').

Paper I suggested that the reverberation mapping of strongly lensed quasars could be used to put constraints on an angular size distance ratio. The method was advertised as a purely geometric probe of cosmology which would be independent of the lensing potential, and thereby would not suffer from lens modelling challenges such as the mass-sheet degeneracy. It suggested that differential time delays from features within the broad emission lines originating from spatially separated signals in the Broad Line Region (BLR) of a quasar, could be distinguished and measured from the spectroscopy of the images. 

In Paper II, I re-examined the ability of this method to recover cosmological information. A fundamental assumption was that the signals under consideration are discrete, spatially separated, and lack any physical correlation. If there is no physical correlation between discrete signals, then the shape of the flux corresponding to each is unique. However, this assumption is not applicable when utilising reverberation mapping of the BLR. Instead, reverberation mapping involves the response of an \textit{extended region} to the \textit{same} continuum signal, giving only general constraints on the BLR structure and kinematics. The technique is ill-motivated: reverberation mapping does not uniquely map the observed time delay and inferred line-of-sight velocity to the three-dimensional location of individual points in the BLR, meaning that the problem is underdetermined. The determination of the parameters describing the BLR structure is also not a priori independent of the source position.

I also gave in Paper II a detailed analytic description of the effect of the differential lensing on the emission line spectral flux for axisymmetric BLR geometries, with the inclined ring or disk, spherical shell, and double cone as examples. I also included a Bayesian parameter estimation, using a Markov chain Monte Carlo (MCMC) method to additionally illustrate the flux from each image is largely insensitive to differential lensing, and hence to cosmology (such as ratios of angular size distances).


\appendix 





\printbibliography[heading=bibintoc]


\end{document}